\documentclass[12pt,letterpaper,hidelinks,usenames,dvipsnames]{article}
\usepackage{jheppub}\usepackage{amsmath,amssymb,amsfonts,amsthm,amscd}
\usepackage{bm} %boldmath = \bm
\usepackage[acronym,symbols]{glossaries}
\usepackage{bbm} %\mathbbm = doublestruck math for upper, lower, numbers, \mathbbmss=sans serif, \mathbbmtt=typewriter
\usepackage{eufrak} %gothic
\usepackage{mathrsfs} %caligraphic
\usepackage{lscape} 
\usepackage{setspace}
\usepackage{caption}
\usepackage{cancel}
\usepackage{enumitem}
\usepackage{tcolorbox}
\usepackage{multirow} % multiple row elements in a table
\usepackage{colortbl} % colored rows and columns in tables
\usepackage{array,hhline,arydshln,multirow}
\usepackage{floatrow,subfig}
\usepackage[export]{adjustbox}
\usepackage{rotating} % rotate text
\usepackage{trimclip}
\allowdisplaybreaks[2]  %allow page breaks in multi-line eqns, [1]=as little as poss; ... [4]=most permissive
\usepackage{pifont,dsfont}% http://ctan.org/pkg/pifont
\usepackage{bm} %boldmath

\usepackage{graphicx,tikz}
\usepackage[framemethod=TikZ]{mdframed} % fancy boxes
\usepackage{tikz-cd} %for complicated commutative diagrams

\usepackage{xcolor}
\def\red#1{{\color{red}{#1}}}
\def\blue#1{{\color{blue}{#1}}}
\def\green#1{{\color{black!30!green}{#1}}}

\def\bM{\begin{matrix}}
\def\eM{\end{matrix}}
\newcommand{\bpm}{\begin{pmatrix}}
\newcommand{\epm}{\end{pmatrix}}
\newcommand{\bsm}{\begin{smallmatrix}}
\newcommand{\esm}{\end{smallmatrix}}
\newcommand{\bspm}{\left(\begin{smallmatrix}}
\newcommand{\espm}{\end{smallmatrix}\right)}
\newcommand{\beq}{\begin{equation}}
\newcommand{\eeq}{\end{equation}}
\def\bar{\overline}

\def\hat{\widehat}

\def\^{\wedge}

\def\O{{\rm O}}

\def\su{\mathfrak{su}}

\def\sp{\mathfrak{sp}}

\def\C{\mathbbm{C}} 
\def\cC{{\mathcal C}}

\def\H{\mathbbm{H}}
\def\cH{{\mathcal H}}

\def\cM{{\mathcal M}}

\def\cN{{\mathcal N}}

\def\R{\mathbbm{R}} 

\def\cS{{\mathcal S}}

\def\cT{{\mathcal T}}

\def\cV{{\mathcal V}}

\def\Z{\mathbbm{Z}}
\def\G{{\Gamma}}
\def\d{{\delta}}
\def\D{{\Delta}}

\def\l{{\lambda}}
\def\L{{\Lambda}}

\def\vf{{\varphi}}

\usepackage{hyperref}
 \hypersetup{
     colorlinks=true,
     linkcolor=Black!60!ForestGreen,
     filecolor=blue,
     citecolor = MidnightBlue,      
     urlcolor=cyan,
     }

\usepackage{tikz}
\usetikzlibrary{arrows,snakes,shapes.arrows,decorations.markings}
     \tikzset{>=triangle 90}
     \tikzstyle{bbc}=[draw,circle,fill=black,scale=.75]
     \tikzstyle{rc}=[circle,fill=red,scale=.6]
     \tikzstyle{wc}=[draw,circle,scale=.75]

\newcommand{\Mtwo}{%
  \mbox{%
    \def\Mhalf##1{\dimexpr.5\width ##1 0.025em\relax}
    \textcolor{red}{\clipbox{0 0 {\Mhalf+} 0}{$\mathcal{M}$}}%
    \textcolor{blue}{\clipbox{{\Mhalf-} 0 0 0}{$\mathcal{M}$}}%
  }%
}

\def\doub{\hyperref[DR]{doubling rule}}

\def\ie{\emph{i.e.}}

\def\sd{\emph{6d}\ }
\def\fid{\emph{5d}\ }
\def\fod{\emph{4d}\ }

\def\blue#1{{\color{blue}{#1}}}

\def\green#1{{\color{black!25!green}{#1}}}

\def\red#1{{\color{red}{#1}}}

\def\bar{\overline}

\def\hat{\widehat}

\def\^{\wedge}

\def\tTE{\widetilde{\cT}_{E_6,2}}
\def\hTE{\widehat{\cT}_{E_6,2}}
\def\hTD{\widehat{\cT}_{D_4,3}}
\def\hTA{\widehat{\cT}_{A_2,4}}

\def\rC{\red{\mathcal{C}}}
\def\bH{\blue{\mathcal{H}}}
\def\rTf{\red{\mathfrak{T}}}
\def\bTf{\blue{\mathfrak{T}}}

\def\G{{\Gamma}}
\def\d{{\delta}}

\def\D{{\Delta}}

\def\l{{\lambda}}

\def\L{{\Lambda}}

\def\vf{{\varphi}}

\def\bR{{\bf R}}
\def\Bb{{\bar B}}

\def\bh{{\boldsymbol h}}

\def\bu{{\boldsymbol u}}

\def\af{\mathfrak{a}}
\def\bff{\mathfrak{b}}
\def\cf{\mathfrak{c}}
\def\df{\mathfrak{d}}
\def\ef{\mathfrak{e}}
\def\ff{\mathfrak{f}}
\def\nff{\ff^\natural}

\def\ffIR{\ff_{\rm IR}}
\def\ffUV{\ff_{\rm UV}}
\def\gf{\mathfrak{g}}

\def\hf{\mathfrak{h}}

\def\nf{\mathfrak{n}}

\def\Rf{\mathfrak{R}}
\def\bRf{\bf \Rf}

\def\Sf{\mathfrak{S}}
\def\bbSf{\bar{\mathfrak{S}}}
\def\rbbSf{\bar{\red{\mathfrak{S}}}}
\def\bbbSf{\bar{\blue{\mathfrak{S}}}}
\def\bSf{\blue{\mathfrak{S}}}
\def\rSf{\red{\mathfrak{S}}}
\def\slf{\mathfrak{sl}}
\def\sof{\mathfrak{so}}
\def\spf{\mathfrak{sp}}
\def\suf{\mathfrak{su}}

\def\Tf{\mathfrak{T}}
\def\uf{\mathfrak{u}}

\def\mT{\mathsf{T}}

\def\cC{{\mathcal C}}

\def\rCrg{\red{\cC}_{\rm reg}}

\def\cH{{\mathcal H}}

\def\cM{{\mathcal M}}

\def\cN{{\mathcal N}}

\def\cS{{\mathcal S}}

\def\cT{{\mathcal T}}

\def\cV{{\mathcal V}}

\def\C{\mathbb{C}} 
 
\def\H{\mathbb{H}}
 
\def\O{\mathbb{O}}

\def\R{\mathbb{R}} 
 
\def\Z{\mathbb{Z}} 

\def\beq{\begin{equation}}
\def\eeq{\end{equation}}

\newcommand{\bpmat}{\begin{pmatrix}}
\newcommand{\epmat}{\end{pmatrix}}
\newcommand{\bsmat}{\begin{smallmatrix}}
\newcommand{\esmat}{\end{smallmatrix}}
\newfloatcommand{capbtabbox}{table}[][\FBwidth]
\newtheorem{fact}{Fact}

\def\rcb{\rowcolor{blue!07}}
\def\rcg{\rowcolor{green!07}}
\def\rcy{\rowcolor{black!25!yellow!10}}

\newcounter{ex}[section]
\renewcommand{\theex}{\arabic{section}.\arabic{ex}}

\newcounter{example}[section]
\renewcommand{\theexample}{\arabic{section}.\arabic{example}}

\newcounter{dig}[section]
\renewcommand{\thedig}{\arabic{section}.\arabic{dig}}

\newcounter{definition}[section]
\renewcommand{\thedefinition}{\arabic{section}.\arabic{definition}}

\newcounter{solution}[section]
\renewcommand{\thesolution}{\arabic{section}.\arabic{solution}}

\makenoidxglossaries
\newglossaryentry{ECBg}{name={ECB},
 description={The Enhanced Coulomb Branch (ECB) is a branch of the moduli space where the SCFT $\Tf$ is Higgsed to a theory of the same rank}}

\newglossaryentry{ScActg}{name={scaling action},
 description={$\C^*$ action defined on the moduli space of vacua, arising from the spontaneous breaking of the $\R^+\times \uf(1)^*$ symmetry, with $\uf(1)^*=\uf(1)_r$ or the cartan of the $\suf(2)_R$ depending on whether we are on the \gls{CB} or the \gls{HB}}}

\newglossaryentry{CBg}{name={CB},
 description={The Coulomb Branch (CB) of a $\cN=2$ SCFT is a branch of the moduli space where only the $\uf(1)_r$ component of the theory's R-symmetry is spontaneously broken}}

\newglossaryentry{HBg}{name={HB},
 description={The Higgs Branch (HB) of a $\cN=2$ SCFT is a branch of the moduli space where only the $\suf(2)_R$ component of the theory's R-symmetry is spontaneously broken}}

\newglossaryentry{MBg}{name={MB},
 description={The Mixed Branch (MB) of a $\cN=2$ SCFT is a branch of the moduli space where the whole theory's R-symmetry is spontaneously broken.}}

\newglossaryentry{UVIR}{name={UV-IR simple flavor condition},
 description={The UV-IR simple flavor condition \cite{Martone:2020nsy}, or F-condition for brevity, states that simple flavor factors $\ff$ of SCFTs of arbitrary rank, are realized (with possible rank-preserving enhancement) as the flavor symmetries of (at least one) rank-1 theory \gls{rTf}}}

\newglossaryentry{gHWg}{name={gHW}, description={We conjecture that these are the Higgsings which are realized on the \gls{CB} as highest weight Higgsings. A generalized highest weight Higgsing is a Higgsing which respects \eqref{gHWeq}.}}

\newglossaryentry{higgs}{name={Higgsable \gls{CB} parameter}, description={A \gls{CB} parameter $u_i$ is higgsable if it can support a non-trivial (either IR-free or \gls{SCFT}) rank-1 theory $\rTf_{u_i}$ giving rise to an unknotted stratum $u_i=0$ of the \gls{CB} stratification}}

\newglossaryentry{totally}{name={totally higgsable}, description={An \gls{SCFT} is totally higgsable if all its \gls{CB} parameters are \gls{higgs}.Because of the \gls{UVIR}, totally higgsable SCFTs have in general non-simple flavor factors and intricate \gls{HB} Hasse diagrams}}

\newglossaryentry{Hlin}{name={linear}, description={Symplectic leaves of a symplectic singularities (\emph{e.g.} \gls{HB}s of $\cN=2$ \gls{SCFT}s) form a partially order set, were the ordering is given by the inclusion of their closures, and therefore can be represented by a Hasse diagram. When the symplectic leaves are instead totally ordered the corresponding Hasse diagram can be depicted as a straight line. In this case we call the \gls{HB} a linear \gls{HB}}}

\newglossaryentry{AS}{name={AS}, description={hypermultiplet in the antisymmetric representation}}

\newglossaryentry{S}{name={S}, description={hypermultiplet in the symmetric representation}}

\newglossaryentry{F}{name={F}, description={hypermultiplet in the fundamental representation}}

\newglossaryentry{V}{name={V}, description={hypermultiplet in the vector representation}}

\newglossaryentry{adj}{name={adj}, description={hypermultiplet in the adjoint representation}}

\newglossaryentry{ccf}{name={central charge formulae}, description={the formulae \eqref{actotaint}-\eqref{actotcint} which allow to compute the $a$, $c$ and $k_\ff$ central charges of an arbitrary \gls{SCFT} from \gls{CB} stratification data}}

\newglossaryentry{CcrgG}{name={CCRG}, description={a crystallographic (point) complex reflection group is a finite group which is generated by complex reflections and which acts on a lattice}}

\newglossaryentry{kstr}{name={knotted stratum}, description={The \gls{ScActg} on the \gls{CB} imposes that the singular locus $\rSf$ has to be closed under it. This in turn tremendously constraints the type of complex co-dimension one singularity which can appear at rank-2. A knotted stratum is one such connected complex co-dimension one stratum which algebraically can be written as $u^p+v^q=0$. Where $p,q\in\Z$ are completely fixed by scaling invariance (\ie homogeneity of the polynomial) and gcd$(p,q)=1$. Any interacting rank-2 \gls{SCFT} must possess at least a knotted stratum}}

\newglossaryentry{ukstr}{name={unknotted stratum}, description={The \gls{ScActg} on the \gls{CB} imposes that the singular locus $\rSf$ has to be closed under it. This in turn tremendously constraints the type of complex co-dimension one singularity which can appear at rank-2. A unknotted stratum is one such connected complex co-dimension one stratum which algebraically can be written either as $u=0$ or $v=0$. Where $u$ ($v$) must be a \gls{higgs}. For any interacting rank-2 \gls{SCFT} there can be at most one such unknotted stratum for each \gls{higgs}}}

\newglossaryentry{elesl}{name={elementary slice}, description={The transverse slice among two adjacent symplectic leaves. Elementary slices represent the individual transition depicted in the Hasse diagram and correspond to minimal higgsings}}

\newglossaryentry{ADt}{name={Argyres-Douglas theory}, description=An {\gls{AD} is characterized by having at least one \gls{CB} parameter of dimension $\D<2$ which implies that these theories have a relevant deformation which does not arise as a flavor symmetry and therefore is not constrained by the \gls{UVIR}. \gls{AD} theories are often described as theories with fractional \gls{CB} scaling dimensions. From a purely \fod perspective, it makes little qualitative difference whether a given scaling dimension is integer or fractional, whereas the existence of a chiral relevant deformation does}}

\newglossaryentry{Hasse}{name={Hasse diagram}, description={A Hasse diagram is a graphical depiction of a partially ordered set (poset). A point is drawn for each element of the poset, and line segments are drawn between these points according to the following two rules:
\begin{itemize}
\item[1.] If $x<y$ in the poset, then the point corresponding to $x$ appears lower in the drawing than the point corresponding to $y$.
\item[2.] The line segment between the points corresponding to any two elements $x$ and $y$ of the poset is included in the drawing iff $x$ covers $y$ or $y$ covers $x$.
\end{itemize}}}

\newglossaryentry{ECB}{type=\acronymtype, name={ECB}, description={Enhanced Coulomb Branch}, first={Enhanced Coulomb Branch (ECB)\glsadd{ECBg}}, see=[Glossary:]{ECBg}}

\newglossaryentry{SCFT}{type=\acronymtype, name={SCFT}, description={Superconformal Field Theory}, first={Superconformal Field Theory (SCFT)}}

\newglossaryentry{CB}{type=\acronymtype, name={CB}, description={Coulomb Branch}, first={Coulomb Branch (CB)\glsadd{CBg}}, see=[Glossary:]{CBg}}

\newglossaryentry{HB}{type=\acronymtype, name={HB}, description={Higgs Branch}, first={Higgs Branch (HB)\glsadd{HBg}}, see=[Glossary:]{HBg}}

\newglossaryentry{MB}{type=\acronymtype, name={MB}, description={Mixed Branch}, first={Mixed Branch (MB)\glsadd{MBg}}, see=[Glossary:]{MBg}}

\newglossaryentry{gHW}{type=\acronymtype, name={gHW}, description={generalized highest weight Higgsing}, first={generalized highest weight Higgsing (gHW)\glsadd{gHWg}}, see=[Glossary:]{gHWg}}

\newglossaryentry{Ccrg}{type=\acronymtype, name={CCRG}, description={crystallographic complex reflection group}, first={crystallographic complex reflection group (CCRG)\glsadd{CcrgG}}, see=[Glossary:]{CcrgG}}

\newglossaryentry{tsk}{type=\acronymtype, name={TSK}, description={triple special K\"ahler \cite{Argyres:2019yyb}}, first={triple special K\"ahler (TSK)}}

\newglossaryentry{AD}{type=\acronymtype, name={AD theory}, description={Argyres-Douglas theory}, first={Argyres-Douglas (AD) theory\glsadd{ADt}}, see=[Glossary:]{ADt}}

\newglossaryentry{SW}{type=\acronymtype, name={SW}, description={Seiberg-Witten}, first={Seiberg-Witten (SW)}}

\newglossaryentry{MN}{type=\acronymtype, name={MN}, description={Minahan-Nemeschansky}, first={Minahan-Nemeschansky (MN)}}

\newglossaryentry{rTf}{name={$\rTf_i$},
 description={Rank-1 theory supported on a CB stratum of co-dimension 1},
 type=symbols}
 
 \newglossaryentry{Tf}{name={T},
 description={Rank-2 SCFT supported at the origin of the moduli space},
 type=symbols}
 
\newglossaryentry{Dsingi}{name={$D_i^{\rm sing}$},
description={Scaling dimension of polynomial defining the algebraic variety on which $\rTf_i$ is supported},
type=symbols}
 
\newglossaryentry{bi}{name={$b_i$},
 description={Contribution of a $\rTf_i$ to the central charge formulae},
 type=symbols}
 
\newglossaryentry{h}{name={$h$},
 description={Total quaternionic dimension of the \gls{ECB} of the theory},
 type=symbols}

\title{Testing our understanding of SCFTs: a catalogue of rank-2 $\cN=2$ theories in four dimensions}
\author[1,2]{Mario Martone}
\affiliation[1]{C.~N.~Yang Institute for Theoretical Physics,  Stony Brook University,Stony Brook, NY 11794-3840, USA}
\affiliation[2]{Simons Center for Geometry and Physics, Stony Brook University, Stony Brook, NY 11794-3840, USA}
\emailAdd{mmartone@scgp.stonybrook.edu}

\abstract{In this paper we begin mapping out the space of rank-2 $\cN=2$ superconformal field theories (SCFTs) in four dimensions. This represents an ideal set of theories which can be potentially classified using purely quantum field-theoretic tools, thus providing a precious case study to probe the completeness of the current understanding of SCFTs, primarily derived from string theory constructions. Here, we collect and systematize a large amount of field theoretic data characterizing each theory. We also provide a detailed description of each case and determine the theories' Coulomb, Higgs and Mixed branch stratification. The theories naturally organize themselves into series connected by RG flows but which have gaps suggesting that our current understanding is not complete. }

\begin{document}
\maketitle 

\section{Introduction}

In this paper, we initiate the bottom-up analysis of rank-2 $\cN=2$ Superconformal Field Theories in four dimensions. We compile a catalogue of all rank-2 $\cN=2$ SCFTs currently known and determine the full moduli space stratification for each theory heavily leveraging the recent advancements of \cite{Argyres:2020wmq,Martone:2020nsy}. This remarkable amount of information already shows interesting patterns suggesting that our current understanding is not complete (for more details see section \ref{sec:overview}). Our results are summarized in table \ref{tab:r21}, \ref{tab:r22}, \ref{tab:r23}, \ref{Higgs:one}, \ref{Higgs:two} and \ref{Higgs:three}\footnote{It is my intention to constantly update the list on known rank-2 theories as our understanding improves. I would therefore be extremely grateful if the reader who is aware of any rank-2 $\cN=2$ \gls{SCFT} not appearing in the tables, could readily communicate this information to me.} which are organized in sets of theories mutually connected by mass deformations. Table \ref{tab:r23} and \ref{Higgs:three} lists in particular the ``isolated'' ones for which no mass deformation connecting them to any other rank-2 \gls{SCFT} is known. This hints at the existence of new theories as it will be further argued below though a detailed discussion of RG-flows among \fod rank-2 theories will be postponed to \cite{Martone:2021drm}. It is also worth notice that some gaps in our understanding of the detailed structure of some theory's moduli space remain  (mostly those in the $\suf(6)$ and $\suf(5)$ series).  %A quick comparison between the tables in this manuscript and the corresponding ones in the rank-1 case, \emph{e.g.} \cite[table 1]{Argyres:2016xmc}, shows the increase in complexity in going from rank-1 to rank-2.

The study of \gls{SCFT}s has been at the center stage of mathematical physics for over two decades. By now, we have a remarkably large amount of information on the space of the allowed theories in different spacetime dimensions and a great deal of understanding of the interconnections among them. This achievement is largely due to string theory which has provided, and still does, absolutely amazing and essential tools to study theories with large amount of supersymmetry, particularly in dimensions larger than four (see below). These results are so inextricably linked with string theory that is often challenging to be sure that our current understanding reflects properties of quantum field theory rather than string theory itself. Assessing the completeness of the string theoretic picture it is then an important priority and the work in this manuscript is a step in this direction.

One way to make this assessment is to develop tools to answer a basic question: can all consistent \gls{SCFT}s, with perhaps appropriate caveat like $d\geq3$ and with eight or more supercharges, be indeed engineered in string theory? It is a standard result that the maximum amount of dimensions where superconformal invariance is even conceivable is six \cite{Nahm:1977tg}. Since the gauge coupling is irrelevant for $d>4$ and all gauge theories are thus IR-free, it was not immediately obvious that \gls{SCFT}s above four dimension could exist at all.  The evidence that this was the case came two decades later, thanks primarily to the improvement in the understanding of string theory \cite{Witten:1995zh,Strominger:1995ac,Seiberg:1996qx,Morrison:1996xf,Seiberg:1996bd,Douglas:1996xp,Intriligator:1997pq}. By now the picture in six dimension is fairly clear with a belief that maximally symmetric (2,0) theories are completely classified by an ADE classification \cite{Witten:1995zh,Henningson:2004dh,Heckman:2013pva} and growing evidence of a complete story in the (1,0) case \cite{DelZotto:2014hpa,Heckman:2015bfa,Bhardwaj:2015xxa,Bhardwaj:2019hhd}\footnote{For a recent review of \gls{SCFT}s in six dimensions and many more references see \cite{Heckman:2018jxk}.}. In five dimensions our understanding is less settled but extraordinary progress has taken place recently leading to some initial attempt at classifying $\cN=1$ \gls{SCFT}s in \fid \cite{Jefferson:2017ahm,Jefferson:2018irk,Apruzzi:2019vpe,Bhardwaj:2019fzv,Apruzzi:2019enx,Apruzzi:2019opn,Bhardwaj:2019jtr,Closset:2020scj}\footnote{The literature on the subject is too vast so I apologize in advance for not providing fair credit to those who deserve it.}. A similar classification of \gls{SCFT}s with eight (or more) supercharges in \fod is instead wide open. A large majority of \fod $\cN=2$ \gls{SCFT}s belong to the so called class-$\cS$ set \cite{Gaiotto:2009we,Gaiotto:2009hg} which directly descends from the compactification of (2,0) theories in \emph{6d}. Thus again much of our current understanding in \fod is derived from string theoretic constructions. %Perhaps a shortcoming of these remarkable advancements, is that overwhelmingly, both in \sd and \fid, the claims on the completeness of the various \gls{SCFT} classification schemes are based on string theoretic arguments with relatively slim bottom-up field theoretic evidence. 

\begin{table}[ht]
\begin{adjustbox}{center,max width=.75\textwidth}
$\def\arraystretch{1.0}
\begin{array}{r|c:c:c:c:cc|c:cc|c}
\multicolumn{11}{c}{\Large\textsc{Table\ of\ Rank-2\ theories:\ CFT\ \&\ Coulomb\ branch\ data \ I}}\\
\hline
\hline
&\multicolumn{6}{l|}{\qquad\qquad\qquad\quad\qquad \text{Moduli Space}} &
\multicolumn{3}{l|}{\quad \text{Flavor and central charges}} & 
\\[1mm]
\multirow{-2}{4mm}{\#}
& \D_{u,v}&\rTf_{\rm knot}&\rTf_u&\rTf_v
&\ \ d_{\text{HB}}\ \  
&\ \  h\ \  &\quad \ff\quad &\ \ 24a\ \ & 12c &\qquad\multirow{-2}{0mm}{\text{Comments}}\hfill
\\[1.5mm]
\hline\hline
\multicolumn{11}{c}{\ef_8-\sof(20)\ \text{series}}\\
\hline
1.
& \{6,12\}
&\blue{\cS^{(1)}_{\varnothing,2}}& \varnothing &\quad \cT^{(1)}_{E_8,1} \quad\ & 59 & 1
& [\ef_8]_{24}\times \suf(2)_{13} & 263 & 161 &\hyperref[sec:E8r2]{\cT^{(2)}_{E_8,1}} \\
2.
& \{6,8\}
&[I_1,\varnothing]&\varnothing & [I_6^*,\sof(20)] & 46 & 0
&\sof(20)_{16} & 202 & 124 &\hyperref[sec:t2]{3^{\rm rd}{\rm\, entry\, at\, page\, 25\, of}}\, \text{\cite{Chacaltana:2014jba}} \\
3.
& \{4,10\}
&[I_1,\varnothing]&\varnothing & \cT^{(1)}_{E_8,1}  & 46 & 0
&[\ef_8]_{20} & 202 & 124 & \hyperref[sec:D20E8]{D^{20}_1(E_8)}\ \text{\cite{Giacomelli:2017ckh}}\\
4.
& \{4,8\}
&\blue{\cS^{(1)}_{\varnothing,2}}& \varnothing & \cT^{(1)}_{E_7,1} & 35 & 1
& [\ef_7]_{16}\times \suf(2)_{9} & 167 & 101 &\hyperref[sec:E7r2]{\cT^{(2)}_{E_7,1}} \\
5.
& \{4,6\}
&[I_1,\varnothing]&[I_2,\suf(2)] &[I_4^*,\sof(16)] & 30 & 0
&\suf(2)_8 \times \sof(16)_{12} & 138 & 84 &\hyperref[sec:t5]{3^{\rm rd}{\rm\, entry\, at\, page\, 26\, of}}\, \text{\cite{Chacaltana:2011ze}} \\
6.
& \{4,5\}
&[I_1,\varnothing]& \varnothing & [I_{10},\suf(10)] & 26 & 0
&\suf(10)_{10} & 122  & 74&\hyperref[sec:S5]{S_5: 3^{\rm rd}{\rm\, entry\, of\, page\, 30\, of}}\, \text{\cite{Chacaltana:2010ks}} \\
7.
& \{3,6\}
&\blue{\cS^{(1)}_{\varnothing,2}}& \varnothing & \cT^{(1)}_{E_6,1} & 23 & 1
& [\ef_6]_{12}\times \suf(2)_{7} & 119 & 71 &\hyperref[sec:E6r2]{\cT^{(2)}_{E_6,1}} \\
8.
& \{3,5\}
&[I_1,\varnothing]&\varnothing & [I_3^*,\sof(14)] & 22 & 0
&\sof(14)_{10}\times \uf(1) & 106  & 64&\hyperref[sec:t8]{R_{2,5}}\ \text{\cite{Chacaltana:2010ks}}  \\
9.
& \{3,4\}
&[I_1,\varnothing]&[I_2,\suf(2)] & [I_8,\suf(8)] & 18 & 0
&\suf(2)_6\times \suf(8)_8 & 90  & 54 &\hyperref[sec:t9]{R_{0,4}}\ \text{\cite{Chacaltana:2010ks}}  \\
\rcy10.
& \{2,4\}
&[I_1,\varnothing]^2& \varnothing & [I_2^*,\sof(12)] & 14 & 0 
& \sof(12)_8 & 74 & 44 & \hyperref[sec:Sp464]{USp(4)+6 F}\\
\rcy 11.
& \{2,4\} 
&\blue{\cS^{(1)}_{\varnothing,2}}& \varnothing & \cT^{(1)}_{D_4,1}  & 11 & 1 
& \sof(8)_8\times\suf(2)_5 & 71 & 42  &\hyperref[sec:D4r2]{USp(4)+4F+V\, {\rm or}\, \cT^{(2)}_{D_4,1}} \\
\rcy12.
& \{2,3\} 
&[I_1,\varnothing]^2& \varnothing & [I_6,\suf(6)] & 10 & 0
& \uf(6)_6 & 58 & 34 &\hyperref[sec:SU3Nf6]{SU(3)+6F}\\
\rcy13.
& \{2,2\}
&[I_2,\suf(2)]^2& [I_2,\suf(2)]^3 & \varnothing & 6 & 0
& \suf(2)_4^5 & 42 & 24 &\hyperref[sec:SU22b]{2 F+SU(2)-SU(2)+2 F} \\
14.
& \{\frac32,3\}
&\blue{\cS^{(1)}_{\varnothing,2}}& \varnothing & \cT^{(1)}_{A_2,1} & 5 & 1
& \suf(3)_{6}\times \suf(2)_{4} & 47 & 26 &\hyperref[sec:H2r2]{\cT^{(2)}_{A_2,1}} \\
15.
& \{\frac32,\frac52\} 
&[I_1,\varnothing]& \varnothing&[I_5,\suf(5)] & 6 & 0  
& \suf(5)_5 & 42 & 24 &\hyperref[sec:D2SU5]{D_2\big(SU(5)\big)}\ \text{\cite{Cecotti:2013lda}}\\
16.
& \{\frac43,\frac83\}
&\blue{\cS^{(1)}_{\varnothing,2}}& \varnothing & \cT^{(1)}_{A_1,1} & 3 & 1
& \suf(2)_{\frac{16}3}\times \suf(2)_{\frac{11}3} & 39 & 21 &\hyperref[sec:H1r2]{\cT^{(2)}_{A_1,1}} \\
17.
& \{\frac43,\frac53\}
&[I_1,\varnothing]& \varnothing & [I_2,\suf(2)] & 2 & 0
&\suf(2)_{\frac{10}3} \times \uf(1) & 26 & 14 &\hyperref[sec:(A1,D6)]{(A_1,D_6) \text{\ AD\ Theory}}\\
18.
& \{\frac65,\frac{12}5\}
&\blue{\cS^{(1)}_{\varnothing,2}}& \varnothing & \cT^{(1)}_{\varnothing,1} & 1 & 1
& \suf(2)_{\frac{17}5} & \frac{163}5 & 17 &\hyperref[sec:H0r2]{\cT^{(2)}_{\varnothing,1}} \\
19.
& \{\frac65,\frac85\}
&[I_1,\varnothing]&\varnothing &[I_2,\suf(2)] & 1 & 0 &
\suf(2)_{\frac{16}5} & \frac{114}5 & 12 &\hyperref[sec:(A1,D5)]{(A_1,D_5) \text{\ AD\ Theory} } \\
20.
& \{\frac54,\frac32\} 
&[I_1,\varnothing]& \varnothing &\varnothing & 1 & 0
& \uf(1) & 22 & \frac{23}2 &\hyperref[sec:(A1,A5)]{(A_1,A_5) \text{\ AD\ Theory}} \\
21.
& \{\frac87,\frac{10}7\} 
&[I_1,\varnothing]&\varnothing&\varnothing & 0 &0  & \varnothing
& \frac{134}7 & \frac{68}7 &\hyperref[sec:(A1,A4)]{(A_1,A_4)\ \text{AD\ Theory}}\\
\cdashline{1-11}

%%End of E_8/SO(20) series.  ------- beginning of the sp(12)-Sp(8)-F_4 series
\multicolumn{11}{c}{\spf(12)-\spf(8)-\ff_4\ \text{series}}\\
\hline
22.
& \{4,6\} 
&[I_1,\varnothing]^2& [I_{12},\suf(12)]_{\Z_2} & \varnothing & 22 & 0
& \spf(12)_8 & 130 & 76 &\hyperref[sec:t22]{66^{\rm th}{\rm\, entry\, at\, page\, 49\, of}}\, \text{\cite{Chacaltana:2013oka}}  \\
23.
& \{4,6\}
&[I_1,\varnothing]&[I_8,\suf(8)]_{\Z_2} &[I_1^*,\spf(4)] & 20 & 2
&\spf(4)_7 \times \spf(8)_8 & 128 & 74 &\hyperref[sec:t23]{5^{\rm th}/6^{\rm th}{\rm\, entry\, at\, page\, 29\, of}}\, \text{\cite{Chacaltana:2011ze}} \\
24.
&\{6,6\} 
&[\blue{\cS^{(1)}_{\varnothing,2}}]^2& [\cT^{(1)}_{E_6,1}]_{\Z_2} & \varnothing  & 24 & 2
& \suf(2)_7^2\times [\ff_4]_{12} & 156 & 90 & \hyperref[sec:TE62]{\cT_{E_6,2}^{(2)}}\ \text{\cite{Giacomelli:2020jel}}  \\
25.
& \{3,4\}
&[I_1,\varnothing]& [I_8,\suf(8)]_{\Z_2} & [I_2,\suf(2)] & 12 & 0
&\suf(2)_8\times \spf(8)_6 & 84  & 48 &\hyperref[sec:t25]{6^{\rm th}{\rm\, entry\, of\, page\, 61\, of}}\, \text{\cite{Chacaltana:2012ch}}  \\
26.
& \{3,4\}
&[I_1,\varnothing]& [I_6,\suf(6)]_{\Z_2}&\blue{\cS^{(1)}_{\varnothing,2}} & 11 & 1
&\suf(2)_5\times \spf(6)_6\times \uf(1) & 83  & 47 &\hyperref[sec:t26]{2^{\rm nd}{\rm\, entry\, of\, page\, 61\, of}}\, \text{\cite{Chacaltana:2012ch}}\\
27.
&\{4,4\} 
&[\blue{\cS^{(1)}_{\varnothing,2}}]^2& [\cT^{(1)}_{D_4,1}]_{\Z_2} & \varnothing  & 12 & 2 
& \suf(2)_5^2\times \sof(7)_8 & 96 & 54 & \hyperref[sec:TD42]{\cT_{D_4,2}^{(2)}}\ \text{\cite{Giacomelli:2020jel}}  \\
28.
& \{4,5\}
&[I_1,\varnothing]& \varnothing & [\cT^{(1)}_{E_6,1}]_{\Z_2} & 16 & 0
&[\ff_4]_{10}\times \uf(1) & 112  & 64 & \hyperref[sec:tE6]{\widehat{\cT}_{E_6,2}}\ \text{\cite{Wang:2018gvb}}\\
29.
& \{\frac52,3\}
&[I_1,\varnothing]& [I_6,\suf(6)]_{\Z_2} &\varnothing & 7 & 0
& \spf(6)_5\times \uf(1) & 61 & 34 &\hyperref[sec:tTE62]{ \widetilde{\cT}_{E_6,2}}\ \text{\cite{Zafrir:2016wkk}}\\
30.
& \{3,3\}
&[\blue{\cS^{(1)}_{\varnothing,2}}]^2&[ \cT^{(1)}_{A_2,1}]_{\Z_2} & \varnothing & 6 & 2
& \suf(3)_6\times\suf(2)^2_4 & 66 & 36 & \hyperref[sec:TA22]{\cT_{A_2,2}^{(2)}}\ \text{\cite{Giacomelli:2020jel}} \\
\rcy31.
& \{2,2\}
&[I_1,\varnothing]^4& [I_4,\suf(4)]_{\Z_2} & \varnothing & 3 & 0
& \spf(4)_4 & 38 & 20 &\hyperref[sec:SU22a]{SU(2)-SU(2)} \\
\rcb32.
& \{2,2\}
&\blue{\cS^{(1)}_{\varnothing,2}}& \blue{\cS^{(1)}_{\varnothing,2}} & \varnothing & 2 & 2
& \suf(2)_3\times\suf(2)_3 & 36 & 18 & \hyperref[sec:N4SU22]{\cN=4\ SU(2)\times SU(2) } \\
[.5mm]
\hline\hline
%% end of sp(12)-Sp(8)-F_4 series
%
%
\end{array}$
\caption{\label{tab:r21}{\small First part of the list of rank-2 theories organized by \emph{series}. Theories in each series are mutually connected by mass deformations. The second column lists the scaling dimension of the \gls{CB} parameters. Columns 3, 4 and 5 instead list the rank-1 theories describing the massless states on the connected components of the \gls{CB} singular locus, see table \ref{Theories} for how to read these entries. The $\Z_\ell$ subscript indicates discrete gauging. Continuing, $d_{\rm HB}$ indicates the quaternionic dimension of the \gls{HB} while \gls{h} that of the \gls{ECB}. Column 8 indicates the flavor symmetry of theory, along with the level, followed by the $a$ and $c$ central charges.}}
\end{adjustbox}
\end{table}

\begin{table}[ht]
\begin{adjustbox}{center,max width=.67\textwidth}
$\def\arraystretch{1.0}
\begin{array}{r|c:c:c:c:cc|c:cc|c}
\multicolumn{11}{c}{\Large\textsc{Table\ of\ Rank-2\ theories:\ CFT\ \&\ Coulomb\ branch\ data \ II }}\\
\hline
\hline
&\multicolumn{6}{l|}{\qquad\qquad\qquad\quad\qquad \text{Moduli Space}} &
\multicolumn{3}{l|}{\quad \text{Flavor and central charges}} & 
%&
%\multicolumn{2}{l}{\text{Invt's:}}
\\[1mm]
\multirow{-2}{4mm}{\#}
& \D_{u,v}&\rTf_{\rm knot}&\rTf_u&\rTf_v
&\ \ d_{\text{HB}}\ \  
&\ \  h\ \  &\quad \ff\quad &\ \ 24a\ \ & 12c &\multirow{-2}{0mm}{\text{Comments}}\hfill
\\[1.5mm]
\hline\hline

%%% Beginning of the SU(6) series:
\multicolumn{11}{c}{\suf(6)\ \text{series}}\\
\hline
33.
& \{6,8\} 
&[I_1,\varnothing]&\varnothing&[I_6^*,\sof(12)\times \suf(2)]_{\Z_2} & 23 &1& 
\suf(6)_{16}{\times}\suf(2)_9 & 179 & 101 &\hyperref[sec:t33]{33^{\rm th}{\rm\, entry\, at\, page\, 16\, of}\,} \text{\cite{Chacaltana:2015bna}}\\
34.
& \{4,6\}
&[I_1,\varnothing]&[I_2,\suf(2)] &[I_4^*,\sof(8)\times\suf(2)]_{\Z_2} & 13 & 1&
\suf(4)_{12}{\times} \suf(2)_7{\times}\uf(1) & 121 & 67 & \hyperref[sec:t34]{10^{\rm th}{\rm\, entry\, at\, page\, 41\, of}\,} \text{\cite{Chacaltana:2013oka}}  \\
35.
& \{4,5\}
&[I_1,\varnothing]&\varnothing &\star\ {\rm w/}\ b=7 & 11 & 0&
\suf(3)_{10}{\times} \suf(3)_{10}{\times}\uf(1) & 107 & 59 & \hyperref[sec:t35]{\emph{5d}\ SCFT\,{\rm on}\,S^1|_{\Z_2}} \text{\cite{Zafrir:2016wkk}} \\
36.
& \{3,5\}
&[I_1,\varnothing]&\varnothing &[I_3^*,\sof(6)\times\suf(2)] & 8 & 1 &
\suf(3)_{10}{\times} \suf(2)_6{\times}\uf(1) & 92 & 50 &\hyperref[sec:t36]{\emph{5d}\ SCFT\,{\rm on}\,S^1|_{\Z_2}} \text{\cite{Zafrir:2016wkk}}  \\
37.
& \{3,4\}
&[I_1,\varnothing]&[I_2,\suf(2)] &\star\ {\rm w/}\ b=5 & 6 & 0 &
\suf(2)_8{\times} \suf(2)_8{\times}\uf(1)^2 & 78 & 42 &\hyperref[sec:t37]{\emph{5d}\ SCFT\,{\rm on}\,S^1|_{\Z_2}} \text{\cite{Zafrir:2016wkk}}  \\
\rcy38.
& \{2,3\} 
&[I_1,\varnothing]^2& \varnothing & [I_6,\suf(2)] & 2 & 0
& \uf(1)\times \uf(1) & 49 & 25 &\hyperref[sec:SU363]{SU(3)+F+S }\\
\cdashline{1-11}

%%% End of the SU(6) series - beginning of the Sp(14) series:
\multicolumn{11}{c}{\spf(14)\ \text{series}}\\
\hline
39.
& \{6,8\}
&[I_1,\varnothing]& \varnothing & [I_6^*,\sp(14)] & 29 & 7
&\spf(14)_9 & 185 & 107& \hyperref[sec:TD4f]{{\rm min}\, (D_7,D_7)\, {\rm on}\, T^2|_{\Z_2}}\ \text{\cite{Ohmori:2018ona}} \\
40.
& \{4,6\}
&[I_1,\varnothing]&[I_2,\suf(2)] & [I_4^*,\spf(10)] & 17 & 5
&\suf(2)_8 \times\spf(10)_7& 125  & 71&\hyperref[sec:t40]{15^{\rm th}{\rm\, entry\, at\, page\, 53\, of}}\, \text{\cite{Chacaltana:2013oka}}  \\
41.
& \{4,6\}
&[I_1,\varnothing]&\blue{\cS^{(1)}_{\varnothing,2}} &[I_3^*,\spf(8)] & 15 & 5
&\suf(2)_5 \times \spf(8)_7 & 123 & 69 &\ \ \hyperref[sec:t64]{5^{\rm th}{\rm\, entry\, at\, page\, 51\, of}}\, \text{\cite{Chacaltana:2013oka}} \ \ \\
42.
& \{3,5\}
&[I_1,\varnothing]&\varnothing & [I_3^*,\spf(8)]  & 11 & 4
&\spf(8)_6\times \uf(1) & 95  & 53 &\hyperref[sec:R24]{ R_{2,4}}\ \text{\cite{Chacaltana:2014nya}} \\
\rcy43.
& \{2,4\}
&[I_1,\varnothing]^2& \varnothing & [I_2^*,\spf(6)] & 6 & 3
& \spf(6)_5 & 65 & 35 &\hyperref[sec:Sp435]{USp(4)+ 3V} \\
\cdashline{1-11}

%%% End of the Sp(14) series - beginning of the SU(5) series:

\multicolumn{11}{c}{\suf(5)\ \text{series}}\\
\hline
44.
& \{6,8\}
&[I_1,\varnothing]&\varnothing & [I_2^*,\sof(12)]_{\mathbb{Z}_3} & 19 & 0
&\suf(5)_{16} & 170 & 92 & \hyperref[sec:t43]{\emph{5d}\ T_5\,{\rm on}\,S^1|_{\Z_3}} \text{\cite{Zafrir:2016wkk}} \\
45.
& \{4,6\}
&[I_1,\varnothing]&[I_2,\suf(2)] &[\cT^{(1)}_{D_4,1}]_{\mathbb{Z}_3} & 6 & 0 &
\suf(3)_{12}{\times} \uf(1) & 114 & 60 &\hyperref[sec:t44]{{\rm\, page\, 39\, of}\,} \text{\cite{Zafrir:2016wkk}}  \\
46.
& \{3,5\}
&[I_1,\varnothing]&\varnothing &[I_\star^*,\sof(6)]_{\mathbb{Z}_3} & 3 & 0 &
\suf(2)_{10}{\times} \uf(1) & 86 & 44 & \hyperref[sec:t45]{{\rm Section\ 5.3.2\ of\ } \text{\cite{Martone:2021drm}}} \\
\cdashline{1-11}

%%% End of the SU(5) series - beginning of the Sp(12) series:

\multicolumn{11}{c}{\spf(12)\ \text{series}}\\
\hline
47.
& \{4,10\}
&[I_1,\varnothing]&\varnothing &[I_5^*,\spf(12)] & 32 & 6
&\spf(12)_{11} & 188 & 110 &\hyperref[sec:t46]{2^{\rm nd}{\rm\, entry\, at\, page\, 29\, of}}\, \text{\cite{Chacaltana:2017boe}}\\
\rcy48.
& \{2,4\}
&[I_2,\suf(2)]^2& \varnothing & [I_1^*,\spf(4)] & 8 & 2
&\spf(4)_5 \times\sof(4)_8 & 68 & 38 & \hyperref[sec:Sp42524]{USp(4)+2F+2V} \\
\rcy49.
& \{2,6\}
&[I_1,\varnothing]^2& \varnothing & [I_3^*,\spf(8)] & 14 & 4
& \spf(8)_7 & 98 & 56 &\hyperref[sec:G247]{G_2+4F }\\
50.
& \{\frac43,\frac{10}3\}
&[I_1,\varnothing]& \varnothing & [I_1^*,\spf(4)] & 4 & 2
& \spf(4)_{\frac{13}3} & 48 & 26 &\hyperref[sec:ADc2]{{\rm AD}(\cf_2)}\, \text{\cite{Kaidi:2021tgr}}\\

\cdashline{1-11}

%%% End of the Sp(12) series - beginning of the Sp(8) series:

\multicolumn{11}{c}{\spf(8)-\suf(2)^2\ \text{series}}\\
\hline
51.
& \{6,12\}
&\blue{\cS^{(1)}_{\varnothing,2}}& \varnothing & \cS^{(1)}_{E_6,2} & 28 & 6
& \spf(8)_{13}\times \suf(2)_{26} & 232 & 130 & \hyperref[sec:SE62]{\cS^{(2)}_{E_6,2}}\ \text{\cite{ Apruzzi:2020pmv}}\\
52.
& \{4,8\}
&\blue{\cS^{(1)}_{\varnothing,2}}& \varnothing & \cS^{(1)}_{D_4,2} & 14 & 4
& \spf(4)_{9}{\times} \suf(2)_{16} {\times}\suf(2)_{18}& 146 & 80 & \hyperref[sec:SD42]{\cS^{(2)}_{D_4,2}}\ \text{\cite{ Apruzzi:2020pmv}}\\
53.
& \{3,6\}
&\blue{\cS^{(1)}_{\varnothing,2}}& \varnothing & \cS^{(1)}_{A_2,2}& 7 & 3
& \suf(2)_{7}\times \suf(2)_{14}\times \uf(1) & 103 & 55 &\hyperref[sec:SA22]{ \cS^{(2)}_{A_2,2}}\ \text{\cite{ Apruzzi:2020pmv}}\\
54.
&\{3,6\}   
&[\blue{\cS^{(1)}_{\varnothing,2}}]^2& [\cT^{(1)}_{A_2,1}]_{\Z_4} & \varnothing & 6 & 2
& \suf(2)_6\times \suf(2)_{14} & 102 & 54  &\hyperref[sec:TA24]{ \cT_{A_2,4}^{(2)}}\ \text{\cite{Giacomelli:2020jel}}  \\

55.
& \{\frac52,4\}
&[I_1,\varnothing]&[\cT^{(1)}_{A_2,1}]_{\Z_4} & \varnothing & 2 & 0
&\suf(2)_5 & 67 & 34 &\hyperref[sec:tA2]{ \widehat{\cT}_{A_2,4}}\ \text{\cite{Giacomelli:2020gee}} \\

\rcb56.
& \{2,4\}
&\blue{\cS^{(1)}_{\varnothing,2}}& \varnothing & \blue{\cS^{(1)}_{\varnothing,2}} & 2 & 2
& \suf(2)_{10} & 60 & 30 &\hyperref[sec:N4Sp4]{ \cN=4\ USp(4) }\\

\cdashline{1-11}

%%% End of the Sp(8) series - beginning of the G2 series:

\multicolumn{11}{c}{\gf_2\ \text{series}}\\
\hline
57.
&\{4,6\} 
&\blue{\cS^{(1)}_{\varnothing,2}}& [\cT^{(1)}_{D_4,1}]_{\Z_3} & \varnothing & 12 & 2
& [\gf_2]_8\times \suf(2)_{14} & 120 & 66 &\hyperref[sec:TD43]{ \cT_{D_4,3}^{(2)}}\ \text{\cite{Giacomelli:2020jel}}  \\
58.
&\{\frac83,4\} 
&\blue{\cS^{(1)}_{\varnothing,2}}& [\cT^{(1)}_{A_1,1}]_{\Z_3} & \varnothing    & 4 & 2
& \suf(2)_{\frac{16}3}\times \suf(2)_{10}  & 72 & 38 & \hyperref[sec:TA13]{\cT_{A_1,3}^{(2)}}\ \text{\cite{Giacomelli:2020jel}}  \\
59.
& \{\frac{10}3,4\}
&[I_1,\varnothing]&[\cT^{(1)}_{D_4,1}]_{\Z_3} & \varnothing & 6& 0
&[\gf_2]_{\frac{20}3} & 82  & 44 &\hyperref[sec:tD4]{ \widehat{\cT}_{D_4,3}}\ \text{\cite{Giacomelli:2020gee}}\\
\rcb60.
& \{2,3\}
&\blue{\cS^{(1)}_{\varnothing,2}}&\varnothing & \varnothing & 2 & 2
& \suf(2)_8 & 48 & 24 & \hyperref[sec:N4SU3]{\cN=4\ SU(3)} \\
\cdashline{1-11}

%%% End of the G2 series - beginning of the SU(3) series:

\multicolumn{11}{c}{\suf(3)\ \text{series}}\\
\hline
61.
& \{6,12\}
&\blue{\cS^{(1)}_{\varnothing,2}}& \varnothing & \cS^{(1)}_{D_4,3} & 15 & 5
& \suf(3)_{26}\times \uf(1) & 219 & 117 &\hyperref[sec:SD43]{ \cS^{(2)}_{D_4,3}}\ \text{\cite{ Apruzzi:2020pmv}}\\
62.
& \{4,8\}
&\blue{\cS^{(1)}_{\varnothing,2}}& \varnothing & \cS^{(1)}_{A_1,3}& 5 & 3
& \uf(1)^2 & 137 & 71 &\hyperref[sec:SA13]{ \cS^{(2)}_{A_1,3}}\ \text{\cite{ Apruzzi:2020pmv}}\\
\rcg63.
& \{3,6\}
&\blue{\cS^{(1)}_{\varnothing,2}}& \varnothing & \green{\cS^{(1)}_{\varnothing,3}} & 2 & 2
& \uf(1) & 96 & 48 &\hyperref[sec:G312]{\cN=3\ G(3,1,2)}\ \text{\cite{Aharony:2016kai}}
\\
\cdashline{1-11}
%%% End of the SU(3) series - beginning of the SU(2) series:
\multicolumn{11}{c}{\suf(2)\ \text{series}}\\
\hline
64. 
& \{6,12\}
&\blue{\cS^{(1)}_{\varnothing,2}}& \varnothing & \cS^{(1)}_{A_2,4} & 8 & 4
& \suf(2)_{16}\times \uf(1) & 212&  110 & \hyperref[sec:SA24]{\cS^{(2)}_{A_2,4}}\ \text{\cite{ Apruzzi:2020pmv}}\\
\rcg65.
& \{4,8\}
&\blue{\cS^{(1)}_{\varnothing,2}}& \varnothing & \green{\cS^{(1)}_{\varnothing,4}} & 2 & 2
& \uf(1) & 132 & 66 &\hyperref[sec:G412]{\cN=3\ G(4,1,2)}\ \text{\cite{Aharony:2016kai}}\\[.5mm]
\hline\hline

%%% End of the SU(2) series

\end{array}$
\caption{\label{tab:r22}{\small Second part of the list of rank-2 theories organized by \emph{series}. Theories in each series are mutually connected by mass deformations. The second column lists the scaling dimension of the \gls{CB} parameters. Columns 3, 4 and 5 instead list the rank-1 theories describing the massless states on the connected components of the \gls{CB} singular locus, see table \ref{Theories} for how to read these entries. The $\Z_\ell$ subscript indicates discrete gauging. Continuing, $d_{\rm HB}$ indicates the quaternionic dimension of the \gls{HB} while \gls{h} that of the \gls{ECB}. Column 8 indicates the flavor symmetry of theory, along with the level, followed by the $a$ and $c$ central charges.}}
\end{adjustbox}
\end{table}

\begin{table}[ht]
\begin{adjustbox}{center,max width=.75\textwidth}
$\def\arraystretch{1.0}
\begin{array}{r|c:c:c:c:cc|c:cc|c}
\multicolumn{11}{c}{\Large\textsc{Rank-2:\ CFT\ \&\ Coulomb\ branch\ data \ III (Isolated theories)}}\\
\hline
\hline
&\multicolumn{6}{l|}{\quad\qquad\qquad\quad\qquad \text{Moduli Space}} &
\multicolumn{3}{l|}{\ \ \text{Flavor and central charges}} & 
\\[1mm]
\multirow{-2}{4mm}{\#}
& \D_{u,v}&\rTf_{\rm knot}&\rTf_u&\rTf_v
&\ \ d_{\text{HB}}\ \  
&\ \  h\ \  &\quad \ff\quad &\ \ 24a\ \ & 12c &\quad\multirow{-2}{0mm}{\text{Comments}}\hfill
\\[1.5mm]
\hline\hline

%% Isolated
66.
& \{4,6\}
&[I_1^*,\spf(4)]&  [I_2,\suf(2)] &\varnothing & 10 & 4 
& \spf(4)_{14}\times\suf(2)_8 & 118 & 64 &\hyperref[sec:t65]{17^{\rm th}{\rm\, entry\, at\, page\, 7\, of}}\, \text{\cite{Chacaltana:2016shw}}   \\
\cdashline{1-11}

67.
& \{\frac{12}5,6\}
&\blue{\cS^{(1)}_{\varnothing,2}}& [\cT^{(1)}_{\varnothing,1}]_{\Z_5}& \varnothing  & 2 & 2
& \suf(2)_{14} & \frac{456}5 & \frac{234}5 &\hyperref[sec:Tvar]{ \cT^{(2)}_{\varnothing,5}}\ \text{\cite{Giacomelli:2020gee}}\\
\cdashline{1-11}

\rcb68.
& \{2,6\}
&\blue{\cS^{(1)}_{\varnothing,2}}& \varnothing & \blue{\cS^{(1)}_{\varnothing,2}} & 2 & 2
& \suf(2)_{14} & 84 & 42 &\hyperref[sec:N4G2]{ \cN=4\ G_2 }\\
\cdashline{1-11}

\multicolumn{11}{c}{\text{Theory with no known string theory realization}}\\
\hline
\rcy69.
& \{2,4\}
&[I_1,\varnothing]^5& \varnothing & [I_1,\varnothing] & 0 & 0 
& \varnothing & 58 & 28 & \hyperref[sec:Sp416]{USp(4)\, \rm{w/}\  \frac12\,{\bf16} }\\

\hline\hline
\end{array}$
\caption{\label{tab:r23}{\small Third part of the list of rank-2 theories. In this table we list the \emph{isolated} theories, those for which no mass deformation connecting them to other rank-2 \gls{SCFT}s is known. The second column lists the scaling dimension of the \gls{CB} parameters. Columns 3, 4 and 5 instead list the rank-1 theories describing the massless states on the connected components of the \gls{CB} singular locus, see table \ref{Theories} for how to read these entries. The $\Z_\ell$ subscript indicates discrete gauging. Continuing, $d_{\rm HB}$ indicates the quaternionic dimension of the \gls{HB} while \gls{h} that of the \gls{ECB}. Column 8 indicates the flavor symmetry of theory, along with the level, followed by the $a$ and $c$ central charges.}}
\end{adjustbox}
\end{table}

\begin{table}
\begin{adjustbox}{center,max width=.8\textwidth}
\renewcommand{\arraystretch}{1.3}
$\begin{array}{|c|c|c|c|c|c|c|c|c|c|c}
\cline{1-9}
\multicolumn{9}{|c|}{\text{\bf Summary of rank-1 theories $\Tf_i$ supported on $\Sf_i$}}\\
\hhline{=========~}
\text{Name} & \multicolumn{1}{c|}{\ \ 12\, c\ \ } &\quad \D_u\quad{} &\quad h\quad{}&\, {\boldsymbol R_{2h}}\,{}&\quad b\quad{} &\quad \ff\quad{} &\quad k_\ff\quad{}&{\rm Comments}&\, {}&\quad{} \\
\cline{1-9}
\ \cT^{(1)}_{E_8,1} &62&6&0&{\bf1}&10&\ef_8&12&{[II^*,\ef_8]}\ {\rm in}\ \text{\cite{Argyres:2016xmc}}&\multicolumn{2}{c}{\parbox[t]{2mm}{\multirow{16}{*}{\rotatebox[origin=c]{90}{SCFTs}}}}\\
\ \cT^{(1)}_{E_7,1} &38&4&0&{\bf1}&9&\ef_7&8&{[III^*,\ef_7]}\ {\rm in}\ \text{\cite{Argyres:2016xmc}}\\
\ \cT^{(1)}_{E_6,1} &26&3&0&{\bf1}&8&\ef_6&6&{[IV^*,\ef_6]}\ {\rm in}\ \text{\cite{Argyres:2016xmc}}\\
\ \cT^{(1)}_{D_4,1} &14&2&0&{\bf1}&6&\sof(8)&4&{[I_0^*,\sof(8)]}\ {\rm in}\ \text{\cite{Argyres:2016xmc}}\\
\ \cT^{(1)}_{A_2,1} &8&\frac32&0&{\bf1}&4&\suf(3)&3&{[IV,\suf(3)]}\ {\rm in}\ \text{\cite{Argyres:2016xmc}}\\
\ \cT^{(1)}_{A_1,1} &6&\frac43&0&{\bf1}&3&\suf(2)&\frac83&{[III,\suf(2)]}\ {\rm in}\ \text{\cite{Argyres:2016xmc}}\\
\ \cT^{(1)}_{\varnothing,1} &\frac{22}5&\frac65&0&{\bf1}&2&\varnothing&-&{[II,\varnothing]}\ {\rm in}\ \text{\cite{Argyres:2016xmc}}\\
\cdashline{1-9}
\ \cS^{(1)}_{E_6,2}&49&6&5&{\bf10}&7&\spf(10)&7&[II^*,\spf(10)]\ {\rm in}\ \text{\cite{Argyres:2016xmc}}&\\
\ \cS^{(1)}_{D_4,2}&29&4&3&({\bf6,1})&6&\spf(6){\times} \suf(2)&(5,8)&[III^*,\spf(6){\times}\suf(2)]\ {\rm in}\ \text{\cite{Argyres:2016xmc}}&\\
\ \cS^{(1)}_{A_2,2}&19&3&2&{\bf4}_0&5&\spf(4){\times} \uf(1)&(4,\star)&[IV^*,\spf(4){\times}\uf(1)]\ {\rm in}\ \text{\cite{Argyres:2016xmc}}\\
\ \blue{\cS^{(1)}_{\varnothing,2}}&9&2&1&{\bf2}&3&\suf(2)&3&\cN=4 \ \suf(2)\\
\cdashline{1-9}
\ \cS^{(1)}_{D_4,3}&42&6&4&{\bf4}\oplus\bar{\bf 4}&6&\suf(4)&14&[II^*,\suf(4)]\ {\rm in}\ \text{\cite{Argyres:2016xmc}}\\
\ \cS^{(1)}_{A_1,3}&24&4&3&{\bf2}_+\oplus{\bf 2}_-&5&\suf(2){\times}\uf(1)&(10,\star)&{[III^*,\suf(2){\times}\uf(1)]}\ {\rm in}\ \text{\cite{Argyres:2016xmc}}\\
\ \green{\cS^{(1)}_{\varnothing,3}}&15&3&1&{\bf1}_+\oplus{\bf 1}_-&4&\uf(1)&\star&\cN=3 \ S\text{-}fold\ \text{\cite{Aharony:2016kai}}\ \\
\cdashline{1-9}
\ \cS^{(1)}_{A_2,4}&38&6&3&{\bf3}\oplus\bar{\bf 3}&\frac{11}2&\suf(3)&14&{[IV^*,\suf(3)]}\ {\rm in}\ \text{\cite{Argyres:2016xmc}}\\
\ \green{\cS^{(1)}_{\varnothing,4}}&21&4&1&{\bf1}_+\oplus{\bf 1}_-&\frac92&\uf(1)&\star&\cN=3 \ S\text{-}fold\ \text{\cite{Aharony:2016kai}}\\
\hhline{=========~}
{[I_1,\varnothing]} &3&1&0&{\bf 1}&1&\uf(1)&1&\uf(1)\,{\rm Theory\, w/ 1\, hyper}&\multicolumn{2}{c}{\parbox[t]{2mm}{\multirow{4}{*}{\rotatebox[origin=c]{90}{IR-free}}}}\\ 
{[I_n,\suf(n)]} &n+2&1&0&{\bf 1}&n&\suf(n){\times}\uf(1)&2&\uf(1)\,{\rm Theory\, w/ n\, hyper}&\\
{[I_{2n},\suf(2n)]_{\Z_2}} &2n+2&1&0&{\bf 1}&2n&\spf(2n)&2&[\uf(1)\,{\rm Theory\, w/\, 2n\, hyper}]_{\Z_2}&\\
{[I_n^*,\sof(2n+8)]} &2(7+n)&2&0&{\bf1}&n+6&\sof(2n+8)&4&\suf(2)\,{\rm w/}\,(n+4)\,{\bf2}&\\
{[I_n^*,\spf(2n+2)]} &3(3+n)&2&n+1&\boldsymbol{2n+2}&n+3&\spf(2n+2)&3&\suf(2)\,{\rm w/}\,(n+1)\,{\bf3}^\clubsuit\\
\cline{1-9}
\multicolumn{9}{l}{\qquad^\clubsuit={\rm \, the\, beta\, function\, is\, renormalized\, by\,}\frac14,\text{see discussion in \cite[Section 4.2]{Argyres:2015ffa}}.}
\end{array}$
\caption{\label{Theories}{\small For the convenience of the reader, we list the properties of the rank-1 theories which can describe the low energy physics on the \gls{CB} co-dimension one singular locus as well as that of rank-decreasing \gls{HB} strata. The list of $\Z_n$ gauging of SCFTs, and their CFT data, can be found in \cite{Argyres:2016yzz}. The discretely gauged theories appearing on the \gls{CB} singular locus of theories in the $\suf(6)$ series are instead discussed explicitly in the text in the appropriate sections.}}
\end{adjustbox}
\end{table}

Despite its richness, the \fod situation is qualitatively different from the higher dimensional one. There are in fact \gls{SCFT}s which have no known string theory realization \cite{Bhardwaj:2013qia} and the lowest rank at which they appear is precisely two (\emph{i.e.} $USp(4)+\frac12{\bf 16}$). Furthermore in this case a variety of tools \cite{Seiberg:1994rs,Seiberg:1994pq,Donagi:1995cf,Beem:2013sza} which are constraining enough to conceive a bottom-up approach, are available. Potentially, this could give a way to probe the completeness of the string theoretic description of the space of supersymmetric field theories or, at least, expand it. The approach most dear to the author is primarily focused on the systematic study of the consistency of moduli space geometries \cite{Argyres:2020nrr}. 

This philosophy came already to fruition after completing the classification of the simplest set of four dimensional $\cN=2$ \gls{SCFT}s, the so-called rank-1 theories, \cite{Argyres:2015ffa,Argyres:2015gha,Argyres:2016xua,Argyres:2016xmc}. This work highlighted the incompleteness of our understanding at the time and led to the discovery of many new theories \cite{Chacaltana:2016shw,Giacomelli:2020jel} as well as new insights into string constructions \cite{Aharony:2016kai,Apruzzi:2020pmv,Heckman:2020svr,Giacomelli:2020gee} and compactification of higher dimensional \gls{SCFT}s \cite{Ohmori:2018ona}. Thanks in particular to \cite{Ohmori:2018ona,Apruzzi:2020pmv}, the question of whether all \fod rank-1 $\cN=2$ \gls{SCFT}s are realizable in string theory was settled with an affirmative answer. Nevertheless there are too many simplifications which take place in the rank-1 case (\emph{e.g.} all scale invariant $\cN=2$ rank-1 \gls{CB} geometries are flat) and it is therefore unwise to extrapolate this positive result to all ranks.

It is perhaps appropriate to comment on the hopes of achieving a bottom-up complete classification for the rank-2 case. The methods used in \cite{Argyres:2015ffa,Argyres:2015gha,Argyres:2016xua,Argyres:2016xmc} are certainly insufficient. The main roadblock currently being the absence of a rank-2 equivalent of the \emph{Kodaira classification} \cite{KodairaI,KodairaII} which provided a list of consistent geometries interpretable as \gls{CB}s of rank-1 \gls{SCFT}s\footnote{This statement requires some clarification. Strictly speaking the analog of the Kodaira classification at rank-2, \emph{i.e.} the study of the possible degeneration of a genus two curve over a one dimensional base, was already performed in \cite{Ogg:1965,Namikawa:1973}. Since the rank-2 \gls{SCFT} ``lives'' at the origin of the moduli space, which is complex co-dimension two, we need here to understand the possible degenerations of genus two curves on a two-dimensional base also satisfying special K\"ahler constraint (\emph{i.e.} exists SW differential).}. At rank-$r$ this task entails determining the possible ways a rank-$r$ abelian variety can be fibered over a $r$ complex dimensional base compatibly with the constraints from scale invariant special K\"ahler geometry \cite{Freed:1997dp}. An enormous simplification takes place for $r=2$ when all polarized abelian varieties can be written as Jacobian tori of (possibly singular) hyperelliptic curves and therefore \gls{CB} geometries can be expressed in a simple algebraic form:
\beq\label{hypergenus2}
y^2=f(x,u,v)
\eeq
where $u$ and $v$ are the (globally defined) coordinate of the \gls{CB} and $f(x,u,v)$ is a either a six or fifth order polynomial in $x$ with coefficient meromorphic in $(u,v)$.  An attempt to perform a study along these lines was made over a decade ago \cite{Argyres:2005pp,Argyres:2005wx} producing many new geometries but falling short of providing a complete picture. 

A second obstacle is to develop appropriate tools which can translate the geometric information into field theory data. For example in \cite{Argyres:2005pp,Argyres:2005wx} only a very limited amount of physical information was provided regarding the \gls{SCFT}s realizing the many new geometries making it hard to make clear predictions testable with other methods. The recent results in \cite{Argyres:2020wmq,Martone:2020nsy} as well as the techniques implemented here, de-facto overcome this problem altogether. We will report on the progress on the Kodaira classification at rank-2 in a separate publication \cite{Argyres:2020}.

The paper is organized as follows. In the next section we will present an overview of the current status of rank-2 theories. In section \ref{sec:background} we will provide some useful background on the geometry of the moduli space of vacua of $\cN=2$ theories highlighting in particular the notion of their stratification, the de-facto non-perturbative generalization of partial higgsing, which plays a central role in our analysis. In section \ref{sec:theories} we will delve instead into the detailed description of the rank-2 theories listed in table \ref{tab:r21}, \ref{tab:r22} and \ref{tab:r23}. We kept the discussion of each theory as self-contained as possible. The paper also has four appendices. In appendix \ref{sec:TandS} we will summarize the properties of the recently discovered $\cT$ and $\cS$ theories which will not be discussed individually. Appendix \ref{sec:N34} summarizes the properties of theories with enhanced supersymmetry which again will not be discussed in much detail individually. In appendix \ref{genfreef} we provide some rudimental information on the generalized free field constructions of the VOA of the rank-2 theories. Finally appendix \ref{glo} is dedicated to a glossary, a list of acronyms and symbols which appear throughout the manuscript\footnote{I would like to thank Jason Pollack, Patrick Rall and Andrea Rocchetto, for the stimulating weekly discussions on the quantum error correcting code interpretation of holography. It is these interactions that inspired the idea of including a glossary.}. \vspace{1em}

\noindent\emph{A note on notation}: throughout the paper, capital letter will denote the gauge group of the various theories (though we won't be extremely careful on the global structure) \emph{e.g.} $SU(3)+N_f \gls{F}$ will indicate a theory with gauge group $SU(3)$ and matter in the fundamental representation. Conversely lower case bold letter, will instead indicate the global flavor symmetry group (again we will be somewhat sloppy on questions regarding the global structure), \emph{e.g.} $\suf(3)$ will denote a theory with global flavor symmetry $\suf(3)$. We will overwhelmingly indicate Lagrangian theories by their gauge group plus matter representation data while non-Lagrangian theories, with the exception of the $\cT$ and $\cS$ theories \cite{Apruzzi:2020pmv,Giacomelli:2020jel,Giacomelli:2020gee}, by their flavor symmetry group as well as their level.

\section{Overview of the space of rank-2 theories}\label{sec:overview}

Before delving into a detailed description of the structure of the moduli space of vacua of $\cN=2$ theories, let us provide an update on the current state of the classification of rank-2 theories. 

First, nearly all the theories in table \ref{tab:r21}, \ref{tab:r22} and \ref{tab:r23} descend from \sd theories, thus nearly all have a string theory construction. The lone exception being the lagrangian theory $USp(4)+\frac12{\bf16}$ which is one of the numerous theories (the only one with rank-2!) found in \cite{Bhardwaj:2013qia} and which still lack a string theory realization, we'll discuss this theory more below. This is perhaps not surprising. As I discussed at length in the previous section, most of our current understanding \emph{descends} from higher dimensions. This knowledge can also provide extremely useful insights in developing the tools for a bottom-up classification directly in \emph{4d}. For example, overwhelmingly \fid \gls{SCFT}s have at least one IR-free gauge theory description associated to it, namely (one of) the gauge theory of which they are the infinite coupling limit. This allows to easily study the space of mass deformations of a given theory. If the realization of \fod \gls{SCFT}s from \fid is understood, this information can also be extended to \fod and the RG-flows trajectories mapped. This approach, which has numerous subtleties, will be further developed in \cite{Martone:2021drm} where a detailed discussion of the RG-flows among the rank-2 theories will be presented.

With this knowledge in mind, the structure of the currently known theories can be nicely organized by RG-flows, and in particular mass deformations. We will call a set of theories which are interconnected by mass deformations a \emph{series}. This allows us to see patterns naturally generalizing what observed in rank-1. Each series has a (not unique) top theory, from which the rest can be obtained, and a (again not unique) bottom theory, which cannot be mass deformed to any interacting rank-2 \gls{SCFT}. We name each series by the largest flavor symmetry factor of the top theories (\emph{e.g.} the top theory of the $\ef_8-\sof(20)$ series have flavor symmetries $[\ef_8]_{24}{\times}\suf(2)_{13}$ and $\sof(20)_{16}$). Overwhelmingly the bottom theories are either lagrangian or $\cN=3$ theories \cite{Martone:2021drm}. An analogous hierarchy exists at rank-1. An important difference between rank-1 and rank-2 is that for the latter we have little understanding of how the scaling dimensions of the \gls{CB} parameters change along mass deformations (it is not clear at all that a definite rule even exists). Conversely, the extremely constrained structure of the allowed geometries in rank-1 made this unambiguous and it was of tremendous help in performing a systematic analysis.

Of course for some of the lagrangian $\cN=2$ \gls{SCFT}s there exist special mass deformations which land you on an \gls{AD} and allow to continue mass deforming beyond a lagrangian theory. It is curious that the current list of known rank-2 \gls{AD} only descends from two of the ten $\cN=2$ lagrangian \gls{SCFT}s (namely $SU(3)+6\gls{F}$ and $USp(4)+4\gls{F}+1\gls{V}$) both belonging to the same, $\ef_8-\sof(20)$ series. The appearance of a \gls{AD} seems instead a completely generic feature of theories with a large enough flavor symmetry. It is therefore not unconceivable that more theories are awaiting to be discovered.

To further support this point, it is worth noticing the following. Special K\"ahler geometry, and in particular the fact that particular monodromies which correspond to special paths encircling the singualr locus on the \gls{CB}, have to be elliptic (\emph{i.e.} their eigenvalues have to \emph{all} lie on the unit circle) provides extremely strong constraints on the allowed scaling dimensions for \gls{CB} parameters. So much so, that there is a finite set of permitted values at any given rank and there is a closed formula to compute them \cite{Argyres:2018zay,Caorsi:2018zsq,Argyres:2018urp}. Of course there is no guarantee that all allowed scaling dimensions have to be realized but it is interesting that this happens at rank-1. At rank-2 all nine integer \gls{CB} scaling dimensions are also realized by theories in our tables. Among the 15 fractional ones instead, there are three ($\frac{12}{11},\frac{10}{9},\frac{12}7$) which appear nowhere and four of those are smaller than two. At general rank, the number of allowed fractional scaling dimensions dramatically exceeds the integer ones yet in our current understanding of \fod $\cN=2$ \gls{SCFT}s, theories with only integer scaling dimensions dramatically exceed the number of AD theories. It is then tempting to suggest that rather than a fundamental feature of \fod \gls{SCFT}s this is to be blamed on the techniques currently available to construct such theories. In particular AD theories are harder to construct from higher dimension and impossible for untwisted class-$\cS$ with regular punctures which, thanks to the Herculaneum effort of the tinkertoys program \cite{Chacaltana:2010ks,Chacaltana:2011ze,Chacaltana:2012zy,Chacaltana:2014jba,Chacaltana:2017boe,Chacaltana:2018zag}, has provided a large chunk of currently known \fod $\cN=2$ theories,.

There is one more reason which make it plausible that there might be new \gls{SCFT}s of the AD type ``hiding'' on the \gls{CB} of some of the lagrangian theories, those highlighted in yellow in the tables. And that is that a systematic search of all loci where non-mutually local particles could coincide is prohibitive at rank-2 and a search of this kind necessarily makes some initial assumptions introducing biases which limit the scope of the search itself. 

Let's now also comment on the list of isolated theories in table \ref{tab:r23}. Looking at the higher dimensional construction, it is plausible that two of these theories (entry 66 and 68) are indeed isolated. For the remaining theories, it is not unconceivable that there might be other theories which connect via mass deformation and which are not yet known. In particular the somewhat curious lagrangian theory $USp(4)+\frac12{\bf 16}$ has no flavor symmetry nor \gls{HB} and it is also the only lagrangian theory which does not appear as a bottom of an RG-flow. Coincidentally, it is also the only theory with no known realization in string theory. It is again tentative to speculate that there might exist a tower of new $\cN=2$ \gls{SCFT}s which can be then mass deformed to this lagrangian theory. 

There is of course the possibility of the existence of entire new series which don't flow to any of the known theories. But perhaps the most concrete indication of the incompleteness of our current understanding of $\cN=2$ theories at rank-2, is provided by the fact that many (depending on the degree of optimism from three to nine) seemingly consistent rank-2 $\cN=3$ theories \cite{Argyres:2019ngz} have yet to find a physical realization. Given the organizational structure described above, each new $\cN=3$ might be the bottom component of a series and therefore bring along many new theories.

\section{Background on $\cN=2$ moduli space}\label{sec:background}

Before moving to a detailed discussion of the each rank-2 theory individually, let us start from a quick review of the general structure of the moduli space of vacua of $\cN=2$ field theories;  see, \emph{e.g.}, \cite{Argyres:1996eh,Argyres:2016xmc}, focusing on the various branches and on their stratification as either Special K\"ahler or Hyperkhaler varieties.  

\subsection{Different branches of the moduli space}

The presence of supersymmetry allows in general for ground state configurations parametriezed by a set of continuous variables which can in turn be interpreted as coordinates of a space called the moduli space of vacua. The relation between the structure of the operator algebra and the moduli space of vacua of four dimensional SCFTs seems special. In particular the problem of when a given operator can acquire a vev can be conjecturally formulated in terms of a set of precise conditions on the operator algebra in the four dimensional case. These heavily rely on the complex structure that these moduli spaces inherit by virtue of supersymmetry, and, relatedly, on the shortening conditions satisfied by the BPS operators whose vevs parametrize the space. 

We define the moduli space of vacua of an $\cN=2$ \gls{SCFT}, which henceforth we will generically label as $\Tf$, as the space of vevs of Lorentz scalar chiral primaries of BPS operators. Depending on the $\suf(2)_R{\times}\uf(1)_r$ R-symmetry charges of these operators, their vevs are interpreted as complex coordinates of various branches of the moduli space. Specifically, labeling as $\bR$ and $r$ their $\suf(2)_R$ and $\uf(1)_r$ charges respectively, we have a \emph{Coulomb branch}, which will be indicated as $\cC$, if $\bR=0$, a \emph{Higgs branch}, indicated as $\cH$, if $r=0$, or a \emph{mixed branch}, indicated as $\cM$, if $\bold{R} \, r \neq0$. We also have a projection from the \gls{MB} into \gls{CB} (\gls{HB}) by simply setting to zero all the vevs of operators with $\bR\neq0$ ($r\neq0$).

Supersymmetry induces different structures on the various branches. The \gls{CB} is Special K\"ahler \cite{Freed:1997dp} and its complex dimension is called the \emph{rank} of the \gls{SCFT} while the \gls{HB} is a hyperk\"ahler cone \cite{Hitchin:1986ea} and therefore a symplectic form on its smooth locus (more below). A \gls{MB} intersects the \gls{CB} along an in general singular special K\"ahler subvariety.  It can likewise intersect a \gls{HB}, along an, again in general singular, hyperk\"ahler subvariety.  (Also, \gls{MB}s can intersect each other in both special K\"ahler and hyperk\"ahler directions.)

The operators whose vevs parametrize each branch, form a corresponding chiral ring which are therefore called \emph{Coulomb}, \emph{Higgs} and \emph{mixed chiral rings}. Even though we are not going to use it, it might be useful to connect with the nomenclature introduced in \cite{Cordova:2016emh}. The $\mathbf{\mathbf{B\Bb}}$ multiplets with general $\bR$ are the as Higgs branch operators and their OPEs contain the Higgs branch chiral ring.  The Coulomb branch chiral ring is generated by those scalar $\mathbf{L\Bb}$ chiral multiplet primaries $\vf_a^{[0,0]}$ with $\bR=0$. Finally the mixed branch chiral ring is generated by scalar primaries of the $\mathbf{L\Bb}$ chiral multiplet with $\bR\neq0$. Explicit  examples of chiral rings of theories containing mixed branches were worked out in \cite{Argyres:2016xmc}.

There is a special case of \gls{MB} which deserves a separate discussion and will play a important role in our construction. This is the case when the projection of the \gls{MB} into the \gls{CB} described above, gives back the entire \gls{CB}. In this case the \gls{CB} is a subvariety of the \gls{MB} and for this reason we will call such \gls{MB} an \emph{Enhanced Coulomb branch} and label its quaternionic dimensionality as \gls{h}.

Not all directions in the moduli space of vacua are equivalent. In fact there are special Higgsings which do not Higgs completely the theory but perhaps take the theory to another one of close enough complexity. If the theory has a weakly coupled gauge description, these Higgsings correspond to those vacuum expectation values of the microscopic fields for which the gauge group is minimally broken. Iterating this process, we see an interesting pattern of partial Higgsing, which can be characterised by the various subspaces of the \gls{CB} or \gls{HB}. These subspaces are naturally partially ordered by inclusion of their closures, and as such can be arranged into a \gls{Hasse}. A \gls{Hasse} simply represents a finite partially ordered set, in the form of a drawing of its transitive reduction.

The majority of $\cN=2$ \gls{SCFT}s \emph{do not} have such a gauge description and a fundamental understanding on how to reformulate the problem of Higgsing in the general case is still lacking. What helps is that the presence of charged massless states makes the metric on the moduli space of vacua singular on the loci where the low-energy theory is not described by free-fields. The moduli space of vacua is therefore in general a singular space and studying the singular locus, which we will label $\bbSf$, gives insights into interesting Higgsing directions. This singular structure induces on $\bbSf$ a \emph{stratification}. The \emph{type} of this stratification, depends on the specific branch we focus on. The study of the various branches of the moduli space of vacua as stratified spaces is extremely helpful to characterize theories and to extend the notion of minimal Higgsings to theories with no weakly coupled lagrangian description. This will play a central role in the analysis below.

Many of the properties just outlined apply more generally to any $\cN=2$ field theory. But one of the key properties which distinguish apart the conformal case, is that they carry a $\C^*$ action which arises as the combination of the spontaneously broken $\R^+$ dilatation transformation and the (also spontaneously broken) $\uf(1)_r$, on the \gls{CB}, or the Cartan of the $\suf(2)_R$, on the \gls{HB}. We will commonly refer to this action as the \gls{ScActg}. The geometry of the moduli space transforms homogenously under this transformation and we will refer to this property as \emph{scale invariance} of the moduli space.

\paragraph{Note on color coding} Throughout the manuscript we will adopt the color coding proposed in \cite{Grimminger:2020dmg} and use the color \blue{blue} for \gls{HB} related quantities $\cH\to\bH$ and \red{red} for \gls{CB} ones, $\cC\to\rC$. Since \gls{MB}s can be seen as either extension of the \gls{CB} or of the \gls{HB}, depending on the context, we will use both color $\cM\to$\Mtwo\footnote{The double coloring was introduced at an earlier stage of the draft. At the end, we made limited to no use of it, nevertheless we left it to show off our coding abilities.}.

\subsection{Coulomb branch}\label{sec:CBstr}

Let us start now with a summary of the structure of the \gls{CB} of $\cN=2$ \gls{SCFT}s and in particular a review of the notion of \emph{\gls{CB} stratification} \cite{Argyres:2020wmq} which is particularly effective for rank-2 theories. For more details on \gls{CB} geometry see for example \cite{Seiberg:1994pq,Seiberg:1994rs,Argyres:2015ffa,Argyres:2020wmq}, or more pedagogical reviews \cite{AlvarezGaume:1996mv,Lerche:1996xu,Argyres:1996nonp,Martone:2020hvy}.

The low-energy theory on a generic point of the \gls{CB} $\rC$ is almost as boring as it gets; a free $\cN=2$ supersymmetric $U(1)^r$ gauge theory with no massless charged states. $r$ is called the \emph{rank} of the theory and coincides with the complex dimensionality of $\rC$, ${\rm dim}_\C\rC=r$; we will indicate the global collective coordinates of $\rC$ as $\bu$. As we said above, $\rC$ is a singular space and its singular locus, which is a closed subset of $\rC$, will be denoted as $\rbbSf$.  The \gls{CB} singularities can be of two types. The metric singularities arise when charged states become massless while singularities of the complex structure denote non-trivial relations among \gls{CB} chiral ring generators \cite{Argyres:2017tmj,Bourget:2018ond,Argyres:2018wxu}. In what follows we will always make the simplifying assumption that $\rC$ is a non singular complex variety and thus we are only interested in the singularities of the first kind. This in turn implies that $\rC$ is topologically $\C^r$.

The smooth part of the \gls{CB} is $\rCrg := \rC \setminus \rbbSf$ and thus $\rCrg$ is an open subset of $\rC$. When the $\cN=2$ theory is superconformal the symmetry group includes an $\R^+ \times \uf(1)_R$ (we are neglecting the $\suf(2)_R$ factor as it acts trivially on $\rC$) which is in general spontaneously broken and combines to give a $\C^*$ action on the \gls{CB}.  The entire structure of $\rC$ has to be compatible with this $\C^*$ action and in particular $\rbbSf$ and $\rCrg$ have to be closed under it and the \gls{CB} coordinates $\bu$ have definite scaling dimension, which will be label by the letter $\D_{u_i}$ with the subscript indicating the specific coordinate we refer to. From here onwards we will only focus on the rank-2 case, in which case scale invariance is particularly constraining \cite{Argyres:2018zay} as there will be only two type of singular loci which we will call \gls{kstr} and \gls{ukstr}. These will be defined shortly. Henceforth we will use the following convention $\bu:=(u,v)$, where $u$ has the lowest scaling dimension of the two \gls{CB} coordinates.

%It is well-known that the \gls{CB} of $\cN=2$ theories in four dimensions is singular, where the loci of the \gls{CB} corresponding to the singular locus are interpreted physically as loci where charged states are becoming massless. The singular locus has a stratified structure which induced by the increasing rank of the lattice of charges of the BPS states becoming massless or, in other words, by the intersections of the various higher complex co-dimensions components and the loci where they are themselves singular.

General argument on the physical interpretation of the \gls{CB} singularities \cite{Seiberg:1994pq,Seiberg:1994rs,Argyres:2018zay}, show that $\rbbSf$ has to be a complex co-dimension one algebraic subvariety of $\rC$. Thus $\rbbSf$ can be defined as the zero locus of a single polynomial in $u$ and $v$. The fact that $\rbbSf$ has to be closed under the \gls{ScActg} implies that the polyonomial has to be homogenous, which in turn implies that it can always be brought to the form:
\beq\label{PolSf}
\rbbSf:=\Big\{(u,v)\in\rC\Big| P(u,v)=0\Big\},\qquad P(u,v)=u\cdot v\cdot\prod_{i\in I} (u^p+\l_iv^q)
\eeq
where $\l_i\in\C$, and $p$ and $q$ are integers fixed by the relative scaling dimension of $u$ and $v$ and by requiring that gcd$(p,q)=1$. Each factor in \eqref{PolSf}, $P_i(u,v)$, identifies a connected component of $\rbbSf$ (which is then the union of a bunch of disconnected pieces). The homogenous dimension of the $P_i(u,v)$ identifying each connected component plays an important role in what follows. We will label it as:
    \beq\label{Dsing}
    \D_i^{\rm sing}:=\D\Big(P_i(u,v)\Big).
    \eeq
We also adopt the following nomenclature:\vspace{.5cm}

\begin{tcolorbox}
\begin{itemize}
\item[1)] $u=0$: $u$ \gls{ukstr} or $\rbbSf_u$. $\D_u^{\rm sing}=\D_u$.

\item[2)] $v=0$: $v$ \gls{ukstr} or $\rbbSf_v$. $\D_v^{\rm sing}=\D_v$.

\item[3)] $u^p+v^q=0$: \gls{kstr} or $\rbbSf_{u^p+v^q}$. $\D_{u^p+v^q}^{\rm sing}=p\D_u=q \D_v$.

\end{itemize}
\end{tcolorbox}\vspace{0.5em}

\noindent The nomenclature knotted/unknotted is explained in detail in \cite{Argyres:2018zay}

It can be proven that the following facts apply \cite{Argyres:2018urp,Argyres:2020wmq}:\vspace{.5cm}

\begin{tcolorbox}
\begin{fact}\label{knot}
A four dimensional rank 2 $\cN=2$ \gls{SCFT} which cannot be decomposed into the product of two rank-1 theories has at least one \gls{kstr}. 
\end{fact}
\end{tcolorbox}\vspace{0.5em}

\begin{tcolorbox}
\begin{fact}\label{scaling}
A two complex dimensional \gls{CB} $\rC$, parametrized by \gls{CB} coordinates $(u,v)$, only admits a stratum corresponding to the $\C^*$ orbit $\rSf_v$ ($\rSf_u$) if $\D_u$ ($\D_v$) is a scaling dimension allowed at rank 1. The corresponding \gls{CB} parameter is called \gls{higgs}.
\end{fact}
\end{tcolorbox}\vspace{0.5em}

\noindent Both statements can be relatively straightforwardly generalized to higher ranks.

Now comes one of the key point. Each connected stratum in \eqref{PolSf} supports an either IR-free or superconformal rank-1 low energy theory which describes precisely the charged states which are becoming massless there\footnote{Technically this statement is incorrect as the origin of each stratum is the origin of the moduli space where our rank-2 \gls{SCFT} is supported. Thus a rank-1 theory is supported on a dense open subset of the stratum which is called the \emph{component} associated to the stratum \cite{Argyres:2020wmq}. In order to keep things as intuitive as possible, we will be sloppy and not make this distinction here.}. Understanding these rank-1 theories is a central piece of our analysis. So let's formalize this point a bit more.

Again, call the rank-$2$ theory at the superconformal vacuum $\Tf$ and call $\rTf_\bu$ the low-energy effective description of $\Tf$  at the generic point of the \gls{CB} $\bu$. For example we have:
    \beq
    \rTf_\bu\equiv \text{free $\cN{=}2$\ \  $\uf(1)^2$},\qquad \bu\in \rCrg.
    \eeq
If instead $\bu\in \rbbSf$, extra charged states become massless and the effective theory in the IR is no longer $\uf(1)^2$ but, rather, an either IR-free or superconfrmal rank-1 theory.  $\rTf_\bu$ is identified precisely with this theory describing the low-energy degrees of freedom which plays a special role in what follows. We will call:
    \beq\label{pairT}
    \rTf_i\equiv\left\{
    \begin{array}{l}
    \rTf_u, \ {\rm for}\ (u,v)\in\rbbSf_u\\
    \rTf_v, \ {\rm for}\ (u,v)\in\rbbSf_v\\
    \rTf_{u^p+v^q}, \ {\rm for}\ (u,v)\in\rbbSf_{u^p+v^q}\\
    \end{array}
    \right.
    \eeq
    and the quantities indexed by $i\in I$, $(c_i,k_i,h_i)$, label the central charges of these rank-1 theories \gls{rTf} and will be used to compute the central charges of the \gls{SCFT} at the superconformal vacuum $\Tf$ (see below). We also use $u_i$ to label the coordinate parametrizing the one complex dimensional \gls{CB} of \gls{rTf} and define:
    \beq\label{Di}
    \D_i:=\D(u_i)
    \eeq
    which defines the last quantity entering the central charge formulae which we will shortly define. If this discussion is a bit too abstract, many many explicit examples can be found in \cite{Argyres:2020wmq} or below in section \ref{sec:theories}.

Before introducing the \gls{ccf} \cite{Martone:2020nsy} let's discuss another very constraining property of \gls{CB} geometries which was proven in the same paper and which will be extremely useful in our analysis below. The \gls{UVIR} states that simple flavor factors of \gls{SCFT}s of arbitrary rank, and thus in particular of rank-2 \gls{SCFT}s, act on the massless BPS spectrum which arise on singular complex co-dimension one strata of the \gls{CB}. In other words any simple flavor factor $\ff$ of an \gls{SCFT} $\Tf$, is realized (with possible rank-preserving enhancement) as the flavor symmetries of (at least one) rank-1 theory \gls{rTf} defined in \eqref{pairT}. This observation allows to then study the structure of the \gls{HB} from the \gls{CB} perspective and gain new insights on allowed Higgsing.

 The rank-1 theories \gls{rTf} carry more information than just the flavor symmetry of the SCFT. In fact, generalizing \cite{Shapere:2008zf}, it is possible to derive explicit formulae expressing the central charges of an arbitrary $\cN=2$ \gls{SCFT} in terms of corresponding quantities of the rank-1 theories \gls{rTf}'s \cite{Martone:2020nsy}:
    \begin{subequations}
    \begin{align}
    \label{actotaint}
    24 a &= 5r + h + 6 \left(\sum_{\ell=1}^r\D_{\bu_\ell}-2\right) +\sum_{i\in I}\D^{\rm sing}_i b_i ,
    \\\label{actotbint}
    12 c &= 2r + h + \sum_{i\in I}\D^{\rm sing}_{i} b_i,\\\label{actotcint}
    k_\ff&=\sum_{i\in I_{\ff}}\frac{\D_i^{\rm sing}}{d_i\D_i} \left(k^i-T({\bf2}\bh_i)\right)+T({\bf2}\bh).
    \end{align}
    \end{subequations}
    Here, $r$ is the rank of the SCFT, $h$ is the quaternionic dimension of the theory's \gls{ECB} and $\D_{\bu_\ell}$ is the scaling dimension of the theory's $\ell$-th component of the \gls{CB} coordinate vector $\bu$. The sums indexed by $i$ are performed over all the singular strata $\rbbSf_i$ and the \gls{bi} are defined to be:
 \beq
b_i :=\frac{12 c_i-2-h_i}{\D_i}
 \eeq   
where $\D^{\rm sing}_i$ and $\D_i$ are defined in \eqref{Dsing} and \eqref{Di},  all the remaining quantities  indexed by $i$ (except $d_i$) refer to corresponding quantities of \gls{rTf} defined in \eqref{pairT}. Finally $d_i$ is the embedding index of the flavor symmetry.  We call these formulae \emph{central charge formulae} and their great service is that they allow to re-write the \gls{SCFT} data of a rank-$r$ \gls{SCFT} in terms of easily accessible geometric data (e.g. the scaling dimension of their \gls{CB} parameter or dimension of its \gls{ECB}) and the \gls{SCFT} data of rank-1 theories which have been fully classified.
    
A warm-up example, which will also enable us to derive a rule which will turn very handy in what follows, is to compute the level $k_\ff$ for a simple flavor factor realized on a stratum identified by a polynomial of dimension $\tilde{\D}$ by any of the entry in table \ref{Theories} with \gls{h}=0. Plugging the appropriate values in \eqref{actotcint} we derive the following:\vspace{.3cm}
    
\begin{tcolorbox}
\paragraph{Doubling rule}\label{DR} Any entry in table \ref{Theories} with \gls{h}=0 realizes a flavor symmetry factor with
\beq
k_\ff=2\tilde{\D}
\eeq
where $\tilde{\D}$ is the homogenous dimension of the polynomial identifying the singular stratum ($\tilde{\D}=\D_u/\D_v$ is the theory is supported on an \gls{ukstr} or $\tilde{\D}=p\D_u=q\D_v$ if is supported on a \gls{kstr}). The converse is nearly always true as well.
\end{tcolorbox}\vspace{0.5em}

The stratification of the \gls{CB} singular locus is even richer than what is discussed above; in fact, the strata themselves have to be scale invariant special K\"ahler varieties. But we won't review this here and refer the interested reader to the original paper for more details \cite{Argyres:2020wmq}. 

Let's conclude with a remark which will be used in a few cases below. As discussed in \cite{Martone:2020nsy}, if a form of the \gls{SW} curve is known where the curve is written as a fibration of an hyperelliptic curve over $\rC$, that is in the form
\beq
y^2=f(u,v,x)
\eeq
where $f(u,v,x)$ is at most of degree six in $x$ with meromorphic coefficient in $(u,v)$, there is an easy way to gain many information about the singular locus of the \gls{CB} by taking the $x$ discriminant of $f(u,v,x)$. This is called the quantum discriminant of the geometry \cite{Martone:2020nsy} which allows in most cases to characterize the entire \gls{Hasse}, not just $\bar{\rSf}$. The relation between the quantum discriminant and the \gls{CB} stratification will be further investigated in \cite{Argyres:2020}.

\subsection{Higgs branch stratification}

A wonderful recent discussion of the structure of \gls{HB}s of \gls{SCFT}s with eight supercharges (thus in particular $\cN=2$ in \emph{4d}) was recently presented in \cite{Bourget:2019aer} where many lagrangian examples are also explicitly discussed. It is hard to do a better job and in fact we most likely won't. But to keep the paper as self-contained as possible we nevertheless present a brief discussion of the \gls{HB} stratitification.

Supeconformal invariance implies that the \gls{HB}, is a hyper-K\"ahler cone \cite{Hitchin:1986ea} which in particular implies that it is a symplectic singularity \cite{beauville1999symplectic}. Like the \gls{CB}, the \gls{HB} is also a singular space. In analogy with what we did in the previous section, we call $\bH^{\rm reg}$ the set of points where the symplectic structure is non-degenerate. This symplectic form naturally induces a symplectic structure on the singular points which we will label as $\bbbSf_{\cH}$ \cite{brieskorn1970singular, slodowy1980simple,beauville2000symplectic}. A powerful and general result is that symplectic singularities admit a finite stratification \cite{kaledin2006symplectic}: 
\beq
\bH \equiv \bigsqcup_{i=0}^\nf \bH_i
\eeq
where $\bigsqcup$ indicates the disjoint union, the $\bH_i$'s are irreducible and connected and are called \emph{symplectic leaves}. The normalization of their closure are symplectic singularities. Importantly symplectic leaves are partially ordered by the operation of inclusion in the closure of other symplectic leaves and they can be represented by a \gls{Hasse}. A leaf $\bH_a$ \emph{covers} $\bH_b$ and notated $a\gtrdot b$ iff $a>b$ and there is no $\bH_c$ such that $a>c>b$, we will also say that $\bH_a$ and $\bH_b$ are neighboring leaves. In addition each pair of symplectic leaves, $(\bH_a,\bH_b)$, defines a subvariety, $\cS_{(a,b)}$, which is transverse in the sense of \cite{slodowy1980} and whose dimension is equal to the codimension of the smaller leaf into the closure of the larger one. We will call $\cS_{(a,b)}$ the \textit{transverse slice} of $\bH_a$ into $\bH_b$ if $a<b$. A transverse slice between two neighboring leaves is called an \emph{\gls{elesl}} and the edges of the \gls{Hasse} are precisely labeled by those. Standard examples of \gls{Hasse} representations of symplectic singularities stratification are given by the Kraft-Procesi transition between nilpotent orbits \cite{kraft1980minimal,kraft1982geometry}. A full list of possible \gls{elesl}s is still unknown and it is unclear whether there is an answer for this question. Initially there was a hope that these could be restricted to minimal nilpotent orbits of classical and exceptional Lie algebras (see table \ref{NilOrbits}) and Du Val or Kleinian singularities \cite{Bourget:2019aer}. But  as our understanding of \gls{HB} of \gls{SCFT}s with eight of more supercharges improves, new \gls{elesl}s are discovered \cite{AffineAJ}. In this work we will also conjecture the existence of new mysterious ones.

The structure of symplectic singularities is illuminated by its physical interpretation.  The smallest and largest symplectic leaves are naturally identified with the origin of the moduli space, $\bH_0=\{0\}$, and the non-singular part of the \gls{HB}, $\bH_\nf=\bH^{\rm reg} $ respectively. The other leaves $\bH_i$, since singularities of moduli spaces corresponds to loci where extra interacting degrees of freedom become massless, precisely identify the subvarieties of $\bH$ where a \gls{SCFT} $\bTf_i$ is supported. For a lagrangian SCFT, the $\bH_i$ are spanned by pattern of partial Higgsing where subgroups of lower and lower rank, as $i$ increases, are left unbroken. In the non-lagrangian set up, moving from one leaf to the other involves turning on vevs of some \gls{HB} operator. Nevertheless the physical intuition remains the same and therefore we will henceforth refer to the action of moving from one symplectic leaf to neighboring ones as \emph{partial Higgsing}. If the higgsing associated to a given leaf is of \gls{gHW} type, then there is an extremely useful relation which we will be used extensively below to reconstruct the \gls{HB} structure of \gls{SCFT}s \cite{Giacomelli:2020jel,CCLMW2020}:\vspace{.2cm}
    
\begin{tcolorbox}
\paragraph{GHW central charge formula}\label{gHWbox} For a leaf associated to a higgsing of \gls{gHW} type, the difference between the $c$ central charge of an \gls{SCFT} $\bTf_0$ supported at the origin of the (closure of the) leaf and that of  \gls{SCFT}, $\bTf_{\rm gHW}$ supported on a generic point of the leaf
\beq\label{gHWeq}
12 c_{\bTf_0}-12 c_{\bTf_{\rm gHW}}=2\left(\frac32k_\ff-1\right)+\d {\rm dim}_\H \bH-1
\eeq
where $\d {\rm dim}_\H \bH$ is the variation in \gls{HB} dimension induced by the Higgsing or the quaternionic dimension of the leaf.
\end{tcolorbox}

\begin{table}
\begin{adjustbox}{center,max width=.9\textwidth}
\renewcommand{\arraystretch}{1.3}
$\begin{array}{|c|c|c|c|c|}
\hline
\hline
\multicolumn{5}{|c|}{\text{\bf Minimal nilpotent orbit of Lie algebras}}\\
\hline
\hline
\ff & {\rm dim}_\H &\ff^\natural & \pi_{\bRf}&I_{\ff^\natural\hookrightarrow\ff}\\
\hline
\ \ \af_N\ \  & N & \af_{N-2}\oplus \C &\ \  ({\bf N-2})_+\oplus (\bar{{\bf N-2}})_-\ \ &\ (1,1) \ \\
\bff_N & 2N-2 & \af_1\oplus \bff_{N-2} & ({\bf2},{\bf2N-3})&\ (1,1) \ \\
\cf_N & N & \cf_{N-1} & {\bf 2(N-1)}&\ (1,1) \ \\
\df_N &\ \  2N-3\ \  &\ \ \af_1\oplus \df_{N-2}\ \ & ({\bf2},{\bf 2N-4})&\ (1,1) \ \\
\ef_6 & 11 & \af_5 & {\bf 20}&1\\
\ef_7 & 17 & \df_6 & {\bf 32}&1\\
\ef_8 & 29 & \ef_7 & {\bf 56}&1\\
\gf_2 & 3 & \af_1 & {\bf 4}&3\\
\ff_4 & 8 & \cf_3 & {\bf 14}' &1\\
\hline
\hline
\end{array}$
\caption{\label{NilOrbits}For the convenience of the reader, we summarize the properties of minimal nilpotent orbit for all classical and exceptional Lie algebras which appear copiously as \gls{elesl}s on the \gls{HB}. This table is almost verbatim taken from \cite{Beem:2019tfp}.}
\end{adjustbox}
\end{table}
\vspace{-.5cm}

\section{Detailed description of rank-2 theories}\label{sec:theories}

In this section we will delve into the details of the results reported in table \ref{tab:r21}, \ref{tab:r22}, \ref{tab:r23}, \ref{Higgs:one}, \ref{Higgs:two} and \ref{Higgs:three}. The first three tables collect the CFT data of the $\cN=2$ rank-2 \gls{SCFT}s currently known to the author along with some information about the \gls{CB} stratification, while the latter three specifically list information about the theories' \gls{HB}s and information about how the flavor symmetry is realized along the various higgsings\footnote{This is also useful to guess the basics of the generalized free field construction of their VOAs \cite{Beem:2019tfp,Beem:2019snk} which will be sketched in appendix \ref{genfreef}.}. We organize the known rank-2 theories in ten series of theories which are mutually connected by mass deformations and five isolated ones. A more detailed study of RG-flows of rank-2 theories will be presented elsewhere \cite{Martone:2021drm}. We will do our best in referencing the literature and clarify the results which have appeared elsewhere and apologize in advance for those who don't get the credit they certainly deserve. For example, it is worth mentioning that the class-$\cS$ construction \cite{Gaiotto:2009we,Gaiotto:2009hg} of $\cN=2$ \gls{SCFT}s gives often remarkable information about the \gls{HB} structure of the theory, thus parts of the \gls{HB} \gls{Hasse}s of the theories below could be derived this way and it was certainly known before. We won't follow this path and use instead primarily the techniques in \cite{Argyres:2020wmq,Martone:2020nsy}.  At the cost of being slightly repetitive, we will write the description of each theory in such a way that it can be read somewhat independently from the rest.

The upshot of our analysis is that most rank-2 theories have been completely understood and reveal general patterns; one, or multiple, \gls{kstr} supporting either a $[I_1,\varnothing]$ or a $\blue{\cS^{(1)}_{\varnothing,2}}$, and \gls{ukstr} supporting one of the rank-1 theories in table \ref{Theories} (see below). But some entries don't quite fit these patterns. Specifically, \hyperref[sec:t65]{$\spf(4)_{14}{\times}\suf(2)_8$} is the only theory with an IR free theory with a non-trivial \gls{HB}, and a non-trivial \gls{ECB}, supported over a \gls{kstr}. The \gls{ECB} structure of this theory is also particularly complicated and involved. Also theories in the $\suf(6)$ and $\suf(5)$ series have IR-free theories supported on \gls{CB} strata with semi-simple flavor symmetries and a \gls{HB} stratification presenting mysterious elementary transitions which are labeled with a question mark and a \blue{blue} dashed line in the corresponding \gls{Hasse}s. 

\begin{landscape}
\begin{table}[ht]
\begin{adjustbox}{center,max width=.64\textwidth}
$\def\arraystretch{1.0}
\begin{array}{r|c|ccc:cc|ccc:cc|}
\multicolumn{12}{c}{\Large\textsc{Rank-2:\ Higgs\ branch\ data\ I}}\\
\hline
\hline
\#&\ff&
\bSf_u&([\ff^\natural]_{k^\natural},I_{\ff^\natural\hookrightarrow\ff_{\rm UV}})&\pi_{\bRf}&\bTf_u
&([\ff_{\rm IR}]_{k_{\rm IR}},I_{\ff^\natural\hookrightarrow\ff_{\rm IR}}) 
&\blue{\Sf}_v&( [\ff^\natural]_{k^\natural},I_{\ff^\natural\hookrightarrow\ff_{\rm UV}})&\pi_{\bRf}&\bTf_v
&([\ff_{\rm IR}]_{k_{\rm IR}},I_{\ff^\natural\hookrightarrow\ff_{\rm IR}}) 
\\
\hline\hline

\multicolumn{12}{c}{\ef_8-\sof(20)\ \text{series}}\\
\hline

%Beginning of the E8-SO(20) series

1.&[\ef_8]_{24}{\times}\suf(2)_{13}
&\varnothing&\text{-}&\text{-}&\text{-}&\text{-}
&\ef_8&([\ef_7]_{24}{\times}\suf(2)_{13},(1,1))&({\bf56},{\bf1})&\cT^{(1)}_{E_8,1}{\times}\H&([\ef_7]_{12}{\times}\suf(2)_{12}{\times}\suf(2)_1,(1,1,1))
\\
2.&\sof(20)_{16}
&\varnothing&\text{-}&\text{-}&\text{-}&\text{-}
&\df_{10}&(\suf(2)_{16}{\times}\sof(16)_{16},(1,1))&({\bf2},{\bf16})&\cT^{(1)}_{E_8,1}&(\sof(16)_{12},1)
\\
3.&[\ef_8]_{20}
&\varnothing&\text{-}&\text{-}&\text{-}&\text{-}
&\ef_8&([\ef_7]_{20},1)&{\bf56}&\cT^{(1)}_{E_7,1}&([\ef_7]_8,1)
\\
4.&[\ef_7]_{16}{\times}\suf(2)_9
&\varnothing&\text{-}&\text{-}&\text{-}&\text{-}
&\ef_7&(\sof(12)_{16}{\times}\suf(2)_9,(1,1))&({\bf 32},{\bf1})&\cT^{(1)}_{E_7,1}{\times}\H&(\sof(12)_8{\times}\suf(2)_8{\times}\suf(2)_1,(1,1,1))
\\
5.&
 \suf(2)_8{\times}\sof(16)_{12}
&\af_1 & (\sof(16)_{12},1) & - & \cT^{(1)}_{E_8,1}&(\sof(16)_{12},1)&
\df_8& (\suf(2)_8{\times}\suf(2)_{12}{\times}\sof(12)_{12},(1,1,1)) &  ({\bf 2},{\bf12}) & \cT^{(1)}_{E_7,1} & (\suf(2)_8{\times}\sof(12)_8,(1,1)) 
\\
6.&\
\suf(10)_{10}
&\varnothing&\text{-}&\text{-}&\text{-}&\text{-}
&\af_9&(\suf(8)_{10}{\times}\uf(1),1)&{\bf8}\oplus\bar{{\bf8}}&\cT^{(1)}_{E_7,1}&(\suf(8)_8,1)
\\
7.&
[\ef_6]_{12}{\times}\suf(2)_7
&\varnothing&\text{-}&\text{-}&\text{-}&\text{-}
&\ef_6&(\suf(6)_{12}{\times}\suf(2)_7,(1,1))&({\bf20},{\bf1})&\cT^{(1)}_{E_6,1}{\times}\H&(\suf(6)_6{\times}\suf(2)_6{\times}\suf(2)_1,(1,1,1))
\\
8.&
\sof(14)_{10}{\times}\uf(1)
&\varnothing&\text{-}&\text{-}&\text{-}&\text{-}
&\df_7&(\suf(2)_{10}{\times}\sof(10)_{10},(1,1))&({\bf2},{\bf10})&\cT^{(1)}_{E_6,1}&(\sof(10)_6{\times}\uf(1),1)
\\
9.&
\suf(2)_6{\times}\suf(8)_8&
\af_1&(\suf(8)_8,1)&-&\cT^{(1)}_{E_7,1}&(\suf(8)_8,1)
&\af_7&(\suf(6)_8{\times}\suf(2)_6{\times}\uf(1),(1,1))&{\bf6}\oplus\bar{{\bf6}}&\cT^{(1)}_{E_6,1}&(\suf(6)_6{\times}\suf(2)_6,(1,1))
\\
\rcy10.&
\sof(12)_8
&\varnothing&\text{-}&\text{-}&\text{-}&\text{-}
&\df_6&(\suf(2)_8{\times}\sof(8)_8,(1,1))&({\bf2},{\bf8})&\cT^{(1)}_{D_4,1}&(\sof(8)_4,1)
\\
\rcy11.&
\sof(8)_8{\times}\suf(2)_5
&\varnothing&\text{-}&\text{-}&\text{-}&\text{-}
&\df_4&(\suf(2)^3_8{\times}\suf(2)_5,(1,1))&({\bf 2}^3,{\bf1})&\cT^{(1)}_{D_4,1}\times \H&(\suf(2)^3_4{\times}\suf(2)_1,(1,1))
\\
\rcy12.&
\uf(6)_6
&\varnothing&\text{-}&\text{-}&\text{-}&\text{-}
&\af_5&(\suf(4)_6{\times}\uf(1),1)&{\bf4}\oplus\bar{{\bf4}}&\cT^{(1)}_{D_4,1}&(\suf(4)_5{\times}\uf(1),1)
\\
\rcy13.&
\suf(2)^5_4&
\af_1&(\suf(2)^4_5,1)&-&\cT^{(1)}_{D_4,1}&(\suf(2)^4_4,1)
&\varnothing&\text{-}&\text{-}&\text{-}&\text{-}
\\
14.&
\suf(3)_6{\times}\suf(2)_4
&\varnothing&\text{-}&\text{-}&\text{-}&\text{-}
&\af_2&(\uf(1){\times}\suf(2)_4,(-,1))&{\bf1}_+\oplus{\bf1}_-&\cT^{(1)}_{A_2,1}{\times}\H&(\suf(2)_3{\times}\suf(2)_1,(1,1))
\\
15.&
\suf(5)_5
&\varnothing&\text{-}&\text{-}&\text{-}&\text{-}
&\af_4&(\suf(3)_5{\times}\uf(1),1)&{\bf3}\oplus\bar{{\bf3}}&\cT^{(1)}_{A_2,1}&(\suf(3)_3,1)
\\
16.&
\suf(2)_{\frac{16}3}{\times}\suf(2)_{\frac{11}3}
&\varnothing&\text{-}&\text{-}&\text{-}&\text{-}
&\af_1&(\suf(2)_{\frac{11}3},1)&{\bf1}&\cT^{(1)}_{A_1,1}{\times}\H&(\suf(2)_{\frac83},\suf(2)_1,(1,1))
\\
17.&
\suf(2)_{\frac{10}3}{\times}\uf(1)
&\varnothing&\text{-}&\text{-}&\text{-}&\text{-}
&\af_1&(\varnothing,\text{-})&-&\cT^{(1)}_{A_1,1}&\uf(1)
\\
18.&
\suf(2)_{\frac{17}5}
&\varnothing&\text{-}&\text{-}&\text{-}&\text{-}
&\varnothing&\text{-}&\text{-}&\text{-}&\text{-}
\\
19.&
\uf(1)
&\varnothing&\text{-}&\text{-}&\text{-}&\text{-}
&\varnothing&\text{-}&\text{-}&\text{-}&\text{-}
\\
20.&
\suf(2)_{\frac{16}5}
&\varnothing&\text{-}&\text{-}&\text{-}&\text{-}
&\af_1&(\varnothing,\text{-})&-&\cT^{(1)}_{\varnothing,1}&\varnothing
\\
21.&
\varnothing
&\varnothing&\text{-}&\text{-}&\text{-}&\text{-}
&\varnothing&\text{-}&\text{-}&\text{-}&\text{-}
\\
\cdashline{1-12}
%End of the E8-SO(20) series -- beginning of the Sp(12)-Sp(8)-F4 series

\multicolumn{12}{c}{\spf(12)-\spf(8)-\ff_4\ \text{series}}\\
\hline

22.&
\spf(12)_8&
\cf_6&(\spf(10)_8,1)&{\bf10}&\cS^{(1)}_{E_6,2}&(\spf(10)_7,1)
&\varnothing&\text{-}&\text{-}&\text{-}&\text{-}
\\
23.&
 \spf(4)_7{\times}\spf(8)_8
&\cf_4 & (\spf(4)_7{\times}\spf(6)_8,(1,1)) & {\bf6} & \cS^{(1)}_{E_6,2}&(\spf(4)_7{\times}\spf(6)_7,(1,1))&
\cf_2& (\suf(2)_7{\times}\spf(8)_8,(1,1)) &  {\bf 2} & \hyperref[sec:t9]{\suf(2)_6{\times}\suf(8)_8} & (\suf(2)_6{\times}\spf(8)_8,(1,1)) 
\\
24.&
[\ff_4]_{12}{\times}\suf(2)_7^2
&\ff_4&(\spf(6){\times}\suf(2)^2,(1,1))&({\bf14}',{\bf1})&\cS^{(1)}_{E_6,2}&(\spf(6)_7{\times}\suf(2)_7^2,(1,1))
&\varnothing&\text{-}&\text{-}&\text{-}&\text{-}
\\
25.&
\suf(2)_8{\times}\spf(8)_6&
\cf_4&(\spf(6)_6{\times}\suf(2)_8,(1,1))&{\bf6}&\cS^{(1)}_{D_4,2}&(\spf(6)_5{\times}\suf(2)_8,(1,1))&
\af_1&(\spf(8)_6,1)&{\bf1}&\cT^{(1)}_{E_6,1}&(\spf(8)_6,1)
\\
26.&
\suf(2)_5{\times}\spf(6)_6{\times}\uf(1)&
\cf_3&(\spf(4)_6{\times}\suf(2)_5{\times}\uf(1),(1,1))&{\bf4}&\cS^{(1)}_{D_4,2}&(\spf(4)_5{\times}\suf(2)_5{\times}\uf(1),(1,1))&
\af_1&(\spf(6)_6,1)&-&\hyperref[sec:SU3Nf6]{SU(3)+6\gls{F}}&(\spf(6)_6,1)
\\
27.&
\sof(7)_8{\times}\suf(2)^2_5
&\bff_3&(\suf(2){\times}\suf(2){\times}\suf(2)^2,(1,2,1))&({\bf 2},{\bf3},{\bf1})&\cS^{(1)}_{D_4,2}&(\suf(2)_5^3{\times}\suf(2)_8,(1,1))
&\varnothing&\text{-}&\text{-}&\text{-}&\text{-}
\\
28.&
[\ff_4]_{10}{\times}\uf(1)
&\varnothing&\text{-}&\text{-}&\text{-}&\text{-}
&\ff_4&(\spf(6)_{10}{\times}\uf(1),1)&{\bf14}'&\cS^{(1)}_{D_4,2}&(\spf(6)_5{\times}\uf(1),1)
\\
29.&
\spf(6)_5{\times}\uf(1)&
\cf_3&(\spf(4)_5{\times}\uf(1),1)&{\bf4}&\cS^{(1)}_{A_2,2}&(\spf(4)_4,1)
&\varnothing&\text{-}&\text{-}&\text{-}&\text{-}
\\
30.&
\suf(3)_6{\times}\suf(2)^2_4
&\af_2&(\uf(1){\times}\suf(2)^2,(-,1))&{\bf1}_+\oplus{\bf1}_-&\cS^{(1)}_{A_2,2}&(\suf(2)_4^2{\times}\uf(1),(1,\text{-}))
&\varnothing&\text{-}&\text{-}&\text{-}&\text{-}
\\
\rcy31.&
\spf(4)_4&
\cf_2&(\suf(2)_4,1)&{\bf2}&\blue{\cS^{(1)}_{\varnothing,2}}&(\suf(2)_3,1)
&\varnothing&\text{-}&\text{-}&\text{-}&\text{-}
\\
\rcb32.&
 \suf(2)_6&
\af_1&(\suf(2)_3,1)&-&\blue{\cS^{(1)}_{\varnothing,2}}&(\suf(2)_3,1)
&\varnothing&\text{-}&\text{-}&\text{-}&\text{-}
\\

%% End of the Sp(12)-Sp(8)-F4 series

\hline\hline
\end{array}$
\caption{\label{Higgs:one}{\footnotesize This table summarizes the \gls{HB} data for the theories in table \ref{tab:r21}. The second column lists the flavor symmetry of the \gls{SCFT}, column three to seven, lists the information of the higgsing of the flavor symmetry realized on the \gls{CB} \gls{ukstr} $u=0$, while the last four columns present the same information for $v=0$. For more details and an explanation of the connection with generalized free fields realization of the theory's VOA, see appendix \ref{genfreef}. $\uf(1)$ factors in this table will be mostly omitted.}}
\end{adjustbox}
\end{table}
\end{landscape}

\begin{landscape}
\begin{table}[ht]
\begin{adjustbox}{center,max width=.65\textwidth}
$\def\arraystretch{1.0}
\begin{array}{r|c|ccc:cc|ccc:cc|}
\multicolumn{12}{c}{\Large\textsc{Rank-2:\ Higgs\ branch\ data\ II}}\\
\hline
\hline
\#&\ff&
\bSf_u&([\ff^\natural]_{k^\natural},I_{\ff^\natural\hookrightarrow\ff_{\rm UV}})&\pi_{\bRf}&\bTf_u
&([\ff_{\rm IR}]_{k_{\rm IR}},I_{\ff^\natural\hookrightarrow\ff_{\rm IR}}) 
&\blue{\Sf}_v&( [\ff^\natural]_{k^\natural},I_{\ff^\natural\hookrightarrow\ff_{\rm UV}})&\pi_{\bRf}&\bTf_v
&([\ff_{\rm IR}]_{k_{\rm IR}},I_{\ff^\natural\hookrightarrow\ff_{\rm IR}}) 
\\
\hline\hline
\multicolumn{12}{c}{\suf(6)\ \text{series}}\\
\hline
33.&\red{\suf(6)_{16}{\times}\suf(2)_9}
&\varnothing&\text{-}&\text{-}&\text{-}&\text{-}
&\af_1&(\suf(6)_{16},1)&\text{-}&\hyperref[sec:t22]{\spf(12)_8}&(\suf(6)_8,2)
\\
34.&\red{\suf(4)_{12}{\times}\suf(2)_7{\times}\uf(1)}
&\red{?}&\red{?}&\red{?}&\red{?}&\red{?}
&\af_1&(\suf(4)_{12}{\times}\uf(1),1)&\text{-}&\hyperref[sec:t25]{\suf(2)_8{\times}\spf(8)_6}&(\suf(2)_8{\times}\suf(4)_6,(1,2))
\\
35.&\red{\suf(3)_{10}{\times}\suf(3)_{10}{\times}\uf(1)}
&\varnothing&\text{-}&\text{-}&\text{-}&\text{-}
&\bar{h}_{2,3}&\red{\suf(3)_{10}{\times}\suf(2)_{10}}&\red{{\bf2}\oplus\bar{\bf 2}} &\cS^{(1)}_{D_4,2}&(\suf(3)_5{\times}\suf(2)_8,(2,1))
\\
36.&\red{\suf(3)_{10}{\times}\suf(2)_6{\times}\uf(1)}
&\varnothing&\text{-}&\text{-}&\text{-}&\text{-}
&\af_1&(\suf(3)_{10}{\times}\uf(1),1)&\text{-}&\hyperref[sec:tTE62]{\spf(6)_5{\times}\uf(1)}&(\suf(3)_5{\times}\uf(1),2)
\\
37.&\red{\suf(2)_{8}{\times}\suf(2)_8{\times}\uf(1)^2}
&A_3&(\suf(2)_8{\times}\suf(2)_8,(1,1))&\text{-}&\cT^{(1)}_{D_4,1}&(\suf(2)_4{\times}\suf(2)_4,(2,2))
&\bar{h}_{2,2}&\red{\suf(2)_8{\times}\uf(1)^3}&\text{-}&\cS^{(1)}_{A_2,2}&(\suf(2)_4{\times}\uf(1),2)
\\
\rcy38.&\uf(1){\times}\uf(1)
&\varnothing&\text{-}&\text{-}&\text{-}&\text{-}
&\af_1&(\varnothing,\text{-})&\text{-}&\blue{\cS^{(1)}_{\varnothing,2}}&\uf(1)
\\
\cdashline{1-12}

%%%End SU(6) series beginning of the Sp(14)

\multicolumn{12}{c}{\spf(14)\ \text{series}}\\
\hline
39.& \spf(14)_9
&\varnothing&\text{-}&\text{-}&\text{-}&\text{-}
&\cf_7&(\spf(12)_9,1)&{\bf12}&\hyperref[sec:t22]{\spf(12)_8}&(\spf(12)_8,1)
\\
40.&\suf(2)_8{\times}\spf(10)_7&
\af_1&(\spf(10)_7,1)&-&\cS^{(1)}_{E_6,2}&(\spf(10)_7,1)
&\cf_5&(\spf(8)_7{\times}\suf(2)_8,(1,1))&{\bf8}&\hyperref[sec:t25]{\suf(2){\times}\spf(8)_6}&(\spf(8)_6{\times}\suf(2)_8,(1,1)
\\
41.&
\suf(2)_5{\times}\spf(8)_7&
\af_1&(\spf(8)_7,1)&-&\hyperref[sec:G247]{G_2+4F}&(\spf(8)_7,1)
&\cf_4&(\suf(2)_5{\times}\spf(6)_7,(1,1))&{\bf6}&\hyperref[sec:t26]{\suf(2)_5{\times}\spf(6)_6{\times}\uf(1)}&(\suf(2)_5{\times}\spf(6)_6,(1,1))
\\
42.&\spf(8)_6{\times}\uf(1)
&\varnothing&\text{-}&\text{-}&\text{-}&\text{-}
&\cf_4&(\spf(6)_6{\times}\uf(1),1)&{\bf6}&\hyperref[sec:tTE62]{\spf(6)_5{\times}\uf(1)}&(\spf(6)_5{\times}\uf(1),1)
\\
\rcy43.&\spf(6)_5
&\varnothing&\text{-}&\text{-}&\text{-}&\text{-}
&\cf_3&(\spf(4)_5,1)&{\bf4}&\hyperref[sec:SU22a]{SU(2)\text{-}SU(2)}&(\spf(4)_4,1)
\\

\cdashline{1-12}

%%%End Sp(14) series beginning of the SU(5)

\multicolumn{12}{c}{\suf(5)\ \text{series}}\\
\hline

44.&\suf(5)_{16}
&\varnothing&\text{-}&\text{-}&\text{-}&\text{-}
&\bar{h}_{5,3}&\suf(4)_{16}&{\bf 4}\oplus\bar{\bf4}&\cS^{(1)}_{D_4,3}&(\suf(4)_{14},1)
\\
45.&\suf(3)_{12}{\times}\uf(1)
&\red{?}&\red{?}&\text{-}&\cS^{(1)}_{A_2,4}&(\suf(3)_{14},1)
&\bar{h}_{3,3}&(\suf(2)_{12}{\times}\uf(1),1)&{\bf 2}\oplus\bar{\bf2}&\cS^{(1)}_{A_1,3}&(\suf(2)_{10}{\times}\uf(1),1)
\\
46.&\suf(2)_{10}\uf(1)
&\varnothing&\text{-}&\text{-}&\text{-}&\text{-}
&\bar{h}_{5,3}&\suf(4)_{16}&{\bf 4}\oplus\bar{\bf4}&\cS^{(1)}_{D_4,3}&(\suf(4)_{14},1)
\\

\cdashline{1-12}

%%%End SU(5) series beginning of the Sp(12)

\multicolumn{12}{c}{\spf(12)\ \text{series}}\\
\hline

47.&
\spf(12)_{11}
&\varnothing&\text{-}&\text{-}&\text{-}&\text{-}
&\cf_6&(\spf(10)_{11},1)&{\bf10}&\hyperref[sec:S5]{\suf(10)_{10}}&(\spf(10)_{10},1)
\\
\rcy48.&
\spf(4)_5{\times}\sof(4)_4
&\varnothing&\text{-}&\text{-}&\text{-}&\text{-}
&\cf_2&(\suf(2)_5{\times}\sof(4)_4,(1,1))&{\bf2}&\hyperref[sec:SU22b]{2F+SU(2)-SU(2)+F}&(\suf(2)_4{\times}\sof(4)_4{\times}\uf(1)^2,(1,1))
\\
\rcy49.&
\spf(8)_7
&\varnothing&\text{-}&\text{-}&\text{-}&\text{-}
&\cf_4&(\spf(6)_7,1)&{\bf6}&\hyperref[sec:SU3Nf6]{SU(3)+6F}&(\spf(6)_6,1)
\\
50.&
\spf(4)_{\frac{13}3}
&\varnothing&\text{-}&\text{-}&\text{-}&\text{-}
&\cf_2&(\spf(2)_{\frac{13}3},1)&{\bf2}&\hyperref[sec:(A1,D6)]{(A_1,D_6)}&(\suf(2)_{\frac{10}3}{\times} \uf(1),1)
\\

\cdashline{1-12}

%%%End Sp(12) series beginning of the Sp(8)-SU(2)^2

\multicolumn{12}{c}{\spf(8)-\suf(2)^2\ \text{series}}\\
\hline
51.&
\spf(8)_{13}{\times}\suf(2)_{26}
&\varnothing&\text{-}&\text{-}&\text{-}&\text{-}
&\cf_4&(\spf(6){\times}\suf(2),(1,1))&({\bf 6},{\bf1})&\cT^{(2)}_{E_6,2}&(\spf(6)_{12}{\times}\suf(2)_{12}{\times}\suf(2)_7^2,(1,1,1))
\\
52.&
\spf(4)_9{\times}\suf(2)_{16}{\times}\suf(2)_{18}
&\varnothing&\text{-}&\text{-}&\text{-}&\text{-}
&\cf_2&(\suf(2){\times}\suf(2),(1,1))&({\bf2},{\bf1})&\cT^{(2)}_{D_4,2}&(\suf(2)_{8}{\times}\suf(2)^2_{8}{\times}\suf(2)_5^2,(2,1,1))
\\
53.&
\suf(2)_7{\times}\suf(2)_{14}\times \uf(1)
&\varnothing&\text{-}&\text{-}&\text{-}&\text{-}
&\af_1&(\suf(2),1)&{\bf1}&\cT^{(2)}_{A_2,2}&(\suf(2)_6{\times}\suf(2)_4^2{\times}\uf(1),(1,1,\text{-})
\\
54.&\suf(2)_6{\times}\suf(2)_8
&\af_1&(\suf(2)_8,1)&{\bf1}&\cS^{(1)}_{A_2,4}&\uf(1){\times}\uf(1)
&\varnothing&\text{-}&\text{-}&\text{-}&\text{-}
\\
55.&\suf(2)_5&
\af_1&(\varnothing,\text{-})&-&\green{\cS^{(1)}_{\varnothing,4}}&(\uf(1))
&\varnothing&\text{-}&\text{-}&\text{-}&\text{-}
\\
\rcb56.&\suf(2)_{10}
&\varnothing&\text{-}&\text{-}&\text{-}&\text{-}
&\af_1&(\suf(2)_3,1)&\text{-}&\blue{\cS^{(1)}_{\varnothing,2}}{\times}\uf(1)&(\suf(2)_3,1)
\\

\cdashline{1-12}

%%%End Sp(8)-SU(2)^2 series beginning of the G2

\multicolumn{12}{c}{\gf_2\ \text{series}}\\
\hline

57.&[\gf_2]_8{\times}\suf(2)_{14}
&\gf_2&(\suf(2){\times}\suf(2),(3,1))&({\bf 4},{\bf 1})&\cS^{(1)}_{D_4,3}&(\suf(2)_{14}{\times}\suf(2)_{14}{\times}\uf(1),(1,1))
&\varnothing&\text{-}&\text{-}&\text{-}&\text{-}
\\
58.&\suf(2)_{\frac{16}3}{\times}\suf(2)_{10}
&\af_1&(\suf(2),1)&{\bf1}&\cS^{(1)}_{A_1,3}&(\suf(2)_{10}{\times}\uf(1),1)
&\varnothing&\text{-}&\text{-}&\text{-}&\text{-}
\\
59.&[\gf_2]_{\frac{20}3}&
\gf_2&(\suf(2)_{\frac{20}3},3)&{\bf4}&\cS^{(1)}_{A_1,3}&(\suf(2)_{10},1)
&\varnothing&\text{-}&\text{-}&\text{-}&\text{-}
\\
\rcb60.&\suf(2)_8
&\varnothing&\text{-}&\text{-}&\text{-}&\text{-}
&\af_1&(\suf(2)_3,1)&\text{-}&\blue{\cS^{(1)}_{\varnothing,2}}{\times}\uf(1)&(\suf(2)_3,1)
\\

\cdashline{1-12}

%%%End G2 series beginning of the SU(3)

\multicolumn{12}{c}{\suf(3)\ \text{series}}\\
\hline

61.&\suf(3)_{26}{\times}\uf(1)
&\varnothing&\text{-}&\text{-}&\text{-}&\text{-}
&h_{2,4}&(\suf(2)_{26},1)&{\bf2}\oplus\bar{\bf 2}&\cT^{(2)}_{D_4,3}&(\suf(2)_8{\times}\suf(2)_{14},(3,1))
\\
62.&\uf(1)^2
&\varnothing&\text{-}&\text{-}&\text{-}&\text{-}
&h_{2,3}&\uf(1)&\text{-}&\cT^{(2)}_{A_1,3}&(\suf(2)_{\frac{16}3}{\times}\suf(2)_{10},(1,1))
\\
\rcg63.&\uf(1)
&\varnothing&\text{-}&\text{-}&\text{-}&\text{-}
&A_4&\text{-}&\text{-}&\blue{\cS^{(1)}_{\varnothing,2}}{\times}\uf(1)&(\suf(2)_3,1)
\\

\cdashline{1-12}

%%%End SU(3) series beginning of the SU(2)

\multicolumn{12}{c}{\suf(2)\ \text{series}}\\
\hline

64.&\suf(2)_{16}{\times}\uf(1)
&\varnothing&&\text{-}&\text{-}&\text{-}
&h_{3,4}&\uf(1)^2&\text{-}&\cT^{(2)}_{A_2,4}&(\suf(2)_6{\times}\suf(2)_8,(1,1))
\\
\rcg65.&\uf(1)
&\varnothing&\text{-}&\text{-}&\text{-}&\text{-}
&A_5&\text{-}&\text{-}&\blue{\cS^{(1)}_{\varnothing,2}}{\times}\uf(1)&(\suf(2)_3,1)\\
\hline\hline
\end{array}$
\caption{\label{Higgs:two}{\footnotesize This table summarizes the \gls{HB} data for the theories in table \ref{tab:r22}. The second column lists the flavor symmetry of the \gls{SCFT}, column three to seven, lists the information of the higgsing of the flavor symmetry realized on the \gls{CB} \gls{ukstr} $u=0$, while the last four columns present the same information for $v=0$. The entries in red are uncertain as discussed in the corresponding sections. For more details and an explanation of the connection with generalized free fields realization of the theory's VOA, see appendix \ref{genfreef}. $\uf(1)$ factors in this table will be mostly omitted.}}
\end{adjustbox}
\end{table}
\end{landscape}

\begin{table}[ht]
\begin{adjustbox}{center,max width=.7\textwidth}
$\def\arraystretch{1.0}
\begin{array}{r|c|ccc:cc|ccc:cc|}
\multicolumn{12}{c}{\Large\textsc{Rank-2:\ Higgs\ branch\ data\ III\ (Isolated)}}\\
\hline
\hline
\#&\ff&
\bSf_u&([\ff^\natural]_{k^\natural},I_{\ff^\natural\hookrightarrow\ff_{\rm UV}})&\pi_{\bRf}&\bTf_u
&([\ff_{\rm IR}]_{k_{\rm IR}},I_{\ff^\natural\hookrightarrow\ff_{\rm IR}}) 
&\blue{\Sf}_v&( [\ff^\natural]_{k^\natural},I_{\ff^\natural\hookrightarrow\ff_{\rm UV}})&\pi_{\bRf}&\bTf_v
&([\ff_{\rm IR}]_{k_{\rm IR}},I_{\ff^\natural\hookrightarrow\ff_{\rm IR}}) 
\\
\hline\hline

66.&\spf(4)_{14}{\times}\suf(2)_8&
\af_1&(\spf(4)_{14},1)&-&\cS^{(1)}_{D_4,3}&(\spf(4)_{14},1)
&\varnothing&\text{-}&\text{-}&\text{-}&\text{-}
\\
\cdashline{1-12}
67.&\suf(2)_{14}
&\af_1&\text{-}&\text{-}&\cT^{(2)}_{\varnothing,1}&\text{-}
&\varnothing&\text{-}&\text{-}&\text{-}&\text{-}
\\
\cdashline{1-12}
\rcb68.&\suf(2)_{14}
&\varnothing&\text{-}&\text{-}&\text{-}&\text{-}
&\af_1&(\suf(2)_3,1)&\text{-}&\blue{\cS^{(1)}_{\varnothing,2}}{\times}\uf(1)&(\suf(2)_3,1)
\\

\cdashline{1-12}
\multicolumn{12}{c}{\text{Theory with no known string theory realization}}\\
\hline
\rcy69.&\varnothing
&\varnothing&\text{-}&\text{-}&\text{-}&\text{-}
&\varnothing&\text{-}&\text{-}&\text{-}&\text{-}
\\
\hline\hline
\end{array}$
\caption{\label{Higgs:three}{\small This table summarizes the \gls{HB} data for the isolated theories. The second column lists the flavor symmetry of the \gls{SCFT}, column three to seven, lists the information of the higgsing of the flavor symmetry realized on the \gls{CB} \gls{ukstr} $u=0$, while the last four columns present the same information for $v=0$. For more details and an explanation of the connection with generalized free fields realization of the theory's VOA, see appendix \ref{genfreef}. $\uf(1)$ factors in this table will be mostly omitted.}}
\end{adjustbox}
\end{table}

A final note is that the discussion of the $\cS$ and $\cT$ theories, as well as the theories with extended supersymmetry, will be far less detailed than the rest. The former have been studied in depth recently and it would be redundant to present the same results here, only less eloquently. The latter are instead extremely constrained so there is a limited number of moving parts. We therefore made the choice of collecting the main results for theories in these two classes, as well as their CFT data, in Appendix \ref{sec:TandS} and \ref{sec:N34}.

\subsection{$\ef_8-\sof(20)$ series}

This is the largest series with two \gls{SCFT}s at the top from which descend a total of twenty one theories. This series includes all the \gls{AD} known to the author. It is also worth mentioning that no discretely gauged rank-1 theories appear on \gls{CB} singular strata. 

\subsubsection*{$\boldsymbol{\cT^{(2)}_{E_8,1}}$}\label{sec:E8r2} This is the rank-2 theory with the largest \gls{HB} and central charges and sits at the top of the $\ef_8-\sof(20)$ series. This theory can be engineered in type \emph{II}B string theory as the worldvolume theory of two \emph{D3} branes probing an $E_8$ 7brane singularity \cite{Banks:1996nj,Douglas:1996js,Sen:1996vd,Dasgupta:1996ij}. It is also commonly known as the rank-2 $E_8$ \gls{MN} theory \cite{Minahan:1996fg,Minahan:1996cj}. Recently this theory has been shown to belong to a larger class of $\cN=2$ theories dubbed $\cT$-theories which have been studied in detail and their properties are summarized in appendix \ref{sec:TandS}. We collect the relevant CFT data as well the stratification in table \ref{Fig:T2G1}.

\subsubsection*{\boldsymbol{$\sof(20)_{16}$}}\label{sec:t2}

This theory is the other top theory of the series and can be obtained, for example, in the untwisted $E_6$ class-$\cS$ series \cite{Chacaltana:2014jba}. This study allows to fill in most of the CFT data reported in table \ref{CchTh23} which will be used below to fill in all the details of the full moduli space of vacua which we now discuss.

\begin{figure}[h!]
\ffigbox{
\begin{subfloatrow}
\ffigbox[8.5cm][]{
\begin{tikzpicture}[decoration={markings,
mark=at position .5 with {\arrow{>}}}]
\begin{scope}[scale=1.5]
\node[bbc,scale=.5] (p0a) at (0,-1) {};
\node[scale=.5] (p0b) at (0,-3) {};
\node[scale=.8] (t0b) at (0,-3.1) {$\sof(20)_{16}$};
\node[scale=.8] (p3) at (.7,-2) {\ \ $[I_1,\varnothing]$};
\node[scale=.8] (p1) at (-.7,-2) {$[I_6^*,\sof(20)]$\ \ };
%\node[scale=.8] (t1a) at (-.6,-.4) {$I_0^*$};
%\node[scale=.8] (t2c) at (-.7,-1.6) {$K_{\D_7}$};
\node[scale=.8] (t1c) at (.85,-2.65) {{\scriptsize$\big[u^4+v^3=0\big]$}};
%\node[scale=.8] (t2c) at (.7,-1.6) {$K_{2\D_7}$};
\node[scale=.8] (t3c) at (-.7,-2.6) {{\scriptsize$\big[v=0\big]$}};
\draw[red] (p0a) -- (p1);
\draw[red] (p0a) -- (p3);
\draw[red] (p1) -- (p0b);
\draw[red] (p3) -- (p0b);
\end{scope}
\begin{scope}[scale=1.5,xshift=2.85cm]
\node[scale=.5] (p2) at (0,-1) {};
\node[scale=.8] (t0a) at (0,-.9) {$\H^{\rm d_{HB}}$};
\node[scale=.8] (tp2) at (-.2,-1.5) {$\ef_8$};
\node[scale=.8] (p1) at (0,-2) {$\cT^{(1)}_{E_8,1}$};
\node[scale=.8] (tp1) at (-.2,-2.5) {$\df_{10}$};
\node[scale=.5] (p0) at (0,-3) {};
\node[scale=.8] (t0b) at (0,-3.1) {$\sof(20)_{16}$};
\draw[blue] (p0) -- (p1);
\draw[blue] (p1) -- (p2);
\end{scope}
\end{tikzpicture}}
{\caption{\label{CBhTh23}The Coulomb and Higgs stratification of $\sof(20)_{16}$.}}
\end{subfloatrow}\hspace{1cm}%
\begin{subfloatrow}
\capbtabbox[5cm]{%
  \renewcommand{\arraystretch}{1.1}
  \begin{tabular}{|c|c|} 
  \hline
  \multicolumn{2}{|c|}{$\sof(20)_{16}$}\\
  \hline\hline
  $(\D_u,\D_v)$  &\quad $\left(6,8\right)$\quad{} \\
  $24a$ & 202\\  
  $12c$ & 124\\
$\ff_k$ & $\sof(20)_{16}$ \\ 
$d_{\rm HB}$& 46\\
$h$&0\\
$T({\bf2}\bh)$&0\\
\hline\hline
\end{tabular}
}{%
  \caption{\label{CchTh23}Central charges, \gls{CB} parameters and \gls{ECB} dimension.}%
}
\end{subfloatrow}}{\caption{\label{TothTh23}Information about the $\sof(20)_{16}$.}}
\end{figure}

The flavor symmetry of this theory is simple, expectedly so given it has only one \gls{higgs}, $v$. Because of the \gls{UVIR}, the $\sof(20)$ factor must be realized on the \gls{CB} as the flavor symmetry of a rank-1 theory supported on a singular stratum, and since the only allowed \gls{ukstr} is $v=0$, we can start with such an option. An encouraging fact is that  the level of the $\sof(20)$ precisely doubles $\D_v$ so we can use the \doub. A quick look at table \ref{Theories} makes it obvious that the right guess is $\rTf_v\equiv [I_6^*,\sof(20)]$ where this latter theory is nothing but an $\cN=2$ $SU(2)$ gauge theory with ten fundamental hypers. Using \eqref{actotcint}, it is immediate to check that this guess does reproduce the correct level $k_{\sof(20)}=16$. Since this $\cN=2$ gauge theory has no \gls{ECB}, we also conclude that \gls{h}=0. That the guess we just made is correct, can be also checked by reproducing the $a$ and $c$ central charges of this theory, shown in table \ref{CchTh23}, plugging the \gls{bi} for the $[I_6^*,\sof(20)]$ (which the reader can check to be 12) in \eqref{actotaint}-\eqref{actotbint}. With $\rTf_{u^4+v^3}\equiv [I_1,\varnothing]$, everything works beautifully.

The \gls{Hasse} of the \gls{HB} is \gls{Hlin} and it involves only two transitions, the first one being associated with the \gls{HB} of the theory supported on the \gls{CB} which has a $\df_{10}$ as its \gls{HB}. To identify the (rank-1) theory supported on $\df_{10}$ we can use the property that the total \gls{HB} dimension of the rank-2 theory is 46 from which, subtracting the 17 dimension of the $\df_{10}$, we obtain a prediction for the dimension of the \gls{HB} of the rank-1 theory: 19. This immediately singles out $\bTf_{\df_{10}}\equiv \cT^{(1)}_{E_8,1}$. There is another way of going about determining the theory supported on various strata and which will be used copiously below. Using \eqref{gHWeq} we can directly predict the central charge of the theory after higgsing. This formula only applies to higgings of \gls{gHW} type. If the \gls{CB} realization of the flavor symmetry which is getting spontaneously broken is known, it is easy to assess whether or not a given higgsing has this property. For the case of the $\df_{10}$ transition this is indeed the case. \eqref{gHWeq} then gives $12c_{\bTf_{\df_{10}}}=62$ which matches our previous guess. As it is explained in section \ref{genfreef}, in this case the matching of the moment map along the higgsing works in a non-trivial and somewhat interesting way.

\subsubsection*{$\boldsymbol{[\ef_8]_{20}}$}\label{sec:D20E8}

This theory was discussed in \cite{Giacomelli:2017ckh} where it is also pointed out that it can be obtained in the $E_7$ class-$\cS$ \cite{Chacaltana:2017boe}. The nomenclature $[\ef_8]_{20}$ was introduced in \cite{Cecotti:2012jx,Cecotti:2013lda} where the geometric engineering realization of this theory in type \emph{II}B string theory is also discussed. Most of the CFT data in table \ref{CchD1E8} is taken from \cite{Giacomelli:2017ckh} and leveraged here to complete the study of the full moduli space.

\begin{figure}[h!]
\ffigbox{
\begin{subfloatrow}
\ffigbox[8.5cm][]{
\begin{tikzpicture}[decoration={markings,
mark=at position .5 with {\arrow{>}}}]
\begin{scope}[scale=1.5]
\node[bbc,scale=.5] (p0a) at (0,-1) {};
\node[scale=.5] (p0b) at (0,-3) {};
\node[scale=.8] (t0b) at (0,-3.1) {$[\ef_8]_{20}$};
\node[scale=.8] (p3) at (.7,-2) {\ \ $[I_1,\varnothing]$};
\node[scale=.8] (p1) at (-.7,-2) {$\cT^{(1)}_{E_8,1}$\ \ };
%\node[scale=.8] (t1a) at (-.6,-.4) {$I_0^*$};
%\node[scale=.8] (t2c) at (-.7,-1.6) {$K_{\D_7}$};
\node[scale=.8] (t1c) at (.85,-2.65) {{\scriptsize$\big[u^5+v^2=0\big]$}};
%\node[scale=.8] (t2c) at (.7,-1.6) {$K_{2\D_7}$};
\node[scale=.8] (t3c) at (-.7,-2.6) {{\scriptsize$\big[v=0\big]$}};
\draw[red] (p0a) -- (p1);
\draw[red] (p0a) -- (p3);
\draw[red] (p1) -- (p0b);
\draw[red] (p3) -- (p0b);
\end{scope}
\begin{scope}[scale=1.5,xshift=2.85cm]
\node[scale=.5] (p0a) at (0,-1) {};
\node[scale=.8] (t0a) at (0,-.9) {$\H^{\rm d_{HB}}$};
\node[scale=.8] (tp2) at (-.2,-1.5) {$\ef_7$};
\node[scale=.8] (p2a) at (0,-2) {$\cT^{(1)}_{E_7,1}$};
\node[scale=.8] (tp1) at (-.2,-2.5) {$\ef_8$};
\node[scale=.5] (p0b) at (0,-3) {};
\node[scale=.8] (t0b) at (0,-3.1) {$[\ef_8]_{20}$};
\draw[blue] (p0a) -- (p2a);
\draw[blue] (p2a) -- (p0b);
\end{scope}
\end{tikzpicture}}
{\caption{\label{CBhD1E8}The Coulomb and Higgs stratification of $[\ef_8]_{20}$}}
\end{subfloatrow}\hspace{1cm}%
\begin{subfloatrow}
\capbtabbox[5cm]{%
  \renewcommand{\arraystretch}{1.1}
  \begin{tabular}{|c|c|} 
  \hline
  \multicolumn{2}{|c|}{$[\ef_8]_{20}$}\\
  \hline\hline
  $(\D_u,\D_v)$  &\quad $\left(4,10\right)$\quad{} \\
  $24a$ &  202\\  
  $12c$ & 124\\
$\ff_k$ & $[\ef_8]_{20}$ \\ 
$d_{\rm HB}$& 46\\
$h$&0\\
$T({\bf2}\bh)$&0\\
\hline\hline
\end{tabular}
}{%
  \caption{\label{CchD1E8}Central charges, \gls{CB} parameters and \gls{ECB} dimension.}%
}
\end{subfloatrow}}{\caption{\label{TothD1E8}Information about the $[\ef_8]_{20}$}}
\end{figure}

Since the theory has a single \gls{higgs} we expect a relatively simple structure. The fact that the flavor symmetry is simple, and furthermore exceptional, makes our life quite easy. In fact the only natural guess for the realization of the $\ef_8$ on the \gls{CB} is $\rTf_v\equiv \cT^{(1)}_{E_8,1}$. This is further confirmed from the fact that the level of the $\ef_8$ flavor factor is indeed double of $\D_v$. This guess can be checked in two ways. Firstly, as we have done in the previous case, we can apply \eqref{actotaint}-\eqref{actotbint} to match the central charges of this theory. This works well and in turns allows to determine the theory on the \gls{kstr}: $\rTf_{u^5+v^2}\equiv [I_1,\varnothing]$. The second approach is insightful. This is one of the few lucky cases where the \gls{CB} geometry is known in terms of a hyperelleptic fibration of a two dimensional base \cite{Argyres:2005pp,Argyres:2005wx}. Therefore we have a way to extract the \gls{CB} stratification by studying the discriminant locus of the fibration as discussed at the end of section \ref{sec:CBstr}. This philosophy is described in more details, for example, in \cite{Martone:2020nsy}. 

From \cite{Argyres:2005wx} the \gls{CB} geometry of this theory can be written as:
\beq
y^2=x^5+(u x+v)^3.
\eeq
Taking the discriminant of the right hand side we obtain:
\beq
D_{x^5}\sim v^{10}(c_1 u^5+ c_2 v^2)
\eeq
where $c_{1,2}$ are irrelevant numerical factor. This result implies that the \gls{CB} geometry is only singular at $v=0$ and $u^5+v^2=0$, which matches nicely with our previous guess. But this is not all. In fact the order of the zero of the discriminant carries extra information which can be used to characterize the theory supported on two singular strata. Performing the analysis we find that the $v=0$ singularity (we are taking implicitly $u\neq0$) is a $II^*$ singularity while the one at $u^5+v^2=0$ is an $I_1$. This perfectly match with what we find using the \gls{UVIR} and the \gls{ccf}.

We are now ready to study the \gls{HB} which we expect to be \gls{Hlin}. Furthermore this theory has \gls{h}=0 and therefore to identify the theory supported on the $\ef_8$ stratum suffices to impose the constraint that the total dimension of the \gls{HB} should add up to 46. This singles out $\bTf_{\ef_8}\equiv \cT^{(1)}_{E_7,1}$. As we did before, we can confirm this guess recognizing that the spontaneous breaking of the $\ef_8$ gives a higgsing of \gls{gHW} type and apply \eqref{gHWeq} to find $12c_{\bTf_{\ef_8}}=38$.

\subsubsection*{$\boldsymbol{\cT^{(2)}_{E_7,1}}$}\label{sec:E7r2} This theory can be engineered in type \emph{II}B string theory as the worldvolume theory of two \emph{D3} branes probing an $E_7$ \emph{7}brane exceptional singularity \cite{Banks:1996nj,Douglas:1996js,Sen:1996vd,Dasgupta:1996ij}. It is also commonly known as the rank-2 $E_7$ \gls{MN} theory \cite{Minahan:1996fg,Minahan:1996cj}. It is well-known that this theory can be obtained by mass deforming the $\cT^{(2)}_{E_8,1}$ and the mass deformation is geometric in the sense that it corresponds to ``peel'' away a \emph{D7} brane which makes the $E_8$ \emph{7}brane singularity a $E_7$ one. This theory has been shown to belong to a larger class of $\cN=2$ theories dubbed $\cT$-theories which have been studied in detail and their properties are summarized in appendix \ref{sec:TandS}. We collect the relevant CFT data as well the stratification in table \ref{Fig:T2G1}.

\subsubsection*{\boldsymbol{$\suf(2)_8{\times}\sof(16)_{12}$}}\label{sec:t5}

This theory appears in numerous class-$\cS$ constructions, one example is the untwisted $D_4$ \cite{Chacaltana:2011ze}. This is where most of the CFT data in table \ref{CcTh19} is taken from.

\begin{figure}[h!]
\ffigbox{
\begin{subfloatrow}
\ffigbox[7cm][]{
\begin{tikzpicture}[decoration={markings,
mark=at position .5 with {\arrow{>}}}]
\begin{scope}[scale=1.5]
\node[bbc,scale=.5] (p0a) at (0,0) {};
\node[scale=.5] (p0b) at (0,-2) {};
\node[scale=.8] (t0b) at (0,-2.1) {$\suf(2)_8{\times}\sof(16)_{12}$};
\node[scale=.8] (p3) at (0,-1) {$[I_2,\suf(2)]$\ \ };
\node[scale=.8] (p2) at (1.,-1) {\ \ $[I_4^*,\sof(16)]$};
\node[scale=.8] (p1) at (-1,-1) {$[I_1,\varnothing]$\ \ };
%\node[scale=.8] (t1a) at (-.6,-.4) {$I_0^*$};
%\node[scale=.8] (t2c) at (-.7,-1.6) {$K_{\D_7}$};
\node[scale=.8] (t1c) at (-.9,-1.7) {{\scriptsize$\big[u^3+v^2=0\big]$}};
%\node[scale=.8] (t2c) at (.7,-1.6) {$K_{2\D_7}$};
\node[scale=.8] (t1c) at (.7,-1.7) {{\scriptsize$\big[v=0\big]$}};
\node[scale=.8] (t1c) at (.3,-1.35) {{\scriptsize$\big[u=0\big]$}};
\draw[red] (p0a) -- (p1);
\draw[red] (p0a) -- (p2);
\draw[red] (p0a) -- (p3);
\draw[red] (p1) -- (p0b);
\draw[red] (p2) -- (p0b);
\draw[red] (p3) -- (p0b);
\end{scope}
\begin{scope}[scale=1.5,xshift=3cm]
\node[scale=.5] (p0a) at (0,0) {};
\node[scale=.5] (p0b) at (0,-2) {};
\node[scale=.8] (t0a) at (0,.1) {$\H^{\rm d_{HB}}$};
\node[scale=.8] (t0b) at (0,-2.1) {$\suf(2)_8{\times}\sof(16)_{12}$};
\node[scale=.8] (p1a) at (.8,-1) {$\cT^{(1)}_{E_7,1}$};
\node[scale=.8] (p1b) at (-.8,-1) {$\cT^{(1)}_{E_8,1}$};
\node[scale=.8] (t1a) at (0.6,-.5) {$\ef_7$};
\node[scale=.8] (t1b) at (-.6,-0.5) {$\ef_8$};
\node[scale=.8] (t2a) at (.6,-1.5) {$\df_8$};
\node[scale=.8] (t2b) at (-.6,-1.5) {$\af_1$};
\draw[blue] (p0a) -- (p1a);
\draw[blue] (p0a) -- (p1b);
\draw[blue] (p1a) -- (p0b);
\draw[blue] (p1b) -- (p0b);
\end{scope}
\end{tikzpicture}}
{\caption{\label{CBTh19}The Coulomb and Higgs stratification of $\suf(2)_8{\times}\sof(16)_{12}$.}}
\end{subfloatrow}\hspace{1cm}%
\begin{subfloatrow}
\capbtabbox[7cm]{%
  \renewcommand{\arraystretch}{1.1}
  \begin{tabular}{|c|c|} 
  \hline
  \multicolumn{2}{|c|}{$\suf(2)_8{\times}\sof(16)_{12}$}\\
  \hline\hline
  $(\D_u,\D_v)$  &\quad $\left(4,6\right)$\quad{} \\
  $24a$ &  138\\  
  $12c$ & 84 \\
$\ff_k$ & $\suf(2)_8{\times}\sof(16)_{12}$ \\ 
$d_{\rm HB}$& 30\\
$h$&0\\
$T({\bf2}\bh)$&0\\
\hline\hline
\end{tabular}
}{%
  \caption{\label{CcTh19}Central charges, \gls{CB} parameters and \gls{ECB} dimension.}%
}
\end{subfloatrow}}{\caption{\label{TotTh19}Information about the $\suf(2)_8{\times}\sof(16)_{12}$ theory.}}
\end{figure}

This is the first case we encounter of a \gls{totally} theory. This property is also reflected in the fact that the flavor symmetry has two simple flavor factors. Given the value of the levels, we can use the \doub\ to determine how both factors are realized on the \gls{CB}. The $\sof(16)_{12}$ is easily identified as the flavor symmetry of a $\cN=2$ $SU(2)$ gauge theory with eight fundamental flavors, therefore leading to the identification $\rTf_v\equiv [I_4^*,\sof(16)]$. The $\suf(2)$ is more ambiguous as it might be the isometry of an \gls{ECB} but again the fact that the level is twice $\D_u$ convincingly suggests that $\rTf_u\equiv[I_2,\suf(2)]$. This conclusions are confirmed by the computing the $a$ and $c$ central charges of the theory using \eqref{actotaint}-\eqref{actotbint} which also allows to fix the last ambiguity $\rTf_{u^3+v^2}\equiv [I_1,\varnothing]$ thus concluding our analysis of the \gls{CB}.

The \gls{CB} perspective indicates that \gls{h}=0 which also implies that the same constraints applies for the rank-1 theories supported on the $\af_1$ and $\df_8$ higgsings. This information, along with the Ricci-flatness of the \gls{HB} and the constraint that the total \gls{HB} dimension should add up to 30, is sufficient to make the identification $\bTf_{\af_1}\equiv \cT^{(1)}_{E_8,1}$ and  $\bTf_{\df_8}\equiv \cT^{(1)}_{E_7,1}$. This guess can be checked by exploiting the fact that both higgsings are of \gls{gHW} type and that \eqref{gHWeq} applied to these cases gives $12c_{\bTf_{\af_1}}=62$ and $12c_{\bTf_{\df_8}}=38$. Thus concluding our analysis.

\subsubsection*{\boldsymbol{$\suf(10)_{10}$}}\label{sec:S5}

This a rank-2 theories, belongs to an infinite series of $\cN=2$ \gls{SCFT}s discussed in \cite{Chacaltana:2010ks}. The generic $S_N$ theories have flavor symmetry $\suf(N+2)_{2N}{\times}\suf(3)_{10}{\times}\uf(1)$ and precisely for $N=5$ there is a possibility of an enhancement to $\suf(10)_{10}$ which in fact happens. The S-duality property of these theories for any $N$ are discussed in the original paper along with the computation of many CFT data which, for $N=5$ are reported in table \ref{CcS5}. Let's start with the analysis of the full moduli space.

\begin{figure}[h!]
\ffigbox{
\begin{subfloatrow}
\ffigbox[8cm][]{
\begin{tikzpicture}[decoration={markings,
mark=at position .5 with {\arrow{>}}}]
\begin{scope}[scale=1.5]
\node[bbc,scale=.5] (p0a) at (0,-1) {};
\node[scale=.5] (p0b) at (0,-3) {};
\node[scale=.8] (t0b) at (0,-3.1) {$\suf(10)_{10}$};
\node[scale=.8] (p2) at (.7,-2) {\ \ $[I_{10},\suf(10)]$};
\node[scale=.8] (p1) at (-.7,-2) {$[I_1,\varnothing]$\ \ };
%\node[scale=.8] (t1a) at (-.6,-.4) {$I_0^*$};
%\node[scale=.8] (t2c) at (-.7,-1.6) {$K_{\D_7}$};
\node[scale=.8] (t1c) at (-.85,-2.7) {{\scriptsize$\big[u^5+v^4=0\big]$}};
%\node[scale=.8] (t2c) at (.7,-1.6) {$K_{2\D_7}$};
\node[scale=.8] (t1c) at (.7,-2.7) {{\scriptsize$\big[v=0\big]$}};
\draw[red] (p0a) -- (p1);
\draw[red] (p0a) -- (p2);
\draw[red] (p1) -- (p0b);
\draw[red] (p2) -- (p0b);
\end{scope}
\begin{scope}[scale=1.5,xshift=3cm]
\node[scale=.5] (p0a) at (0,-1) {};
\node[scale=.5] (p0b) at (0,-3) {};
\node[scale=.8] (t0a) at (0,-.9) {$\H^{\rm d_{HB}}$};
\node[scale=.8] (t0b) at (0,-3.1) {$\suf(10)_{10}$};
\node[scale=.8] (p1) at (0,-2) {$\cT^{(1)}_{E_7,1}$};
\node[scale=.8] (t01) at (.3,-1.5) {$\ef_7$};
\node[scale=.8] (t02) at (0.3,-2.5) {$\af_9$};
\draw[blue] (p0a) -- (p1);
\draw[blue] (p1) -- (p0b);
\end{scope}
\end{tikzpicture}}
{\caption{\label{CBTh14}The Coulomb and Higgs stratification of $\suf(10)_{10}$.}}
\end{subfloatrow}\hspace{1cm}%
\begin{subfloatrow}
\capbtabbox[5cm]{%
  \renewcommand{\arraystretch}{1.1}
  \begin{tabular}{|c|c|} 
  \hline
  \multicolumn{2}{|c|}{$\suf(10)_{10}$}\\
  \hline\hline
  $(\D_u,\D_v)$  &\quad $\left(4,5\right)$\quad{} \\
  $24a$ &  122\\  
  $12c$ & 74\\
$\ff_k$ & $\suf(10)_{10}$ \\ 
$d_{\rm HB}$& 26\\
$h$&0\\
$T({\bf2}\bh)$&0\\
\hline\hline
\end{tabular}
}{%
  \caption{\label{CcS5}Central charges, \gls{CB} parameters and \gls{ECB} dimension.}%
}
\end{subfloatrow}}{\caption{\label{TotS5}Information about the $\suf(10)_{10}$ theory.}}
\end{figure}

The fact that the flavor symmetry is simple is consistent with the fact that this theory has a single \gls{higgs}, $v$. Furthermore, since the level of the flavor symmetry is precisely doubled the scaling dimension of the \gls{higgs}, the \doub\  immediately suggests the identifcation $\rTf_v=[I_{10},\suf(10)]$. The rest of the \gls{CB} stratification can be easily filled in by matching the central charge using \eqref{actotbint} and we therefore conclude that $\rTf_{u^5+v^4}=[I_1,\varnothing]$. 

The analysis of the \gls{HB} is also straightforward; the $\af_9$ strata is mandated by the \gls{CB} analysis which also shows that this higgsing is of \gls{gHW} type. Since \gls{h}=0, the theory supported there is a rank-1 theory which could be immediately identified from matching the unbroken flavor symmetry along the Higgsing and imposing that the total \gls{HB} of the theory matches what is found in the original class-$\cS$ construction. This leads us to the conclusion that $\bTf_{\af_9}=\cT^{(1)}_{E_7,1}$. It is a useful exercise to check that the result from \eqref{gHWeq} are consistent with this identification.

\subsubsection*{$\boldsymbol{\cT^{(2)}_{E_6,1}}$}\label{sec:E6r2} This theory can be engineered in type \emph{II}B string theory as the worldvolume theory of two \emph{D3} branes probing an $E_6$ \emph{7}brane exceptional singularity \cite{Banks:1996nj,Douglas:1996js,Sen:1996vd,Dasgupta:1996ij}. It is also commonly known as the rank-2 $E_6$ \gls{MN} theory \cite{Minahan:1996fg,Minahan:1996cj}. Again, it is well-known that this theory can be obtained by mass deforming the $\cT^{(2)}_{E_7,1}$ and the mass deformation is again geometric corresponding to making the $E_7$ \emph{7}brane singularity a $E_6$ one. This theory has been shown to belong to a larger class of $\cN=2$ theories dubbed $\cT$-theories which have been studied in detail and their properties are summarized in appendix \ref{sec:TandS}. We collect the relevant CFT data as well the stratification in table \ref{Fig:T2G1}. 

\subsubsection*{\boldsymbol{$\sof(14)_{10}{\times}\uf(1)$}}\label{sec:t8}

This theory was first introduced in the context of $\Z_2$ twisted $E_6$ class-$\cS$ \cite{Chacaltana:2015bna}. The class-$\cS$ construction gives access to most of the CFT data reported in figure \ref{CcTh10} which we will leverage here to fully solve the moduli space structure of the theory.

\begin{figure}[h!]
\ffigbox{
\begin{subfloatrow}
\ffigbox[7cm][]{
\begin{tikzpicture}[decoration={markings,
mark=at position .5 with {\arrow{>}}}]
\begin{scope}[scale=1.5]
\node[bbc,scale=.5] (p0a) at (0,0) {};
\node[scale=.5] (p0b) at (0,-2) {};
\node[scale=.8] (t0b) at (0,-2.1) {$\sof(14)_{10}{\times}\uf(1)$};
\node[scale=.8] (p1) at (-.7,-1) {$[I_1,\varnothing]$\ \ };
\node[scale=.8] (p2) at (.7,-1) {\ \ $[I_3^*,\sof(14)]$};
%\node[scale=.8] (t1a) at (-.6,-.4) {$I_0^*$};
%\node[scale=.8] (t2c) at (-.7,-1.6) {$K_{\D_7}$};
\node[scale=.8] (t1c) at (-.8,-1.6) {{\scriptsize$\big[u^5+v^3=0\big]$}};
%\node[scale=.8] (t2c) at (.7,-1.6) {$K_{2\D_7}$};
\node[scale=.8] (t1c) at (.7,-1.55) {{\scriptsize$\big[v=0\big]$}};
\draw[red] (p0a) -- (p1);
\draw[red] (p0a) -- (p2);
\draw[red] (p1) -- (p0b);
\draw[red] (p2) -- (p0b);
\end{scope}
\begin{scope}[scale=1.5,xshift=2.5cm]
\node[scale=.5] (p0a) at (0,0) {};
\node[scale=.5] (p0b) at (0,-2) {};
\node[scale=.8] (t0a) at (0,.1) {$\H^{\rm d_{HB}}$};
\node[scale=.8] (t0b) at (0,-2.1) {$\sof(14)_{10}{\times}\uf(1)$};
\node[scale=.8] (p1) at (0,-1) {$\cT^{(1)}_{E_6,1}$};
\node[scale=.8] (t2c) at (.3,-0.5) {$\ef_6$};
\node[scale=.8] (t1c) at (.3,-1.5) {$\df_7$};
\draw[blue] (p0b) -- (p1);
\draw[blue] (p1) -- (p0a);
\end{scope}
\end{tikzpicture}}
{\caption{\label{CBtTh10}The Coulomb and Higgs stratification of $\sof(14)_{10}{\times}\uf(1)$.}}
\end{subfloatrow}\hspace{1cm}%
\begin{subfloatrow}
\capbtabbox[7cm]{%
  \renewcommand{\arraystretch}{1.1}
  \begin{tabular}{|c|c|} 
  \hline
  \multicolumn{2}{|c|}{$\sof(14)_{10}{\times}\uf(1)$}\\
  \hline\hline
  $(\D_u,\D_v)$  &\quad $\left(3,5\right)$\quad{} \\
  $24a$ &  106\\  
  $12c$ & 64 \\
$\ff_k$ & $\sof(14)_{10}{\times}\uf(1)$ \\ 
$d_{\rm HB}$& 22\\
$h$&0\\
$T({\bf2}\bh)$&0\\
\hline\hline
\end{tabular}
}{%
  \caption{\label{CcTh10}Central charges, \gls{CB} parameters and \gls{ECB} dimension.}%
}
\end{subfloatrow}}{\caption{\label{TotTh10}Information about the $\sof(14)_{10}{\times}\uf(1)$ theory.}}
\end{figure}

Since the theory has a single simple flavor symmetry factor, we expect an easy \gls{HB} structure. This is a reflection that the theory is not \gls{totally} and only $v$ is a \gls{higgs}. The \gls{CB} realization of the $\sof(14)_{10}$ can be easily and readily identified as $\rTf_v\equiv [I_3^*,\sof(14)]$. This identification predicts a $\df_7$ transition on the \gls{HB} side of things but imposes no constraints on the subsequent transitions while restricting the \gls{ECB} of the theory supported over it to be zero. As a further check that the \gls{CB} identification which we just made is correct, we can check that, plugging the \gls{bi} corresponding to $[I_3^*,\sof(14)]$ supported over $v=0$ and a $[I_1,\varnothing]$ on the \gls{kstr} in \eqref{actotaint}-\eqref{actotbint}, we perfectly reproduce the $a$ and $c$ central charges of the theory.

Completing the \gls{HB} analysis is straightforward. From the \gls{CB} analysis, we could notice that the spontaneous breaking of the $\sof(14)$ factor is of \gls{gHW} type and use \eqref{gHWeq} to compute the central charge of the theory supported over $\df_7$. But a possibly even simpler way to complete the study of the \gls{HB} stratification is to notice that the theory $\bTf_{\df_7}$ has to have no \gls{ECB} (coming from the \gls{CB} analysis) and an 11 dimensional \gls{HB}. Using the Ricci-flatness of the \gls{HB} we are left with only one possibility: $\bTf_{\df_7}\equiv \cT^{(1)}_{E_6,1}$. We leave it for the reader to check that the prediction from \eqref{gHWeq} indeed perfectly match with the value of the central charge of the rank-1 MN $E_6$ theory.

\subsubsection*{\boldsymbol{$\suf(2)_6{\times}\suf(8)_8$}}\label{sec:t9}

This theory was first introduced in the context of twisted $\Z_2$ $A_3$ class-$\cS$ theories \cite{Chacaltana:2012ch}. In the original paper interesting S-dualities of this theory are discussed as well as most of the CFT data reported in table \ref{CcTh8} computed.

The analysis will be similar to the previous cases. The theory is \gls{totally} and the two simple factors of the flavor symmetry show that the \gls{HB} contains two disconnected transitions. It is easy to argue that the $\suf(8)_8$ is realized as the flavor symmetry of an $[I_8,\suf(8)]$ supported on an unknotted $v=0$ stratum leading to the identification $\rTf_v\equiv[I_8,\suf(8)]$. We therefore expect that one of the two \gls{HB} transition is an $\af_7$. The $\suf(2)_6$ instead could potentially give rise to an \gls{ECB} but a more careful look at the level of this symmetry, which precisely doubles $\D_u$,  suggests that it should be realized as the flavor symmetry of a $[I_2,\suf(2)]$ on a $v=0$ stratum. Thus $\rTf_v\equiv [I_2,\suf(2)]$ and \gls{h}=0. As usual the calculus of the $c$ central charge via \eqref{actotbint}, inputting the known \gls{bi}s of the already identified \gls{CB} components, allow us to both check that the those are indeed correct, and determine the theory supported on the unknotted strata. This complets the analysis of the stratification of the \gls{CB}.

\begin{figure}[h!]
\ffigbox{
\begin{subfloatrow}
\ffigbox[7cm][]{
\begin{tikzpicture}[decoration={markings,
mark=at position .5 with {\arrow{>}}}]
\begin{scope}[scale=1.5]
\node[bbc,scale=.5] (p0a) at (0,0) {};
\node[scale=.5] (p0b) at (0,-2) {};
\node[scale=.8] (t0b) at (0,-2.1) {$\suf(2)_6{\times}\suf(8)_8$};
\node[scale=.8] (p3) at (0,-1) {$[I_2,\suf(2)]$\ \ };
\node[scale=.8] (p2) at (1.,-1) {\ \ $[I_8,\suf(8)]$};
\node[scale=.8] (p1) at (-1,-1) {$[I_1,\varnothing]$\ \ };
%\node[scale=.8] (t1a) at (-.6,-.4) {$I_0^*$};
%\node[scale=.8] (t2c) at (-.7,-1.6) {$K_{\D_7}$};
\node[scale=.8] (t1c) at (-.9,-1.7) {{\scriptsize$\big[u^4+v^3=0\big]$}};
%\node[scale=.8] (t2c) at (.7,-1.6) {$K_{2\D_7}$};
\node[scale=.8] (t1c) at (.7,-1.7) {{\scriptsize$\big[v=0\big]$}};
\node[scale=.8] (t1c) at (.3,-1.35) {{\scriptsize$\big[u=0\big]$}};
\draw[red] (p0a) -- (p1);
\draw[red] (p0a) -- (p2);
\draw[red] (p0a) -- (p3);
\draw[red] (p1) -- (p0b);
\draw[red] (p2) -- (p0b);
\draw[red] (p3) -- (p0b);
\end{scope}
\begin{scope}[scale=1.5,xshift=3cm]
\node[scale=.5] (p0a) at (0,0) {};
\node[scale=.5] (p0b) at (0,-2) {};
\node[scale=.8] (t0a) at (0,.1) {$\H^{\rm d_{HB}}$};
\node[scale=.8] (t0b) at (0,-2.1) {$\suf(2)_6{\times}\suf(8)_8$};
\node[scale=.8] (p1a) at (.8,-1) {$\cT^{(1)}_{E_7,1}$};
\node[scale=.8] (p1b) at (-.8,-1) {$\cT^{(1)}_{E_6,1}$};
\node[scale=.8] (t1a) at (0.6,-.5) {$\ef_7$};
\node[scale=.8] (t1b) at (-.6,-0.5) {$\ef_6$};
\node[scale=.8] (t2a) at (.6,-1.5) {$\af_1$};
\node[scale=.8] (t2b) at (-.6,-1.5) {$\af_7$};
\draw[blue] (p0a) -- (p1a);
\draw[blue] (p0a) -- (p1b);
\draw[blue] (p1a) -- (p0b);
\draw[blue] (p1b) -- (p0b);
\end{scope}
\end{tikzpicture}}
{\caption{\label{CBTh8}The Coulomb and Higgs stratification of $\suf(2)_6{\times}\suf(8)_8$.}}
\end{subfloatrow}\hspace{1cm}%
\begin{subfloatrow}
\capbtabbox[7cm]{%
  \renewcommand{\arraystretch}{1.1}
  \begin{tabular}{|c|c|} 
  \hline
  \multicolumn{2}{|c|}{$\suf(2)_6{\times}\suf(8)_8$}\\
  \hline\hline
  $(\D_u,\D_v)$  &\quad $\left(3,4\right)$\quad{} \\
  $24a$ &  90\\  
  $12c$ & 54 \\
$\ff_k$ & $\suf(2)_6{\times}\suf(8)_8$ \\ 
$d_{\rm HB}$& 18\\
$h$&0\\
$T({\bf2}\bh)$&0\\
\hline\hline
\end{tabular}
}{%
  \caption{\label{CcTh8}Central charges, \gls{CB} parameters and \gls{ECB} dimension.}%
}
\end{subfloatrow}}{\caption{\label{TotTh8}Information about the $\suf(2)_6{\times}\suf(8)_8$ theory.}}
\end{figure}

Let's move now to the analysis of the \gls{HB}. The absence of an \gls{ECB} and the fact that both Higgsing are of \gls{gHW} type, make things fairly easy to work out. Indeed using \eqref{gHWeq} we can right away identify the theories supported on the $\af_7$ and $\af_1$ as the rank-1 MN $E_6$ ($\cT^{(1)}_{E_6,1}$) and $E_7$ ($\cT^{(1)}_{E_7,1}$) theory (notice that the difference in \gls{HB} dimensions among these two theories precisely makes up for the difference in dimension of the strata over which they are supported to give rise to a total \gls{HB} of dimension 12). This is enough to reproduce the \gls{HB} stratification in figure \ref{CcTh8}.

\subsubsection*{\boldsymbol{$USp(4)+$}$6\gls{F}$}\label{sec:Sp464}

Let's now discuss the first lagrangian case. We will be somewhat brief since most of this, is standard material. The huge advantage of the lagrangian case is that to determine the \gls{CB} singular structure we can directly study the masses induced by the vev of the adjoint vector multiplet scalar for the various hypers present in the theory. The extra charged states which can become massless are either W-bosons (where there is an unbroken $SU(2)$ gauge factors) or charged matter, \emph{i.e.} specific components of the hypermultiplets which become massless.

\begin{figure}[h!]
\ffigbox{
\begin{subfloatrow}
\ffigbox[7cm][]{
\begin{tikzpicture}[decoration={markings,
mark=at position .5 with {\arrow{>}}}]
\begin{scope}[scale=1.5]
\node[bbc,scale=.5] (p0a) at (0,0) {};
\node[scale=.5] (p0b) at (0,-2) {};
\node[scale=.8] (t0b) at (0,-2.1) {$USp(4)+6\gls{F}$};
\node[scale=.8] (p1) at (-.8,-1) {$[I_1,\varnothing]$};
\node[scale=.8] (p2) at (.8,-1) {$[I_2^*,\sof(12)]$};
\node[scale=.8] (p3) at (0,-1) {$[I_1,\varnothing]$};
\node[scale=.8] (t2b) at (-.3,-1.5) {{\scriptsize$\big[u^2+v=0\big]$}};
\node[scale=.8] (t3b) at (.7,-1.6) {{\scriptsize$\big[v=0\big]$}};
\draw[red] (p0a) -- (p1);
\draw[red] (p0a) -- (p2);
\draw[red] (p0a) -- (p3);
\draw[red] (p1) -- (p0b);
\draw[red] (p2) -- (p0b);
\draw[red] (p3) -- (p0b);
\end{scope}
\begin{scope}[scale=1.5,xshift=2.85cm]
\node[scale=.8] (p0a) at (0,0) {$\H^{d_{\rm HB}}$};
\node[scale=.8] (t1a) at (-0.2,-.5) {$\df_4$};
\node[scale=.8] (p1) at (-0,-1) {$\cT^{(1)}_{D_4,1}$};
\node[scale=.8] (t1b) at (-.2,-1.5) {$\df_6$};
\node[scale=.8] (p0b) at (0,-2) {$USp(4)+6\gls{F}$};
\draw[blue] (p0a) -- (p1);
\draw[blue] (p1) -- (p0b);
\end{scope}
\end{tikzpicture}}
{\caption{\label{CBSp464}The \gls{Hasse} for the \gls{CB} and the \gls{HB} of the $USp(4)$ gauge theory with six hypermultiplets in the ${\bf 4}$.}}
\end{subfloatrow}\hspace{1cm}%
\begin{subfloatrow}
\capbtabbox[7cm]{%
  \renewcommand{\arraystretch}{1.1}
  \begin{tabular}{|c|c|} 
  \hline
  \multicolumn{2}{|c|}{$USp(4)+6\gls{F}$}\\
  \hline\hline
  $(\D_u,\D_v)$  &\quad (2,4)\quad{} \\
  $24a$ & 74\\  
  $12c$ & 44 \\
$\ff_k$ & $\sof(12)_8$ \\ 
$d_{\rm HB}$&14\\
$h$&0\\
$T({\bf2}\bh)$&0\\
\hline\hline
 \end{tabular}
}{%
  \caption{\label{CcSp464}Central charges, \gls{CB} parameters and \gls{ECB} dimension.}%
}
\end{subfloatrow}}{\caption{\label{TotSp464}Information about the $USp(4)$ $\cN=2$ theory with six hypermultiplet in the fundamental.}}
\end{figure}
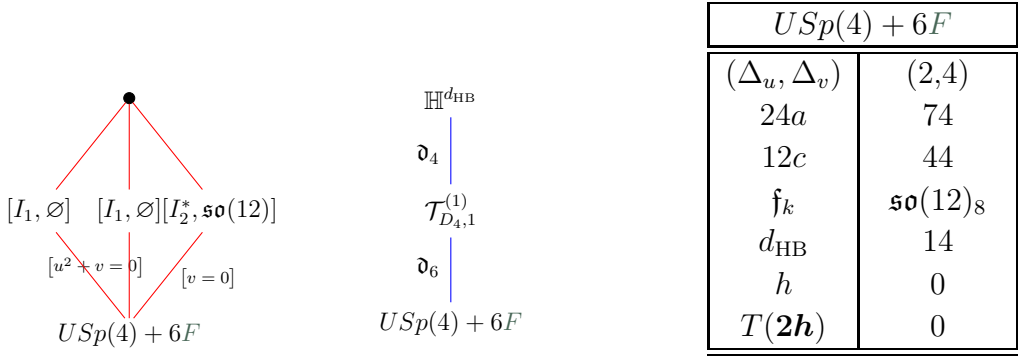

In the $USp(4)$ case there are two inequivalent directions, up to Weyl tranformation, where an $SU(2)$ is left unbroken (corresponding to the long and short simple roots) and which therefore give surely rise to singularities. In one case each hypermultiplet in the ${\bf 4}$ contributes a massless flavor while in the other it contributes no massless matter. It therefore implies that along these two interesting directions we find in one case an $\cN=2$ $SU(2)$ theory with $N_f=6$ and in the other a pure $SU(2)$ theory. The latter theory is asymptotically free and the result of strong coupling is to ``split'' the singularity into two knotted strata each supporting a $[I_1,\varnothing]$. The other low-energy theory is instead IR-free and contributes a \gls{kstr} supporting a $[I_2^*,\sof(12)]$ reproducing the \gls{CB} stratification in figure \ref{CBSp464}. This theory has no \gls{ECB}.

This result can be confirmed both by reproducing the central charges of this theory from \eqref{actotaint}-\eqref{actotbint} and by studying the discriminant locus of the Seiberg-Witten curve which has been worked out explicitly \cite{Argyres:1995fw}. We leave both checks as an exercise for the reader.

The analysis of the \gls{HB} can also be performed in a straightforward manner by analyzing the possible vevs of the hypermultiplets. Rather than performing the group theory analysis, it is quicker to impose our ``non-lagrangian'' constraints. In fact the $\df_6$ Higgsing should support a rank-1 theory with a five dimensional \gls{HB} and no \gls{ECB}. This leads to the only consistent guess $\bTf_{\df_6}\equiv \cT^{(1)}_{D_4,1}$ which can also be checked by carefully turning a minimal vev for the mesons of this theory.

\subsubsection*{\boldsymbol{$USp(4)+4F+AS\ {\rm or} \cT^{(2)}_{D_4,1}$}}\label{sec:D4r2} This is a lagrangian theory which belong to an infinite series of $USp(2n)+4\gls{F}+1\gls{AS}$ gauge theories with  which can be engineered in type \emph{II}B string theory as the worldvolume theory of two \emph{D3} branes probing an exceptional $D_4$ \emph{7}brane singularity \cite{Banks:1996nj,Douglas:1996js,Sen:1996vd,Dasgupta:1996ij}. For $n=2$ $USp(4)\cong SO(5)$ and thus we label the two indices traceless antisymmetric of $USp(4)$ simply as \gls{V}. Because of the string theoretic realization, it is well-known that this theory can be obtained by mass deforming the $\cT^{(2)}_{E_6,1}$ with the mass deformation geometric realized as moving a \emph{D7} away from the $E_6$ exceptional \emph{7}brane to make it a $D_4$ one. This theory has been shown to belong to a larger class of $\cN=2$ theories dubbed $\cT$-theories which have been studied in detail and their properties are summarized in appendix \ref{sec:TandS}. We collect the relevant CFT data as well the stratification in table \ref{Fig:T2G1}. 

\subsubsection*{\boldsymbol{$SU(3)+6F$}}\label{sec:SU3Nf6}

\begin{figure}[h!]
\ffigbox{
\begin{subfloatrow}
\ffigbox[7cm][]{
\begin{tikzpicture}[decoration={markings,
mark=at position .5 with {\arrow{>}}}]
\begin{scope}[scale=1.5]
\node[bbc,scale=.5] (p0a) at (0,0) {};
\node[scale=.5] (p0b) at (0,-2) {};
\node[scale=.8] (t0b) at (0,-2.1) {$SU(3)+6\gls{F}$};
\node[scale=.8] (p1) at (-.8,-1) {$[I_1,\varnothing]$};
\node[scale=.8] (p2) at (.8,-1) {$[I_6,\suf(6)]$};
\node[scale=.8] (p3) at (0,-1) {$[I_1,\varnothing]$};
\node[scale=.8] (t2b) at (-.3,-1.5) {{\scriptsize$\big[u^3+v^2=0\big]$}};
\node[scale=.8] (t3b) at (.7,-1.6) {{\scriptsize$\big[v=0\big]$}};
\draw[red] (p0a) -- (p1);
\draw[red] (p0a) -- (p2);
\draw[red] (p0a) -- (p3);
\draw[red] (p1) -- (p0b);
\draw[red] (p2) -- (p0b);
\draw[red] (p3) -- (p0b);
\end{scope}
\begin{scope}[scale=1.5,xshift=2.85cm]
\node[scale=.8] (p0a) at (0,0) {$\H^{d_{\rm HB}}$};
\node[scale=.8] (p0b) at (0,-2) {$SU(3)+6\gls{F}$};
\node[scale=.8] (p1) at (-0,-1) {$\cT^{(1)}_{D_4,1}$};
\node[scale=.8] (t1a) at (-0.2,-.5) {$\df_4$};
\node[scale=.8] (t1b) at (-.2,-1.5) {$\af_5$};
\draw[blue] (p0a) -- (p1);
\draw[blue] (p1) -- (p0b);
\end{scope}
\end{tikzpicture}}
{\caption{\label{CBSU363}The \gls{Hasse} for the \gls{CB} and \gls{HB} of the $SU(3)$ gauge theory with 6 ${\bf 3}$s.}}
\end{subfloatrow}\hspace{1cm}%
\begin{subfloatrow}
\capbtabbox[7cm]{%
  \renewcommand{\arraystretch}{1.1}
  \begin{tabular}{|c|c|} 
  \hline
  \multicolumn{2}{|c|}{$SU(3)+6\gls{F}$}\\
  \hline\hline
  $(\D_u,\D_v)$  &\quad (2,3)\quad{} \\
  $24a$ &  58\\  
  $12c$ & 34 \\
$\ff_k$ & $\suf(6)_6$ \\ 
$d_{\rm HB}$&10\\
$h$&0\\
$T({\bf2}\bh)$&0\\
\hline\hline
  \end{tabular}
}{%
  \caption{\label{CcSU363}Central charges, \gls{CB} parameters and \gls{ECB} dimension.}%
}
\end{subfloatrow}}{\caption{\label{TotSU363} Information about the $SU(3)$ $\cN=2$ theory with 6 hypermultiplets in the ${\bf 3}$.}}
\end{figure}
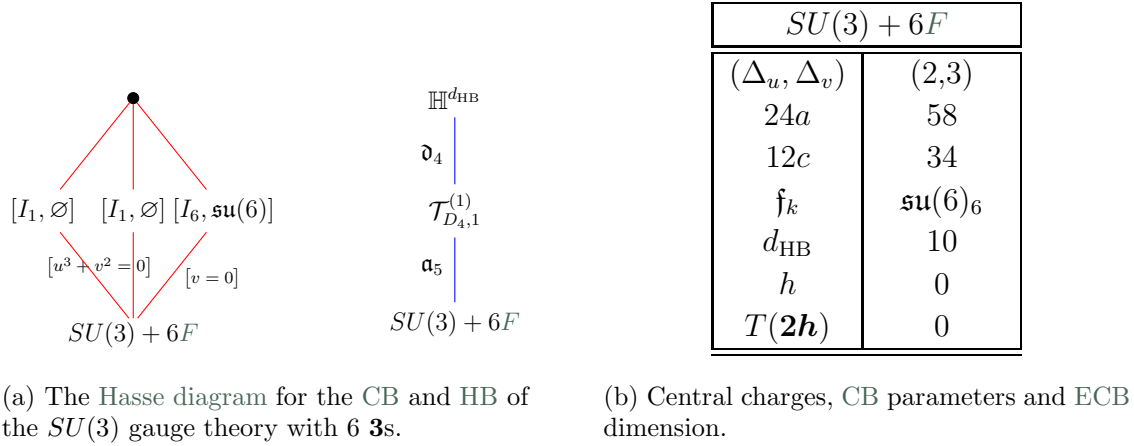

The \gls{CB} stratification of this theory is discussed explicitly, for example, in \cite{Argyres:2018zay,Argyres:2020wmq} while the \gls{HB} in \cite{Bourget:2019aer}. Let us simply discuss a few interesting points. Firstly the double \gls{kstr}, is the ``mark'' of the dyon-monopole singularity of the pure $\cN=2$ $SU(2)$ \gls{CB} solution \cite{Argyres:2018zay}. Secondly the $a$ and $c$ central charges in table \ref{CBSU363} can be readily matched using \eqref{actotaint}-\eqref{actotbint}. Thirdly the $\af_5$ transition, is immediately associated with the \gls{HB} of $\rTf_v$ and it is a useful exercise to check that \eqref{gHWeq} precisely reproduces the central charge of the $\cN=2$ $SU(2)$ theory with $N_f=4$. Our more abstract way of going about characterizing the full moduli space structure, perfectly reproduces the result expected from the lagrangian analysis and does it perhaps even more straightforwardly than the standard way of working with gauge variant fields and equations of motions.

\subsubsection*{\boldsymbol{$2F+SU(2){-}SU(2)+2F$}}\label{sec:SU22b}

Let's start the analysis of this theory from the \gls{CB} perspective. Here there are three higgsing directions which we need to consider. We can turn on a vev for the scalar component of the vector multiplet corresponding to each separate $SU(2)$ or the two combined. In the first case, all the components of the ${(\bf2,2})$ are massive while the other hypermultiplets contribute two flavors for each separate $SU(2)$. Therefore, each one of these Higgsings gives rise to a $SU(2)$ with $N_f=2$ which is reflected each by two strata supporting a $[I_2,\suf(2)]$ (which realize the $\sof(4)$ symmetry of the gauge theory on the \gls{CB}). The other higgsing instead, breaks $SU(2){\times}SU(2)\to U(1){\times}U(1)$, makes massive all the components of the hypers in the $({\bf 2,1})\oplus({\bf1,2})$, but the hyper in the bifundamental contributes two hypers with charge one under an appropriate linear combination of the $U(1)$ which survives. This contributes a singular locus supporting yet another $[I_2,\suf(2)]$. This analysis reproduces the intricate \gls{CB} stratification depicted in figure \ref{CBSU2b} and immediately implies that \gls{h}=0.

\begin{figure}[h!]
\ffigbox{
\begin{subfloatrow}
\ffigbox[7.5cm][]{
\begin{tikzpicture}[decoration={markings,
mark=at position .5 with {\arrow{>}}}]
\begin{scope}[scale=1.5]
\node[bbc,scale=.5] (p0a) at (0,0) {};
\node[scale=.5] (p0b) at (0,-2) {};
\node[scale=.8] (t0b) at (0,-2.1) {$2\gls{F}+SU(2){-}SU(2)+2\gls{F}$};
\node[scale=.5] (p1) at (-1.2,-1) {$[I_2,\suf(2)]$};
\node[scale=.5] (p2) at (1.2,-1) {$[I_2,\suf(2)]$};
\node[scale=.5] (p4) at (.6,-1) {$[I_2,\suf(2)]$};
\node[scale=.5] (p5) at (-.6,-1) {$[I_2,\suf(2)]$};
\node[scale=.5] (p3) at (0,-1) {$[I_2,\suf(2)]$};
\node[scale=.7] (t2b) at (-.3,-1.5) {{\scriptsize$\big[u+v=0\big]$}};
\node[scale=.7] (t3b) at (.7,-1.6) {{\scriptsize$\big[v=0\big]$}};
\draw[red] (p0a) -- (p1);
\draw[red] (p0a) -- (p2);
\draw[red] (p0a) -- (p3);
\draw[red] (p0a) -- (p4);
\draw[red] (p0a) -- (p5);
\draw[red] (p1) -- (p0b);
\draw[red] (p2) -- (p0b);
\draw[red] (p3) -- (p0b);
\draw[red] (p4) -- (p0b);
\draw[red] (p5) -- (p0b);
\end{scope}
\begin{scope}[scale=1.5,xshift=3cm]
\node[scale=.8] (p0a) at (0,0) {$\H^{d_{\rm HB}}$};
\node[scale=.6] (t1a) at (-0.1,-.5) {$\df_4$};
\node[scale=.5] (p1) at (-0,-1) {$\cT^{(1)}_{D_4,1}$};
\node[scale=.5] (p1a) at (-1.2,-1) {$\cT^{(1)}_{D_4,1}$};
\node[scale=.5] (p2) at (1.2,-1) {$\cT^{(1)}_{D_4,1}$};
\node[scale=.5] (p4) at (.6,-1) {$\cT^{(1)}_{D_4,1}$};
\node[scale=.5] (p5) at (-.6,-1) {$\cT^{(1)}_{D_4,1}$};
\node[scale=.6] (t1a) at (-0.1,-1.5) {$\af_1$};
\node[scale=.8] (pb) at (-0,-2) {$2\gls{F}+SU(2){-}SU(2)+2\gls{F}$};
\draw[blue] (p0a) -- (p1);
\draw[blue] (p1) -- (pb);
\draw[blue] (p0a) -- (p1a);
\draw[blue] (p1a) -- (pb);
\draw[blue] (p0a) -- (p2);
\draw[blue] (p2) -- (pb);
\draw[blue] (p0a) -- (p4);
\draw[blue] (p4) -- (pb);
\draw[blue] (p0a) -- (p5);
\draw[blue] (p5) -- (pb);
\end{scope}
\end{tikzpicture}}
{\caption{\label{CBSU2b}The \gls{Hasse} for the \gls{CB} and the \gls{HB} of the $2\gls{F}+SU(2){-}SU(2)+2\gls{F}$.}}
\end{subfloatrow}\hspace{1cm}%
\begin{subfloatrow}
\capbtabbox[7cm]{%
  \renewcommand{\arraystretch}{1.1}
  \begin{tabular}{|c|c|} 
  \hline
  \multicolumn{2}{|c|}{$2\gls{F}{+}SU(2){-}SU(2){+}2\gls{F}$}\\
  \hline\hline
  $(\D_u,\D_v)$  &\quad (2,2)\quad{} \\
  $24a$ & 42\\  
  $12c$ & 24 \\
$\ff_k$ & $\suf(2)_4^5$ \\ 
$d_{\rm HB}$&6\\
$h$&0\\
$T({\bf2}\bh)$&0\\
\hline\hline
 \end{tabular}
}{%
  \caption{\label{CcSU2b}Central charges, \gls{CB} parameters and \gls{ECB} dimension.}%
}
\end{subfloatrow}}{\caption{\label{TotSU22b} Information about the $SU(2)$ $\cN=2$ theory with two hypermultiplets in the $({\bf 2},{\bf1})\oplus({\bf 1},{\bf2})$ and one in the $({\bf2},{\bf2})$}}
\end{figure}

To analyze the \gls{HB} we only consider one of the five possible Higgsing, since there is a symmetry among all of them and will give rise to the same structure (in the \gls{Hasse} in the figure we only depict one of the five transitions). Rather than explicitly doing the calculation, let's use shortcut to identify the rest of the \gls{HB} \gls{Hasse}. From the \gls{CB} analysis we concluded that the theory supported on $\af_1$ should be a rank-1 theory with no \gls{ECB}. Reproducing the total \gls{HB} dimension of the theory readily implies that the \gls{HB} of the rank-1 theory should be five quaternionic dimension, which in turn immediately singles out the $\cN=2$ $SU(2)$ with $N_f=4$ completing our analysis.

\subsubsection*{$\boldsymbol{\cT^{(2)}_{A_2,1}}$}\label{sec:H2r2} This is an \gls{AD} which can be engineered in type \emph{II}B string theory as the worldvolume theory of two \emph{D3} branes probing an $A_2$ \emph{7}brane  singularity \cite{Banks:1996nj,Douglas:1996js,Sen:1996vd,Dasgupta:1996ij}. It is also commonly known as the rank-2 $H_2$ theory. It is well-known that this theory can be obtained by mass deforming the $\cT^{(2)}_{D_4,1}$ and the mass deformation is geometric and corresponds to moving away two \emph{D7} branes to make the $D_4$ \emph{7}brane singularity a $A_2$ one. This theory has been shown to belong to a larger class of $\cN=2$ theories dubbed $\cT$-theories which have been studied in detail and their properties are summarized in appendix \ref{sec:TandS}. We collect the relevant CFT data as well the stratification in table \ref{Fig:T2G1}.

\subsubsection*{$\boldsymbol{D_2(\suf(5))}$}\label{sec:D2SU5}

This theory appears on the \gls{CB} of $\suf(3)$ theory with $N_f=5$ by appropriately tuning their mass parameters \cite{Eguchi:1996vu} and is the rank-2 entry of an infinite series $D_2(\suf(2N+1))$, see below. A useful expression for its \gls{SW} curve was derived in \cite{Argyres:2005wx} but the derivation of its central charges were discussed in \cite{Cecotti:2012jx,Cecotti:2013lda} in the context of geometric engineering were the name was also coined. Recently \cite{Beem:2020pry}, this theory was shown to also arise in twisted $A_4$ class-$\cS$ (while the twisted $A_{2N}$ engineers the $D_2(\suf(2N+1)$). 

Since the hyperelliptic form of the \gls{SW} curve is known in this case, we use this route to characterize the moduli space structure:
\beq
y^2=x^5+(ux+v)^2
\eeq
taking the $x$ discriminant of the RHS is it is straightforward to identify its \gls{CB} stratification which is depicted in figure \ref{CBD2SU5} while the CFT data is reported in \ref{CcD2SU5}.

\begin{figure}[h!]
\ffigbox{
\begin{subfloatrow}
\ffigbox[7cm][]{
\begin{tikzpicture}[decoration={markings,
mark=at position .5 with {\arrow{>}}}]
\begin{scope}[scale=1.5]
\node[bbc,scale=.5] (p0a) at (0,0) {};
\node[scale=.5] (p0b) at (0,-2) {};
\node[scale=.8] (t0b) at (0,-2.1) {$D_2\big[\suf(5)\big]$};
\node[scale=.8] (p1) at (-.5,-1) {$[I_1,\varnothing]$};
\node[scale=.8] (p2) at (.5,-1) {$[I_5,\suf(5)]$};
\node[scale=.8] (t1c) at (.5,-1.7) {{\scriptsize$\big[v=0\big]$}};
\node[scale=.8] (t1c) at (-.7,-1.7) {{\scriptsize$\big[u^5+v^3=0\big]$}};
\draw[red] (p0a) -- (p1);
\draw[red] (p0a) -- (p2);
\draw[red] (p1) -- (p0b);
\draw[red] (p2) -- (p0b);
\end{scope}
\begin{scope}[scale=1.5,xshift=2.5cm]
\node[scale=.5] (p0a) at (0,0) {};
\node[scale=.5] (p0b) at (0,-2) {};
\node[scale=.8] (t0a) at (0,.1) {$\H^{\rm d_{HB}}$};
\node[scale=.8] (t0b) at (0,-2.1) {$D_2\big[\suf(5)\big]$};
\node[scale=.8] (p1) at (-0,-1) {$\cT^{(1)}_{A_2,1}$};
\node[scale=.8] (t1a) at (-0.2,-.5) {$\af_2$};
\node[scale=.8] (t1b) at (-.2,-1.5) {$\af_4$};
\draw[blue] (p0a) -- (p1);
\draw[blue] (p1) -- (p0b);
\end{scope}
\end{tikzpicture}}
{\caption{\label{CBD2SU5}The \gls{Hasse} for the \gls{CB} and \gls{HB} of the $D_2\big[\suf(5)\big]$ theory.}}
\end{subfloatrow}\hspace{1cm}%
\begin{subfloatrow}
\capbtabbox[7cm]{%
  \renewcommand{\arraystretch}{1.1}
  \begin{tabular}{|c|c|} 
  \hline
  \multicolumn{2}{|c|}{$D_2\big[\suf(5)\big]$}\\
  \hline\hline
  $(\D_u,\D_v)$  &\quad $\left(\frac32,\frac52\right)$\quad{} \\
  $24a$ &  42\\  
  $12c$ & 24 \\
$\ff_k$ & $\suf(5)_{5}$ \\ 
$d_{\rm HB}$& 6\\
$h$&0\\
$T({\bf2}\bh)$&0\\
\hline\hline
\end{tabular}
}{%
  \caption{\label{CcD2SU5}Central charges, \gls{CB} parameters and \gls{ECB} dimension.}%
}
\end{subfloatrow}}{\caption{\label{TotD2SU5}Information about the $D_2\big[\suf(5)\big]$ AD theory.}}
\end{figure}

As said, this theory is part of an infinite series dubbed $D_2\big[\suf(2N+1)\big]$. This series is characterized by the following properties:

\begin{itemize}

\item It appears on the \gls{CB} of a $SU(N)$ $\cN=2$ gauge theory with $N_f=2N+1$ flavors.

\item The $D_2\big[\suf(2N+1)\big]$ has rank $N$ and \gls{CB} scaling dimension $\D_i=\frac{2i+1}2$, $i=1,...,N$.

\item The $c$ and $a$ central charges are given by $24a=7 N(N+1)$ and $12 c= 4N(N+1)$.

\item The flavor symmetry is a $\suf(2N+1)_{2N+1}$.

\item The associated vertex operator algebra is conjectured to be the affine current algebra \cite{Xie:2016evu}:
\beq
\mathbb{V}\Big[D_2\big[\suf(2N+1)\big]\Big]=\widehat{\suf(2N+1)}_{-\frac{2N+1}2}
\eeq

\end{itemize}

\subsubsection*{$\boldsymbol{\cT^{(2)}_{A_1,1}}$}\label{sec:H1r2}  This is an \gls{AD} which can be engineered in type II\emph{B} string theory as the worldvolume theory of two \emph{D3} branes probing an $A_1$ \emph{7}brane  singularity \cite{Banks:1996nj,Douglas:1996js,Sen:1996vd,Dasgupta:1996ij}. It is also commonly known as the rank-2 $H_1$ theory. This theory can be obtained by mass deforming the $\cT^{(2)}_{A_2,1}$ by moving away a single \emph{D7} brane to make the $A_2$ \emph{7}brane singularity a $A_1$ one. The mass deformation is thus geometric. This theory has been shown to belong to a larger class of $\cN=2$ theories dubbed $\cT$-theories which have been studied in detail and their properties are summarized in appendix \ref{sec:TandS}. We collect the relevant CFT data as well the stratification in table \ref{Fig:T2G1}.

\subsubsection*{$\boldsymbol{(A_1,D_6)}$}\label{sec:(A1,D6)}

This theory belongs to an infinite series of rank-$n$ AD theory: $(A_1,D_{2n+2})$. We describe their somewhat involved \gls{HB} structure below. This theory was first found on the \gls{CB} of a pure $SO(12)$ $\cN=2$ theory (as we will further discuss below, in general an $(A_1,G)$ theory, where $G$ is a simply laced Lie algebra, arise on special loci of pure $G$ $\cN=2$ gauge theory) and it can be engineered both in class-$\cS$ and type \emph{II}B on a Calabi-Yau threefold hypersurface singularity. 

\begin{figure}[h!]
\ffigbox{
\begin{subfloatrow}
\ffigbox[7cm][]{
\begin{tikzpicture}[decoration={markings,
mark=at position .5 with {\arrow{>}}}]
\begin{scope}[scale=1.5]
\node[bbc,scale=.5] (p0a) at (0,0) {};
\node[scale=.5] (p0b) at (0,-2) {};
\node[scale=.8] (t0b) at (0,-2.1) {$(A_1,D_6)$};
\node[scale=.8] (p1) at (-.5,-1) {$[I_1,\varnothing]$};
\node[scale=.8] (p2) at (.5,-1) {$[I_2,\suf(2)]$};
\node[scale=.8] (t1c) at (.5,-1.7) {{\scriptsize$\big[v=0\big]$}};
\node[scale=.8] (t1c) at (-.7,-1.7) {{\scriptsize$\big[u^5+v^4=0\big]$}};
\draw[red] (p0a) -- (p1);
\draw[red] (p0a) -- (p2);
\draw[red] (p1) -- (p0b);
\draw[red] (p2) -- (p0b);
\end{scope}
\begin{scope}[scale=1.5,xshift=2.5cm]
\node[scale=.5] (p0a) at (0,0) {};
\node[scale=.5] (p0b) at (0,-2) {};
\node[scale=.8] (t0a) at (0,.1) {$\H^{\rm d_{HB}}$};
\node[scale=.8] (t0b) at (0,-2.1) {$(A_1,D_6)$};
\node[scale=.8] (p1) at (-0,-1) {$\cT^{(1)}_{A_1,1}$};
\node[scale=.8] (t1a) at (-0.2,-.5) {$\af_1$};
\node[scale=.8] (t1b) at (-.2,-1.5) {$\af_1$};
\draw[blue] (p0a) -- (p1);
\draw[blue] (p1) -- (p0b);
\end{scope}
\end{tikzpicture}}
{\caption{\label{CBA1D6}The \gls{Hasse} for the \gls{CB} and \gls{HB} of the $(A_1,D_6)$ AD theory.}}
\end{subfloatrow}\hspace{1cm}%
\begin{subfloatrow}
\capbtabbox[7cm]{%
  \renewcommand{\arraystretch}{1.1}
  \begin{tabular}{|c|c|} 
  \hline
  \multicolumn{2}{|c|}{$(A_1,D_6)$}\\
  \hline\hline
  $(\D_u,\D_v)$  &\quad $\left(\frac43,\frac53\right)$\quad{} \\
  $24a$ &  26\\  
  $12c$ & 14 \\
$\ff_k$ & $\suf(2)_{\frac{10}3}\times \uf(1)$ \\ 
$d_{\rm HB}$& 2\\
$h$&0\\
$T({\bf2}\bh)$&0\\
\hline\hline
\end{tabular}
}{%
  \caption{\label{CcA1D6}Central charges, \gls{CB} parameters and \gls{ECB} dimension.}%
}
\end{subfloatrow}}{\caption{\label{TotA1D6}Information about the $(A_1,D_6)$ AD theory.}}
\end{figure}
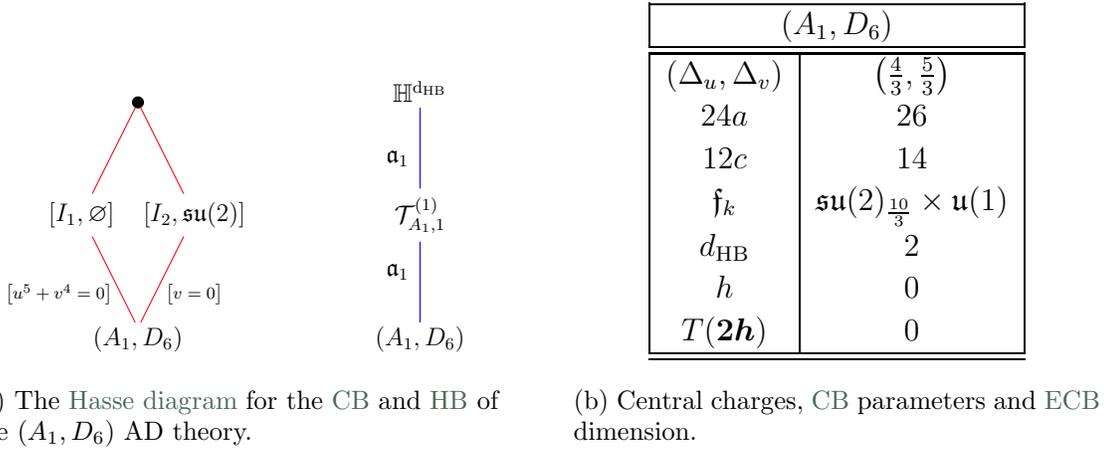

The stratification of the \gls{CB} singular locus can again be read off straightforwardly from the expression of the \gls{SW} curve reported in \cite{Argyres:2005pp}:
\beq\label{disA1D6}
{\rm \gls{SW}\ curve}:y^2=x^6 +x( u x + v)\quad \Rightarrow\quad D^\L_x=v^2\left(256 u^5+3125 v^4\right),
\eeq
and then we conclude that there is a (4,5) \gls{kstr} ($\rTf_{u^5+v^4}\equiv[I_1,\varnothing]$) as well as an \gls{ukstr} at $v=0$. \eqref{disA1D6} leads to the identification $\rTf_v\equiv[I_2,\suf(2)]$. The analysis which leads to the reproduction of the central charges as well as the full characterization of the \gls{Hasse} in figure \ref{CBA1D6}, is largely similar to the one above therefore we won't discuss it and instead focus on discussing the \gls{HB} of this theory.

The \gls{HB} of the $(A_1,D_{2n+2})$ can be elegantly written as the intersection of symplectic varieties \cite{Beem:2017ooy}
\beq
\bar{\O_{[n+1,1]}}\cap \cS_{[n,1,1]},
\eeq
where $\O_{[n+1,1]}$ is the subregular nilpotent orbit of $\slf(n+2)$ and $\cS_{[n,1,1]}$ is the Slodowy slice of the nilpotent orbit associated to the $[n,1,1]$ partition. Adapting the notation that we have used to label the stratification on the \gls{CB} to the stratification of nilpotent orbits: $\cS_{[n,1,1]}\cong\mT(\O_{[n,1,1]},\O_{[n+2]})$, that is, $\cS_{[n,1,1]}$ can be identified as the transverse slice of the nilpotent orbit associated to the $[n,1,1]$ into the principal nilpotent orbit of $\slf(n+2)$. As $n$ increases this space can be quite complicated but for $n=2$ is relatively simple. It is two quaternionic dimensional, and it has only three strata with \gls{elesl}s $\C^2/\Z_2$ (see figure \ref{CBA1D6}). 

From the \gls{CB} analysis, it is obvious that the higgsing is of \gls{gHW} type and therefore we can use \eqref{gHWeq} to compute the central charge of the theory supported on the second \gls{HB} stratum finding $12 c_{\bTf_{\af_1}}=6$ which immediately singles out the $\cT^{(1)}_{A_1,1}$ as depicted in figure \ref{CBA1D6}. And indeed the stratification that we find is compatible with the fact the lower leaf of the \gls{HB} extends into the \gls{MB} of this theory and the low-energy theory living on the second stratum of the \gls{HB} is the rank-1 AD theory $(A_1,D_3)$ which is the same as $(A_1,A_3)$.

\subsubsection*{$\boldsymbol{\cT^{(2)}_{\varnothing,1}}$}\label{sec:H0r2} This \gls{AD} is the first of three bottom theories of the $\ef_8-\sof(20)$ series. It can be engineered in type \emph{II}B string theory as the worldvolume theory of two \emph{D3} branes probing two \emph{D7} branes with mutually non-local charges \cite{Banks:1996nj,Douglas:1996js,Sen:1996vd,Dasgupta:1996ij}. It is also commonly known as the rank-2 $H_0$ theory. This theory can be obtained by mass deforming the $\cT^{(2)}_{A_1,1}$ by moving away a single \emph{D7} brane from the $A_1$ \emph{7}brane singularity. The mass deformation is thus geometric. This theory has been shown to belong to a larger class of $\cN=2$ theories dubbed $\cT$-theories which have been studied in detail and their properties are summarized in appendix \ref{sec:TandS}. We collect the relevant CFT data as well the stratification in table \ref{Fig:T2G1}.

\subsubsection*{$\boldsymbol{(A_1,D_5)}$} \label{sec:(A1,D5)}
This theory appears on a special locus of a $\cN=2$ $SO(10)$ pure gauge theory \cite{Eguchi:1996vu,Eguchi:1996ds} and it belongs instead to an infinite series of AD theory with a $\C^2/\Z_2$ \gls{HB}: $(A_1,D_{2n+1})$. This is the second bottom theory of the $\ef_8-\sof(20)$ series. These can be geometrically engineered in type \emph{II}B on a Calabi-Yau three-fold hypersurface singularity specified by the two simply laced Lie algebras $A_1$ and $D_{2n+1}$ \cite{Cecotti:2010fi}. It can also be obtained in $A_N$ class-$\cS$ with a full puncture and a $I_{2,3}$ irregular puncture \cite{Xie:2012hs} (in general $(A_1,D_N)$ theories can be obtained in $A_N$ class-$\cS$ with a full puncture and an irregular $I_{2,N-2}$ irregular puncture). Although the class-$\cS$ description gives a formulation for the theory's \gls{CB} geometry, we instead read off the stratification of the \gls{CB} singular locus, following the remarks at the end of section \ref{sec:CBstr}, using a different form for the \gls{SW} curve reported in \cite{Argyres:2005wx},
\beq\label{disA1D5}
{\rm \gls{SW}\ curve}:y^2=x^5 +x( u x + v)\quad \Rightarrow\quad D_x^\L=v^2\left(27 u^4-256 v^3\right).
\eeq
This implies that there is a (3,4) \gls{kstr} as well as an \gls{ukstr} at $v=0$.

\begin{figure}[h!]
\ffigbox{
\begin{subfloatrow}
\ffigbox[7cm][]{
\begin{tikzpicture}[decoration={markings,
mark=at position .5 with {\arrow{>}}}]
\begin{scope}[scale=1.5]
\node[bbc,scale=.5] (p0a) at (0,0) {};
\node[scale=.8] (p0b) at (0,-2) {};
\node[scale=.8] (t0b) at (0,-2.1) {$(A_1,D_5)$};
\node[scale=.8] (p1) at (-.5,-1) {$[I_1,\varnothing]$};
\node[scale=.8] (p2) at (.5,-1) {$[I_2,\suf(2)]$};
\node[scale=.8] (t1c) at (-.7,-1.7) {{\scriptsize$\big[u^4+v^3=0\big]$}};
\node[scale=.8] (t1c) at (.5,-1.7) {{\scriptsize$\big[v=0\big]$}};
\draw[red] (p0a) -- (p1);
\draw[red] (p0a) -- (p2);
\draw[red] (p1) -- (p0b);
\draw[red] (p2) -- (p0b);
\end{scope}
\begin{scope}[scale=1.5,xshift=2.5cm]
\node[scale=.8] (p0b) at (0,-2) {$(A_1,D_5)$};
\node[scale=.8] (p1) at (-0,-1) {$\cT^{(1)}_{\varnothing,1}$};
\node[scale=.8] (t1b) at (-.2,-1.5) {$\af_1$};
\draw[blue] (p1) -- (p0b);
\end{scope}
\end{tikzpicture}}
{\caption{\label{CBA1D5}The \gls{Hasse} for the Coulomb and Higgs branch of the $(A_1,D_5)$ AD theory.}}
\end{subfloatrow}\hspace{1cm}%
\begin{subfloatrow}
\capbtabbox[7cm]{%
  \renewcommand{\arraystretch}{1.1}
  \begin{tabular}{|c|c|} 
  \hline
  \multicolumn{2}{|c|}{$(A_1,D_5)$}\\
  \hline\hline
  $(\D_u,\D_v)$  &$\quad \left(\frac65,\frac85\right)\quad{}$ \\
  $24a$ & $ \frac{114}5$\\  
  $12c$ & 12 \\
$\ff_k$ & $\suf(2)_{\frac{16}5}$ \\ 
$d_{\rm HB}$& 1\\
$h$&0\\
$T({\bf2}\bh)$&0\\
\hline\hline
  \end{tabular}
}{%
  \caption{\label{CcA1D5}Central charges, \gls{CB} parameters and \gls{ECB} dimension.}%
}
\end{subfloatrow}}{\caption{\label{TotA1D5}Information about the $(A_1,D_5)$ AD theory.}}
\end{figure}
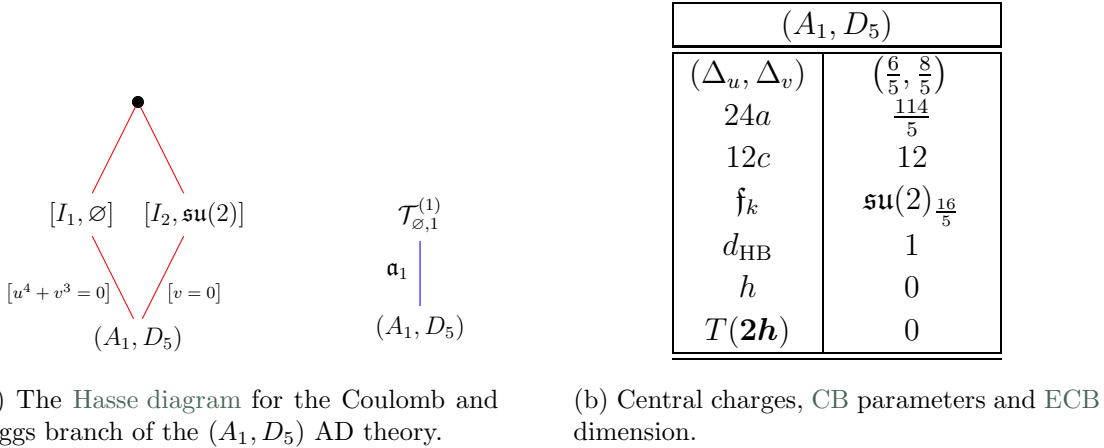

As we did in other cases, we can use the extra information provided by the order of the zeros of the discriminant to further characterize the \gls{CB}. As before we infer that $\rTf_{u^4+v^3}\equiv [I_1,\varnothing]$. There is now an ambiguity in identifying $\rTf_v$ as both an $[I_2,\suf(2)]$ and a $\cT^{(1)}_{\varnothing,1}$ would be compatible with the discriminant \eqref{disA1D5}. This ambiguity can be resolved using the \gls{UVIR} which implies that the $\suf(2)$ flavor symmetry has to be realized as a flavor symmetry of a rank-1 theory on the \gls{CB} and we therefore conclude that $\rTf_v\equiv[I_2,\suf(2)]$. This perfectly reproduces the level $k_{\suf(2)}=\frac{16}5$ (via \eqref{actotcint}) and reproduces the \gls{CB} stratification shown in figure \ref{CBA1D5}. 

We observe that this \gls{CB} stratification has two implications which are also consistent with known facts about this theory \cite{Beem:2019tfp}. First, the entire \gls{HB} of the ($A_1,D_5$) extends over its \gls{CB} and it is therefore a \gls{MB}, and secondly the low-energy theory on the generic point of the \gls{MB}, of the theory is the rank-1 ($A_1,A_2$).

\subsubsection*{$\boldsymbol{(A_1,A_5)}$}\label{sec:(A1,A5)}

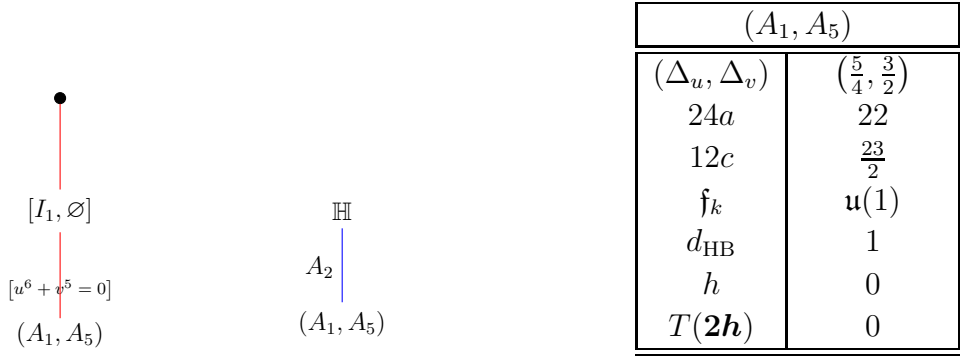
\begin{figure}[h!]
\ffigbox{
\begin{subfloatrow}
\ffigbox[7cm][]{
\begin{tikzpicture}[decoration={markings,
mark=at position .5 with {\arrow{>}}}]
\begin{scope}[scale=1.5]
\node[bbc,scale=.5] (p0a) at (0,0) {};
\node[scale=.5] (p0b) at (0,-2) {};
\node[scale=.8] (t0b) at (0,-2.1) {$(A_1,A_5)$};
\node[scale=.8] (p1) at (-0,-1) {$ [I_1,\varnothing]$};
\node[scale=.8] (t1c) at (0,-1.7) {{\scriptsize$\big[u^6+v^5=0\big]$}};
\draw[red] (p0a) -- (p1);
\draw[red] (p1) -- (p0b);
\end{scope}
\begin{scope}[scale=1.5,xshift=2.5cm]
\node[scale=.8] (p0b) at (0,-2) {$(A_1,A_5)$};
\node[scale=.8] (p1) at (-0,-1) {$\H$};
\node[scale=.8] (t1b) at (-.2,-1.5) {$A_2$};
\draw[blue] (p1) -- (p0b);
\end{scope}
\end{tikzpicture}}
{\caption{\label{CBA1A5}The \gls{Hasse} for the \gls{CB} and \gls{HB} of the $(A_1,A_5)$ AD theory.}}
\end{subfloatrow}\hspace{1cm}%
\begin{subfloatrow}
\capbtabbox[7cm]{%
  \renewcommand{\arraystretch}{1.1}
  \begin{tabular}{|c|c|} 
  \hline
  \multicolumn{2}{|c|}{$(A_1,A_5)$}\\
  \hline\hline
  $(\D_u,\D_v)$  &\quad $\left(\frac54,\frac32\right)$\quad{} \\
  $24a$ &  22\\  
  $12c$ & $\frac{23}2$ \\
$\ff_k$ &$ \uf(1) $\\ 
$d_{\rm HB}$& 1\\
$h$&0\\
$T({\bf2}\bh)$&0\\
\hline\hline
  \end{tabular}
}{%
  \caption{\label{CcA1A5}Central charges, \gls{CB} parameters and \gls{ECB} dimension.}%
}
\end{subfloatrow}}{\caption{\label{TotA1A5}Information about the $(A_1,A_5)$ AD theory.}}
\end{figure}

This theory, appears on a special locus of the pure $\cN=2$ $SU(6)$ gauge theory. It can be engineered in class-$\cS$ and type \emph{II}B string theory like other \gls{AD}. It also belongs to an infinite series of rank-$n$ AD theory, called $(A_1,A_{2n+1})$, whose \gls{HB} can be written somewhat homogeneously as $\C^2/\Z_{n+1}$. The stratification of the \gls{CB} singular locus can be again read off straightforwardly from the expression of the \gls{SW} curve reported in \cite{Argyres:2005pp}:
\beq\label{disA1A5}
{\rm \gls{SW}\ curve}:y^2=x^6 + u x + v\quad \Rightarrow\quad D^\L_x=\left(3125 u^6-46656 v^5\right),
\eeq
and again it implies that $\rTf_{u^6+v^5}\equiv[I_1,\varnothing]$. Performing an analysis analogous to the one above, the rest of the \gls{Hasse}, shown in figure \ref{CBA1A5}, can be completely characterized and the central $(a,c)$ correctly reproduced via the \gls{ccf}. The theory supported on the \gls{CB} singular locus has no \gls{HB} but the rank-2 theory at the origin has a non-trivial \gls{HB}.  This in turn implies that the low-energy theory on the generic point of the \gls{MB}, is trivial.

\subsubsection*{$\boldsymbol{(A_1,A_4)}$}\label{sec:(A1,A4)}

Finally the last bottom theory of the $\ef_8-\sof(20)$ series. This theory is the rank-2 entry of an infinite series of rank-$n$ $\cN=2$ \gls{SCFT}s with trivial \gls{HB}, which is often labeled as $(A_1,A_{2n})$. It appears on a special locus on the \gls{CB} of a pure $SU(5)$ $\cN=2$ gauge theory \cite{Eguchi:1996vu,Eguchi:1996ds}. It can also be geometrically engineered in type \emph{II}B string theory on a Calabi-Yau 3-fold hypersurface singularity \cite{Cecotti:2010fi}. This is where the name $(A_1,A_{2n})$ comes from as these two labels uniquely identify the polynomial cutting the 3-fold singularity. Finally, this theory can also be obtained in class-$\mathcal{S}$ compactifying an 6d $A_1$ (2,0) theory on a sphere with a single, type I, irregular puncture \cite{Xie:2012hs}. Its VOA is conjectured to be the (2,5) Virasoro minimal model \cite{Cordova:2015nma,Beem:2017ooy}.

\begin{figure}[h!]
\ffigbox{
\begin{subfloatrow}
\ffigbox[7cm][]{
\begin{tikzpicture}[decoration={markings,
mark=at position .5 with {\arrow{>}}}]
\begin{scope}[scale=1.5]
\node[bbc,scale=.5] (p0a) at (0,0) {};
\node[bbc,scale=.5] (p0b) at (0,-2) {};
\node[scale=.8] (t0b) at (0,-2.3) {($A_1,A_4$)};
\node[scale=.8] (p1) at (-0,-1) {$[I_1,\varnothing]$};
\node[scale=.8] (t1c) at (0,-1.7) {{\scriptsize$\big[u^5+v^4=0\big]$}};\draw[red] (p0a) -- (p1);
\draw[red] (p1) -- (p0b);
\end{scope}
\begin{scope}[scale=1.5,xshift=2.5cm]
\node[bbc,scale=.5] (p0b) at (0,-2) {};
\node[scale=.8] (t0b) at (0,-2.3) {($A_1,A_4$)};
\end{scope}
\end{tikzpicture}}
{\caption{\label{CBA1A4}The \gls{Hasse} for the \gls{CB} and the, trivial, \gls{HB} of the $(A_1,A_4)$ AD theory.}}
\end{subfloatrow}\hspace{1cm}%
\begin{subfloatrow}
\capbtabbox[7cm]{%
  \renewcommand{\arraystretch}{1.1}
  \begin{tabular}{|c|c|} 
  \hline
  \multicolumn{2}{|c|}{$(A_1,A_4)$}\\
  \hline\hline
  $(\D_u,\D_v)$  &\quad $\left(\frac87,\frac{10}7\right)$\quad{} \\
  $24a$ &  $\frac{134}7$\\  
  $12c$ &$\frac{68}7 $\\
$\ff_k$ & $\varnothing$ \\ 
$d_{\rm HB}$& 0\\
$h$&0\\
$T({\bf2}\bh)$&0\\
\hline
\hline
  \end{tabular}
}{%
  \caption{\label{CcA1A4}Central charges, \gls{CB} parameters and \gls{ECB} dimension.}%
}
\end{subfloatrow}}{\caption{\label{TotA1A4} Information about the $(A_1,A_4)$ AD theory.}}
\end{figure}
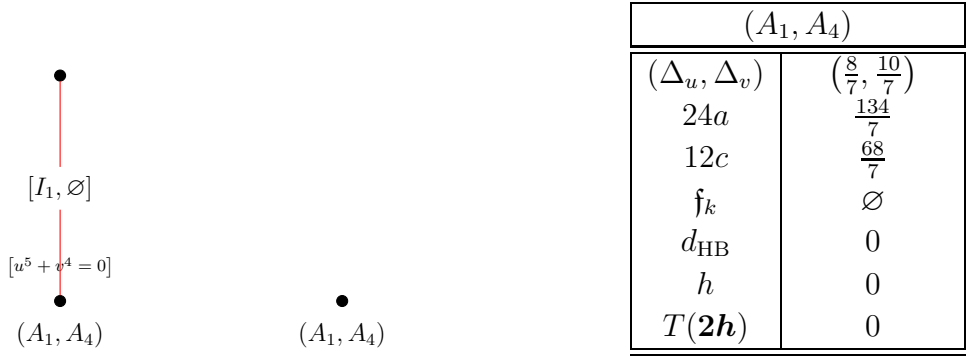

The stratification of the \gls{CB} can be read off directly from the discriminant of the curve presented in \cite{Argyres:2005wx}, as remarked at the end of section \ref{sec:CBstr}.  For the $(A_1,A_4)$ we have:
\beq
{\rm \gls{SW}\ curve}:y^2=x^5 + u x + v\quad \Rightarrow\quad D^\L_x=256 u^5+3125 v^4,
\eeq
and therefore we readily conclude that $\rTf_{u^5+v^4}\equiv[I_1,\varnothing]$. It is an instructive exercise to check that  with this information we can reproduce the $a$ and $c$ central charges in table \ref{CcA1A4} using our \gls{ccf}.

A curious phenomenon is that the \gls{kstr} is an \emph{irregular geometry} ($\bar{\rSf}_{u^5+v^4}\equiv I_0^{*(3)}$) as it can be read off by analyzing more closely the special K\"ahler structure of the \gls{kstr}. The uniformizing parameter \cite{Argyres:2020wmq} for this hypersurface is
\beq
t_{(A_1,A_4)} \sim (u^5)^{\frac1{20}} \sim (v^4)^{\frac1{20}}
\quad \Rightarrow\quad 
\D_t=\frac27.
\eeq
For more details on irregular geometries and how to determine the uniformizing parameter for a one-complex dimensional stratum see \cite{Argyres:2017tmj,Argyres:2020wmq}.

\begin{center}
\rule[1mm]{2cm}{.4pt}\hspace{1cm}$\circ$\hspace{1cm} \rule[1mm]{2cm}{.4pt}
\end{center}

\subsection{$\spf(12)-\spf(8)-\ff_4$ series}

This is the second largest series with again multiple top theories (three) to which connect a total of eleven $\cN=2$ \gls{SCFT}s. This series does not contain any \gls{AD}.

\subsubsection*{\boldsymbol{$\spf(12)_8$}}\label{sec:t22}

This theory is one of the theories sitting at the top of the $\spf(12)-\spf(8)-\ff_4$ series and can be obtained, for example, in the $\Z_2$ twisted $D_4$ class-$\cS$ \cite{Chacaltana:2013oka} where most of the CFT data reported below is computed.

\begin{figure}[h!]
\ffigbox{
\begin{subfloatrow}
\ffigbox[8.5cm][]{
\begin{tikzpicture}[decoration={markings,
mark=at position .5 with {\arrow{>}}}]
\begin{scope}[scale=1.5]
\node[bbc,scale=.5] (p0a) at (0,0) {};
\node[scale=.5] (p0b) at (0,-2) {};
\node[scale=.8] (t0b) at (0,-2.1) {$\spf(12)_8$};
\node[scale=.8] (p1) at (-1,-1) {$[I_1,\varnothing]$};
\node[scale=.8] (p2) at (1,-1) {$[I_{12},\suf(12)]_{\Z_2}$};
\node[scale=.8] (p3) at (0,-1) {$[I_1,\varnothing]$};
\node[scale=.8] (t2b) at (-.3,-1.5) {{\scriptsize$\big[u^3+v^2=0\big]$}};
\node[scale=.8] (t3b) at (.75,-1.6) {{\scriptsize$\big[u=0\big]$}};
\draw[red] (p0a) -- (p1);
\draw[red] (p0a) -- (p2);
\draw[red] (p0a) -- (p3);
\draw[red] (p1) -- (p0b);
\draw[red] (p2) -- (p0b);
\draw[red] (p3) -- (p0b);
\end{scope}
\begin{scope}[scale=1.5,xshift=2.85cm]
\node[scale=.5] (p3) at (0,1) {};
\node[scale=.8] (t0a) at (0,1.1) {$\H^{\rm d_{HB}}$};
\node[scale=.8] (tp2) at (-.2,.5) {$\ef_6$};
\node[scale=.8] (p2) at (0,0) {$\cT^{(1)}_{E_6,1}$};
\node[scale=.8] (tp2) at (-.2,-.5) {$\cf_5$};
\node[scale=.8] (p1) at (0,-1) {$\cS^{(1)}_{E_6,2}$};
\node[scale=.8] (tp1) at (-.2,-1.5) {$\cf_6$};
\node[scale=.5] (p0) at (0,-2) {};
\node[scale=.8] (t0b) at (0,-2.1) {$\spf(12)_8$};
\draw[blue] (p0) -- (p1);
\draw[blue] (p1) -- (p2);
\draw[blue] (p2) -- (p3);
\end{scope}
\end{tikzpicture}}
{\caption{\label{CBhTh30}The Coulomb and Higgs stratification of $\spf(12)_8$.}}
\end{subfloatrow}\hspace{1cm}%
\begin{subfloatrow}
\capbtabbox[5cm]{%
  \renewcommand{\arraystretch}{1.1}
  \begin{tabular}{|c|c|} 
  \hline
  \multicolumn{2}{|c|}{$\spf(12)_8$}\\
  \hline\hline
  $(\D_u,\D_v)$  &\quad $\left(4,6\right)$\quad{} \\
  $24a$ & 130\\  
  $12c$ & 76\\
$\ff_k$ & $\spf(12)_8$ \\ 
$d_{\rm HB}$& 22\\
$h$&0\\
$T({\bf2}\bh)$&0\\
\hline\hline
\end{tabular}
}{%
  \caption{\label{CchTh29}Central charges, \gls{CB} parameters and \gls{ECB} dimension.}%
}
\end{subfloatrow}}{\caption{\label{TothTh30}Information about the $\spf(12)_8$.}}
\end{figure}

This theory is \gls{totally} and therefore we would expect an at least semi-simple flavor symmetry. Surprisingly $\ff$ is instead simple and its \gls{HB} \gls{Hlin} and it is therefore reasonable to guess that one of the two allowed \gls{ukstr} associated with each \gls{higgs}, simply supports no low-energy rank-1 theory. By looking at the level of the $\spf(12)$ flavor symmetry, the natural guess is that $\rTf_u\equiv [I_{12},\suf(12)]_{\Z_2}$, where the subscript refer to a $\Z_2$ discrete gauging, while $\rTf_v\equiv I_0$, which is another way of saying that there are no charged states becoming massless at $v=0$. The fact that discretely gauged theories can appear on singular strata of the \gls{CB}, was already noticed in \cite{Argyres:2020wmq} but a closer analysis reveals that when this happens the \gls{CB} analysis is particularly constrained. You can for example derive, in this case, not only that \gls{h}=0 but also that the theory $\bTf_{\cf_6}$ supported on the $\cf_6$ transition associated to the \gls{HB} of this discretely gauged theory, should support a rank-1 theory with \gls{h}=5. Before confirming this fact directly with a \gls{HB} analysis, let's confirm that the \gls{CB} analysis we just performed is correct. For that we can as usual apply the \gls{ccf} \eqref{actotaint}-\eqref{actotbint} which works out nicely and furthermore suggests that there are two knotted strata each supporting a $[I_1,\varnothing]$. Unfortunately there is another possible solution which is compatible with everything that we know, \emph{i.e.} a single \gls{kstr} with $\rTf_{u^3+v^2}=[II,\varnothing]$. In figure \ref{CBhTh30}, we pick the former but for not better reason than symmetries with the other \gls{ukstr}, also of $I_n$ type. It is important to clarify, that these two choices are indistinguishable as far as the analysis we are performing here goes but there is of course an easy way to distinguish them as the two rank-1 theories are associated with different monodromies. In fact one is parabolic (the $I_1$s) and the other is elliptic. Thus we expect that a more careful analysis of the global structure of the \gls{CB} could very likely distinguish the two cases.

With all this information at hand, the \gls{HB} is straightforward. In fact the prediction of the higgsing from the \gls{CB} are enough to uniquely identify $\bTf_{\cf_6}\equiv \cS^{(1)}_{E_6,2}$ from which we can determine the rest of the Higgsings and thus the full \gls{HB} stratification. We leave it up to the reader to check that this identification is indeed consistent with \eqref{gHWeq} but it reassuring that the entire structure holds up together so well.

\subsubsection*{\boldsymbol{$\spf(4){\times}\spf(8)_8$}}\label{sec:t23}

This theory is also at top of the series and was initially obtained in the untwisted $D_4$ class-$\cS$ series \cite{Chacaltana:2011ze} where most of the CFT data reported in figure \ref{CchTh18} was initially computed. In the original paper, various dual description of this theory are also discussed.

\begin{figure}[h!]
\ffigbox{
\begin{subfloatrow}
\ffigbox[8.5cm][]{
\begin{tikzpicture}[decoration={markings,
mark=at position .5 with {\arrow{>}}}]
\begin{scope}[scale=1.5]
\node[bbc,scale=.5] (p0a) at (0,-1) {};
\node[scale=.5] (p0b) at (0,-3) {};
\node[scale=.8] (t0b) at (0,-3.1) {$\spf(4){\times}\spf(8)_8$};
\node[scale=.8] (p3) at (1,-2) {\ \ $[I_1,\varnothing]$};
\node[scale=.8] (p2) at (0,-2) {\ \ $[I_8,\suf(8)]_{\Z_2}$};
\node[scale=.8] (p1) at (-1,-2) {$[I_1^*,\spf(4)]$\ \ };
%\node[scale=.8] (t1a) at (-.6,-.4) {$I_0^*$};
%\node[scale=.8] (t2c) at (-.7,-1.6) {$K_{\D_7}$};
\node[scale=.8] (t1c) at (.85,-2.75) {{\scriptsize$\big[u^3+v^2=0\big]$}};
\node[scale=.8] (t2c) at (-.25,-2.4) {{\scriptsize$\big[u=0\big]$}};
%\node[scale=.8] (t2c) at (.7,-1.6) {$K_{2\D_7}$};
\node[scale=.8] (t3c) at (-.7,-2.7) {{\scriptsize$\big[v=0\big]$}};
\draw[red] (p0a) -- (p1);
\draw[red] (p0a) -- (p2);
\draw[red] (p0a) -- (p3);
\draw[red] (p1) -- (p0b);
\draw[red] (p2) -- (p0b);
\draw[red] (p3) -- (p0b);
\end{scope}
\begin{scope}[scale=1.5,xshift=2.85cm]
\node[scale=.5] (p0a) at (0.7,0) {};
\node[scale=.8] (t0a) at (.7,.1) {$\H^{\rm d_{HB}}$};
\node[scale=.8] (tp2) at (.2,-.4) {$\ef_6$};
\node[scale=.8] (tp2) at (1.25,-.45) {$\ef_7$};
\node[scale=.8] (p1b) at (0,-1) {$\cT^{(1)}_{E_6,1}$};
\node[scale=.8] (p1a) at (1.4,-1) {$\cT^{(1)}_{E_7,1}$};
\node[scale=.8] (tp2) at (-.5,-1.4) {$\cf_5$};
\node[scale=.8] (tp2) at (.25,-1.6) {$\af_7$};
\node[scale=.8] (tp2) at (1.25,-1.5) {$\af_1$};
\node[scale=.8] (p2a) at (.7,-2) {\hyperref[sec:t9]{$\suf(2)_6{\times}\suf(8)_8$}};
\node[scale=.8] (p2b) at (-.7,-2) {$\cS^{(1)}_{E_6,2}$};
\node[scale=.8] (tp1) at (-.5,-2.6) {$\cf_4$};
\node[scale=.8] (tp2) at (.5,-2.6) {$\cf_2$};
\node[scale=.5] (p0b) at (0,-3) {};
\node[scale=.8] (t0b) at (0,-3.1) {$\spf(4){\times}\spf(8)_8$};
\draw[blue] (p0a) -- (p1a);
\draw[blue] (p0a) -- (p1b);
\draw[blue] (p1a) -- (p2a);
\draw[blue] (p1b) -- (p2a);
\draw[blue] (p1b) -- (p2b);
\draw[blue] (p2a) -- (p0b);
\draw[blue] (p2b) -- (p0b);
\end{scope}
\end{tikzpicture}}
{\caption{\label{CBhTh18}The Coulomb and Higgs stratification of $\spf(4){\times}\spf(8)_8$}}
\end{subfloatrow}\hspace{1cm}%
\begin{subfloatrow}
\capbtabbox[5cm]{%
  \renewcommand{\arraystretch}{1.1}
  \begin{tabular}{|c|c|} 
  \hline
  \multicolumn{2}{|c|}{$\spf(4){\times}\spf(8)_8$}\\
  \hline\hline
  $(\D_u,\D_v)$  &\quad $\left(4,6\right)$\quad{} \\
  $24a$ &  128\\  
  $12c$ & 74\\
$\ff_k$ & $\spf(4)_7{\times}\spf(8)_8$ \\ 
$d_{\rm HB}$& 20\\
$h$&2\\
$T({\bf2}\bh)$&1\\
\hline\hline
\end{tabular}
}{%
  \caption{\label{CchTh18}Central charges, \gls{CB} parameters and \gls{ECB} dimension.}%
}
\end{subfloatrow}}{\caption{\label{TothTh18}Information about the $\spf(4){\times}\spf(8)_8$.}}
\end{figure}

This theory is \gls{totally}, we therefore expect and interesting moduli space structure. First notice that we can immediately identify how one of the two $\spf$ factor, the $\spf(8)$, is realized on the \gls{CB}. In fact we notice that it has a level which doubles $\D_u$ and using the \doub\ directly leads to the identification $\rTf_u\equiv[I_8,\suf(8)]_{\Z_2}$. On the other hand the relation between $k_{\ff_{\spf(4)}}$ and $\D_v$ suggests that the $\spf(4)$ factor is instead realized as isometry of an \gls{ECB} leading to $\rTf_v\equiv [I_1^*,\spf(4)]$. As usual we can cross-check this identification by plugging the corresponding \gls{bi} into \eqref{actotaint}-\eqref{actotbint} to correctly reproduce the values of the $a$ and $c$ central charges in table \ref{CchTh18}. This also fixes the theory supported on the knotted strata thus completing our analysis of the \gls{CB} stratification. 

Let's start our analysis of the \gls{HB} from the rank decreasing transition. This is a $\cf_4$ and the \gls{CB} analysis suggests that this stratum should support a theory with a five dimensional \gls{ECB}. The theory supported on this stratum has to be a rank-1 theory, thus the \gls{ECB} information just derived uniquely identify $\bTf_{\cf_4}\equiv \cS^{(1)}_{E_6,2}$. This identification can be of course cross-checked by applying \eqref{gHWeq} and reproducing $12c_{\bTf_{\cf_4}}=49$. The higgsing corresponding to the \gls{ECB} is also of \gls{gHW} type and therefore \eqref{gHWeq} can be used to derive $12c_{\bTf_{\cf_4}}=54$. Unfortunately this is information is not enough to identify the rank-2 theory supported on the stratum. As it is often the case, the degeneracy can be lifted by imposing the simple condition that the total \gls{HB} should be 20 dimensional. This leads to the unique identification $\bTf_{\cf_2}\equiv \sof(14)_{10}{\times}\uf(1)$. The subsequent Higgsings can be reproduced by studying the \gls{HB} of $\bTf_{\cf_2}$ and $\bTf_{\cf_4}$ which give rise to the intricate \gls{Hasse} in figure \ref{CBhTh18}.

\subsubsection*{$\boldsymbol{\cT^{(2)}_{E_6,2}}$}\label{sec:TE62} This theory is one of the top theories of the $\spf(12)-\spf(8)-\ff_4$ series. It can be obtained in class-$\cS$, for example, in the untwisted $D_4$ \cite{Chacaltana:2011ze}. It was also recently shown to be obtainable by higgsing the $\cN=2$ $\cS$-fold $\cS^{(1)}_{E_6,2}$ or by a twisted compactification of a \sd (1,0) theory \cite{Giacomelli:2020jel} or as a wordvolume theory of two \emph{D3} branes probing an exceptional $E_6$ \emph{7}brane singularity in the presence of an $\cS$-fold without flux \cite{Giacomelli:2020gee}. Theories probing an exceptional \emph{7}brane plus an $\cS$-fold without fluxes are more generally dubbed $\cN=2$ $\cT$-theories and their properties are summarized in appendix \ref{sec:TandS}. The CFT properties as well as the stratification of this particular theory can be found in table \ref{Fig:T2Gl}.

\subsubsection*{\boldsymbol{$\suf(2)_8{\times}\spf(8)_6$}}\label{sec:t25}

This theory can be obtained by mass deforming both theory \hyperref[sec:t22]{$\spf(12)_8$} and \hyperref[sec:t23]{$\spf(4){\times}\spf(8)_8$} \cite{Martone:2021drm}. It was introduced for the first time in the context of twisted $\Z_2$ $A_3$ class-$\cS$ theories \cite{Chacaltana:2012ch}. In the original paper interesting S-dualities of this theory are discussed as well as most of the CFT data reported in table \ref{CcTh9} computed. This information will be leveraged here to fully understand the moduli space structure of this theory. 

This theory is again characterized by having two simple flavor symmetry factors which reflect the fact that the theory is \gls{totally}. This in turn implies that the \gls{HB} is not \gls{Hlin}. Looking at the levels we can use the \doub\ to identify both theories realizing the two simple flavor symmetry factors. In this particular case this would imply that the $\suf(2)_6$ is realized by $\rTf_u\equiv [I_2,\suf(2)]$ while the $\spf(8)_6$ as a $\rTf_v\equiv [I_8,\suf(8)]_{\Z_2}$. We can immediately support this guess by checking that with an $[I_1,\varnothing]$ supported on the unknotted strata, the $c$ central charge of the theory is perfectly reproduced by \eqref{actotbint}.

\begin{figure}[h!]
\ffigbox{
\begin{subfloatrow}
\ffigbox[8cm][]{
\begin{tikzpicture}[decoration={markings,
mark=at position .5 with {\arrow{>}}}]
\begin{scope}[scale=1.5]
\node[bbc,scale=.5] (p0a) at (0,0) {};
\node[scale=.5] (p0b) at (0,-2) {};
\node[scale=.8] (t0b) at (0,-2.1) {$\suf(2)_8{\times}\spf(8)_6$};
\node[scale=.8] (p3) at (0,-1) {$[I_2,\suf(2)]$\ \ };
\node[scale=.8] (p2) at (1.,-1) {\ \ $[I_8,\suf(8)]_{\Z_2}$};
\node[scale=.8] (p1) at (-1,-1) {$[I_1,\varnothing]$\ \ };
%\node[scale=.8] (t1a) at (-.6,-.4) {$I_0^*$};
%\node[scale=.8] (t2c) at (-.7,-1.6) {$K_{\D_7}$};
\node[scale=.8] (t1c) at (-.85,-1.7) {{\scriptsize$\big[u^4+v^3=0\big]$}};
%\node[scale=.8] (t2c) at (.7,-1.6) {$K_{2\D_7}$};
\node[scale=.8] (t1c) at (.7,-1.7) {{\scriptsize$\big[u=0\big]$}};
\node[scale=.8] (t1c) at (.28,-1.35) {{\scriptsize$\big[v=0\big]$}};
\draw[red] (p0a) -- (p1);
\draw[red] (p0a) -- (p2);
\draw[red] (p0a) -- (p3);
\draw[red] (p1) -- (p0b);
\draw[red] (p2) -- (p0b);
\draw[red] (p3) -- (p0b);
\end{scope}
\begin{scope}[scale=1.5,xshift=3cm]
\node[scale=.5] (p0a) at (0,1) {};
\node[scale=.5] (p0b) at (0,-2) {};
\node[scale=.8] (t0a) at (0,1.1) {$\H^{\rm d_{HB}}$};
\node[scale=.8] (t0b) at (0,-2.1) {$\suf(2)_8{\times}\spf(8)_6$};
\node[scale=.8] (p1a) at (.8,0) {$\cT^{(1)}_{D_4,1}$};
\node[scale=.8] (p2a) at (.8,-1) {$\cS^{(1)}_{D_4,2}$};
\node[scale=.8] (p1b) at (-.8,-.5) {$\cT^{(1)}_{E_6,1}$};
\node[scale=.8] (t01) at (-.6,.3) {$\ef_6$};
\node[scale=.8] (t01) at (0.6,.6) {$\df_4$};
\node[scale=.8] (t1a) at (0.6,-.5) {$\cf_3$};
\node[scale=.8] (t2a) at (.6,-1.6) {$\cf_4$};
\node[scale=.8] (t2b) at (-.6,-1.3) {$\af_1$};
\draw[blue] (p0a) -- (p1a);
\draw[blue] (p0a) -- (p1b);
\draw[blue] (p1a) -- (p2a);
\draw[blue] (p1b) -- (p0b);
\draw[blue] (p2a) -- (p0b);
\end{scope}
\end{tikzpicture}}
{\caption{\label{CBTh9}The Coulomb and Higgs stratification of $\suf(2)_8{\times}\spf(8)_6$.}}
\end{subfloatrow}\hspace{1cm}%
\begin{subfloatrow}
\capbtabbox[5cm]{%
  \renewcommand{\arraystretch}{1.1}
  \begin{tabular}{|c|c|} 
  \hline
  \multicolumn{2}{|c|}{$\suf(2)_8{\times}\spf(8)_6$}\\
  \hline\hline
  $(\D_u,\D_v)$  &\quad $\left(3,4\right)$\quad{} \\
  $24a$ &  84\\  
  $12c$ & 48\\
$\ff_k$ & $\suf(2)_8{\times}\spf(8)_6$ \\ 
$d_{\rm HB}$& 12\\
$h$&0\\
$T({\bf2}\bh)$&0\\
\hline\hline
\end{tabular}
}{%
  \caption{\label{CcTh9}Central charges, \gls{CB} parameters and \gls{ECB} dimension.}%
}
\end{subfloatrow}}{\caption{\label{TotTh9}Information about the $\suf(2)_8{\times}\spf(8)_6$ theory.}}
\end{figure}

The \gls{CB} analysis implies that: $a$) all the Higgsing of this theory decrease the rank, therefore \gls{h}=0 $b$) both Higgsing are of \gls{gHW} type $c$) we expect that the theory supported on the $\cf_4$, which arises because of the $[I_8,\suf(8)]_{\Z_2}$, to have a three dimensional \gls{ECB}. This latter property is immediately verified by using \eqref{gHWeq} which singles out the $\cS^{(1)}_{D_4,2}$ theory as the one supported on $\cf_4$. The other Higgsing instead leads to the rank-1 MN $E_6$ ($\cT^{(1)}_{E_6,1}$) as it can be verified easily using again \eqref{gHWeq}.

\subsubsection*{\boldsymbol{$\suf(2)_5{\times}\spf(6)_6{\times}\uf(1)$}}\label{sec:t26}

This theory was first introduced in the context of the $\Z_2$ twisted $A_3$ \cite{Chacaltana:2012ch}. The original reference also discusses interesting S-duality of this theory. 

The presence of two simple flavor factors signals that the \gls{HB} should have two transitions stemming out of the superconformal vacuum. This is expected as the theory is \gls{totally}. Let's study them in turn. First the $\spf(6)_6$ factor. Since the level is twice the $u$ scaling dimension, it is tempting to use the \doub\ and associate this factor to a theory supported on the \gls{ukstr} $u=0$ and make the following identification $\rTf_u\equiv [I_6,\suf(6)]_{\Z_2}$. This in turn implies that one of the first \gls{HB} transition is a $\cf_3$. The level of the $\suf(2)$ factor suggests instead that the $\suf(2)_5$ is realized on the other \gls{ukstr} by $\rTf_v\equiv\blue{\cS^{(1)}_{\varnothing,2}}$. This in turn implies two things: 1) that the theory should has a one dimensional \gls{ECB} and 2) that the theory supported on the $\cf_3$ stratum has to have a three dimensional \gls{ECB}. These two facts turn out to be completely consistent with what we find from the \gls{HB} analysis. But before turning to that, we can plug the \gls{bi} corresponding to the $\rTf_u$ and $\rTf_v$ into \eqref{actotbint} and solve for the theory supported on the \gls{kstr} which completes our analysis of the \gls{CB} stratification in figure \ref{CBTh7}

\begin{figure}[h!]
\ffigbox{
\begin{subfloatrow}
\ffigbox[7.5cm][]{
\begin{tikzpicture}[decoration={markings,
mark=at position .5 with {\arrow{>}}}]
\begin{scope}[scale=1.5]
\node[bbc,scale=.5] (p0a) at (0,0) {};
\node[scale=.5] (p0b) at (0,-2) {};
\node[scale=.8] (t0b) at (0,-2.1) {$\suf(2)_5{\times}\spf(6)_6{\times}\uf(1)$};
\node[scale=.8] (p3) at (0,-1) {$\blue{\cS^{(1)}_{\varnothing,2}}$\ \ };
\node[scale=.8] (p2) at (1.,-1) {\ \ $[I_6,\suf(6)]_{\mathbb{Z}_2}$};
\node[scale=.8] (p1) at (-1,-1) {$[I_1,\varnothing]$\ \ };
%\node[scale=.8] (t1a) at (-.6,-.4) {$I_0^*$};
%\node[scale=.8] (t2c) at (-.7,-1.6) {$K_{\D_7}$};
\node[scale=.8] (t1c) at (-.8,-1.7) {{\scriptsize$\big[u^4{+}v^3{=}0\big]$}};
%\node[scale=.8] (t2c) at (.7,-1.6) {$K_{2\D_7}$};
\node[scale=.8] (t1c) at (.7,-1.7) {{\scriptsize$\big[u=0\big]$}};
\node[scale=.8] (t1c) at (.25,-1.35) {{\scriptsize$\big[v=0\big]$}};
\draw[red] (p0a) -- (p1);
\draw[red] (p0a) -- (p2);
\draw[red] (p0a) -- (p3);
\draw[red] (p1) -- (p0b);
\draw[red] (p2) -- (p0b);
\draw[red] (p3) -- (p0b);
\end{scope}
\begin{scope}[scale=1.5,xshift=3cm]
\node[scale=.5] (p0a) at (0,1) {};
\node[scale=.5] (p0b) at (0,-2) {};
\node[scale=.8] (t0a) at (0,1.1) {$\H^{\rm d_{HB}}$};
\node[scale=.8] (t0b) at (0,-2.1) {$\suf(2)_5{\times}\spf(6)_6{\times}\uf(1)$};
\node[scale=.8] (p1) at (0,0) {$\cT^{(1)}_{D_4,1}$};
\node[scale=.8] (p2a) at (.8,-1) {\hyperref[sec:SU3Nf6]{$SU(3)+6\gls{F}$}};
\node[scale=.8] (p2b) at (-.8,-1) {$\cS^{(1)}_{D_4,2}$};
\node[scale=.8] (t01) at (0.2,.5) {$\df_4$};
\node[scale=.8] (t1a) at (0.6,-.5) {$\af_5$};
\node[scale=.8] (t1b) at (-.6,-0.5) {$\cf_3$};
\node[scale=.8] (t2a) at (.6,-1.5) {$\af_1$};
\node[scale=.8] (t2b) at (-.6,-1.5) {$\cf_3$};
\draw[blue] (p0a) -- (p1);
\draw[blue] (p1) -- (p2a);
\draw[blue] (p1) -- (p2b);
\draw[blue] (p2a) -- (p0b);
\draw[blue] (p2b) -- (p0b);
\end{scope}
\end{tikzpicture}}
{\caption{\label{CBTh7}The Coulomb and Higgs stratification of $\suf(2)_5{\times}\spf(6)_6{\times}\uf(1)$}}
\end{subfloatrow}\hspace{1cm}%
\begin{subfloatrow}
\capbtabbox[7cm]{%
  \renewcommand{\arraystretch}{1.1}
  \begin{tabular}{|c|c|} 
  \hline
  \multicolumn{2}{|c|}{$\suf(2)_5{\times}\spf(6)_6{\times}\uf(1)$}\\
  \hline\hline
  $(\D_u,\D_v)$  &\quad $\left(3,4\right)$\quad{} \\
  $24a$ &  83\\  
  $12c$ & 47 \\
$\ff_k$ & $\suf(2)_5{\times}\spf(6)_6{\times}\uf(1)$ \\ 
$d_{\rm HB}$& 11\\
$h$&1\\
$T({\bf2}\bh)$&1\\
\hline\hline
\end{tabular}
}{%
  \caption{\label{CcTh7}Central charges, \gls{CB} parameters and \gls{ECB} dimension.}%
}
\end{subfloatrow}}{\caption{\label{TotTh7}Information about the $\suf(2)_5{\times}\spf(6)_6{\times}\uf(1)$ theory.}}
\end{figure}

To identify the theories supported on the two strata which stem out of the superconformal vacuum we notice that both of these Higgsings are of \gls{gHW} type and therefore we can compute the corresponding central charges using \eqref{gHWeq}. Performing this calculation we find that the theory supported on the $\cf_3$ stratum is the $\cS^{(1)}_{D_4,2}$, thus $\bTf_{\cf_3}\equiv \cS^{(1)}_{D_4,2}$. This rank-1 theory has indeed \gls{h}=3 compatibly with the prediction arising from the \gls{CB} analysis. To identify the rank-2 theory supported on the \gls{ECB} we can use the extra information that the total \gls{HB} of the \gls{SCFT} in exam is known (that is 11) and therefore this singles out the lagrangian theory $\suf(3)$ with $N_f=6$ as our candidate. The rest of the \gls{HB} is determined by following the higgsings of both $\bTf_{\cf_3}$ and $\bTf_{\af_1}$ and the final result is summarized in figure \ref{CBTh7}.

\subsubsection*{$\boldsymbol{\cT^{(2)}_{D_4,2}}$}\label{sec:TD42} This theory can be straightforwardly obtained by mass deformation of the $\cT^{(1)}_{E_6,2}$. It was one of the recently discovered $\cT$ theories. It can be obtained by higgsing the $\cN=2$ $\cS$-fold $\cS^{(1)}_{A_2,2}$ or by a twisted compactification of a \sd (1,0) theory \cite{Giacomelli:2020jel}. It can also be realized in type \emph{II}B string theory as a wordvolume theory of two \emph{D3} branes probing an exceptional $E_6$ \emph{7}brane singularity in the presence of an $\cS$-fold without flux \cite{Giacomelli:2020gee}. Finally, it can also be obtained in class-$\cS$, for example in the untwisted $A_7$ case \cite{Giacomelli:2020jel}. The properties of $\cN=2$ $\cT$-theories are summarized in appendix \ref{sec:TandS}. The CFT properties as well as the stratification of this particular theory can be found in table \ref{Fig:T2Gl}.

\subsubsection*{$\boldsymbol{\hat{\cT}_{E_6,2}}$}\label{sec:tE6}

This theory was first constructed in twisted $E_6$ class-$\cS$ \cite{Chacaltana:2015bna} but in the initial reference the flavor symmetry symmetry was wrongly identified as $\sof(9){\times}\uf(1)$. The $\ff_4$ enhancement was instead realized in \cite{Wang:2018gvb}. Finally this theory can also be obtained as mass deformation of the $\cT^{(2)}_{E_6,2}$ \cite{Giacomelli:2020gee} which can most easily seen from the $5d$ construction. $\cT^{(2)}_{E_6,2}$ can be obtained as a $\Z_2$ twisted compactification of a $5d$ \gls{SCFT} which UV completes both a $SU(4)_0+2\gls{AS}+6\gls{F}$ and $1\gls{F}-SU(2)-SU(2)-SU(2)-1\gls{F}$, where the latter description also includes two fundamentals for the middle $\suf(2)$\footnote{It is customary to call two $5d$ gauge theories which have the same UV-completion as \emph{UV dual}. We believe that this terminology is misleading therefore we will refrain from using it.}. The $\hat{\cT}_{E_6,2}$ theory can be obtained as a $\Z_2$ twisted compactification of the $\cN=1$ $5d$ \gls{SCFT} which UV completes $SU(2)-SU(2)-\suf(2)$ with the $\theta$ angle of the two edge $\suf(2)$s set to 0 \cite{Giacomelli:2020gee} (where the middle $\suf(2)$ flavor factor still has two flavors attached to it).

\begin{figure}[h!]
\ffigbox{
\begin{subfloatrow}
\ffigbox[8cm][]{
\begin{tikzpicture}[decoration={markings,
mark=at position .5 with {\arrow{>}}}]
\begin{scope}[scale=1.5]
\node[bbc,scale=.5] (p0a) at (0,-1) {};
\node[scale=.5] (p0b) at (0,-3) {};
\node[scale=.8] (t0b) at (0,-3.1) {$\hTE$};
\node[scale=.8] (p2) at (.7,-2) {\ \ $[\cT^{(1)}_{E_6,1}]_{\Z_2}$};
\node[scale=.8] (p1) at (-.7,-2) {$[I_1,\varnothing]$\ \ };
%\node[scale=.8] (t1a) at (-.6,-.4) {$I_0^*$};
%\node[scale=.8] (t2c) at (-.7,-1.6) {$K_{\D_7}$};
\node[scale=.8] (t1c) at (-.85,-2.7) {{\scriptsize$\big[u^5+v^4=0\big]$}};
%\node[scale=.8] (t2c) at (.7,-1.6) {$K_{2\D_7}$};
\node[scale=.8] (t1c) at (.7,-2.7) {{\scriptsize$\big[v=0\big]$}};
\draw[red] (p0a) -- (p1);
\draw[red] (p0a) -- (p2);
\draw[red] (p1) -- (p0b);
\draw[red] (p2) -- (p0b);
\end{scope}
\begin{scope}[scale=1.5,xshift=2.5cm]
\node[scale=.5] (p0a) at (0,0) {};
\node[scale=.8] (tp2) at (0.3,-.5) {$\mathfrak{d}_4$};
\node[scale=.5] (p0b) at (0,-3) {};
\node[scale=.8] (t0a) at (0,.1) {$\H^{\rm d_{HB}}$};
\node[scale=.8] (t0b) at (0,-3.1) {$\hTE$};
\node[scale=.8] (p2) at (0,-2) {$\cS^{(1)}_{D_4,2}$\ \ };
\node[scale=.8] (tp2) at (0.3,-1.5) {$\cf_3$};
\node[scale=.8] (p1) at (0,-1) {$\cT^{(1)}_{D_4,1}$};
\node[scale=.8] (tp1) at (0.3,-2.5) {$\ff_4$};
\draw[blue] (p0a) -- (p1);
\draw[blue] (p1) -- (p2);
\draw[blue] (p2) -- (p0b);
\end{scope}
\end{tikzpicture}}
{\caption{\label{CBhTE6}The Coulomb and Higgs stratification of $\hTE$.}}
\end{subfloatrow}\hspace{1cm}%
\begin{subfloatrow}
\capbtabbox[5cm]{%
  \renewcommand{\arraystretch}{1.1}
  \begin{tabular}{|c|c|} 
  \hline
  \multicolumn{2}{|c|}{$\hTE$}\\
  \hline\hline
  $(\D_u,\D_v)$  &\quad $\left(4,5\right)$\quad{} \\
  $24a$ &  112\\  
  $12c$ & 65\\
$\ff_k$ & $[\ff_4]_{10}{\times}\uf(1)$ \\ 
$d_{\rm HB}$& 16\\
$h$&0\\
$T({\bf2}\bh)$&0\\
\hline\hline
\end{tabular}
}{%
  \caption{\label{CchTE6}Central charges, \gls{CB} parameters and \gls{ECB} dimension.}%
}
\end{subfloatrow}}{\caption{\label{TothTE6}Information about the $\hTE$. theory.}}
\end{figure}

Since $\D=5$ is not an allowed \gls{CB} scaling dimension at rank-1, the theory is not \gls{totally} which simplifies our lives. The theory has in fact a single simple flavor factor and the \gls{HB} is \gls{Hlin}. Identifying which theory on the \gls{CB} realizes an exceptional flavor symmetry is straightforward and this leads to the following $\rTf_v=[\cT^{(1)}_{E_6,1}]_{\Z_2}$ which is also consistent with the level of the $\ff_4$ flavor symmetry being doubled $\D_v$. The \gls{HB} structure of this rank-1 theory readily implies that \gls{h}=0 and that the total \gls{HB} of the rank-2 \gls{SCFT} should start with a $\ff_4$ stratum followed by a rank preserving $\cf_3$ transition. It is straightforward to complete the \gls{CB} \gls{Hasse} by using \eqref{actotaint}-\eqref{actotbint} to reproduce the $a$ and $c$ central charges listed in table \ref{CchTE6}. This exercise implies that $\rTf_{v^4+u^5}\equiv[I_1,\varnothing]$.

Leveraging all the information which we have gathered from the \gls{CB} analysis will immediately allow us to characterize the \gls{HB} side. In fact the rank-1 theory supported on the $\ff_4$ stratum is uniquely fixed by requiring that such theory has a eight dimensional \gls{HB} (to reproduce the total dimension of the \gls{HB} of $\hat{\cT}_{E_6,2}$) and \gls{h}=3 (to reproduce the subsequent rank-preserving $\cf_3$ transition). Thus we conclude that $\bTf_{\ff_4}\equiv\cS^{(1)}_{D_4,2}$. This result can be also checked by matching the central charge of $\bTf_{\ff_4}$ with what we obtain from \eqref{gHWeq}. 

\subsubsection*{$\boldsymbol{\spf(6)_5{\times}\uf(1)}$}\label{sec:tTE62}

This theory was first discussed in \cite{Zafrir:2016wkk} in the context of twisted compactification of 5d \gls{SCFT}s and then further analyzed in \cite{Giacomelli:2020gee}. It can be obtained via mass deformation of the \hyperref[sec:t25]{$\suf(2)_8{\times}\spf(8)_6$} which was discussed above. In particular the $\Z_2$ twisted compactification of a $5d$ \gls{SCFT} which completes a $SU(4)_0+1\gls{AS}+6\gls{F}$, where the subscript indicates the Chern-Simons level of the $5d$ gauge theory, gives the $\spf(6)_5{\times}\uf(1)$ \cite{Zafrir:2016wkk,Giacomelli:2020gee}.

\begin{figure}[h!]
\ffigbox{
\begin{subfloatrow}
\ffigbox[7cm][]{
\begin{tikzpicture}[decoration={markings,
mark=at position .5 with {\arrow{>}}}]
\begin{scope}[scale=1.5]
\node[bbc,scale=.5] (p0a) at (0,0) {};
\node[scale=.5] (p0b) at (0,-2) {};
\node[scale=.8] (t0b) at (0,-2.1) {$\tTE$};
\node[scale=.8] (p1) at (-.7,-1) {$[I_1,\varnothing]$\ \ };
\node[scale=.8] (p2) at (.7,-1) {\ \ $[I_6,\suf(6)]_{\mathbb{Z}_2}$};
%\node[scale=.8] (t1a) at (-.6,-.4) {$I_0^*$};
%\node[scale=.8] (t2c) at (-.7,-1.6) {$K_{\D_7}$};
\node[scale=.8] (t1c) at (-1,-1.7) {{\scriptsize$\big[u^6+v^5=0\big]$}};
%\node[scale=.8] (t2c) at (.7,-1.6) {$K_{2\D_7}$};
\node[scale=.8] (t1c) at (.7,-1.7) {{\scriptsize$\big[u=0\big]$}};
\draw[red] (p0a) -- (p1);
\draw[red] (p0a) -- (p2);
\draw[red] (p1) -- (p0b);
\draw[red] (p2) -- (p0b);
\end{scope}
\begin{scope}[scale=1.5,xshift=2.5cm]
\node[scale=.5] (p0a) at (0,1) {};
\node[scale=.5] (p0b) at (0,-2) {};
\node[scale=.8] (t0a) at (0,1.1) {$\H^{\rm d_{HB}}$};
\node[scale=.8] (t0b) at (0,-2.1) {$\tTE$};
\node[scale=.8] (p1) at (0,0) {$\cT^{(1)}_{A_2,1}$};
\node[scale=.8] (p2) at (0,-1) {$\cS^{(1)}_{A_2,2}$};
\node[scale=.8] (t1c) at (.3,.5) {$\mathfrak{a}_2$};
\node[scale=.8] (t2c) at (.3,-0.5) {$\mathfrak{c}_2$};
\node[scale=.8] (t1c) at (.3,-1.5) {$\mathfrak{c}_3$};
\draw[blue] (p0a) -- (p1);
\draw[blue] (p1) -- (p2);
\draw[blue] (p2) -- (p0b);
\end{scope}
\end{tikzpicture}}
{\caption{\label{CBtTE}The Coulomb and Higgs stratification of $\tTE$}}
\end{subfloatrow}\hspace{1cm}%
\begin{subfloatrow}
\capbtabbox[7cm]{%
  \renewcommand{\arraystretch}{1.1}
  \begin{tabular}{|c|c|} 
  \hline
  \multicolumn{2}{|c|}{$\tTE$}\\
  \hline\hline
  $(\D_u,\D_v)$  &\quad $\left(\frac52,4\right)$\quad{} \\
  $24a$ &  61\\  
  $12c$ & 34 \\
$\ff_k$ & $\spf(6)_5{\times}\uf(1)$ \\ 
$d_{\rm HB}$& 7\\
$h$&0\\
$T({\bf2}\bh)$&0\\
\hline\hline
\end{tabular}
}{%
  \caption{\label{CctTE}Central charges, \gls{CB} parameters and \gls{ECB} dimension.}%
}
\end{subfloatrow}}{\caption{\label{TottTE}Information about the $\tTE$ theory.}}
\end{figure}

The moduli space of this theory is fairly straightforward to analyze. In fact the theory is not \gls{totally} and it only has a single \gls{higgs} ($u$). As usual this is reflected in the fact that there is a single simple flavor factor which in turn implies that the \gls{HB} is \gls{Hlin}, see figure \ref{CBtTE}. The rank-1 theories which realizes the $\spf(6)_5$  can be easily identified as  the $[I_6,\suf(6)]_{\Z_2}$ and which, by using the \doub\ to match the level via \eqref{actotcint}, it has to be supported on the \gls{ukstr} $u=0$ which leads to the identification $\rTf_u\equiv[I_6,\suf(6)]_{\Z_2}$ and readily implies \gls{h}=0. To determine the theory supported on the \gls{kstr} it is enough to match the $c$ central charge via \eqref{actotbint}.

Moving on the \gls{HB} side, the \gls{HB} of the $\rTf_u$ should give rise to a $\cf_3$ followed by a $\cf_2$ transition. The central charge of the rank-1 theory supported on the $\cf_3$ stratum can be easily identified noticing that this Higgsing is of \gls{gHW} type, thus using \eqref{gHWeq} singles out $\cS^{(1)}_{A_2,2}$. This  also reproduces the subsequent $\cf_2$ Higgsing. The rest of the \gls{Hasse} is obtained straightforwardly from the Higgsings of $\cS^{(1)}_{A_2,2}$. The way in which the flavor symmetry of the theory is reproduced from the Higgsing is straightforward using the properties of $\cf_3$ summarized in table \ref{NilOrbits}.

\subsubsection*{$\boldsymbol{\cT^{(2)}_{A_2,2}}$}\label{sec:TA22}  This theory can be obtained mass deforming the $\cT^{(1)}_{D_4,2}$ and belongs to the recently discovered $\cT$ theories. It can be obtained in class-$\cS$, for example in the recent study of twisted $A_2$ \cite{Beem:2020pry}\footnote{In \cite{Beem:2020pry} $\cT^{(2)}_{A_2,2}$ was named $\tilde{T}_3$.}, as well as by higgsing the $\cN=2$ $\cS$-fold $\cS^{(1)}_{A_2,2}$. As usual, the $\cT$-theories also allow a twisted construction from \sd (1,0) theories \cite{Giacomelli:2020jel} and they arise as a wordvolume theory of two \emph{D3} branes probing an exceptional $E_6$ \emph{7}brane singularity in the presence of an $\cS$-fold without flux \cite{Giacomelli:2020gee}. The properties of $\cN=2$ $\cT$-theories are summarized in appendix \ref{sec:TandS}. The CFT properties as well as the stratification of this particular theory can be found in table \ref{Fig:T2Gl}.

\subsubsection*{\boldsymbol{$SU(2)-SU(2)$}}\label{sec:SU22a}

\begin{figure}[h!]
\ffigbox{
\begin{subfloatrow}
\ffigbox[7.5cm][]{
\begin{tikzpicture}[decoration={markings,
mark=at position .5 with {\arrow{>}}}]
\begin{scope}[scale=1.5]
\node[bbc,scale=.5] (p0a) at (0,0) {};
\node[scale=.5] (p0b) at (0,-2) {};
\node[scale=.8] (t0b) at (0,-2.1) {$SU(2)-SU(2)$)};
\node[scale=.8] (p1) at (-1.5,-1) {$[I_1,\varnothing]$};
\node[scale=.8] (p4) at (-.9,-1) {$[I_1,\varnothing]$};
\node[scale=.8] (p5) at (.3,-1) {$[I_1,\varnothing]$};
\node[scale=.8] (p2) at (1.2,-1) {$[I_4,\suf(4)]_{\Z_2}$};
\node[scale=.8] (p3) at (-.3,-1) {$[I_1,\varnothing]$};
\node[scale=.8] (t2b) at (-.3,-1.5) {{\scriptsize$\big[u+v=0\big]$}};
\node[scale=.8] (t3b) at (.7,-1.6) {{\scriptsize$\big[u=0\big]$}};
\draw[red] (p0a) -- (p1);
\draw[red] (p0a) -- (p2);
\draw[red] (p0a) -- (p3);
\draw[red] (p0a) -- (p4);
\draw[red] (p0a) -- (p5);
\draw[red] (p5) -- (p0b);
\draw[red] (p2) -- (p0b);
\draw[red] (p4) -- (p0b);
\draw[red] (p1) -- (p0b);
\draw[red] (p2) -- (p0b);
\draw[red] (p3) -- (p0b);
\end{scope}
\begin{scope}[scale=1.5,xshift=3.2cm]
\node[scale=.8] (p0a) at (0,0) {$\uf(1)$};
\node[scale=.8] (t1a) at (-0.2,-.5) {$\af_1$};
\node[scale=.8] (p1) at (-0,-1) {$\blue{\cS^{(1)}_{\varnothing,2}}$};
\node[scale=.8] (t1a) at (-0.2,-1.5) {$\cf_2$};
\node[scale=.8] (p2) at (-0,-2) {$SU(2)-SU(2)$};
\draw[blue] (p0a) -- (p1);
\draw[blue] (p1) -- (p2);
\end{scope}
\end{tikzpicture}}
{\caption{\label{CBSU2a}The \gls{Hasse} for the \gls{CB} of the  $SU(2)$ $\cN=2$ theory with two hypermultiplets in the $({\bf2},{\bf2})$.}}
\end{subfloatrow}\hspace{1cm}%
\begin{subfloatrow}
\capbtabbox[7cm]{%
  \renewcommand{\arraystretch}{1.1}
  \begin{tabular}{|c|c|} 
  \hline
  \multicolumn{2}{|c|}{$SU(2)-SU(2)$}\\
  \hline\hline
  $(\D_u,\D_v)$  &\quad (2,2)\quad{} \\
  $24a$ & 38\\  
  $12c$ & 20 \\
$\ff_k$ & $\spf(4)_4$ \\ 
$d_{\rm HB}$&3\\
$h$&0\\
$T({\bf2}\bh)$&0\\
\hline\hline
 \end{tabular}
}{%
  \caption{\label{CcSU2a}Central charges, \gls{CB} parameters and \gls{ECB} dimension.}%
}
\end{subfloatrow}}{\caption{\label{TotSU22a} Information about the $SU(2)$ $\cN=2$ theory with two hypermultiplets in the $({\bf2},{\bf2})$}}
\end{figure}

Let's start from the analysis of the \gls{CB}. Two obvious strata correspond to turning on separately a \gls{CB} vev for each $SU(2)$. This makes all the hypers massive, and semi-classically leaves a pure $SU(2)$ which breaks quantum mechanically into a dyon and monopole giving rise to a total of four knotted singularities each supporting an $[I_1,\suf(2)]$. There is also another semi-classical locus where massless charged matter arises. I can in fact tune the vevs of the two $SU(2)$ and make them equal. This of course will break $SU(2)\times SU(2)\to U(1)\times U(1)$ but each bifundamental hypermultiplet will contribute two massless hyper with charge 1 giving rise to an effective $U(1)$ theory with four massless hypers. Since the $[I_4,\suf(4)]$ is discretely gauged, we expect a $\cf_2$ transition on the \gls{HB} side of things which supports a rank-1 theory with a one dimensional \gls{ECB}.

Let's now focus on the \gls{HB}. The information that we gathered from the \gls{CB} analysis are already enough to make the identification $\bTf_{\cf_2}\equiv\blue{\cS^{(1)}_{\varnothing,2}}$ (the $SU(2)$ $\cN=4$ theory). It is possible to use \eqref{gHWeq} to explicitly confirm this guess or else perform the higgsing explicitly solving for both the $F$ and $D$ term conditions.

\subsubsection*{\boldsymbol{\blue{$SU(2){\times}SU(2)$}}}\label{sec:N4SU22}  This is an $\cN=4$ theory. The moduli space of these theories is extremely constrained and it is basically entirely specified by the Weyl group of the gauge algebra which in this case is $\Z_2\times \Z_2$. More details on the moduli space structure of theories with extended supersymmetry can be found in appendix \ref{sec:N34}. The CFT data of this theory, as well as the explicit \gls{Hasse} of both the \gls{CB} and \gls{HB} stratification are depicted in figure \ref{Fig:N4}.

\begin{center}
\rule[1mm]{2cm}{.4pt}\hspace{1cm}$\circ$\hspace{1cm} \rule[1mm]{2cm}{.4pt}
\end{center}

\subsection{$\suf(6)$ series}

All the theories but one (the lagrangian entry) in this series present some intriguing mysteries. Specifically it appears that unknown \gls{elesl} appear on the \gls{HB} which seem to be realized in highly non-trivial way on the \gls{CB}. In the \gls{Hasse}s below there are many question marks, which characterize our current ignorance on the details of these theories. Most of the \gls{HB} results that we do understand are obtained by quiver subtractions, for more details see \emph{e.g.} \cite{Bourget:2021csg}.

\subsubsection*{\boldsymbol{$\suf(6)_{16}{\times}\suf(2)_9$}}\label{sec:t33}

\begin{figure}[h!]
\ffigbox{
\begin{subfloatrow}
\ffigbox[8.5cm][]{
\begin{tikzpicture}[decoration={markings,
mark=at position .5 with {\arrow{>}}}]
\begin{scope}[scale=1.5]
\node[bbc,scale=.5] (p0a) at (0,0) {};
\node[scale=.5] (p0b) at (0,-2) {};
\node[scale=.8] (t0b) at (0,-2.1) {$\suf(6)_{16}{\times}\suf(2)_9$};
\node[scale=.8] (p1) at (-.7,-1) {$[I_6^*,\suf(2){\times}\sof(12)]_{\Z_2}$};
\node[scale=.8] (p2) at (.7,-1) {$[I_1,\varnothing]$};
\node[scale=.8] (t2b) at (-.7,-1.5) {{\scriptsize$\big[v=0\big]$}};
\node[scale=.8] (t3b) at (.7,-1.6) {{\scriptsize$\big[u^4+v^3=0\big]$}};
\draw[red] (p0a) -- (p1);
\draw[red] (p0a) -- (p2);
\draw[red] (p1) -- (p0b);
\draw[red] (p2) -- (p0b);
\end{scope}
\begin{scope}[scale=1.5,xshift=3cm]
\node[scale=.5] (p4) at (0,2) {};
\node[scale=.8] (t0a) at (0,2.1) {$\H^{\rm d_{HB}}$};
\node[scale=.8] (tp2) at (-.2,1.5) {$\ef_6$};
\node[scale=.8] (p2) at (0,1) {$\cT^{(1)}_{E_6,1}$};
\node[scale=.8] (tp2) at (-.2,.5) {$\cf_5$};
\node[scale=.8] (p1) at (0,0) {$\cS^{(1)}_{E_6,2}$};
\node[scale=.8] (tp1) at (-.6,-.5) {$\cf_6$};
\node[scale=.8] (p1b) at (-.7,-1) {\hyperref[sec:t22]{$\spf(12)_8$}};
\node[scale=.8] (tp1) at (-.6,-1.6) {$\af_1$};
\node[scale=.8] (tp1) at (.7,-1) {?};
\node[scale=.5] (p0) at (0,-2) {};
\node[scale=.8] (t0b) at (0,-2.1) {$\suf(6)_{16}{\times}\suf(2)_9$};
\draw[blue] (p0) -- (p1b);
\draw[blue] (p1b) -- (p1);
\draw[blue] (p1) -- (p2);
\draw[blue] (p2) -- (p4);
\draw[blue,dashed]  (p0) to[out=30,in=-50] (p1);
\end{scope}
\end{tikzpicture}}
{\caption{\label{CBhTh35}The Coulomb and Higgs stratification of $\suf(6)_{16}{\times}\suf(2)_9$.}}
\end{subfloatrow}\hspace{1cm}%
\begin{subfloatrow}
\capbtabbox[5cm]{%
  \renewcommand{\arraystretch}{1.1}
  \begin{tabular}{|c|c|} 
  \hline
  \multicolumn{2}{|c|}{$\suf(6)_{16}{\times}\suf(2)_9$}\\
  \hline\hline
  $(\D_u,\D_v)$  &\quad $\left(6,8\right)$\quad{} \\
  $24a$ & 179\\  
  $12c$ & 101\\
$\ff_k$ & $\suf(6)_{16}{\times}\suf(2)_9$ \\ 
$d_{\rm HB}$& 23\\
$h$&1\\
$T({\bf2}\bh)$&1\\
\hline\hline
\end{tabular}
}{%
  \caption{\label{CchTh35}Central charges, \gls{CB} parameters and \gls{ECB} dimension.}%
}
\end{subfloatrow}}{\caption{\label{TothTh35}Information about the $\suf(6)_{16}{\times}\suf(2)_9$}}
\end{figure}
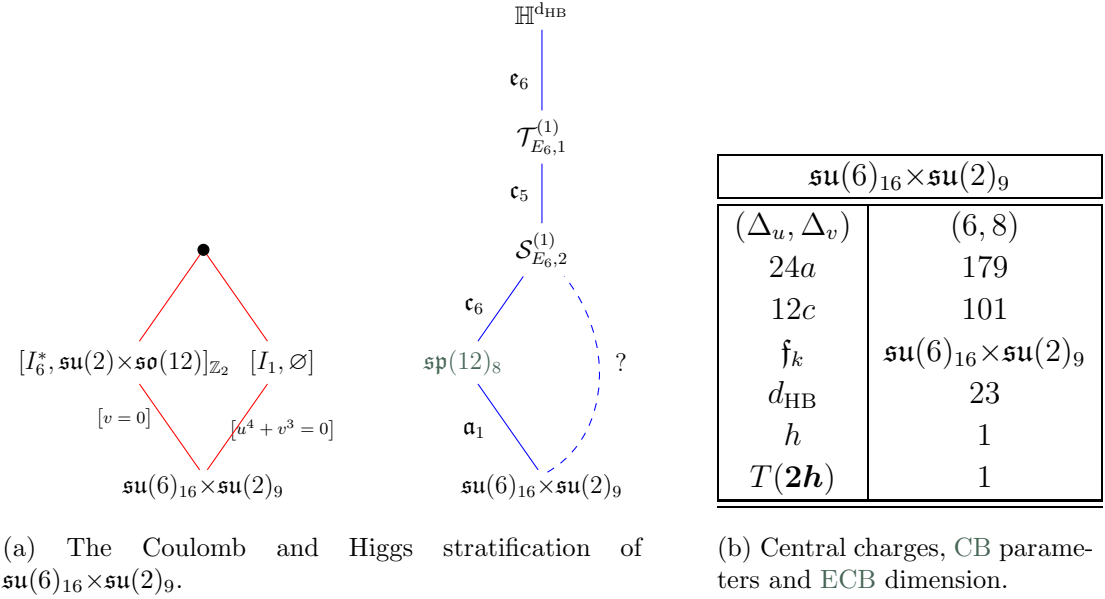

This theory, which sits at the top of this series, was first realized in \cite{Chacaltana:2015bna} but it can also be realized by a twisted $\Z_2$ compactification with non commuting holonomies of a $6d$ \gls{SCFT} completing the $SU(4)+1\gls{AS}+12\gls{F}$ \cite{Ohmori:2018ona}. Its CFT data is summarized in table \ref{TothTh35}.

Immediately we are faced with a puzzle. In fact the flavor symmetry is semi-simple but the theory is not \gls{totally}. A closer look at the levels of the flavor symmetries present a possible resolution: $k_{\suf(6)}=2\D_v$ and $k_{\suf(2)}=\D_v+1$ suggesting that both flavor symmetries are realized on the $v=0$ stratum. A further indication comes from reproducing the central charges in \ref{CchTh35} from the \gls{ccf}. This instructive exercise does not provide a unique solution but the most reasonable one assigns a $b=1$ to \gls{kstr} and $b=9$ to the $v=0$ unknotted one. We are therefore led to the following identification: $\rTf_v\equiv [I_6^*,\suf(2){\times}\sof(12)]_{\Z_2}$ and $\rTf_{u^4+v^3}\equiv[I_1,\varnothing]$. The latter is now standard but let's discuss the former.

$[I_6^*,\suf(2){\times}\sof(12)]$ is nothing but an $\cN=2$ $SU(2)+6\gls{F}+1\gls{adj}$ gauge theory. This theory has a twelve dimensional \gls{HB} of which one dimension is an \gls{ECB}. Clearly the \gls{UVIR} tells us that its flavor symmetry is too large for being the correct identification but $\sof(12)$ has a $\Z_2$ (inner) automorphism whose commutant is $\suf(6)$, for more details on automorphisms of Lie algebras see \cite[Theorem 8.6]{Kac} or for a more physical discussion \cite[Sec.~3.3]{Tachikawa:2011ch}. We therefore claim that the theory that correctly realizes the flavor symmetry of this theory is a $\Z_2$ discretely gauged version of this gauge theory, $[I_6^*,\suf(2){\times}\sof(12)]_{\Z_2}$, where the $\Z_2$ acts trivially on the adjoint hyper and therefore leaves the $\suf(2)$ untouched and implies that \gls{h}=1.

We can now analyze the \gls{HB} of this theory. Great insights on the overall structure of this 29 dimensional symplectic variety can be gained by first analyzing the \gls{ECB} of the theory. From our \gls{CB} analysis we learned that this corresponds to higgsing the adjoint multiplet and therefore this higgsing is of \gls{gHW} type. Using the \eqref{gHWeq} we readily obtain that the central charge of the theory supported on this $\af_1$ stratum is $12c_{\bTf_{\af_1}}=76$ which by itself leads to the identification $\bTf_{\af_1}\equiv \spf(12)_8$. The \gls{HB} of this theory, see figure \ref{CBhTh30}, determines most of the remaining structure. To complete the analysis we need to understand what is the transition carrying the $\suf(6)$ action and which corresponds to giving a vev to the fundamentals of the $\rTf_v$. This transition is depicted by a dashed line in figure \ref{CBhTh35}. Because of the structure of the \gls{Hasse}, this cannot correspond to an \gls{elesl}. A possibility to answer this question is to directly analyze the magnetic quiver of the theory which can be derived from the higher dimensional realization of the theory. Unfortunately, ``subtracting off'' the $\ef_6$ and the $\cf_5$ transition we are left with an unknown transition. Perhaps the discretely gauged realization which we identified on the \gls{CB} side will provide interesting insights allowing to resolve this puzzle.

\subsubsection*{\boldsymbol{$\suf(4)_{12}{\times}\suf(2)_7{\times}\uf(1)$}}\label{sec:t34}

This theory was first realized in class-$\cS$ within the twisted $D$-series \cite{Chacaltana:2013oka}. But it can also be realized as $\Z_2$ compactification of a \emph{5d} brane web which is obtained after mass deformation of the circle compactification of the $6d$ \gls{SCFT} completing a $SU(4)+1\gls{AS}+12\gls{F}$\footnote{The author is deeply thankful to Gabi Zafrir for his patient and clarifying explanation of the brane web deformations which lead to this and the following three theories.}. This implies that this theory can be obtained by mass deforming the theory discussed in the previous section. Most of the CFT data summarized in figure \ref{TothTh35} was computed in the original class-$\cS$ realization.

\begin{figure}[h!]
\ffigbox{
\begin{subfloatrow}
\ffigbox[8.5cm][]{
\begin{tikzpicture}[decoration={markings,
mark=at position .5 with {\arrow{>}}}]
\begin{scope}[scale=1.5]
\node[bbc,scale=.5] (p0a) at (0,0) {};
\node[scale=.5] (p0b) at (0,-2) {};
\node[scale=.8] (t0b) at (0,-2.1) {$\suf(4)_{12}{\times}\suf(2)_7{\times}\uf(1)$};
\node[scale=.8] (p1) at (-1.15,-1) {$[I_4^*,\suf(2){\times}\sof(8)]_{\Z_2}$};
\node[scale=.8] (p2) at (1,-1) {$[I_1,\varnothing]$};
\node[scale=.8] (p3) at (.2,-1) {$[I_2,\suf(2)]$};
\node[scale=.8] (t2b) at (-1,-1.5) {{\scriptsize$\big[v=0\big]$}};
\node[scale=.8] (t2b) at (-.2,-1.5) {{\scriptsize$\big[u=0\big]$}};
\node[scale=.8] (t3b) at (1,-1.6) {{\scriptsize$\big[u^3+v^2=0\big]$}};
\draw[red] (p0a) -- (p1);
\draw[red] (p0a) -- (p2);
\draw[red] (p0a) -- (p3);
\draw[red] (p3) -- (p0b);
\draw[red] (p1) -- (p0b);
\draw[red] (p2) -- (p0b);
\end{scope}
\begin{scope}[scale=1.5,xshift=3cm]
\node[scale=.5] (p4) at (-.7,2) {};
\node[scale=.8] (t0a) at (-.7,2.1) {$\H^{\rm d_{HB}}$};
\node[scale=.8] (p1a) at (.1,1) {$\cT^{(1)}_{D_4,1}$};
\node[scale=.8] (p2a) at (.1,0) {$\cS^{(1)}_{D_4,2}$};
\node[scale=.8] (p1b) at (-1.5,.5) {$\cT^{(1)}_{E_6,1}$};
\node[scale=.8] (t01) at (-1.3,1.3) {$\ef_6$};
\node[scale=.8] (t01) at (-.1,1.6) {$\df_4$};
\node[scale=.8] (t1a) at (-.1,.5) {$\cf_3$};
\node[scale=.8] (t2a) at (-.1,-.6) {$\cf_4$};
\node[scale=.8] (t2b) at (-1.3,-.3) {$\af_1$};
\node[scale=.8] (p0a) at (-.7,-1) {\hyperref[sec:t25]{$\suf(2)_8{\times}\spf(8)_6$}};
\node[scale=.8] (p0d) at (0,-1) {};
\node[scale=.8] (tp1) at (-.7,-1.5) {$\af_1$};
\node[scale=.8] (tp1) at (.7,-1) {?};
\node[scale=.8] (tp1) at (.1,-1.5) {?};
\node[scale=.5] (p0) at (0,-2) {};
\node[scale=.8] (t0b) at (0,-2.1) {$\suf(4)_{12}{\times}\suf(2)_7{\times}\uf(1)$};
\draw[blue,dashed]  (p0) to[out=30,in=-50] (p2a);
\draw[blue] (p0) -- (p0a);
\draw[blue,dashed] (p0) -- (p0d);
\draw[blue] (p0a) -- (p1b);
\draw[blue] (p0a) -- (p2a);
\draw[blue] (p1b) -- (p4);
\draw[blue] (p2a) -- (p1a);
\draw[blue] (p1a) -- (p4);
\end{scope}
\end{tikzpicture}}
{\caption{\label{CBhTh36}The Coulomb and Higgs stratification of $\suf(4)_{12}{\times}\suf(2)_7{\times}\uf(1)$.}}
\end{subfloatrow}\hspace{1cm}%
\begin{subfloatrow}
\capbtabbox[5cm]{%
  \renewcommand{\arraystretch}{1.1}
  \begin{tabular}{|c|c|} 
  \hline
  \multicolumn{2}{|c|}{$\suf(4)_{12}{\times}\suf(2)_7{\times}\uf(1)$}\\
  \hline\hline
  $(\D_u,\D_v)$  &\quad $\left(4,6\right)$\quad{} \\
  $24a$ & 121\\  
  $12c$ & 67\\
$\ff_k$ & $\suf(4)_{12}{\times}\suf(2)_7{\times} \uf(1)$ \\ 
$d_{\rm HB}$& 13\\
$h$&1\\
$T({\bf2}\bh)$&1\\
\hline\hline
\end{tabular}
}{%
  \caption{\label{CchTh36}Central charges, \gls{CB} parameters and \gls{ECB} dimension.}%
}
\end{subfloatrow}}{\caption{\label{TothTh36}Information about the $\suf(4)_{12}{\times}\suf(2)_7{\times}\uf(1)$.}}
\end{figure}

The analysis of this case is largely analogous to the previous one and thus quite involved. The theory is \gls{totally} but again the level of the two simple flavor symmetry factors are both compatible to be realized by a theory living on the same \gls{ukstr}. Indeed $k_{\suf(2)}=\D_v+1$ and $k_{\su(4)}=2\D_v$. This again suggests that the low energy theory $\rTf_v$ will have a semi-simple flavor symmetry. To make progress in this identification we leverage the \gls{ccf} to reproduce the $c$ and $a$ central charges in table \ref{CchTh36}. This exercise does not produce a single solution but again the most reasonable one implies $b_{u^3+v^2}=1$, $b_u=2$ and $b_v=7$. This immediately suggests the identification $\rTf_{u^3+v^2}\equiv[I_1,\varnothing]$, $\rTf_u\equiv[I_2,\suf(2)]$ and $\rTf_v\equiv[I_4^*,\suf(2){\times}\sof(8)]_{\Z_2}$ which is nothing but a $\cN=2$ $SU(2)+1\gls{adj}+4\gls{F}$ gauge theory. This solution is satisfactory in some senses and puzzling in others. Let's elaborate.

The $\Z_2$ gauging that we conjecture is implemented on the $v=0$ \gls{ukstr}, only acts on the fundamental hypers as a inner automorphism of $\sof(8)$ with commutant $\suf(4)$. This implies in turn that the one dimensional \gls{ECB} obtained by turning on a vev for the adjoint hyper is left unchanged and can be leverage to great extent to learn about the \gls{HB} of this theory. Running this anlaysis with our usual tools, which include using \eqref{gHWeq}, we find $\bTf_{\af_1}\equiv \hyperref[sec:t25]{\suf(2)_8{\times}\spf(8)_8}$. This identification allows to fill in the left side of the \gls{Hasse} in figure \ref{CBhTh36}. The higgsing of the fundamental is as usual more complicated. Indeed we are not able to identify the stratum which is acted upon by the $\suf(4)$. We can only conclude that this cannot be and \gls{elesl}. Since the \fid realization of the theory is known, so is the magnetic quiver of the theory. We hope that an in-depth study of this object might help identifying the ``question mark transition'' in the figure. But there is one more puzzle in this case. And that is that the $\rTf_u$ also has a Higgs branch and thus we expect another branch of the total \gls{HB}. The fact that the $\suf(2)$ realizes the $\uf(1)$ with an IR enhancement, suggests that there is an extra discrete identification at the origin of the moduli space acting on the $\af_1$. But even allowing for that, our analysis is not powerful enough to say anything more about this branch nor even conclusively determine whether it is actually there. This is reflected in the figure by another dashed branch which connects nowhere. Again an in-depth study of the magnetic quiver might help resolve this puzzle.

\subsubsection*{\boldsymbol{$\suf(3)_{10}{\times}\suf(3)_{10}{\times}\uf(1)$}}\label{sec:t35}

\begin{figure}[h!]
\ffigbox{
\begin{subfloatrow}
\ffigbox[8.5cm][]{
\begin{tikzpicture}[decoration={markings,
mark=at position .5 with {\arrow{>}}}]
\begin{scope}[scale=1.5]
\node[bbc,scale=.5] (p0a) at (0,0) {};
\node[scale=.5] (p0b) at (0,-2) {};
\node[scale=.8] (t0b) at (0,-2.1) {$\suf(3)_{10}{\times}\suf(3)_{10}{\times}\uf(1)$};
\node[scale=.8] (p1) at (-.7,-1) {$[\star\,{\rm w/}b=7]$};
\node[scale=.8] (p2) at (.7,-1) {$[I_1,\varnothing]$};
\node[scale=.8] (t2b) at (-.7,-1.5) {{\scriptsize$\big[v=0\big]$}};
\node[scale=.8] (t3b) at (.7,-1.6) {{\scriptsize$\big[u^5+v^4=0\big]$}};
\draw[red] (p0a) -- (p1);
\draw[red] (p0a) -- (p2);
\draw[red] (p1) -- (p0b);
\draw[red] (p2) -- (p0b);
\end{scope}
\begin{scope}[scale=1.5,xshift=3cm]
\node[scale=.5] (p4) at (0,1) {};
\node[scale=.8] (t0a) at (0,1.1) {$\H^{\rm d_{HB}}$};
\node[scale=.8] (tp2) at (-.2,.5) {$\df_4$};
\node[scale=.8] (p1) at (0,0) {$\cT^{(1)}_{D_4,1}$};
\node[scale=.8] (tp1) at (-.55,-.5) {$\cf_3$};
\node[scale=.8] (tp1) at (.5,-.5) {$\cf_3$};
\node[scale=.8] (p1b) at (-.7,-1) {$\cS^{(1)}_{D_4,2}$};
\node[scale=.8] (p1a) at (.7,-1) {$\cS^{(1)}_{D_4,2}$};
\node[scale=.8] (tp1) at (-.6,-1.5) {$\bar{h}_{3,2}$};
\node[scale=.8] (tp1) at (.7,-1.5) {$\bar{h}_{3,2}$};
\node[scale=.5] (p0) at (0,-2) {};
\node[scale=.8] (t0b) at (0,-2.1) {$\suf(3)_{10}{\times}\suf(3)_{10}{\times}\uf(1)$};
\draw[blue] (p0) -- (p1b);
\draw[blue] (p0) -- (p1a);
\draw[blue] (p1b) -- (p1);
\draw[blue] (p1a) -- (p1);
\draw[blue] (p1) -- (p4);
\end{scope}
\end{tikzpicture}}
{\caption{\label{CBhTh37}The Coulomb and Higgs stratification of $\suf(3)_{10}{\times}\suf(3)_{10}{\times}\uf(1)$.}}
\end{subfloatrow}\hspace{1cm}%
\begin{subfloatrow}
\capbtabbox[5cm]{%
  \renewcommand{\arraystretch}{1.1}
  \begin{tabular}{|c|c|} 
  \hline
  \multicolumn{2}{|c|}{$\suf(3)_{10}{\times}\suf(3)_{10}{\times}\uf(1)$}\\
  \hline\hline
  $(\D_u,\D_v)$  &\quad $\left(4,5\right)$\quad{} \\
  $24a$ & 107\\  
  $12c$ & 59\\
$\ff_k$ & $\suf(3)_{10}{\times}\suf(3)_{10}{\times}\uf(1)$ \\ 
$d_{\rm HB}$& 11\\
$h$&0\\
$T({\bf2}\bh)$&0\\
\hline\hline
\end{tabular}
}{%
  \caption{\label{CchTh37}Central charges, \gls{CB} parameters and \gls{ECB} dimension.}%
}
\end{subfloatrow}}{\caption{\label{TothTh37}Information about the $\suf(3)_{10}{\times}\suf(3)_{10}{\times}\uf(1)$.}}
\end{figure}

This theory can be obtained by the twisted $\Z_2$ compactification of the \emph{5d} brane web with three \emph{D5} and three \emph{NS5} all intersection at one point described in \cite{Zafrir:2016wkk}. 

Our understanding of this theory suffers of many of the same problems encountered in the previous cases. The theory is not \gls{totally} and $v$ is the only \gls{higgs}. Furthermore even though the flavor symmetry is not simple, the semi-simple component has the right level to be interpreted as flavor symmetry of an IR-free rank-1 gauge theory supported on $v=0$. To gain a better feeling about what this theory could be, it is useful to trying in reproducing the $a$ and $c$ central charges in table \ref{CchTh37} using the \gls{ccf}. Performing this exercise we don't get a unique answer but the most reasonable one predicts a $\rTf_{u^5+v^4}\equiv[I_1,\varnothing]$ while the $\rTf_v$ is predicted to have $b=7$. It is likely that the correct interpretation will involve a discrete gauging of an IR-free theory but we don't have a sharp guess yet.

Without a complete knowledge of the \gls{CB} stratification it is hard to perform a complete analysis on the \gls{HB} side using purely field theoretic methods. Thankfully the information of the magnetic quiver, which can be derived from the \emph{5d} realization, allows to reproduce the full \gls{HB} \gls{Hasse} which is depicted in figure \ref{CBhTh37}\footnote{We thank Julius Grimminger for performing the quiver subtraction which led to this \gls{HB} \gls{Hasse}.}. The careful reader will have notice that a new elementary slice appeared which is labeled as $\bar{h}_{3,2}$. These are new  \gls{elesl}s which will be introduced in \cite{AffineAJ} and appear in the affine Grassmanian of $\spf(2n)$. These slices have been known for quite some time in the math literature with a different name $\bar{h}_{n,2}=a\cf_n$ \cite{ac21,ac22,ac23}\footnote{ We are grateful to Antoine Bourget and Julius Grimminger for the computation of many of the \gls{Hasse}s in this section, for sharing with me unpublished results of (one of) their upcoming paper(s) and for providing the list of math references on \gls{elesl}s in the affine Grassmanian.}

\subsubsection*{\boldsymbol{$\suf(3)_{10}{\times}\suf(2)_6{\times}\uf(1)$}}\label{sec:t36}

This theory can be obtained as twisted $\Z_2$ compactification of a \emph{5d} \gls{SCFT} obtained from a mass deforming previous brane webs, which in particular implies that this $4d$ \gls{SCFT} comes from deforming  \hyperref[sec:t34]{$\suf(4)_{12}{\times}\suf(2)_7{\times}\uf(1)$} \cite{Zafrir:2016wkk,Martone:2021drm}. 

The analysis of the moduli space of vacua of this theory resembles the previous cases. First notice that despite having a single \gls{higgs} ($v$) the flavor symmetry of the theory is semi-simple. A careful look at the levels reveals that it the entire $\suf(3){\times}\suf(2)$ factor should be realized on the $v=0$ stratum by a discretely gauged $SU(2)$ $\cN=2$ gauge theory. The experience built in the analysis of previous theories suggests $\rTf_v\equiv [I_3^*,\suf(2){\times}\sof(6)]_{\Z_2}$. This identification is perfectly confirmed by the exercise of reproducing the $a$ and $c$ central charges in table \ref{CchTh37} using the \gls{ccf}. The $\Z_2$ that we gauge does not act on the $\suf(2)$ factor, thus implying \gls{h}=1, and instead has only a component in the inner automorphism group of $\sof(6)$ with commutant $\suf(3)$.

\begin{figure}[h!]
\ffigbox{
\begin{subfloatrow}
\ffigbox[8.5cm][]{
\begin{tikzpicture}[decoration={markings,
mark=at position .5 with {\arrow{>}}}]
\begin{scope}[scale=1.5]
\node[bbc,scale=.5] (p0a) at (0,0) {};
\node[scale=.5] (p0b) at (0,-2) {};
\node[scale=.8] (t0b) at (0,-2.1) {$\suf(3)_{10}{\times}\suf(2)_6{\times}\uf(1)$};
\node[scale=.8] (p1) at (-.7,-1) {$[I_3^*,\suf(2){\times}\sof(6)]_{\Z_2}$};
\node[scale=.8] (p2) at (.7,-1) {$[I_1,\varnothing]$};
\node[scale=.8] (t2b) at (-.7,-1.5) {{\scriptsize$\big[v=0\big]$}};
\node[scale=.8] (t3b) at (.7,-1.6) {{\scriptsize$\big[u^5+v^3=0\big]$}};
\draw[red] (p0a) -- (p1);
\draw[red] (p0a) -- (p2);
\draw[red] (p1) -- (p0b);
\draw[red] (p2) -- (p0b);
\end{scope}
\begin{scope}[scale=1.5,xshift=3cm]
\node[scale=.5] (p4) at (0,2) {};
\node[scale=.8] (t0a) at (0,2.1) {$\H^{\rm d_{HB}}$};
\node[scale=.8] (p1a) at (0,1) {$\cT^{(1)}_{A_2,1}$};
\node[scale=.8] (p2a) at (0,0) {$\cS^{(1)}_{A_2,2}$};
\node[scale=.8] (t01) at (-.2,1.5) {$\af_2$};
\node[scale=.8] (t1a) at (-.2,.5) {$\cf_2$};
\node[scale=.8] (t2a) at (-.6,-.4) {$\cf_3$};
\node[scale=.8] (p0a) at (-.7,-1) {$\hyperref[sec:tTE62]{\spf(6)_5{\times}\uf(1)}$};
\node[scale=.8] (tp1) at (-.5,-1.6) {$\af_1$};
\node[scale=.8] (tp1) at (.7,-1) {?};
\node[scale=.5] (p0) at (0,-2) {};
\node[scale=.8] (t0b) at (0,-2.1) {$\suf(3)_{10}{\times}\suf(2)_6{\times}\uf(1)$};
\draw[blue,dashed]  (p0) to[out=30,in=-50] (p2a);
\draw[blue] (p0) -- (p0a);
\draw[blue] (p0a) -- (p2a);
\draw[blue] (p2a) -- (p1a);
\draw[blue] (p1a) -- (p4);
\end{scope}
\end{tikzpicture}}
{\caption{\label{CBhTh38}The Coulomb and Higgs stratification of $\suf(3)_{10}{\times}\suf(2)_6{\times}\uf(1)$.}}
\end{subfloatrow}\hspace{1cm}%
\begin{subfloatrow}
\capbtabbox[5cm]{%
  \renewcommand{\arraystretch}{1.1}
  \begin{tabular}{|c|c|} 
  \hline
  \multicolumn{2}{|c|}{$\suf(3)_{10}{\times}\suf(2)_6{\times}\uf(1)$}\\
  \hline\hline
  $(\D_u,\D_v)$  &\quad $\left(3,5\right)$\quad{} \\
  $24a$ & 92\\  
  $12c$ & 50\\
$\ff_k$ & $\suf(3)_{10}{\times}\suf(2)_{6}{\times}\uf(1)$ \\ 
$d_{\rm HB}$& 8\\
$h$&1\\
$T({\bf2}\bh)$&1\\
\hline\hline
\end{tabular}
}{%
  \caption{\label{CchTh38}Central charges, \gls{CB} parameters and \gls{ECB} dimension.}%
}
\end{subfloatrow}}{\caption{\label{TothTh38}Information about the $\suf(3)_{10}{\times}\suf(2)_6{\times}\uf(1)$.}}
\end{figure}

The analysis of the \gls{HB} proceeds also in a similar manner as previous examples. First let's analyze the \gls{ECB}. Using \eqref{gHWeq} we obtain that $12c_{\af_1}=34$. This information, by itself, it is not enough to identify the theory supported on $\af_1$ yet adding the constraints that the total \gls{HB} should be eight dimensional does obtaining $\bTf_{\af_1}\equiv \spf(6)_5{\times}\uf(1)$. The \gls{HB} of the latter allows to fill in most of the remaining part of the \gls{HB} \gls{Hasse} in figure \ref{CBhTh38} and it remains to characterize the higgs direction corresponding to turning on a vev for the fundantamental hypers on the gauge theory supported on the \gls{CB} \gls{ukstr}. Even with the help of the magnetic quiver, we are unable to determine the precise structure of this four dimensional component, which is indicated with a dashed line and a question mark in figure \ref{CBhTh38}, and leave this task for future work.

\subsubsection*{\boldsymbol{$\suf(2)_8{\times}\suf(2)_8{\times}\uf(1)^2$}}\label{sec:t37}

\begin{figure}[h!]
\ffigbox{
\begin{subfloatrow}
\ffigbox[8.5cm][]{
\begin{tikzpicture}[decoration={markings,
mark=at position .5 with {\arrow{>}}}]
\begin{scope}[scale=1.5]
\node[bbc,scale=.5] (p0a) at (0,0) {};
\node[scale=.5] (p0b) at (0,-2) {};
\node[scale=.8] (t0b) at (0,-2.1) {$\suf(2)_8{\times}\suf(2)_8{\times}\uf(1)^2$};
\node[scale=.8] (p1) at (-1,-1) {$[I_1,\varnothing]$};
\node[scale=.8] (p2) at (1,-1) {$[I_2,\suf(2)]$};
\node[scale=.8] (p3) at (0,-1) {$[\star\,{\rm w/}b=5]$};
\node[scale=.8] (t2b) at (-.7,-1.5) {{\scriptsize$\big[u^3+v^2=0\big]$}};
\node[scale=.8] (t3b) at (.7,-1.6) {{\scriptsize$\big[u=0\big]$}};
\node[scale=.8] (t3b) at (.25,-1.4) {{\scriptsize$\big[v=0\big]$}};
\draw[red] (p0a) -- (p1);
\draw[red] (p0a) -- (p2);
\draw[red] (p0a) -- (p3);
\draw[red] (p1) -- (p0b);
\draw[red] (p2) -- (p0b);
\draw[red] (p3) -- (p0b);
\end{scope}
\begin{scope}[scale=1.5,xshift=3cm]
\node[scale=.5] (p4) at (0,1) {};
\node[scale=.8] (t0a) at (0,1.1) {$\H^{\rm d_{HB}}$};
\node[scale=.8] (tp2) at (-.4,.5) {$\af_2$};
\node[scale=.8] (p3) at (-.4,0) {$\cT^{(1)}_{A_2,1}$};
\node[scale=.8] (tp2) at (.9,.25) {$\df_4$};
\node[scale=.8] (p2) at (1,-.5) {$\cT^{(1)}_{D_4,1}$};
\node[scale=.8] (tp1) at (.9,-1) {$A_3$};
\node[scale=.8] (tp2) at (.1,-.5) {$\cf_2$};
\node[scale=.8] (tp2) at (-.85,-.5) {$\cf_2$};
\node[scale=.8] (p1b) at (-1,-1) {$\cS^{(1)}_{A_2,2}$};
\node[scale=.8] (p1a) at (.3,-1) {$\cS^{(1)}_{A_2,2}$};
\node[scale=.8] (tp1) at (-.75,-1.5) {$\bar{h}_{2,2}$};
\node[scale=.8] (tp1) at (-.1,-1.5) {$\bar{h}_{2,2}$};
\node[scale=.5] (p0) at (0,-2) {};
\node[scale=.8] (t0b) at (0,-2.1) {$\suf(2)_8{\times}\suf(2)_8{\times}\uf(1)^2$};
\draw[blue] (p0) -- (p1a);
\draw[blue] (p0) -- (p1b);
\draw[blue] (p0) -- (p2);
\draw[blue] (p1b) -- (p3);
\draw[blue] (p1a) -- (p3);
\draw[blue] (p2) -- (p4);
\draw[blue] (p3) -- (p4);
\end{scope}
\end{tikzpicture}}
{\caption{\label{CBhTh39}The Coulomb and Higgs stratification of $\suf(2)_8{\times}\suf(2)_8{\times}\uf(1)^2$.}}
\end{subfloatrow}\hspace{1cm}%
\begin{subfloatrow}
\capbtabbox[5cm]{%
  \renewcommand{\arraystretch}{1.1}
  \begin{tabular}{|c|c|} 
  \hline
  \multicolumn{2}{|c|}{$\suf(2)_8{\times}\suf(2)_8{\times}\uf(1)^2$}\\
  \hline\hline
  $(\D_u,\D_v)$  &\quad $\left(3,4\right)$\quad{} \\
  $24a$ & 78\\  
  $12c$ & 42\\
$\ff_k$ & $\suf(2)_8{\times}\suf(2)_8{\times}\uf(1)^2$ \\ 
$d_{\rm HB}$& 6\\
$h$&0\\
$T({\bf2}\bh)$&0\\
\hline\hline
\end{tabular}
}{%
  \caption{\label{CchTh39}Central charges, \gls{CB} parameters and \gls{ECB} dimension.}%
}
\end{subfloatrow}}{\caption{\label{TothTh39}Information about the $\suf(2)_8{\times}\suf(2)_8{\times}\uf(1)^2$.}}
\end{figure}

This theory can be obtained by $\Z_2$ twisted compactification of the \emph{5d} brane web with two \emph{D5}, two \emph{NS5} and one (1,-1) \emph{5}brane all intersection at one point \cite{Zafrir:2016wkk}. This is thus connected by mass deformation to the theory \hyperref[sec:t35]{$\suf(3)_{10}{\times}\suf(3)_{10}{\times}\uf(1)$} discussed previously.

The state of our understanding of this theory is only marginally better than theory \hyperref[sec:t35]{$\suf(3)_{10}{\times}\suf(3)_{10}{\times}\uf(1)$}. In this case the theory is \gls{totally} but a careful analysis of the level of the two simple flavor symmetry factors, suggest that they are both realized as an IR-free rank-1 gauge theory supported on $v=0$. To gain a better feeling about the \gls{CB} stratification, it is useful leverage the \gls{ccf} to reproduce the $a$ and $c$ central charges in table \ref{CchTh39}. Performing this exercise we again don't get a unique answer but the most reasonable one predicts a $\rTf_{u^4+v^3}\equiv[I_1,\varnothing]$, $\rTf_{u}\equiv[I_2,\suf(2)]$ (which enhances one of the $\uf(1)$ factors) while the $\rTf_v$ is predicted to have $b=5$.

Again it is hard to perform a complete analysis of the \gls{HB} without a complete knowledge of the \gls{CB} stratification. Luckly in this case, the magnetic quiver, which can be readily obtained from the \emph{5d} realization, can be used to obtain the full \gls{Hasse} which is reproduced in figure \ref{CBhTh39}. Again we point out the appearance of the new elementary slices $\bar{h}_{2,2}$ \cite{AffineAJ}. 

\subsubsection*{\boldsymbol{$SU(3)+F+S$}}\label{sec:SU363}

This case was discussed before \cite{Landsteiner:1998pb} and more recently in \cite{Argyres:2018zay,Argyres:2020wmq}.

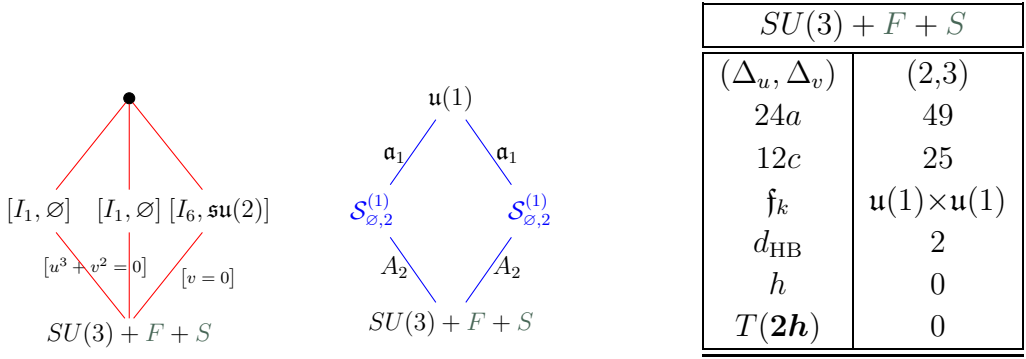
\begin{figure}[h!]
\ffigbox{
\begin{subfloatrow}
\ffigbox[7cm][]{
\begin{tikzpicture}[decoration={markings,
mark=at position .5 with {\arrow{>}}}]
\begin{scope}[scale=1.5]
\node[bbc,scale=.5] (p0a) at (0,0) {};
\node[scale=.5] (p0b) at (0,-2) {};
\node[scale=.8] (t0b) at (0,-2.1) {$SU(3)+\gls{F}+\gls{S}$};
\node[scale=.8] (p1) at (-.8,-1) {$[I_1,\varnothing]$};
\node[scale=.8] (p2) at (.8,-1) {$[I_6,\suf(2)]$};
\node[scale=.8] (p3) at (0,-1) {$[I_1,\varnothing]$};
\node[scale=.8] (t2b) at (-.3,-1.5) {{\scriptsize$\big[u^3+v^2=0\big]$}};
\node[scale=.8] (t3b) at (.7,-1.6) {{\scriptsize$\big[v=0\big]$}};
\draw[red] (p0a) -- (p1);
\draw[red] (p0a) -- (p2);
\draw[red] (p0a) -- (p3);
\draw[red] (p1) -- (p0b);
\draw[red] (p2) -- (p0b);
\draw[red] (p3) -- (p0b);
\end{scope}
\begin{scope}[scale=1.5,xshift=2.85cm]
\node[scale=.8] (p0a) at (0,0) {$\uf(1)$};
\node[scale=.8] (p0b) at (0,-2) {$SU(3)+\gls{F}+\gls{S}$};
\node[scale=.8] (p1) at (.7,-1) {$\blue{\cS^{(1)}_{\varnothing,2}}$};
\node[scale=.8] (t1a) at (0.5,-.5) {$\af_1$};
\node[scale=.8] (t1b) at (.5,-1.5) {$A_2$};
\node[scale=.8] (p2) at (-.7,-1) {$\blue{\cS^{(1)}_{\varnothing,2}}$};
\node[scale=.8] (t1a) at (-0.5,-.5) {$\af_1$};
\node[scale=.8] (t1b) at (-.5,-1.5) {$A_2$};
\draw[blue] (p0a) -- (p1);
\draw[blue] (p1) -- (p0b);
\draw[blue] (p0a) -- (p2);
\draw[blue] (p2) -- (p0b);
\end{scope}
\end{tikzpicture}}
{\caption{\label{CBSU336}The \gls{Hasse} for the \gls{CB} and the \gls{HB} of the $SU(3)$ gauge theory with one ${\bf 3}$ and one ${\bf 6}$.}}
\end{subfloatrow}\hspace{1cm}%
\begin{subfloatrow}
\capbtabbox[7cm]{%
  \renewcommand{\arraystretch}{1.1}
  \begin{tabular}{|c|c|} 
  \hline
  \multicolumn{2}{|c|}{$SU(3)+\gls{F}+\gls{S}$}\\
  \hline\hline
  $(\D_u,\D_v)$  &\quad (2,3)\quad{} \\
  $24a$ &  49\\  
  $12c$ & 25 \\
$\ff_k$ & $\uf(1){\times}\uf(1)$ \\ 
$d_{\rm HB}$&2\\
$h$&0\\
$T({\bf2}\bh)$&0\\
\hline\hline
 \end{tabular}
}{%
  \caption{\label{CcSU336}Central charges, \gls{CB} parameters and \gls{ECB} dimension.}%
}
\end{subfloatrow}}{\caption{\label{TotSU336}Information about the $SU(3)$ $\cN=2$ theory with one hypermultiplet in the ${\bf 3}$ and one in the ${\bf 6}$.}}
\end{figure}

Since this theory has two hypers in two different complex representations, we expect the flavor symmetry of the theory to be $\uf(1)\times \uf(1)$. The \gls{CB} stratification depicted in figure \ref{CBSU336} follows from the lagrangian analysis where again the two \gls{kstr} supporting a $[I_1,\varnothing]$ correspond to the \gls{CB} locus where only a pure $SU(2)$ $\cN=2$ theory arises. In this case, the $\rTf_v\equiv [I_6,\suf(2)]$ corresponds to a $\uf(1)$ theory with three massless hypers, two of which of charge one and the other of charge two. Notice two facts: 1) in this case there is an enhancement of the flavor symmetry on the \gls{CB}: $\uf(1)\to \uf(2)$ 2) the $\rTf_v$ has a one quaternionic dimensional \gls{HB} with a free massless hyper on its generic point. Thus we expect the full theory to have a non-trivial \gls{HB} with a one quaternionic dimensional first transition supporting a theory with a \gls{h}=0. Let's see how this works.

An $SU(3)$ theory with 1 flavor and one symmetric has the following gauge invariant operators:
\begin{align}
M_{\bf 3}&=Q \tilde{Q},\qquad M_{\bf 6}=X\tilde{X}\\
B_1&=X Q^2,\qquad B_3=X^3\\
\tilde{B}_1&=\tilde{X} \tilde{Q}^2,\qquad \tilde{B}_3=\tilde{X}^3,
\end{align}
where we have suppressed the gauge indices, the $X$ and $\tilde X$ are the component of the symmetric hypermultiplet, $Q$ and $\tilde Q$ are the fundamental flavors and the subscript of $B_i$ labels the power of $X$. Also in labeling the representations we chosen the convention of \cite{Yamatsu:2015npn} where the ${\bf 6}$ has Dynkin label $(0,2)$.  When $B_3$ and $\tilde{B}_3$ (which have no fundamental components) are turned on the theory flows precisely to $\blue{\cS^{(1)}_{\varnothing,2}}$, that is $\cN=4$ $SU(2)$ gauge theory. The structure of the first stratum of the \gls{HB} is easily obtained once the F-term conditions are taken into account. For $Q=0$, we find the following relation:
\beq
B_3 \tilde{B}_3=M_{\bf 6}^3\quad\Rightarrow\quad \bar{\blue{\Sf}}_{\uf(1)}=\C^2/\Z_3\equiv A_2.
\eeq
therefore concluding that $\bTf_{A_2}=\blue{\cS^{(1)}_{\varnothing,2}}$ matching our expected \gls{h}=1. The fact that there are two such disconnected components, as reported in figure \ref{CBSU336}, was computed by a careful analysis of the magnetic quiver associated to this theory \cite{Bourget:2021csg}.

It is interesting to notice that this is an example where the transition on the \gls{HB} differs from the generic transition into the \gls{MB}. This is though expected because of the flavor symmetry enhancement. The full \gls{HB} of the theory is reported in figure \ref{CBSU336}.

\begin{center}
\rule[1mm]{2cm}{.4pt}\hspace{1cm}$\circ$\hspace{1cm} \rule[1mm]{2cm}{.4pt}
\end{center}

\subsection{$\spf(14)$ series}

This series has a single top theory and four entry total. The moduli space structure of the theories in this series don't present particular difficulties and have been completely solved.

\subsubsection*{\boldsymbol{$\spf(14)_9$}}\label{sec:TD4f}

\begin{figure}[h!]
\ffigbox{
\begin{subfloatrow}
\ffigbox[8.5cm][]{
\begin{tikzpicture}[decoration={markings,
mark=at position .5 with {\arrow{>}}}]
\begin{scope}[scale=1.5]
\node[bbc,scale=.5] (p0a) at (0,-1) {};
\node[scale=.5] (p0b) at (0,-3) {};
\node[scale=.8] (t0b) at (0,-3.1) {$\spf(14)_9$};
\node[scale=.8] (p3) at (.7,-2) {\ \ $[I_1,\varnothing]$};
\node[scale=.8] (p1) at (-.7,-2) {$[I_6^*,\spf(14)]$\ \ };
%\node[scale=.8] (t1a) at (-.6,-.4) {$I_0^*$};
%\node[scale=.8] (t2c) at (-.7,-1.6) {$K_{\D_7}$};
\node[scale=.8] (t1c) at (.85,-2.65) {{\scriptsize$\big[u^4+v^3=0\big]$}};
%\node[scale=.8] (t2c) at (.7,-1.6) {$K_{2\D_7}$};
\node[scale=.8] (t3c) at (-.7,-2.6) {{\scriptsize$\big[v=0\big]$}};
\draw[red] (p0a) -- (p1);
\draw[red] (p0a) -- (p3);
\draw[red] (p1) -- (p0b);
\draw[red] (p3) -- (p0b);
\end{scope}
\begin{scope}[scale=1.5,xshift=2.85cm]
\node[scale=.5] (p4) at (0,1) {};
\node[scale=.8] (t0a) at (0,1.1) {$\H^{\rm d_{HB}}$};
\node[scale=.8] (tp0) at (-.2,.5) {$\ef_6$};
\node[scale=.8] (p3) at (0,0) {$\cT^{(1)}_{E_6,1}$};
\node[scale=.8] (tp0) at (-.2,-.5) {$\cf_5$};
\node[scale=.8] (p2) at (0,-1) {$\cS^{(1)}_{E_6,2}$};
\node[scale=.8] (tp2) at (-.2,-1.5) {$\cf_6$};
\node[scale=.8] (p1) at (0,-2) {\hyperref[sec:t22]{$\spf(12)_8$}};
\node[scale=.8] (tp1) at (-.2,-2.5) {$\cf_7$};
\node[scale=.5] (p0) at (0,-3) {};
\node[scale=.8] (t0b) at (0,-3.1) {$\spf(14)_9$};
\draw[blue] (p0) -- (p1);
\draw[blue] (p1) -- (p2);
\draw[blue] (p2) -- (p3);
\draw[blue] (p3) -- (p4);
\end{scope}
\end{tikzpicture}}
{\caption{\label{CBhTh22}The Coulomb and Higgs stratification of $\spf(14)_9$}}
\end{subfloatrow}\hspace{1cm}%
\begin{subfloatrow}
\capbtabbox[5cm]{%
  \renewcommand{\arraystretch}{1.1}
  \begin{tabular}{|c|c|} 
  \hline
  \multicolumn{2}{|c|}{$\spf(14)_9$}\\
  \hline\hline
  $(\D_u,\D_v)$  &\quad $\left(6,8\right)$\quad{} \\
  $24a$ & 185\\  
  $12c$ & 107\\
$\ff_k$ & $\spf(14)_9$ \\ 
$d_{\rm HB}$& 29\\
$h$&7\\
$T({\bf2}\bh)$&1\\
\hline\hline
\end{tabular}
}{%
  \caption{\label{CchTh22}Central charges, \gls{CB} parameters and \gls{ECB} dimension.}%
}
\end{subfloatrow}}{\caption{\label{TothTh22}Information about the $\spf(14)_9$}}
\end{figure}

This specific theory was first discussed in \cite{Argyres:2020wmq} but it can be constructed using the general construction presented in \cite{Ohmori:2018ona}. Specifically it can be obtained as a $T^2$ compactification of the minimal $(D_7,D_7)$ conformal matter \cite{DelZotto:2014hpa} with a $\Z_2$ valued non-commuting holonomy\footnote{As explained \cite{Ohmori:2018ona}, for $(D_{2n+1},D_{2n+1})$ there are two possible such $\Z_2$ compactifications, one that presevers the full enhanced $\spf(4n+2)$ flavor symmetry, and one that only preserves a subgroup. We are here interested in the former.}. This theory can also be engineered in twisted $D_5$ class-$\cS$ as it is described in the analysis of \cite{Ohmori:2018ona}.

This theory is not \gls{totally} with $v$ being the only \gls{higgs}. Not surprisingly the flavor symmetry is simple and by noticing that $k_{\spf(14)}=\D_v+1$ we immediately can identify the theory realizing the flavor symmetry on the \gls{CB}: $\rTf_v\equiv [I_6^*,\spf(14)]$. This in turn implies that the theory has \gls{h}=7 and the theory supported on the $\cf_7$ stratum of the \gls{HB} should have no \gls{ECB}. Imposing that the $a$ and $c$ central charges in table \ref{CchTh22} can be reproduced by the \gls{ccf} allows to complete the analysis of the \gls{CB} stratification by identifying the theory supported on the \gls{kstr} as a $[I_1,\varnothing]$. 

The analysis of the \gls{HB} is fairly easy. The fact that the flavor symmetry is simple it immediately implies that the \gls{Hasse} is \gls{Hlin}. Thus the symplectic stratification is basically determined once the theory supported on the first stratum is identified. This identification can be readily performed by noticing that the $\spf(14)$ higgsing is of \gls{gHW} type thus we can use \eqref{gHWeq} and readily derive $12c_{\bTf_{\cf_7}}=76$. This information is enough to identify $\bTf_{\cf_7}\equiv \spf(12)_8$. Following the subsequent higgsings reproduces the \gls{Hasse} in figure \ref{CBhTh22} completing our analysis.

\subsubsection*{\boldsymbol{$\suf(2)_8{\times}\spf(10)_7$}}\label{sec:t40}

This theory was initially obtained in the $\Z_2$ twisted $D_4$ class-$\cS$ series \cite{Chacaltana:2013oka} where most of the CFT data reported in figure \ref{CchTh16}. As usual we will be using this information to fill in the detailed structure of the full moduli space. And as usual we will start our analysis from the \gls{CB} side of things.

\begin{figure}[h!]
\ffigbox{
\begin{subfloatrow}
\ffigbox[8.5cm][]{
\begin{tikzpicture}[decoration={markings,
mark=at position .5 with {\arrow{>}}}]
\begin{scope}[scale=1.5]
\node[bbc,scale=.5] (p0a) at (0,-1) {};
\node[scale=.5] (p0b) at (0,-3) {};
\node[scale=.8] (t0b) at (0,-3.1) {$\suf(2)_8{\times}\spf(10)_7$};
\node[scale=.8] (p3) at (1,-2) {\ \ $[I_1,\varnothing]$};
\node[scale=.8] (p2) at (0,-2) {\ \ $[I_2,\suf(2)]$};
\node[scale=.8] (p1) at (-1,-2) {$[I_4^*,\spf(10)]$\ \ };
%\node[scale=.8] (t1a) at (-.6,-.4) {$I_0^*$};
%\node[scale=.8] (t2c) at (-.7,-1.6) {$K_{\D_7}$};
\node[scale=.8] (t1c) at (.85,-2.75) {{\scriptsize$\big[u^3+v^2=0\big]$}};
\node[scale=.8] (t2c) at (-.25,-2.4) {{\scriptsize$\big[u=0\big]$}};
%\node[scale=.8] (t2c) at (.7,-1.6) {$K_{2\D_7}$};
\node[scale=.8] (t3c) at (-.7,-2.7) {{\scriptsize$\big[v=0\big]$}};
\draw[red] (p0a) -- (p1);
\draw[red] (p0a) -- (p2);
\draw[red] (p0a) -- (p3);
\draw[red] (p1) -- (p0b);
\draw[red] (p2) -- (p0b);
\draw[red] (p3) -- (p0b);
\end{scope}
\begin{scope}[scale=1.5,xshift=2.85cm]
\node[scale=.5] (p0a) at (0.7,1) {};
\node[scale=.8] (t0a) at (.7,1.1) {$\H^{\rm d_{HB}}$};
\node[scale=.8] (tp2) at (1.4,.5) {$\df_4$};
\node[scale=.8] (tp2) at (0.3,.5) {$\ef_6$};
\node[scale=.8] (p0) at (1.4,0) {$\cT^{(1)}_{D_4,1}$};
\node[scale=.8] (tp2) at (1.25,-.45) {$\cf_3$};
\node[scale=.8] (p1b) at (0,-.5) {$\cT^{(1)}_{E_6,1}$};
\node[scale=.8] (p1a) at (1.4,-1) {$\cS^{(1)}_{D_4,2}$};
\node[scale=.8] (tp2) at (-.5,-1.15) {$\cf_5$};
\node[scale=.8] (tp2) at (.2,-1.35) {$\af_1$};
\node[scale=.8] (tp2) at (1.2,-1.5) {$\cf_4$};
\node[scale=.8] (p2a) at (.7,-2) {\hyperref[sec:t25]{$\suf(2)_8{\times}\spf(8)_6$}};
\node[scale=.8] (p2b) at (-.7,-2) {$\cS^{(1)}_{E_6,2}$};
\node[scale=.8] (tp1) at (-.5,-2.6) {$\af_1$};
\node[scale=.8] (tp2) at (.5,-2.6) {$\cf_5$};
\node[scale=.5] (p0b) at (0,-3) {};
\node[scale=.8] (t0b) at (0,-3.1) {$\suf(2)_8{\times}\spf(10)_7$};
\draw[blue] (p0a) -- (p0);
\draw[blue] (p0a) -- (p1b);
\draw[blue] (p0) -- (p1a);
\draw[blue] (p1a) -- (p2a);
\draw[blue] (p1b) -- (p2a);
\draw[blue] (p1b) -- (p2b);
\draw[blue] (p2a) -- (p0b);
\draw[blue] (p2b) -- (p0b);
\end{scope}
\end{tikzpicture}}
{\caption{\label{CBhTh16}The Coulomb and Higgs stratification of $\suf(2)_8{\times}\spf(10)_7$.}}
\end{subfloatrow}\hspace{1cm}%
\begin{subfloatrow}
\capbtabbox[5cm]{%
  \renewcommand{\arraystretch}{1.1}
  \begin{tabular}{|c|c|} 
  \hline
  \multicolumn{2}{|c|}{$\suf(2)_8{\times}\spf(10)_7$}\\
  \hline\hline
  $(\D_u,\D_v)$  &\quad $\left(4,6\right)$\quad{} \\
  $24a$ &  125\\  
  $12c$ & 71\\
$\ff_k$ & $\suf(2)_8{\times}\spf(10)_7$ \\ 
$d_{\rm HB}$& 17\\
$h$&5\\
$T({\bf2}\bh)$&1\\
\hline\hline
\end{tabular}
}{%
  \caption{\label{CchTh16}Central charges, \gls{CB} parameters and \gls{ECB} dimension.}%
}
\end{subfloatrow}}{\caption{\label{TothTh16}Information about the $\suf(2)_8{\times}\spf(10)_7$.}}
\end{figure}

First notice that both of the \gls{CB} parameter are \gls{higgs}, therefore we expect a non-linear \gls{HB} \gls{Hasse} and a somewhat involved \gls{HB} structure. This in fact matches with the fact that the flavor symmetry of the theory contains two simple flavor factors. Giving its level, we can use the \doub\ to identify the theory realizing the $\suf(2)$ factor while the level of the $\spf(10)$ is $k_{\spf(10)}=\D_v+1$. Then there is a natural guess for how these two simple factors are realized on the \gls{CB}: $\rTf_u=[I_2,\suf(2)]$ and $\rTf_v=[I_4^*,\spf(10)]$. This immediately suggests that this rank-2 theory has \gls{h}$=4$ and a $\cf_5$ transition followed by a $\cf_4$ one. As a check that this identification is indeed correct, the reader can easily match the $a$ and $c$ central charges using \eqref{actotaint}-\eqref{actotbint}. This exercise also fixes the theory supported on the \gls{kstr} to be $\rTf_{u^3+v^2}\equiv [I_1,\varnothing]$. 

Now let's analyze the \gls{HB}. Start first from the $\af_1$ transition which is a regular \gls{HB} stratum and as such should support a rank-1 theory. Noticing that the higgsing is of \gls{gHW} type and using \eqref{gHWeq} immediately accomplish the task of making the identification $\bTf_{\af_1}\equiv \cS^{(1)}_{E_6,2}$. To identify the theory supported on the five dimensional \gls{ECB} we can perform a similar analysis. Again the $\cf_5$ higgsing is of \gls{gHW} type and we can use \eqref{gHWeq} to compute the $c$ central charge of the theory. This doesn't quite specify the theory uniquely but either the constraint derived from the \gls{CB} analysis that the $\cf_5$ stratum should be followed by a $\cf_4$ or simply imposing that the total \gls{HB} dimension, is enough to make the identification $\bTf_{\cf_5}\equiv \suf(2)_6{\times}\suf(8)_8$ and conclude our analysis.

\subsubsection*{\boldsymbol{$\suf(2)_5{\times}\spf(8)_7$}}\label{sec:t64}

This theory was initially obtained in the $\Z_2$ twisted $D_4$ class-$\cS$ series \cite{Chacaltana:2013oka} and again the CFT data reported in figure \ref{CcTh17} is for the most part derived from this initial analysis.

\begin{figure}[h!]
\ffigbox{
\begin{subfloatrow}
\ffigbox[8.5cm][]{
\begin{tikzpicture}[decoration={markings,
mark=at position .5 with {\arrow{>}}}]
\begin{scope}[scale=1.5]
\node[bbc,scale=.5] (p0a) at (0,-1) {};
\node[scale=.5] (p0b) at (0,-3) {};
\node[scale=.8] (t0b) at (0,-3.1) {$\suf(2)_5{\times}\spf(8)_7$};
\node[scale=.8] (p3) at (1,-2) {\ \ $[I_1,\varnothing]$};
\node[scale=.8] (p2) at (0,-2) {\ \ $\blue{\cS^{(1)}_{\varnothing,2}}$};
\node[scale=.8] (p1) at (-1,-2) {$[I_3^*,\spf(8)]$\ \ };
%\node[scale=.8] (t1a) at (-.6,-.4) {$I_0^*$};
%\node[scale=.8] (t2c) at (-.7,-1.6) {$K_{\D_7}$};
\node[scale=.8] (t1c) at (.85,-2.75) {{\scriptsize$\big[u^3+v^2=0\big]$}};
\node[scale=.8] (t2c) at (-.25,-2.4) {{\scriptsize$\big[u=0\big]$}};
%\node[scale=.8] (t2c) at (.7,-1.6) {$K_{2\D_7}$};
\node[scale=.8] (t3c) at (-.7,-2.7) {{\scriptsize$\big[v=0\big]$}};
\draw[red] (p0a) -- (p1);
\draw[red] (p0a) -- (p2);
\draw[red] (p0a) -- (p3);
\draw[red] (p1) -- (p0b);
\draw[red] (p2) -- (p0b);
\draw[red] (p3) -- (p0b);
\end{scope}
\begin{scope}[scale=1.5,xshift=2.85cm]
\node[scale=.5] (p0a) at (0.7,1) {};
\node[scale=.8] (t0a) at (.7,1.1) {$\H^{\rm d_{HB}}$};
\node[scale=.8] (tp2) at (0.5,.5) {$\df_4$};
\node[scale=.8] (p0) at (.7,0) {$\cT^{(1)}_{D_4,1}$};
\node[scale=.8] (tp2) at (0.25,-.4) {$\af_5$};
\node[scale=.8] (tp2) at (1.2,-.5) {$\cf_3$};
\node[scale=.8] (p1b) at (0,-1) {\hyperref[sec:SU3Nf5]{$SU(3)+6\gls{F}$}};
\node[scale=.8] (p1a) at (1.4,-1) {$\cS^{(1)}_{D_4,2}$};
\node[scale=.8] (tp2) at (-.5,-1.45) {$\cf_4$};
\node[scale=.8] (tp2) at (.2,-1.55) {$\af_1$};
\node[scale=.8] (tp2) at (1.25,-1.5) {$\cf_3$};
\node[scale=.8] (p2a) at (.7,-2) {\hyperref[sec:t26]{$\suf(2)_5{\times}\spf(6)_6{\times}\uf(1)$}};
\node[scale=.8] (p2b) at (-.7,-2) {\hyperref[sec:G247]{$G_2+4\gls{F}$}};
\node[scale=.8] (tp1) at (-.5,-2.6) {$\af_1$};
\node[scale=.8] (tp2) at (.5,-2.6) {$\cf_4$};
\node[scale=.5] (p0b) at (0,-3) {};
\node[scale=.8] (t0b) at (0,-3.1) {$\suf(2)_5{\times}\spf(8)_7$};
\draw[blue] (p0a) -- (p0);
\draw[blue] (p0) -- (p1a);
\draw[blue] (p0) -- (p1b);
\draw[blue] (p1a) -- (p2a);
\draw[blue] (p1b) -- (p2a);
\draw[blue] (p1b) -- (p2b);
\draw[blue] (p2a) -- (p0b);
\draw[blue] (p2b) -- (p0b);
\end{scope}
\end{tikzpicture}}
{\caption{\label{CBhTh17}The Coulomb and Higgs stratification of $\suf(2)_5{\times}\spf(8)_7$.}}
\end{subfloatrow}\hspace{1cm}%
\begin{subfloatrow}
\capbtabbox[5cm]{%
  \renewcommand{\arraystretch}{1.1}
  \begin{tabular}{|c|c|} 
  \hline
  \multicolumn{2}{|c|}{$\suf(2)_5{\times}\spf(8)_7$}\\
  \hline\hline
  $(\D_u,\D_v)$  &\quad $\left(4,6\right)$\quad{} \\
  $24a$ &  123\\  
  $12c$ & 69\\
$\ff_k$ & $\suf(2)_5{\times}\spf(8)_7$ \\ 
$d_{\rm HB}$& 15\\
$h$&5\\
$T({\bf2}\bh)$&1\\
\hline\hline
\end{tabular}
}{%
  \caption{\label{CcTh17}Central charges, \gls{CB} parameters and \gls{ECB} dimension.}%
}
\end{subfloatrow}}{\caption{\label{TothTh17}Information about the $\suf(2)_5{\times}\spf(8)_7$.}}
\end{figure}

We start our analysis of the full moduli space structure from the \gls{CB}. The fact that this theory has two \gls{higgs} is reflected in the fact that the flavor symmetry has two simple flavor factor. As we will see shortly, this will give rise to an involved \gls{HB} \gls{Hasse}, expectedly. The identification of how the flavor symmetry is realized on the \gls{CB}  is straightforward since the level of both simple flavor factors differ by one from from the \gls{CB} parameters. This observation then readily leads to the following identifications: $\rTf_u\equiv \blue{\cS^{(1)}_{\varnothing,2}}$ and $\rTf_v\equiv[I_3^*,\spf(8)]$ which in turn imply that for this theory \gls{h}$=5$ and it furthermore implies that the $\cf_4$ Higgsing supports a theory with a one dimensional \gls{ECB} and a rank decreasing $\cf_3$ transition while the $\af_1$ component of the \gls{ECB} supports a theory with a four dimensional \gls{ECB}. This guess can be checked by plugging in the \gls{bi} for these theories in \eqref{actotaint}-\eqref{actotbint} and reproducing the $a$ and $c$ reported in table \ref{CcTh17}. This also fixes the theory supported on the \gls{kstr} which is found to be our usual $[I_1,\varnothing]$. 

Let's turn now to the \gls{HB}. Both \gls{ECB} arise from lagrangian theories supported on the \gls{CB} and are therefore of the \gls{gHW} type. This makes our analysis fairly easy. We leave it up to the careful reader to check that the results of \eqref{gHWeq} applied separately to $\af_1$ and $\cf_5$ uniquely lead to the identification $\bTf_{\af_1}\equiv G_2+4\gls{F}$ and $\bTf_{\af_1}\equiv\suf(2)_5{\times}\spf(6)_6{\times}\uf(1)$. The rest of the \gls{Hasse} in figure \ref{CBhTh17} can be filled in by following the \gls{HB}s of these two theories which are worked out in figure \ref{CBG247} and \ref{CBTh7} respectively giving rise to the fairly intricate diagram in figure \ref{CBhTh17}. This theory can be mass deformed into the theory we are describing next.

\subsubsection*{$\boldsymbol{\spf(8)_6{\times}\uf(1)}$}\label{sec:R24}

The $\spf(8)_6{\times}\uf(1)$ is the rank-2 entry of an infinite family of $\cN=2$ \gls{SCFT}s discussed in \cite{Chacaltana:2014nya}. The $R_{2,2N}$ is a rank $N$ \gls{SCFT} with $\spf(4N)_{2N+2}\times \uf(1)$ flavor symmetry whose $USp(2N)$ gauging is S-dual to an $\cN=2$ $SU(2N+1)$ gauge theory with one symmetric and one antisymmetric with $R_{2,2}\equiv \cS^{(1)}_{A_2,2}$, for more details we refer to the original paper.

\begin{figure}[t!]
\ffigbox{
\begin{subfloatrow}
\ffigbox[8cm][]{
\begin{tikzpicture}[decoration={markings,
mark=at position .5 with {\arrow{>}}}]
\begin{scope}[scale=1.5]
\node[bbc,scale=.5] (p0a) at (0,-1) {};
\node[scale=.5] (p0b) at (0,-3) {};
\node[scale=.8] (t0b) at (0,-3.1) {$\spf(8)_6{\times}\uf(1)$};
\node[scale=.8] (p2) at (.7,-2) {\ \ $[I_3^*,\spf(8)]$};
\node[scale=.8] (p1) at (-.7,-2) {$[I_1,\varnothing]$\ \ };
%\node[scale=.8] (t1a) at (-.6,-.4) {$I_0^*$};
%\node[scale=.8] (t2c) at (-.7,-1.6) {$K_{\D_7}$};
\node[scale=.8] (t1c) at (-.85,-2.7) {{\scriptsize$\big[u^5+v^3=0\big]$}};
%\node[scale=.8] (t2c) at (.7,-1.6) {$K_{2\D_7}$};
\node[scale=.8] (t1c) at (.7,-2.7) {{\scriptsize$\big[v=0\big]$}};
\draw[red] (p0a) -- (p1);
\draw[red] (p0a) -- (p2);
\draw[red] (p1) -- (p0b);
\draw[red] (p2) -- (p0b);
\end{scope}
\begin{scope}[scale=1.5,xshift=2.5cm]
\node[scale=.5] (p0a) at (0,1) {};
\node[scale=.8] (t0a) at (0,1.1) {$\H^{\rm d_{HB}}$};
\node[scale=.8] (p0b) at (0,-2) {\hyperref[sec:tTE62]{$\spf(6)_5{\times}\uf(1)$}};
\node[scale=.8] (p1) at (0,0) {$\cT^{(1)}_{A_2,1}$};
\node[scale=.8] (p2) at (0,-1) {$\cS^{(1)}_{A_2,2}$};
\node[scale=.8] (t1c) at (.3,.5) {$\af_2$};
\node[scale=.8] (t2c) at (.3,-0.5) {$\cf_2$};
\node[scale=.8] (t1c) at (.3,-1.5) {$\cf_3$};
\node[scale=.8] (t1c) at (.3,-2.5) {$\cf_4$};
\node[scale=.8] (p0c) at (0,-3) {};
\node[scale=.8] (t0c) at (0,-3.1) {$\spf(8)_6{\times}\uf(1)$};
\draw[blue] (p0a) -- (p1);
\draw[blue] (p1) -- (p2);
\draw[blue] (p2) -- (p0b);
\draw[blue] (p0c) -- (p0b);
\end{scope}
\end{tikzpicture}}
{\caption{\label{CBR24}The Coulomb and Higgs stratification of $\spf(8)_6{\times}\uf(1)$.}}
\end{subfloatrow}\hspace{1cm}%
\begin{subfloatrow}
\capbtabbox[5cm]{%
  \renewcommand{\arraystretch}{1.1}
  \begin{tabular}{|c|c|} 
  \hline
  \multicolumn{2}{|c|}{$\spf(8)_6{\times}\uf(1)$}\\
  \hline\hline
  $(\D_u,\D_v)$  &\quad $\left(3,5\right)$\quad{} \\
  $24a$ & 95\\  
  $12c$ & 53\\
$\ff_k$ & $\spf(8)_6{\times}\uf(1)$ \\ 
$d_{\rm HB}$& 11\\
$h$&4\\
$T({\bf2}\bh)$&1\\
\hline\hline
\end{tabular}
}{%
  \caption{\label{CcR24}Central charges, \gls{CB} parameters and \gls{ECB} dimension.}%
}
\end{subfloatrow}}{\caption{\label{TotR24}Information about the $\spf(8)_6{\times}\uf(1)$ theory.}}
\end{figure}

This theory again is not \gls{totally} and, relatedly, has a single simple flavor factor. This signals that the \gls{HB} is \gls{Hlin}. Since the level of the $\spf(8)$ flavor factor differs precisely by one from the only \gls{higgs} ($v$ in this case), it is immediate to realize the $\spf(8)$ as the flavor symmetry of $\rTf_v\equiv[I_3^*,\spf(8)]$. This in turn implies that \gls{h}=4 and that the $\cf_4$ stratum on the \gls{HB} should be followed by a $\cf_3$. 
 
The analysis of the \gls{HB} is straightforward. In fact the \gls{CB} analysis signals that the $\spf(8)$ higgsing is of \gls{gHW} type and thus we can use \eqref{gHWeq} to compute the central charge of the theory supported on the $\cf_4$ finding $12c_{\bTf_{\cf_4}}=34$. Since we are higgsing along an \gls{ECB} direction, $\bTf_{\cf_4}$ has to be rank-2, thus the $c$ central charge we find does not uniquely specifies it. We can use then either the information coming from the \gls{CB} analysis of the presence of a subsequent $\cf_3$ Higgsing or the constraint coming from the total dimension of the \gls{HB} of $\spf(8)_6{\times}\uf(1)$. Either lift the degeneracy and lead us to the conclusion that $\bTf_{\cf_4}=\spf(6)_5{\times}\uf(1)$. This $ \spf(8)_6{\times}\uf(1)\to \spf(6)_5{\times}\uf(1)$ can also be seen directly from the 5d realization \cite{Giacomelli:2020gee}. The rest of the \gls{HB} \gls{Hasse} in figure \ref{CBR24} can be inferred by following the subsequent higgsing of $\spf(6)_5{\times}\uf(1)$ which are depicted in figure \ref{CBtTE}.

\subsubsection*{\boldsymbol{$USp(4)+3V$}}\label{sec:Sp435}

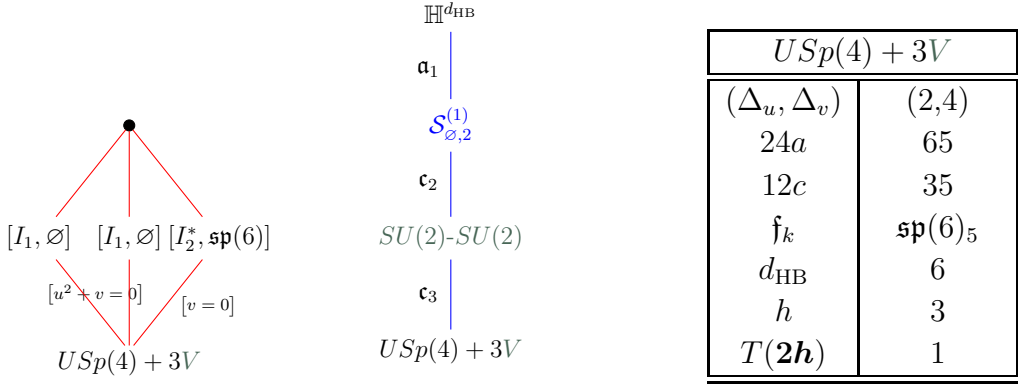
\begin{figure}[h!]
\ffigbox{
\begin{subfloatrow}
\ffigbox[7cm][]{
\begin{tikzpicture}[decoration={markings,
mark=at position .5 with {\arrow{>}}}]
\begin{scope}[scale=1.5]
\node[bbc,scale=.5] (p0a) at (0,0) {};
\node[scale=.5] (p0b) at (0,-2) {};
\node[scale=.8] (t0b) at (0,-2.1) {$USp(4)+3\gls{V}$};
\node[scale=.8] (p1) at (-.8,-1) {$[I_1,\varnothing]$};
\node[scale=.8] (p2) at (.8,-1) {$[I_2^*,\spf(6)]$};
\node[scale=.8] (p3) at (0,-1) {$[I_1,\varnothing]$};
\node[scale=.8] (t2b) at (-.3,-1.5) {{\scriptsize$\big[u^2+v=0\big]$}};
\node[scale=.8] (t3b) at (.7,-1.6) {{\scriptsize$\big[v=0\big]$}};
\draw[red] (p0a) -- (p1);
\draw[red] (p0a) -- (p2);
\draw[red] (p0a) -- (p3);
\draw[red] (p1) -- (p0b);
\draw[red] (p2) -- (p0b);
\draw[red] (p3) -- (p0b);
\end{scope}
\begin{scope}[scale=1.5,xshift=2.85cm]
\node[scale=.8] (p0a) at (0,1) {$\H^{d_{\rm HB}}$};
\node[scale=.8] (t1a) at (-0.2,.5) {$\af_1$};
\node[scale=.8] (p1) at (-0,0) {$\blue{\cS^{(1)}_{\varnothing,2}}$};
\node[scale=.8] (t1a) at (-0.2,-.5) {$\cf_2$};
\node[scale=.8] (p2) at (-0,-1) {\hyperref[sec:SU22a]{$SU(2)\text{-}SU(2)$}};
\node[scale=.8] (t1b) at (-.2,-1.5) {$\cf_3$};
\node[scale=.8] (p0b) at (0,-2) {$USp(4)+3\gls{V}$};
\draw[blue] (p0a) -- (p1);
\draw[blue] (p1) -- (p2);
\draw[blue] (p2) -- (p0b);
\end{scope}
\end{tikzpicture}}
{\caption{\label{CBSp435}The \gls{Hasse} for the \gls{CB} of the $USp(4)$ gauge theory with three hypermultiplets in the ${\bf 5}$.}}
\end{subfloatrow}\hspace{1cm}%
\begin{subfloatrow}
\capbtabbox[7cm]{%
  \renewcommand{\arraystretch}{1.1}
  \begin{tabular}{|c|c|} 
  \hline
  \multicolumn{2}{|c|}{$USp(4)+3\gls{V}$}\\
  \hline\hline
  $(\D_u,\D_v)$  &\quad (2,4)\quad{} \\
  $24a$ & 65\\  
  $12c$ & 35 \\
$\ff_k$ & $\spf(6)_5$ \\ 
$d_{\rm HB}$&6\\
$h$&3\\
$T({\bf2}\bh)$&1\\
\hline\hline
 \end{tabular}
}{%
  \caption{\label{CcSp435}Central charges, \gls{CB} parameters and \gls{ECB} dimension.}%
}
\end{subfloatrow}}{\caption{\label{TotSp435} Information about the $USp(4)$ $\cN=2$ theory with three hypermultiplet in the ${\bf 5}$}}
\end{figure}

As we discussed in the analysis of other $USp(4)$ theories, there are two interesting directions on the \gls{CB} leaving an $SU(2)$ unbroken. Each \gls{V} (${\bf 5}$) contributes a massless hypermultiplet in the ${\bf 3}$ for one $SU(2)$ and none for the other with the result that we expect a \gls{kstr} with an effective $SU(2)$ with 3 adjoint hypermultiplets in the low-energy. The other \gls{kstr} ``splits'' into two, each supporting a $[I_1,\varnothing]$, thus we conclude that \gls{h}=3 and furthermore we expect that the rank-2 theory supported on the \gls{ECB} has either a $\af_3$ or $\cf_2$ \gls{HB} transition. This analysis, which is depicted in figure \ref{CBSp435}, can be confirmed both by reproducing the central charges of the theories using our central charge formulae or explicitly studying the discrimant locus of the Seiberg-Witten curve which is known explicitly \cite{Argyres:1995fw}.

Let's move to the analysis of the \gls{HB}. Since we have an explicit lagrangian description available, we could solve for the higgsging by standard methods \cite{Argyres:1996eh} but we will find it quicker to leverage the geometric constraints which can be derived from the consistency of the full moduli space. In fact from our \gls{CB} analysis we know that the theory supported on the three dimensional \gls{ECB} should have an either $\cf_2$ or $\af_3$ transition. Furthermore, to reproduce the total dimension of the \gls{HB} (which can be computed both from anomaly matching or standard Hyperk\"ahler quotient), implies this rank-2 theory should furthermore have a three dimensional \gls{HB}. This enough information to conclude that $\bTf_{\cf_3}\equiv SU(2)-SU(2)$.

\begin{center}
\rule[1mm]{2cm}{.4pt}\hspace{1cm}$\circ$\hspace{1cm} \rule[1mm]{2cm}{.4pt}
\end{center}

\subsection{$\suf(5)$ series}

This series has a single top theory from which we can reach the remaining three theories. The moduli space structure of these theories is fairly involved and some open questions on the details remain and are explained in the text below. The HB Hasse diagrams below are largely obtained by quiver subtraction, for a discussion see \emph{e.g.} \cite{Bourget:2021csg}.

\subsubsection*{\boldsymbol{$\suf(5)_{16}$}}\label{sec:t43}

\begin{figure}[h!]
\ffigbox{
\begin{subfloatrow}
\ffigbox[8.5cm][]{
\begin{tikzpicture}[decoration={markings,
mark=at position .5 with {\arrow{>}}}]
\begin{scope}[scale=1.5]
\node[bbc,scale=.5] (p0a) at (0,0) {};
\node[scale=.5] (p0b) at (0,-2) {};
\node[scale=.8] (t0b) at (0,-2.1) {$\suf(5)_{16}$};
\node[scale=.8] (p1) at (-.7,-1) {$[I_1,\varnothing]$};
\node[scale=.8] (p2) at (.7,-1) {$[I_2^*,\sof(12)]_{\Z_3}$};
\node[scale=.8] (t2b) at (-.5,-1.5) {{\scriptsize$\big[u^4+v^3=0\big]$}};
\node[scale=.8] (t3b) at (.6,-1.6) {{\scriptsize$\big[v=0\big]$}};
\draw[red] (p0a) -- (p1);
\draw[red] (p0a) -- (p2);
\draw[red] (p1) -- (p0b);
\draw[red] (p2) -- (p0b);
\end{scope}
\begin{scope}[scale=1.5,xshift=2.85cm]
\node[scale=.5] (p3) at (0,1) {};
\node[scale=.8] (t0a) at (0,1.1) {$\H^{\rm d_{HB}}$};
\node[scale=.8] (tp2) at (-.2,.5) {$\df_4$};
\node[scale=.8] (p2) at (0,0) {$\cT^{(1)}_{D_4,1}$};
\node[scale=.8] (tp1) at (-.2,-.5) {$h_{4,3}$};
\node[scale=.8] (p1) at (0,-1) {$\cS^{(1)}_{D_4,3}$};
\node[scale=.8] (tp1) at (-.2,-1.5) {$\bar{h}_{5,3}$};
\node[scale=.5] (p0) at (0,-2) {};
\node[scale=.8] (t0b) at (0,-2.1) {$\suf(5)_{16}$};
\draw[blue] (p0) -- (p1);
\draw[blue] (p1) -- (p2);
\draw[blue] (p2) -- (p3);
\end{scope}
\end{tikzpicture}}
{\caption{\label{CBhTh40}The Coulomb and Higgs stratification of $\suf(5)_{16}$.}}
\end{subfloatrow}\hspace{1cm}%
\begin{subfloatrow}
\capbtabbox[5cm]{%
  \renewcommand{\arraystretch}{1.1}
  \begin{tabular}{|c|c|} 
  \hline
  \multicolumn{2}{|c|}{$\suf(5)_{16}$}\\
  \hline\hline
  $(\D_u,\D_v)$  &\quad $\left(6,8\right)$\quad{} \\
  $24a$ & 170\\  
  $12c$ & 92\\
$\ff_k$ & $\suf(5)_{16}$ \\ 
$d_{\rm HB}$& 14\\
$h$&0\\
$T({\bf2}\bh)$&0\\
\hline\hline
\end{tabular}
}{%
  \caption{\label{CchTh40}Central charges, \gls{CB} parameters and \gls{ECB} dimension.}%
}
\end{subfloatrow}}{\caption{\label{TothTh40}Information about the $\suf(5)_{16}$.}}
\end{figure}

This theory was first proposed in \cite{Zafrir:2016wkk} where it was constructed as circle compactification of the $5d$ $T_5$ \gls{SCFT} with a $\Z_3$ twist along the circle. Much of CFT data was computed a few years later in \cite{Ohmori:2018ona}, here we will present an analysis of its moduli space structure.

In this case we will start from the analysis of the \gls{HB}. Part of its structure was already worked out in the original paper where it was pointed out that the first transition of the \gls{HB} is five quaternionic dimensional and that leads to the rank-1 $\cS^{(1)}_{A_2,3}$ theory. This information, determines the rest of the \gls{HB} \gls{Hasse} but we still need to characterize the first stratum which is five quaternionic dimensional. This can be done by using the magnetic quiver of this theory which corresponds to the $n=3$ entry of the $A_{n+1}$ series in table 11 of \cite{Bourget:2020asf}\footnote{We thank Gabi Zafrir for pointing this out.}. Using the technique of \emph{quiver subtraction} \cite{Cabrera:2018ann} it is possible to determine that this five dimensional stratum is a $\bar{h}_{5,3}$ which is a new \gls{elesl} which further generalizes the $\bar{h}_{2,n}$ series which appeared in the $\suf(6)$ series. These \gls{elesl}s, labeled as $\bar{h}_{n,3}$, do not in general appear in the affine Grasmannian of any lie algebra unless $n=2$ where they appear in the affine Grassmanian of $\gf_2$ and we have the identification $\bar{h}_{2,3}=a\gf_2$\footnote{Again, we are grateful to Julius Grimminger for the explanation of this point.}. In order to further characterize them, let's now turn to the analysis of the \gls{CB}.

The theory is not \gls{totally} with $v$ being the only \gls{higgs}. This is reflected by the fact that the flavor symmetry is simple. Since $k_{\suf(5)}=2\D_v$ one might be tempted to realize the flavor symmetry on the \gls{CB} with a $[I_5,\suf(5)]$. A sign that this cannot be the case is the fact that this identification would imply the existence of a $\af_4$ transition which is four and not five dimensional. Another evidence is that this identification (which implies a $b=5$) is not compatible with reproducing the $a$ and $c$ in table \ref{CchTh38} using the \gls{ccf}. This exercise is the one that provides some hint on how to realize the $\suf(5)$ factor on the \gls{CB}, indeed we find that the \gls{bi} of the theories on the \gls{CB} strata should be one for $\rTf_{u^4+v^3}$ and eight for $\rTf_v$. The former immediately leads to the identification $\rTf_{u^4+v^3}\equiv [I_1,\varnothing]$, let's discuss instead how to interpret the latter.

First let's go back to our \gls{HB} analysis. An extra hint which helps in making the right guess of what theory realizes the $\suf(5)$ on the \gls{CB} is provided by the fact that the rank-1 theory supported on the $\bar{h}_{5,3}$ has \gls{h}=4. This implies that the theory we are after must likely have a nine dimensional \gls{HB} and it is likely a discretely gauged version of a rank-1 theory. With a bit more thinking, the only reasonable guess with these properties is $\rTf_{v}\equiv[I_2^*,\sof(12)]_{\Z_3}$ where the $\Z_3$ acts as an inner automorphism with no outer component and whose commutant  has $\suf(5)$ as its simple component. Therefore we conjecture that modding by this $\Z_3$ action transforms the $\df_6$ (which, from table \ref{NilOrbits} is indeed nine dimensional) to a symplectic.variety with three strata, and a $\bar{h}_{5,3}$ and a $h_{4,3}$ transition.

\subsubsection*{\boldsymbol{$\suf(3)_{12}{\times}\uf(1)$}}\label{sec:t44}

\begin{figure}[h!]
\ffigbox{
\begin{subfloatrow}
\ffigbox[8.5cm][]{
\begin{tikzpicture}[decoration={markings,
mark=at position .5 with {\arrow{>}}}]
\begin{scope}[scale=1.5]
\node[bbc,scale=.5] (p0a) at (0,0) {};
\node[scale=.5] (p0b) at (0,-2) {};
\node[scale=.8] (t0b) at (0,-2.1) {$\suf(3)_{12}{\times}\uf(1)$};
\node[scale=.8] (p1) at (-1,-1) {$[I_1,\varnothing]$};
\node[scale=.8] (p3) at (0,-1) {$[I_2,\suf(2)]$};
\node[scale=.8] (p2) at (1,-1) {$[\cT^{(1)}_{D_4,1}]_{\Z_3}$};
\node[scale=.8] (t2b) at (-.6,-1.55) {{\scriptsize$\big[u^4+v^3=0\big]$}};
\node[scale=.8] (t3b) at (.25,-1.4) {{\scriptsize$\big[u=0\big]$}};
\node[scale=.8] (t3b) at (.75,-1.6) {{\scriptsize$\big[v=0\big]$}};
\draw[red] (p0a) -- (p1);
\draw[red] (p0a) -- (p2);
\draw[red] (p0a) -- (p3);
\draw[red] (p1) -- (p0b);
\draw[red] (p2) -- (p0b);
\draw[red] (p3) -- (p0b);
\end{scope}
\begin{scope}[scale=1.5,xshift=2.85cm]
\node[scale=.5] (p3) at (0,1) {};
\node[scale=.8] (t0a) at (0,1.1) {$\H^{\rm d_{HB}}$};
\node[scale=.8] (tp2) at (.5,.5) {$\af_1$};
\node[scale=.8] (tp2) at (-.5,.5) {$\af_2$};
\node[scale=.8] (p2) at (.6,0) {$\cT^{(1)}_{A_1,1}$};
\node[scale=.8] (p2a) at (-.6,0) {$\cT^{(1)}_{A_2,1}$};
\node[scale=.8] (tp1) at (.8,-.5) {$h_{2,3}$};
\node[scale=.8] (tp1) at (-.8,-.5) {$h_{3,4}$};
\node[scale=.8] (p1) at (.6,-1) {$\cS^{(1)}_{A_1,3}$};
\node[scale=.8] (p1a) at (-.6,-1) {$\cS^{(1)}_{A_2,4}$};
\node[scale=.8] (tp1) at (.6,-1.5) {$\bar{h}_{3,3}$};
\node[scale=.8] (tp1) at (-.5,-1.5) {$A_{?}$};
\node[scale=.5] (p0) at (0,-2) {};
\node[scale=.8] (t0b) at (0,-2.1) {$\suf(3)_{12}{\times}\uf(1)$};
\draw[blue] (p0) -- (p1);
\draw[blue] (p0) -- (p1a);
\draw[blue] (p1) -- (p2);
\draw[blue] (p1a) -- (p2a);
\draw[blue] (p2) -- (p3);
\draw[blue] (p2a) -- (p3);
\end{scope}
\end{tikzpicture}}
{\caption{\label{CBhTh41}The Coulomb and Higgs stratification of $\suf(3)_{12}{\times}\uf(1)$.}}
\end{subfloatrow}\hspace{1cm}%
\begin{subfloatrow}
\capbtabbox[5cm]{%
  \renewcommand{\arraystretch}{1.1}
  \begin{tabular}{|c|c|} 
  \hline
  \multicolumn{2}{|c|}{$\suf(3)_{12}{\times}\uf(1)$}\\
  \hline\hline
  $(\D_u,\D_v)$  &\quad $\left(4,6\right)$\quad{} \\
  $24a$ & 114\\  
  $12c$ & 60\\
$\ff_k$ & $\suf(3)_{12}\times \uf(1)$ \\ 
$d_{\rm HB}$& 6\\
$h$&0\\
$T({\bf2}\bh)$&0\\
\hline\hline
\end{tabular}
}{%
  \caption{\label{CchTh41}Central charges, \gls{CB} parameters and \gls{ECB} dimension.}%
}
\end{subfloatrow}}{\caption{\label{TothTh41}Information about the $\suf(3)_{12}{\times}\uf(1)$.}}
\end{figure}

This theory was first proposed in \cite{Zafrir:2016wkk} as a theory obtained by mass deforming the theory described in the previous section. As we will see, lot of the strange features of the previous analysis carry over to this one. 

We will start again from the \gls{HB} which is more directly accessible from the $5d$ analysis and therefore some of its properties were already pointed out in the original paper. Namely that one of the first \gls{HB} strata is three quaternionic dimensional with a $\cS^{(1)}_{A_1,3}$ supported over it. This identification readily allows to determine the rest of right side of the \gls{HB} \gls{Hasse} yet it leaves the characterization of the three dimensional stratum, as well as the left part, unresolved. The study of the magnetic quiver of this theory which corresponds to the $n=3$ entry in the $A_{n-1}\times \uf(1)$ series in table 11 of \cite{Bourget:2020asf} shows that this stratum should be a $\bar{h}_{3,3}$. It is curious to notice that the magnetic quiver does not give any evidence of the left branch in figure \ref{CBhTh41}.

Now let's start with the \gls{CB} analysis. This theory is \gls{totally}, despite that there is a single simple flavor factor. The relation $k_{\suf(3)}=2\D_v$ it is again suggestive that this $\suf(3)$ could be realized on the \gls{CB} by a $[I_3,\suf(3)]$ supported on the $v=0$ stratum. The study of the \gls{HB} already shows that this identification is not consistent since $[I_3,\suf(3)]$ has a $\af_2$ as its \gls{HB} and thus would give rise to a two dimensional transition and not three which is instead the case we find here. To resolve this puzzle we can again rely on the \gls{ccf} which tells us that $b_u=2$, $b_v=6$ and $b_{u^3+v^2}=1$. The last immediately leads to the identification $\rTf_{u^3+v^2}\equiv [I_1,\varnothing]$, the first to $\rTf_u\equiv[I_2,\suf(2)]$ while the middle one $\rTf_v\equiv [\cT^{(1)}_{D_4,1}]_{\Z_3}$. Let's conclude with a discussion of these last two identifications.

First the higgsing of the $[I_2,\suf(2)]$ is what leads to the left branch of \gls{Hasse} in figure \ref{CBhTh41}. Since we see en enhancement of the flavor symmetry on the \gls{CB} we conjecture that there is a further identification acting on the $\af_1$ at the origin. The question mark in figure \ref{CBhTh41}, signals that we are not able to conclusively determine what this identification is. Conversely the higgsing of the $[\cT^{(1)}_{D_4,1}]_{\Z_3}$ leads to the branch on the right. The $\Z_3$ that we are gauging is the one which has commutant $\suf(3)$ and we conjecture that under this modding the $\df_4$ (which is indeed five dimensional, see table \ref{NilOrbits}) becomes a symplectic variety with three strata, with a $\bar{h}_{3,3}$ transition followed by a $h_{2,3}$.

\subsubsection*{\boldsymbol{$\suf(2)_{10}{\times}\uf(1)$}}\label{sec:t45}

This theory was is obtained as a mass deformation of the theory studied in the previous section but its explicit higher dimensional construction has not appeared anywhere. A possible construction from \fid will be discussed in \cite{Martone:2021drm}. The CFT data in table \ref{CchTh42} is derived using the \fid construction and checked for self-consistency below.

\begin{figure}[h!]
\ffigbox{
\begin{subfloatrow}
\ffigbox[8.5cm][]{
\begin{tikzpicture}[decoration={markings,
mark=at position .5 with {\arrow{>}}}]
\begin{scope}[scale=1.5]
\node[bbc,scale=.5] (p0a) at (0,0) {};
\node[scale=.5] (p0b) at (0,-2) {};
\node[scale=.8] (t0b) at (0,-2.1) {$\suf(2)_{10}{\times}\uf(1)$};
\node[scale=.8] (p1) at (-.7,-1) {$[I_1,\varnothing]$};
\node[scale=.8] (p2) at (.7,-1) {$[I_\star^*,\sof(6)]_{\Z_3}$};
\node[scale=.8] (t2b) at (-.3,-1.5) {{\scriptsize$\big[u^5+v^3=0\big]$}};
\node[scale=.8] (t3b) at (.75,-1.6) {{\scriptsize$\big[v=0\big]$}};
\draw[red] (p0a) -- (p1);
\draw[red] (p0a) -- (p2);
\draw[red] (p1) -- (p0b);
\draw[red] (p2) -- (p0b);
\end{scope}
\begin{scope}[scale=1.5,xshift=2.85cm]
\node[scale=.5] (p3) at (0,0) {};
\node[scale=.8] (t0a) at (0,.1) {$\uf(1)\times \H^3$};
\node[scale=.8] (tp1) at (-.2,-.5) {$A_2$};
\node[scale=.8] (p1) at (0,-1) {$\green{\cS^{(1)}_{\uf(1),3}}$};
\node[scale=.8] (tp1) at (-.2,-1.5) {$\bar{h}_{2,3}$};
\node[scale=.5] (p0) at (0,-2) {};
\node[scale=.8] (t0b) at (0,-2.1) {$\suf(2)_{10}{\times}\uf(1)$};
\draw[blue] (p0) -- (p1);
\draw[blue] (p1) -- (p3);
\end{scope}
\end{tikzpicture}}
{\caption{\label{CBhTh42}The Coulomb and Higgs stratification of $\suf(2)_{10}{\times}\uf(1)$.}}
\end{subfloatrow}\hspace{1cm}%
\begin{subfloatrow}
\capbtabbox[5cm]{%
  \renewcommand{\arraystretch}{1.1}
  \begin{tabular}{|c|c|} 
  \hline
  \multicolumn{2}{|c|}{$\suf(2)_{10}{\times}\uf(1)$}\\
  \hline\hline
  $(\D_u,\D_v)$  &\quad $\left(3,5\right)$\quad{} \\
  $24a$ & 86\\  
  $12c$ & 44\\
$\ff_k$ & $\suf(2)_{10}\times \uf(1)$ \\ 
$d_{\rm HB}$& 3\\
$h$&0\\
$T({\bf2}\bh)$&0\\
\hline\hline
\end{tabular}
}{%
  \caption{\label{CchTh42}Central charges, \gls{CB} parameters and \gls{ECB} dimension.}%
}
\end{subfloatrow}}{\caption{\label{TothTh42}Information about the $\suf(2)_{10}{\times}\uf(1)$.}}
\end{figure}

This theory is not \gls{totally} and we expect that the simple factor of the flavor symmetry to be realized on the $v=0$ \gls{ukstr}. To gain more insight we go through the usual exercise of matching the $a$ and $c$ central charges in table \ref{CBhTh42} with the \gls{ccf}. The most reasonable solution assigns $b_v=5$ and $b_{u^5+v^3}=1$. The latter lead immediately to the identification $\rTf_{u^5+v^3}\equiv [I_1,\varnothing]$ but the former one is a bit more problematic. In fact a natural guess would be a $[I_\star^*,\sof(6)]_{\Z_3}$ with the $\Z_3$ acting as inner automorphism of $\sof(6)\cong \suf(4)$ with commutant $\suf(2)$. And there is a theory which almost has the right properties, that is an $SU(2)+3\gls{F}$ gauge theory. This theory has indeed $b=5$ and the right flavor symmetry. The only shortcoming of this identification is that the theory in question is asymptotically free, thus there is no notion that a single singularity on the \gls{CB} could support all the degree of freedom of that theory. In fact in \cite{Seiberg:1994aj} the geometry of the \gls{CB} of the $SU(2)+3\gls{F}$ is worked out and it presents two singualrities separated by a distance proportional to the strong coupling scale. We don't have at the moment a resolution of this puzzle which makes the evidence for the existence of this theory less solid than other cases. The \gls{HB} depicted in figure \ref{CBhTh42} can be inferred exploiting the mass deformation from other theories but again further checks will be left for the future.

\begin{center}
\rule[1mm]{2cm}{.4pt}\hspace{1cm}$\circ$\hspace{1cm} \rule[1mm]{2cm}{.4pt}
\end{center}

\subsection{$\spf(12)$ series}

This series contains a unique top theory and two lagrangian ones.

\subsubsection*{\boldsymbol{$\spf(12)_{11}$}}\label{sec:t46}

This theory has appeared for the first time in \cite{Chacaltana:2017boe} in the $E_7$ class-$\cS$ and has one \gls{higgs} and therefore we expect the \gls{HB} to \gls{Hlin}.

\begin{figure}[h!]
\ffigbox{
\begin{subfloatrow}
\ffigbox[8.5cm][]{
\begin{tikzpicture}[decoration={markings,
mark=at position .5 with {\arrow{>}}}]
\begin{scope}[scale=1.5]
\node[bbc,scale=.5] (p0a) at (0,-1) {};
\node[scale=.5] (p0b) at (0,-3) {};
\node[scale=.8] (t0b) at (0,-3.1) {$\spf(12)_{11}$};
\node[scale=.8] (p3) at (.7,-2) {\ \ $[I_1,\varnothing]$};
\node[scale=.8] (p1) at (-.7,-2) {$[I_5^*,\spf(12)]$\ \ };
%\node[scale=.8] (t1a) at (-.6,-.4) {$I_0^*$};
%\node[scale=.8] (t2c) at (-.7,-1.6) {$K_{\D_7}$};
\node[scale=.8] (t1c) at (.85,-2.65) {{\scriptsize$\big[u^5+v^2=0\big]$}};
%\node[scale=.8] (t2c) at (.7,-1.6) {$K_{2\D_7}$};
\node[scale=.8] (t3c) at (-.7,-2.6) {{\scriptsize$\big[v=0\big]$}};
\draw[red] (p0a) -- (p1);
\draw[red] (p0a) -- (p3);
\draw[red] (p1) -- (p0b);
\draw[red] (p3) -- (p0b);
\end{scope}
\begin{scope}[scale=1.5,xshift=2.85cm]
\node[scale=.5] (p0a) at (0,0) {};
\node[scale=.8] (t0a) at (0,.1) {$\H^{\rm d_{HB}}$};
\node[scale=.8] (tp2) at (.2,-.5) {$\ef_7$};
\node[scale=.8] (p1a) at (0,-1) {$\cT^{(1)}_{E_7,1}$};
\node[scale=.8] (tp2) at (-.2,-1.5) {$\af_9$};
\node[scale=.8] (p2a) at (0,-2) {\hyperref[sec:S5]{$\suf(10)_{10}$}};
\node[scale=.8] (tp1) at (-.2,-2.5) {$\cf_6$};
\node[scale=.5] (p0b) at (0,-3) {};
\node[scale=.8] (t0b) at (0,-3.1) {$\spf(12)_{11}$};
\draw[blue] (p0a) -- (p1a);
\draw[blue] (p1a) -- (p2a);
\draw[blue] (p2a) -- (p0b);
\end{scope}
\end{tikzpicture}}
{\caption{\label{CBhTh20}The Coulomb and Higgs stratification of $\suf(2)_{10}{\times}\uf(1)$.}}
\end{subfloatrow}\hspace{1cm}%
\begin{subfloatrow}
\capbtabbox[5cm]{%
  \renewcommand{\arraystretch}{1.1}
  \begin{tabular}{|c|c|} 
  \hline
  \multicolumn{2}{|c|}{$\spf(12)_{11}$}\\
  \hline\hline
  $(\D_u,\D_v)$  &\quad $\left(4,10\right)$\quad{} \\
  $24a$ &  188\\  
  $12c$ & 110\\
$\ff_k$ & $\spf(12)_{11}$ \\ 
$d_{\rm HB}$& 32\\
$h$&6\\
$T({\bf2}\bh)$&1\\
\hline\hline
\end{tabular}
}{%
  \caption{\label{CchTh20}Central charges, \gls{CB} parameters and \gls{ECB} dimension.}%
}
\end{subfloatrow}}{\caption{\label{TothTh20}Information about the $\spf(12)_{11}$.}}
\end{figure}

The \gls{CB} stratification can be readily obtained by noticing that the simple flavor symmetry of the theory has a level which is off by one from the scaling dimension of the only \gls{higgs} ($v$). It is therefore reasonable to make the following identification $\rTf_v\equiv [I_5^*,\spf(12)]$. This immediately suggests that \gls{h}=6 and that the rank-2 theory supported on the $\cf_6$ stratum should have a $\af_9$ rank-decreasing transition. Matching the $a$ and $c$ central charges using \eqref{actotaint} and \eqref{actotbint} confirms the validity of our initial guess and in turns fixes the last info needed to complete the \gls{CB} picture: $\rTf_{u^5+v^2}\equiv[I_1,\varnothing]$.

To completely specify the \gls{HB} structure we will employ our usual tactics. The higgsing of $\rTf_v$ is of \gls{gHW} type and therefore we can use \eqref{gHWeq} to compute the $c$ central charge of the rank-2 theory supported on $\cf_6$. This, along with the constraint that this theory should have a 26 dimensional \gls{HB} to account to the remaining dimension, leads to the final identification: $\bTf_{\cf_6}\equiv \suf(10)_{10}$. The \gls{HB} \gls{Hasse} in figure \ref{CBTh14} shows that the constraint that we guessed from the \gls{CB} structure is in fact satisfied providing a fully consistent picture.

\subsubsection*{\boldsymbol{$USp(4)+F+2V$}}\label{sec:Sp42524}

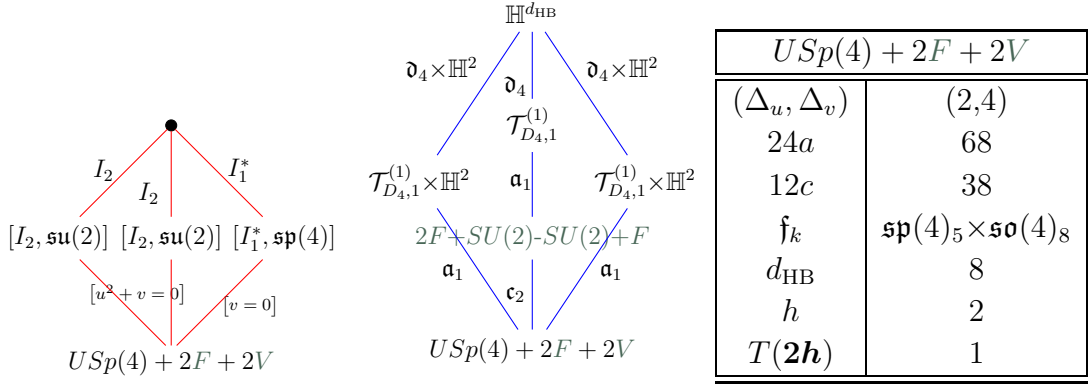
\begin{figure}[h!]
\ffigbox{
\begin{subfloatrow}
\ffigbox[7.5cm][]{
\begin{tikzpicture}[decoration={markings,
mark=at position .5 with {\arrow{>}}}]
\begin{scope}[scale=1.5]
\node[bbc,scale=.5] (p0a) at (0,0) {};
\node[scale=.5] (p0b) at (0,-2) {};
\node[scale=.8] (t0b) at (0,-2.1) {$USp(4)+2\gls{F}+2\gls{V}$};
\node[scale=.8] (p1) at (-1,-1) {$[I_2,\suf(2)]$};
\node[scale=.8] (p2) at (1,-1) {$[I_1^*,\spf(4)]$};
\node[scale=.8] (p3) at (0,-1) {$[I_2,\suf(2)]$};
\node[scale=.8] (t1a) at (-.6,-.4) {$I_2$};
\node[scale=.8] (t2a) at (-.2,-.6) {$I_2$};
\node[scale=.8] (t2b) at (-.3,-1.5) {{\scriptsize$\big[u^2+v=0\big]$}};
\node[scale=.8] (t3a) at (.6,-.4) {$I_1^*$};
\node[scale=.8] (t3b) at (.7,-1.6) {{\scriptsize$\big[v=0\big]$}};
\draw[red] (p0a) -- (p1);
\draw[red] (p0a) -- (p2);
\draw[red] (p0a) -- (p3);
\draw[red] (p1) -- (p0b);
\draw[red] (p2) -- (p0b);
\draw[red] (p3) -- (p0b);
\end{scope}
\begin{scope}[scale=1.5,xshift=3.2cm]
\node[scale=.8] (p0a) at (0,1) {$\H^{d_{\rm HB}}$};
\node[scale=.8] (t1a) at (-0.15,.35) {$\df_4$};
\node[scale=.8] (t2a) at (0.8,.5) {$\df_4{\times}\H^2$};
\node[scale=.8] (t2b) at (-0.8,.5) {$\df_4{\times}\H^2$};
\node[scale=.8] (p2a) at (-1,-.5) {$\cT^{(1)}_{D_4,1}{\times} \H^2$};
\node[scale=.8] (p2b) at (1,-.5) {$\cT^{(1)}_{D_4,1}{\times} \H^2$};
\node[scale=.8] (t2c) at (-0.7,-1.3) {$\af_1$};
\node[scale=.8] (t2d) at (0.7,-1.3) {$\af_1$};
\node[scale=.8] (p1) at (-0,0) {$\cT^{(1)}_{D_4,1}$};
\node[scale=.8] (t1a) at (-0.1,-.5) {$\af_1$};
\node[scale=.8] (p2) at (-0,-1) {\hyperref[sec:SU22b]{$2\gls{F}\text{+}SU(2)\text{-}SU(2)\text{+}\gls{F}$}};
\node[scale=.8] (t1b) at (-.15,-1.5) {$\cf_2$};
\node[scale=.8] (p0b) at (0,-2) {$USp(4)+2\gls{F}+2\gls{V}$};
\draw[blue] (p0a) -- (p2a);
\draw[blue] (p0a) -- (p2b);
\draw[blue] (p0a) -- (p1);
\draw[blue] (p1) -- (p2);
\draw[blue] (p2) -- (p0b);
\draw[blue] (p2a) -- (p0b);
\draw[blue] (p2b) -- (p0b);
\end{scope}
\end{tikzpicture}}
{\caption{\label{CBSp42524}The \gls{Hasse} for the \gls{CB} and the \gls{HB} of the $USp(4)$ gauge theory with two hypermultiplets in the ${\bf 4}\oplus{\bf5}$.}}
\end{subfloatrow}\hspace{1cm}%
\begin{subfloatrow}
\capbtabbox[7cm]{%
  \renewcommand{\arraystretch}{1.1}
  \begin{tabular}{|c|c|} 
  \hline
  \multicolumn{2}{|c|}{$USp(4)+2\gls{F}+2\gls{V}$}\\
  \hline\hline
  $(\D_u,\D_v)$  &\quad (2,4)\quad{} \\
  $24a$ & 68\\  
  $12c$ & 38 \\
$\ff_k$ & $\spf(4)_5{\times}\sof(4)_8$ \\ 
$d_{\rm HB}$&8\\
$h$&2\\
$T({\bf2}\bh)$&1\\
\hline\hline
 \end{tabular}
}{%
  \caption{\label{CcSp42524}Central charges, \gls{CB} parameters and \gls{ECB} dimension.}%
}
\end{subfloatrow}}{\caption{\label{TotSp42524} Information about the $USp(4)$ $\cN=2$ theory with two hypermultiplets in the ${\bf 4}\oplus{\bf 5}$}}
\end{figure}

The analysis of this theory follows the same line as other $USp(4)$ theories. As we mentioned earlier, there are two inequivalent $SU(2)$s (corresponding to a long and a short root) which can be left unbroken by turning on a vev for the scalar component of the $USp(4)$ vector multiplet. For one $SU(2)$ each hypermultiplet in the ${\bf4}$ contributes a massless fundamental flavor while those in the ${\bf 5}$ none. Viceversa for the other $SU(2)$ the hyper in the ${\bf 4}$ has no massless component while those in the ${\bf 5}$ contribute a hyper in the adjoint of $SU(2)$. The result is that we expect three knotted strata, two supporting a $[I_2,\suf(2)]$ (which together carry the $\sof(4)$ flavor symmetry of an $SU(2)$ gauge theory with $N_f=2$) and the other \gls{kstr} supports an $\cN=2$  $SU(2)$ gauge theory with two adjoints. We therefore also conclude that the theory has \gls{h}=2. It is possible to check that this stratification perfectly reproduces the $a$ and $c$ central charges in table \ref{CcSp42524} once the appropriate \gls{bi} are plugged into \eqref{actotaint}-\eqref{actotbint}.

The analysis of the \gls{HB} is even more straightforward. In fact to identify the rank-2 theory supported on $\cf_2$ we can use the constraint on the dimension of the \gls{HB} as well as the central charge obtained using \eqref{gHWeq} (the \gls{CB} analysis clarifies that this Higgsing is indeed of \gls{gHW} type). These two conditions lead to the identification $\bTf_{\cf_2}\equiv 2\gls{F}+SU(2)-SU(2)+2\gls{F}$ which can also be checked by solving the equation of motions and working directly with the vevs of the hypermultiplets.

\subsubsection*{\boldsymbol{$G_2+4F$}}\label{sec:G247}

\begin{figure}[h!]
\ffigbox{
\begin{subfloatrow}
\ffigbox[7cm][]{
\begin{tikzpicture}[decoration={markings,
mark=at position .5 with {\arrow{>}}}]
\begin{scope}[scale=1.5]
\node[bbc,scale=.5] (p0a) at (0,0) {};
\node[scale=.5] (p0b) at (0,-2) {};
\node[scale=.8] (t0b) at (0,-2.1) {$G_2+4\gls{F}$};
\node[scale=.7] (p1) at (-.8,-1) {$[I_1,\varnothing]{\times}\H^4$\quad\ \ };
\node[scale=.7] (p2) at (.8,-1) {\quad\ $[I^*_3,\spf(8)]$};
\node[scale=.7] (p3) at (0,-1) {\ \ $[I_1,\varnothing]{\times}\H^4$};
\node[scale=.8] (t2b) at (-.3,-1.5) {{\scriptsize$\big[u^3+v=0\big]$}};
\node[scale=.8] (t3b) at (.7,-1.6) {{\scriptsize$\big[v=0\big]$}};
\draw[red] (p0a) -- (p1);
\draw[red] (p0a) -- (p2);
\draw[red] (p0a) -- (p3);
\draw[red] (p1) -- (p0b);
\draw[red] (p2) -- (p0b);
\draw[red] (p3) -- (p0b);
\end{scope}
\begin{scope}[scale=1.5,xshift=2.85cm]
\node[scale=.8] (p0a) at (0,1) {$\H^{d_{\rm HB}}$};
\node[scale=.8] (t1a) at (-0.2,.5) {$\df_4$};
\node[scale=.8] (p1) at (0,0) {$\cT^{(1)}_{D_4,1}$};
\node[scale=.8] (t1b) at (-.2,-.5) {$\af_5$};
\node[scale=.8] (p2) at (0,-1) {\hyperref[sec:SU3Nf6]{$SU(3)+6F$}};
\node[scale=.8] (t1b) at (-.2,-1.5) {$\cf_4$};
\node[scale=.8] (p0b) at (0,-2) {$G_2+4\gls{F}$};
\draw[blue] (p0a) -- (p1);
\draw[blue] (p1) -- (p2);
\draw[blue] (p2) -- (p0b);
\end{scope}
\end{tikzpicture}}
{\caption{\label{CBG247}The \gls{Hasse} for the \gls{CB} and the \gls{HB} of the $G_2$ gauge theory with 4 ${\bf 7}$s.}}
\end{subfloatrow}\hspace{1cm}%
\begin{subfloatrow}
\capbtabbox[7cm]{%
  \renewcommand{\arraystretch}{1.1}
  \begin{tabular}{|c|c|} 
  \hline
  \multicolumn{2}{|c|}{$G_2+4\gls{F}$}\\
  \hline\hline
  $(\D_u,\D_v)$  &\quad (2,6)\quad{} \\
  $24a$ &  98\\  
  $12c$ & 56 \\
$\ff_k$ & $\spf(8)_{7}$ \\ 
$d_{\rm HB}$&14\\
$h$&4\\
$T({\bf2}\bh)$&1\\
\hline\hline
  \end{tabular}
}{%
  \caption{\label{CcG247}Central charges, \gls{CB} parameters and \gls{ECB} dimension.}%
}
\end{subfloatrow}}{\caption{\label{TotG247}Information about the $G_2$ $\cN=2$ theory with 4 hypermultiplets in the ${\bf 7}$.}}
\end{figure}
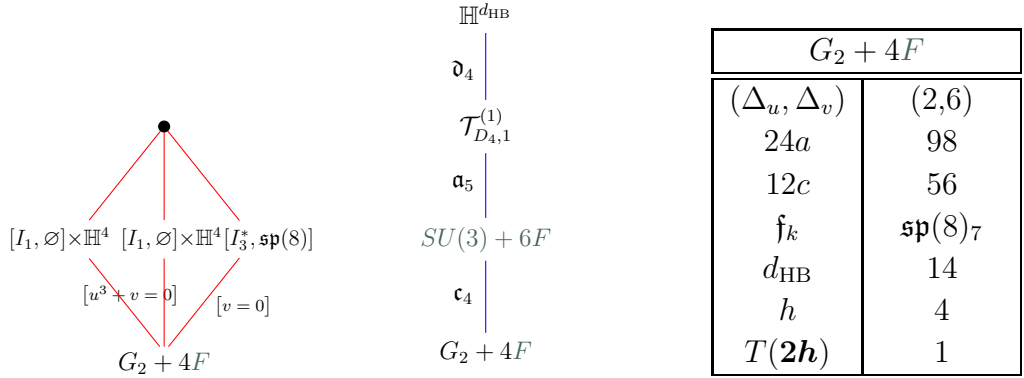

The analysis of the \gls{CB} of this theory follows closely what we have done before. As in the case of $USp(4)$, $G_2$ has two inequivalent $\suf(2)$ factors which can remain unbroken by turning on the vev of the scalar component of the vector multiplet. Each hyper in the ${\bf 7}$ contributes a massless hyper in the ${\bf 3}$ for one $SU(2)$ and none for the other. This readily gives the stratification depicted in figure \ref{CBG247} and sets \gls{h}=4.

The analysis of the \gls{HB} of this theory has been performed, for example, in \cite{Bourget:2019aer} therefore we won't reproduce it here. It is a useful exercise to see that our geometric constraints are perfectly reproduced by the \gls{Hasse} in figure \ref{CBG247}.

\subsubsection*{\boldsymbol{$\spf(4)_{\frac{13}3}$}}\label{sec:ADc2}

\begin{figure}[h!]
\ffigbox{
\begin{subfloatrow}
\ffigbox[7cm][]{
\begin{tikzpicture}[decoration={markings,
mark=at position .5 with {\arrow{>}}}]
\begin{scope}[scale=1.5]
\node[bbc,scale=.5] (p0a) at (0,0) {};
\node[scale=.5] (p0b) at (0,-2) {};
\node[scale=.8] (t0b) at (0,-2.1) {AD($\cf_2$)};
\node[scale=.7] (p1) at (-.7,-1) {$[I_1,\varnothing]{\times}\H^2$\quad\ \ };
\node[scale=.7] (p2) at (.7,-1) {\quad\ $[I^*_1,\spf(4)]$};
\node[scale=.8] (t2b) at (-.3,-1.5) {{\scriptsize$\big[u^5+v^2=0\big]$}};
\node[scale=.8] (t3b) at (.6,-1.6) {{\scriptsize$\big[v=0\big]$}};
\draw[red] (p0a) -- (p1);
\draw[red] (p0a) -- (p2);
\draw[red] (p1) -- (p0b);
\draw[red] (p2) -- (p0b);
\end{scope}
\begin{scope}[scale=1.5,xshift=2.85cm]
\node[scale=.8] (p0a) at (0,1) {$\H^4$};
\node[scale=.8] (t1a) at (-0.2,.5) {$\af_1$};
\node[scale=.8] (p1) at (0,0) {$\cT^{(1)}_{A_1,1}$};
\node[scale=.8] (t1b) at (-.2,-.5) {$\af_1$};
\node[scale=.8] (p2) at (0,-1) {\hyperref[sec:(A1,D6)]{$(A_1,D_6)$}};
\node[scale=.8] (t1b) at (-.2,-1.5) {$\cf_2$};
\node[scale=.8] (p0b) at (0,-2) {AD($\cf_2$)};
\draw[blue] (p0a) -- (p1);
\draw[blue] (p1) -- (p2);
\draw[blue] (p2) -- (p0b);
\end{scope}
\end{tikzpicture}}
{\caption{\label{CBADc2}The \gls{Hasse} for the \gls{CB} and the \gls{HB} of the AD$(\cf_2)$ theory.}}
\end{subfloatrow}\hspace{1cm}%
\begin{subfloatrow}
\capbtabbox[7cm]{%
  \renewcommand{\arraystretch}{1.1}
  \begin{tabular}{|c|c|} 
  \hline
  \multicolumn{2}{|c|}{AD($\cf_2$)}\\
  \hline\hline
  $(\D_u,\D_v)$  &\quad $(\frac43,\frac{10}3)$\quad{} \\
  $24a$ &  48\\  
  $12c$ & 26 \\
$\ff_k$ & $\spf(4)_{\frac{13}3}$ \\ 
$d_{\rm HB}$&4\\
$h$&2\\
$T({\bf2}\bh)$&1\\
\hline\hline
  \end{tabular}
}{%
  \caption{\label{CcADc2}Central charges, \gls{CB} parameters and \gls{ECB} dimension.}%
}
\end{subfloatrow}}{\caption{\label{TotADc2}Information about the AD($\cf_2$) theory.}}
\end{figure}
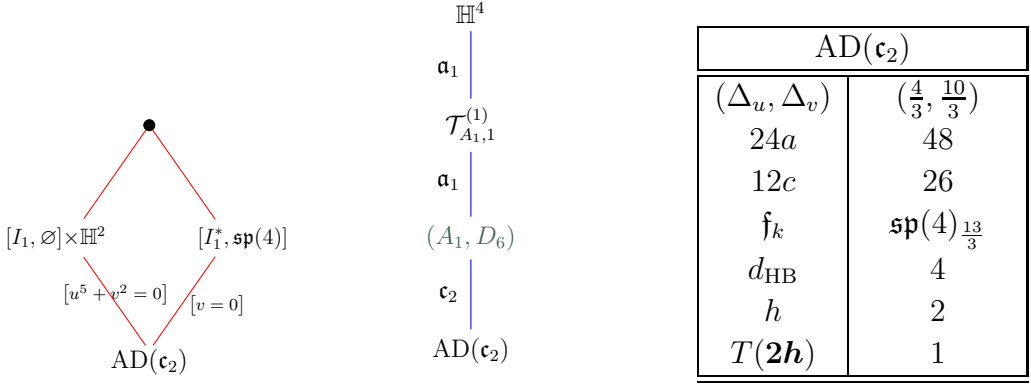

This theory arises most naturally on the \gls{CB} of the $G_2$ theory with three hypermultiplets in the $\bf 7$ \cite{Kaidi:2021tgr}. This theory can also be realized in class-$\cS$ as a twisted compactification of an $A_4$ 6\emph{d} (2,0) theory with a maximum regular puncture and one irregular puncture identified by the value $k'=-4$ of the parameter in \cite{Wang:2018gvb}. The \gls{CB} and \gls{HB} stratification as well as the CFT data can be found figure \ref{CBADc2} and in table \ref{CcADc2} respectively. A detailed discussion of the structure of the moduli space, as well as a conjectural VOA for this theory, namely the affine $\hat{\spf(4)}_{-{13 \over 6}}$ extended by four generators of conformal dimension $h=\frac 32$, is provided in \cite{Kaidi:2021tgr} thus we will not repeat the analysis here.

\begin{center}
\rule[1mm]{2cm}{.4pt}\hspace{1cm}$\circ$\hspace{1cm} \rule[1mm]{2cm}{.4pt}
\end{center}

\subsection{Other series}

We here summarize the rest of the series, many of which had been understood in good detail already in \cite{Apruzzi:2020pmv,Giacomelli:2020gee}.

\subsubsection{$\spf(8)-\suf(2)^2$ series}

This series contain two top theories and a total of six theories and all but one mass deformation can be seen geometrically from the brane realization of the corresponding theories. The remaining one was discussed in \cite{Giacomelli:2020gee}.

\paragraph{$\boldsymbol{\cS^{(2)}_{E_6,2}}$}\label{sec:SE62} This theory, which is the top theory of the $\spf(8)$ series, can be obtained in class-$\cS$, for example in the untwisted $D_7$ case \cite{Chacaltana:2011ze,Giacomelli:2020jel}. More recently, it was shown to belong to an infinite set of theories called $\cN=2$ $\cS$-fold \cite{Apruzzi:2020pmv}, or $\cS$ theories for short. This specific case, can be engineered as worldvolume theory of $2$ \emph{D3} branes probing an exceptional $E_6$ \emph{7}brane singularity in the presence of a $\Z_2$ $\cS$-fold \cite{Aharony:2016kai} with flux. This theory can also be realized as the compactification of a (1,0) \sd theory \cite{Giacomelli:2020jel}. A summary of the properties of $\cS$-theories can be found in appendix \ref{sec:TandS} and the CFT data and depiction of the \gls{Hasse}s of both the \gls{CB} and \gls{HB} stratification, can be found in figure \ref{Fig:S2Gl}.

\paragraph{$\boldsymbol{\cS^{(2)}_{D_4,2}}$}\label{sec:SD42} This theory can be obtained mass deforming the previous one. Specifically this mass deformation is realized in the brane picture as moving away two \emph{D7} branes from the $E_6$ exceptional \emph{7}brane singularity. The $\cS^{D_4,2}$ arises probing the remaining $D_4$ singularity in the presence of a fluxfull $\cS$-fold by two \emph{D3} branes. It can also be obtained in class-$\cS$, for example in the $\Z_2$ twisted $A_7$ case and  as the compactification of a (1,0) \sd theory \cite{Giacomelli:2020jel}. A summary of the properties of $\cS$-theories can be found in appendix \ref{sec:TandS} and the CFT data and depiction of the \gls{Hasse}s of both the \gls{CB} and \gls{HB} stratification, can be found in figure \ref{Fig:S2Gl}.

\paragraph{$\boldsymbol{\cS^{(2)}_{A_2,2}}$}\label{sec:SA22} A similar mass deformation works in this as well. In fact this theory can be obtained by moving away another two \emph{D7} branes from the $E_6$ exceptional \emph{7}brane singularity, or just two from the $D_4$, and probing the remaining $A_2$ singularity plus $\cS$-fold with flux by two \emph{D3} branes. It can also be obtained as compactification of a (1,0) \sd theory \cite{Giacomelli:2020jel}. Currently the author is unaware of any class-$\cS$ realization. A summary of the properties of $\cS$-theories can be found in appendix \ref{sec:TandS} and the CFT data and depiction of the \gls{Hasse}s of both the \gls{CB} and \gls{HB} stratification, can be found in figure \ref{Fig:S2Gl}.

\paragraph{$\boldsymbol{\cT^{(2)}_{A_2,4}}$}\label{sec:TA24}  It was one of the recently discovered $\cT$ theories. It can be obtained higgsing a $\cS^{(2)}_{A_2,4}$ $\cN=2$ $\cS$-fold \cite{Apruzzi:2020pmv,Giacomelli:2020jel} or as a wordvolume theory of two \emph{D3} branes probing an exceptional $A_2$ \emph{7}brane singularity in the presence of a $\Z_4$ $\cS$-fold without flux \cite{Giacomelli:2020gee}. This theory cannot be obtained by mass deformation of the previous three and it is therefore a top theory of the series. Currently the author is not aware of any class-$\cS$ realization of this theory. The properties of $\cN=2$ $\cT$-theories are summarized in appendix \ref{sec:TandS}. The CFT properties as well as the stratification of this particular theory can be found in table \ref{Fig:T2Gl}.

\paragraph{$\boldsymbol{\hat{\cT}_{A_2,4}}$}\label{sec:tA2}

This theory was initially found in \cite{Giacomelli:2020gee} as a mass deformation of the $\cT^{(2)}_{A_2,4}$. This mass deformation is apparent from the $5d$ perspective. Indeed this latter theory can be obtained as a twisted $\Z_4$ compactification of a $5d$ \gls{SCFT} whose brane web is known \cite{Giacomelli:2020gee}. The brane web gives access to the mass deformation that preserves the $\Z_4$ and the twisted compactification of the resulting $5d$ \gls{SCFT} is what gives $\hat{\cT}_{A_2,4}$.

\begin{figure}[t!]
\ffigbox{
\begin{subfloatrow}
\ffigbox[8cm][]{
\begin{tikzpicture}[decoration={markings,
mark=at position .5 with {\arrow{>}}}]
\begin{scope}[scale=1.5]
\node[bbc,scale=.5] (p0a) at (0,-1) {};
\node[scale=.5] (p0b) at (0,-3) {};
\node[scale=.8] (t0b) at (0,-3.1) {$\hTA$};
\node[scale=.8] (p2) at (.7,-2) {\ \ $[\cT^{(1)}_{A_2,1}]_{\Z_4}$};
\node[scale=.8] (p1) at (-.7,-2) {$[I_1,\varnothing]$\ \ };
%\node[scale=.8] (t1a) at (-.6,-.4) {$I_0^*$};
%\node[scale=.8] (t2c) at (-.7,-1.6) {$K_{\D_7}$};
\node[scale=.8] (t1c) at (-.85,-2.7) {{\scriptsize$\big[u^8+v^5=0\big]$}};
%\node[scale=.8] (t2c) at (.7,-1.6) {$K_{2\D_7}$};
\node[scale=.8] (t1c) at (.7,-2.7) {{\scriptsize$\big[u=0\big]$}};
\draw[red] (p0a) -- (p1);
\draw[red] (p0a) -- (p2);
\draw[red] (p1) -- (p0b);
\draw[red] (p2) -- (p0b);
\end{scope}
\begin{scope}[scale=1.5,xshift=3cm]
\node[scale=.5] (p0a) at (0,-1) {};
\node[scale=.5] (p0b) at (0,-3) {};
\node[scale=.8] (t0a) at (0,-.9) {$\H^{\rm d_{HB}}$};
\node[scale=.8] (t0b) at (0,-3.1) {$\hTA$};
\node[scale=.8] (p2) at (0,-2) {$\green{\cS^{(1)}_{\varnothing,4}}$\ \ };
\node[scale=.8] (tp2) at (0.3,-1.5) {$\af_1$};
\node[scale=.8] (tp1) at (0.3,-2.5) {$\af_1$};
\draw[blue] (p0a) -- (p2);
\draw[blue] (p2) -- (p0b);
\end{scope}
\end{tikzpicture}}
{\caption{\label{CBhTA2}The Coulomb and Higgs stratification of $\hTA$.}}
\end{subfloatrow}\hspace{1cm}%
\begin{subfloatrow}
\capbtabbox[5cm]{%
  \renewcommand{\arraystretch}{1.1}
  \begin{tabular}{|c|c|} 
  \hline
  \multicolumn{2}{|c|}{$\hTA$}\\
  \hline\hline
  $(\D_u,\D_v)$  &\quad $\left(\frac52,4\right)$\quad{} \\
  $24a$ &  67\\  
  $12c$ & 34\\
$\ff_k$ & $\suf(2)_5$ \\ 
$d_{\rm HB}$& 2\\
$h$&0\\
$T({\bf2}\bh)$&0\\
\hline\hline
\end{tabular}
}{%
  \caption{\label{CchTA2}Central charges, \gls{CB} parameters and \gls{ECB} dimension.}%
}
\end{subfloatrow}}{\caption{\label{TothTA2}Information about the $\hTA$. theory.}}
\end{figure}

Let's move now to the analysis of the full moduli space structure. The theory has a single simple flavor factor compatibly with the lone \gls{higgs} ($u$). The level of the $\suf(2)$ is double $\D_u$ and therefore a possible realization of the $\suf(2)$ is via a $[I_2,\suf(2)]$ supported on a $u=0$ \gls{ukstr}. This identification would immediately imply that \gls{h}=0 and that the theory supported on the $\af_1$ stratum also has no \gls{ECB}. Right now we cannot exclude this possibility. Though the structure of the $\cT^{(2)}_{A_2,4}$ moduli space, along with general behavior of the moduli space under mass deformation \cite{Martone:2021drm} suggests an alternative realization with $\rTf_u\equiv [\cT^{(1)}_{A_2,4}]_{\Z_4}$. This latter identification also predicts a $\suf(2)_5$ and \gls{h}=0 but now the theory supported on the $\af_1$ stratum should have a one dimensional \gls{ECB}. To check the validity of this working assumption, we can use \eqref{actotaint}-\eqref{actotbint} to match the $a$ and $c$ central charges in figure \ref{CchTA2} to then complete the analysis of the \gls{CB} stratification and reproduce the \gls{Hasse} in figure. Note that the $[I_2,\suf(2)]$ and $[\cT^{(1)}_{A_2,4}]_{\Z_4}$ have different \gls{bi} and contirbute differently to \eqref{actotaint}-\eqref{actotbint}. It is an instructive exercise to check that the former choice simply gives no consistent solution and we cannot reproduce the central charges from the \gls{CB} analysis.

Now let's turn to the analysis of the \gls{HB}. From the previous analysis we expect that the \gls{HB} would start with a $\af_1$ transition and we have also derived that the rank-1 theory supported there should have \gls{h}=1. To reproduce the total dimension of the $\hat{\cT}_{A_2,4}$, we immediately obtain that the entire one dimensional \gls{HB} should be an \gls{ECB}, which suggests that the theory must have $\cN\geq 3$. The \gls{CB} analysis tells us that the $\af_1$ higgsing is of \gls{gHW} type. Applying \eqref{gHWeq} to this stratum then implies $12c_{\bTf_{\af_1}}=21 $ which readily implies that $\bTf_{\af_1}\equiv \green{\cS^{(1)}_{\varnothing,4}}$. This concludes our anlaysis.

\paragraph{\blue{$USp(4)$}}\label{sec:N4Sp4} This is an $\cN=4$ theory. The moduli space of these theories is extremely constrained and it is basically entirely specified by the Weyl group of the gauge algebra which in this case is $D_4$, the dihedral group of order eight. More details on the moduli space structure of theories with extended supersymmetry can be found in appendix \ref{sec:N34}. The CFT data of this theory, as well as the explicit \gls{Hasse} of both the \gls{CB} and \gls{HB} stratification are depicted in figure \ref{Fig:N4}.

\begin{center}
\rule[1mm]{2cm}{.4pt}\hspace{1cm}$\circ$\hspace{1cm} \rule[1mm]{2cm}{.4pt}
\end{center}

\subsubsection{$\gf_2$ series}

This series was basically already discussed in \cite{Apruzzi:2020pmv,Giacomelli:2020gee}. In fact all the theories here belong to the same $\cN=2$ $\cS$-fold class which can be engineered in type \emph{II}B string theory, and all mass deformations connecting the theories in the series are geometrically realized by motion of \emph{D7} branes.

\paragraph{$\boldsymbol{\cT^{(2)}_{D_4,3}}$}\label{sec:TD43} This theory, which sits at the top of the $\gf_2$ series, can be obtained in class-$\cS$ by compactifying a $D_4$ (2,0) theory with a $\Z_3$ twist \cite{Chacaltana:2016shw}. It was also recently recognize to belong to the infinite series of $\cT$ theories \cite{Giacomelli:2020jel}. Specifically it can be realized as the worldvolume theory of two \emph{D3} branes probing an exceptional $D_4$ \emph{7}brane singularity in the presence of a fluxless $\Z_3$ $\cS$-fold. General properties of $\cT$-theories are summarized in appendix \ref{sec:TandS} and the CFT properties as well as the stratification of this particular theory can be found in table \ref{Fig:T2Gl}.

\paragraph{$\boldsymbol{\cT^{(2)}_{A_1,3}}$}\label{sec:TA13} This theory can be obtained by moving away three \emph{D7} branes from the exceptional $D_4$ \emph{7}brane singularity plus a $\Z_3$ fluxless $\cS$-fold and probing the remaining $A_1$ singularity, in the presence of the $\cS$-fold, with two \emph{D3} branes. This action in the brane system corresponds to a mass deformation. It belongs to the recently discovered infinite series of $\cT$ theories \cite{Giacomelli:2020jel} and it can be obtained by mass deforming $\cT^{(2)}_{D_4,3}$. General properties of $\cT$-theories are summarized in appendix \ref{sec:TandS} and the CFT properties as well as the stratification of this particular theory can be found in table \ref{Fig:T2Gl}

\paragraph{$\boldsymbol{\hat{\cT}_{D_4,3}}$}\label{sec:tD4}

This theory was initially found in \cite{Giacomelli:2020gee} as a mass deformation of the $\cT^{(2)}_{D_4,3}$. The $\cT^{(2)}_{D_4,3}$ can be obtained as a twisted $\Z_3$ compactification of a $5d$ \gls{SCFT} with a known brane web \cite{Giacomelli:2020gee}. The brane web then has a deformation which preserves the $\Z_3$ symmetry and the twisted compactification of the resulting $5d$ \gls{SCFT} is what gives $\hat{\cT}_{D_4,3}$. Let's move to the analysis of the moduli space which is anyway performed already in \cite{Giacomelli:2020gee}.

\begin{figure}[h!]
\ffigbox{
\begin{subfloatrow}
\ffigbox[8cm][]{
\begin{tikzpicture}[decoration={markings,
mark=at position .5 with {\arrow{>}}}]
\begin{scope}[scale=1.5]
\node[bbc,scale=.5] (p0a) at (0,-1) {};
\node[scale=.5] (p0b) at (0,-3) {};
\node[scale=.8] (t0b) at (0,-3.1) {$\hTD$};
\node[scale=.8] (p2) at (.7,-2) {\ \ $[\cT^{(1)}_{D_4,1}]_{\Z_3}$};
\node[scale=.8] (p1) at (-.7,-2) {$[I_1,\varnothing]$\ \ };
%\node[scale=.8] (t1a) at (-.6,-.4) {$I_0^*$};
%\node[scale=.8] (t2c) at (-.7,-1.6) {$K_{\D_7}$};
\node[scale=.8] (t1c) at (-.75,-2.7) {{\scriptsize$\big[u^6+v^5=0\big]$}};
%\node[scale=.8] (t2c) at (.7,-1.6) {$K_{2\D_7}$};
\node[scale=.8] (t1c) at (.6,-2.7) {{\scriptsize$\big[u=0\big]$}};
\draw[red] (p0a) -- (p1);
\draw[red] (p0a) -- (p2);
\draw[red] (p1) -- (p0b);
\draw[red] (p2) -- (p0b);
\end{scope}
\begin{scope}[scale=1.5,xshift=2.5cm]
\node[scale=.5] (p0a) at (0,0) {};
\node[scale=.8] (tp2) at (0.3,-.5) {$\af_1$};
\node[scale=.5] (p0b) at (0,-3) {};
\node[scale=.8] (t0a) at (0,.1) {$\H^{\rm d_{HB}}$};
\node[scale=.8] (t0b) at (0,-3.1) {$\hTD$};
\node[scale=.8] (p2) at (0,-2) {$\cS^{(1)}_{A_1,3}$\ \ };
\node[scale=.8] (p1) at (0,-1) {$\cT^{(1)}_{A_1,1}$};
\node[scale=.8] (tp1) at (0.3,-2.5) {$\gf_2$};
\draw[blue] (p0a) -- (p1);
\draw[blue] (p1) -- (p2);
\draw[blue] (p2) -- (p0b);
\end{scope}
\end{tikzpicture}}
{\caption{\label{CBhTD4}The Coulomb and Higgs stratification of $\hTD$.}}
\end{subfloatrow}\hspace{1cm}%
\begin{subfloatrow}
\capbtabbox[5cm]{%
  \renewcommand{\arraystretch}{1.1}
  \begin{tabular}{|c|c|} 
  \hline
  \multicolumn{2}{|c|}{$\hTD$}\\
  \hline\hline
  $(\D_u,\D_v)$  &\quad $\left(\frac{10}3,4\right)$\quad{} \\
  $24a$ &  82\\  
  $12c$ & 44\\
$\ff_k$ & $[\gf_2]_{\frac{20}3}$ \\ 
$d_{\rm HB}$& 6\\
$h$&0\\
$T({\bf2}\bh)$&0\\
\hline\hline
\end{tabular}
}{%
  \caption{\label{CchTD4}Central charges, \gls{CB} parameters and \gls{ECB} dimension.}%
}
\end{subfloatrow}}{\caption{\label{TothTD4}Information about the $\hTD$. theory.}}
\end{figure}

The theory is not \gls{totally} with $u$ representing the only \gls{higgs}. The exceptional flavor symmetry makes it easy to identify the \gls{CB} structure: $\rTf_u\equiv [\cT^{(1)}_{D_4,1}]_{\Z_3}$ which is also compatible with the fact that $k_{\gf_2}=2\D_u$. This immediately predicts that \gls{h}=0, that the \gls{HB} has a $\gf_2$ transition and that the rank-1 theory supported on this stratum has a two dimensional \gls{ECB}. We can conclude the analysis of the \gls{CB} by reproducing the $a$ and $c$ central charges in figure \ref{CchTD4} using \eqref{actotaint}-\eqref{actotbint} which imposes that $\rTf_{u^6+v^5}\equiv [I_1,\varnothing]$.

The analysis of the \gls{HB} is straightforward. The constraints coming from the \gls{CB} analysis as well as the demand that the total \gls{HB} being six quaternionic dimensions uniquely identifies the rank-1 theory and we conclude that $\bTf_{\gf_2}\equiv \cS^{(1)}_{A_1,3}$. This identification is further confirmed by using \eqref{gHWeq} which can be applied to $\gf_2$ and predicts precisely the central charge of $\cS^{(1)}_{A_1,3}$.

\paragraph{\blue{$\boldsymbol{SU(3)}$}}\label{sec:N4SU3}  This is an $\cN=4$ theory. The moduli space of these theories is extremely constrained and it is basically entirely specified by the Weyl group of the gauge algebra which in this case is $S_3$, the symmetric group of order six. More details on the moduli space structure of theories with extended supersymmetry can be found in appendix \ref{sec:N34}. The CFT data of this theory, as well as the explicit \gls{Hasse} of both the \gls{CB} and \gls{HB} stratification are depicted in figure \ref{Fig:N4}.

\begin{center}
\rule[1mm]{2cm}{.4pt}\hspace{1cm}$\circ$\hspace{1cm} \rule[1mm]{2cm}{.4pt}
\end{center}

\subsubsection{$\suf(3)$ series}

This series was basically already discussed in \cite{Apruzzi:2020pmv}. In fact all the theories here belong to the same $\cN=2$ $\cS$-fold class which can be engineered in type \emph{II}B string theory, and all mass deformations connecting the theories in the series are geometrically realized by motion of \emph{D7} branes.

\paragraph{$\boldsymbol{\cS^{(2)}_{D_4,3}}$}\label{sec:SD43}  This theory is the top theory of the $\suf(3)$ series and can be obtained by probing an exceptional $D_4$ \emph{7}brane singularity in the presence of a fluxfull $\Z_3$ $\cS$-fold by two \emph{D3} branes. It can also be obtained as compactification of a (1,0) \sd theory \cite{Giacomelli:2020jel}. Currently the author is unaware of any class-$\cS$ realization. A summary of the properties of $\cS$-theories can be found in appendix \ref{sec:TandS} and the CFT data and depiction of the \gls{Hasse}s of both the \gls{CB} and \gls{HB} stratification, can be found in figure \ref{Fig:S2Gl}.

\paragraph{$\boldsymbol{\cS^{(2)}_{A_1,3}}$}\label{sec:SA13} This theory can be obtained by moving away three \emph{D7} branes from the $D_4$ exceptional \emph{7}brane singularity and probing the remaining $A_1$ singularity plus a $\Z_3$ $\cS$-fold with flux by two \emph{D3} branes. The motion of the three \emph{D7} brane corresponds to a mass deformation in the $\cN=2$ theory. This theory can also be obtained as compactification of a (1,0) \sd theory \cite{Giacomelli:2020jel}. Again, the author is unaware of any class-$\cS$ realization. A summary of the properties of $\cS$-theories can be found in appendix \ref{sec:TandS} and the CFT data and depiction of the \gls{Hasse}s of both the \gls{CB} and \gls{HB} stratification, can be found in figure \ref{Fig:S2Gl}.

\paragraph{\green{$\boldsymbol{G(3,1,2)}$}}\label{sec:G312} This is an $\cN=3$ theory. The moduli space of these theories is as constrained as in the $\cN=4$ case and it is basically entirely specified by a \gls{Ccrg} which in this case is $G(3,1,2)$, a rank-2 \gls{Ccrg} of order eighteen. More details on the moduli space structure of theories with extended supersymmetry can be found in appendix \ref{sec:N34}. The CFT data of this theory, as well as the explicit \gls{Hasse} of both the \gls{CB} and \gls{HB} stratification are depicted in figure \ref{Fig:N3}.

\begin{center}
\rule[1mm]{2cm}{.4pt}\hspace{1cm}$\circ$\hspace{1cm} \rule[1mm]{2cm}{.4pt}
\end{center}

\subsubsection{$\suf(2)$ series}

As it was the case in the previous few series, this set of RG-flows was basically already discussed in \cite{Apruzzi:2020pmv}. Again, all the theories belong to the same $\cN=2$ $\cS$-fold class and all mass deformations are geometrically realized by motion of \emph{D7} branes.

\paragraph{$\boldsymbol{\cS^{(2)}_{A_2,4}}$}\label{sec:SA24} This theory, which sits at the top of the two theories $\suf(2)$ series, can be obtained by probing the $A_2$ exceptional \emph{7}brane singularity with two \emph{D3} branes but in the presence of a $\Z_4$ $\cS$-fold with fluxes. It can also be obtained as compactification of a (1,0) \sd theory \cite{Giacomelli:2020jel}. Currently the author is unaware of any class-$\cS$ realization. A summary of the properties of $\cS$-theories can be found in appendix \ref{sec:TandS} and the CFT data and depiction of the \gls{Hasse}s of both the \gls{CB} and \gls{HB} stratification, can be found in figure \ref{Fig:S2Gl}.

\paragraph{\green{$\boldsymbol{G(4,1,2)}$}}\label{sec:G412}   This is an $\cN=3$ theory. The moduli space of these theories is extremely constrained and it is basically entirely specified by a \gls{Ccrg} which in this case is $G(4,1,2)$, a rank-2 \gls{Ccrg} of order 32. More details on the moduli space structure of theories with extended supersymmetry can be found in appendix \ref{sec:N34}. The CFT data of this theory, as well as the explicit \gls{Hasse} of both the \gls{CB} and \gls{HB} stratification are depicted in figure \ref{Fig:N3}.

\begin{center}
\rule[1mm]{2cm}{.4pt}\hspace{1cm}$\circ$\hspace{1cm} \rule[1mm]{2cm}{.4pt}
\end{center}

\subsection{Isolated theories}

All theories discussed in this section do not belong to any series. While for those which can be realized by brane constructions there are reasons to expect that they are indeed not connected by mass deformations to any other $\cN=2$ theory, for the rest there aren't really strong argument in this direction. Thus one could speculate that there might be theories which are connected to them by RG-flows, awaiting to be discovered. The case of the lagrangian $USp(4)+\frac12{\bf 16}$ is particularly interesting as it currently has no string theory realizations\footnote{This is not the only lagrangian case with no known string theory realization. Many more examples are described in \cite{Bhardwaj:2013qia}. We thank Yuji Tachikawa for pointing this out.}.

\subsubsection*{\boldsymbol{$\spf(4)_{14}{\times}\suf(2)_8$}}\label{sec:t65}

\begin{figure}[h!]
\ffigbox{
\begin{subfloatrow}
\ffigbox[8.5cm][]{
\begin{tikzpicture}[decoration={markings,
mark=at position .5 with {\arrow{>}}}]
\begin{scope}[scale=1.5]
\node[bbc,scale=.5] (p0a) at (0,0) {};
\node[scale=.5] (p0b) at (0,-2) {};
\node[scale=.8] (t0b) at (0,-2.1) {$\spf(4)_{14}{\times}\suf(2)_8$};
\node[scale=.8] (p1) at (-.7,-1) {$[I_1^*,\spf(4)]$};
\node[scale=.8] (p2) at (.7,-1) {$[I_2,\suf(2)]$};
\node[scale=.8] (t2b) at (-.7,-1.5) {{\scriptsize$\big[u^3+v^2=0\big]$}};
\node[scale=.8] (t3b) at (.7,-1.6) {{\scriptsize$\big[u=0\big]$}};
\draw[red] (p0a) -- (p1);
\draw[red] (p0a) -- (p2);
\draw[red] (p1) -- (p0b);
\draw[red] (p2) -- (p0b);
\end{scope}
\begin{scope}[scale=1.5,xshift=3.5cm]
\node[scale=.5] (p4) at (0,2) {};
\node[scale=.8] (t0a) at (0,2.1) {$\H^{\rm d_{HB}}$};
\node[scale=.8] (tp2) at (-.2,1.5) {$\df_4$};
\node[scale=.8] (p3) at (0,1) {$\cT^{(1)}_{D_4,1}$};
\node[scale=.8] (p1a) at (.7,-.5) {$\cS^{(1)}_{D_4,3}$};
\node[scale=.8] (tp2) at (-.6,.5) {$\af_1$};
\node[scale=.8] (p2b) at (-.7,0) {\hyperref[sec:SU22b]{$2\gls{F}+SU(2)-SU(2)+2\gls{F}$}};
\node[scale=.8] (tp1) at (-1,-.5) {$\cf_2$};
\node[scale=.8] (p1b) at (-.7,-1) {\hyperref[sec:Sp42524]{$USp(4)+2\gls{F}+2\gls{V}$}};
\node[scale=.8] (tp1) at (-.7,-1.5) {$\cf_2$};
\node[scale=.8] (tp1) at (.7,-1.5) {$\af_1$};
\node[scale=.5] (p0) at (0,-2) {};
\node[scale=.8] (t0b) at (0,-2.1) {$\spf(4)_{14}{\times}\suf(2)_8$};
\draw[blue] (p0) -- (p1a);
\draw[blue] (p0) -- (p1b);
\draw[blue] (p1b) -- (p2b);
\draw[blue] (p2b) -- (p3);
\draw[blue] (p1a) -- (p3);
\draw[blue] (p3) -- (p4);
\end{scope}
\end{tikzpicture}}
{\caption{\label{CBhTh34}The Coulomb and Higgs stratification of $\spf(4)_{14}{\times}\suf(2)_8$.}}
\end{subfloatrow}\hspace{1cm}%
\begin{subfloatrow}
\capbtabbox[5cm]{%
  \renewcommand{\arraystretch}{1.1}
  \begin{tabular}{|c|c|} 
  \hline
  \multicolumn{2}{|c|}{$\spf(4)_{14}{\times}\suf(2)_8$}\\
  \hline\hline
  $(\D_u,\D_v)$  &\quad $\left(4,6\right)$\quad{} \\
  $24a$ & 118\\  
  $12c$ & 64\\
$\ff_k$ & $\spf(4)_{14}{\times}\suf(2)_8$ \\ 
$d_{\rm HB}$& 10\\
$h$&4\\
$T({\bf2}\bh)$&2\\
\hline\hline
\end{tabular}
}{%
  \caption{\label{CchTh34}Central charges, \gls{CB} parameters and \gls{ECB} dimension.}%
}
\end{subfloatrow}}{\caption{\label{TothTh34}Information about the $\spf(4)_{14}{\times}\suf(2)_8$.}}
\end{figure}

This theory can be obtained, for example, in the $\Z_3$ twisted $D_4$ class-$\cS$ \cite{Chacaltana:2016shw} where most of the CFT data reported below is computed.

This theory is \gls{totally} with a semi-simple flavor symmetry. At first one might be tempted to guess that the two simple flavor symmetry factors are realized each on an allowed \gls{ukstr}, $u=0$ and $v=0$. But a more careful look at the value of the levels immediately reveal that the situation cannot be as simple. The level of the $\suf(2)$ does not create much problem. It is indeed double $\D_u$ thus the most natural guess is the identification $\rTf_u\equiv [I_2,\suf(2)]$. The $\spf(4)$ is instead puzzling. Since the $u=0$ is already ``occupied'' the two other options are either the $v=0$ which has $\D_v=8$ or the \gls{kstr} which instead has $\D_{\rm knot}=12$. The insight comes from the fact that \gls{h}=4 while this theory would naturally support a \gls{h}=2 acted upon by the $\spf(4)$ factor (and for example realized by a $[I_1^*,\spf(4)]$). The resolution is that the four quaternionic (eight complex) dimensional \gls{ECB} transforms as ${\bf 4}\oplus{\bf4}$ of the flavor symmetry realized on the \gls{kstr} by a $\rTf_{u^3+v^2}\equiv[I_1^*,\spf(4)]$. This perfectly reproduces the level but since is a new situation let's do things explicitly. Recall \eqref{actotcint}:
\beq
k_\ff=\sum_{i\in I_{\ff}}\frac{\D_i^{\rm sing}}{d_i\D_i} \left(k^i-T({\bf2}\bh_i)\right)+T({\bf2}\bh).
\eeq
In our case, $\D_i^{\rm sing}=\D_{\rm knot}=12$. The fact that the theory supported on the stratum is a $[I_1^*,\spf(4)]$ implies $d_i=1$, $\D_i=2$, $k^i=3$ and $T({\bf2}\bh_i\equiv{\bf4})=1$. But because of our previous observation on the dimensionality of the \gls{ECB}, $T({\bf2}\bh\equiv{\bf4}\oplus{\bf4})=2$ and thus we reproduce the correct level of the $\spf(4)$ factor. This identification also perfectly reproduces the $a$ and $c$ central charges in figure \ref{CchTh34} using \eqref{actotaint} and \eqref{actotbint}.

Let's now move to analyze the \gls{HB}, this analysis will confirm the validity of our previous guesses. Let us first focus on the stratum associated with the \gls{CB} higgsing of the $[I_2,\suf(2)]$. This is of \gls{gHW} type and thus we can use \eqref{gHWeq} to predict the rank-1 theory supported there which leads to $\bTf_{\af_1}\equiv\cS^{(1)}_{D_4,3}$ (this guess could have also been made by observing that to match the total \gls{HB} dimension, the rank-1 theory supported on $\af_1$ had to have a nine dimensional \gls{HB}). The other ``side'' of the \gls{Hasse} is trickier. Since we have a single $\spf(4)$ factor we expect a single $\cf_2$ transition, yet we know from the \gls{CB} analysis that the \gls{ECB} does not transform irreducibly under the flavor symmetry, signaling that the theory that is supported on the $\cf_2$ should itself have a \gls{h}=2. By matching both the total dimensionality of the \gls{HB} and using the \eqref{gHWeq} it is possible to identify that the rank-2 theory supported on the $\cf_2$ stratum is a $USp(4)+2\gls{F}+2\gls{V}$, which indeed has a two quaternionic dimensional \gls{ECB} as expected (this higgsing can also be guessed from the class-$\cS$ construction and can be then checked independently). Following the subsequent higgsings of the two theories supported on the first two strata, we can readily reproduce the full \gls{Hasse} in figure \ref{CBhTh34}.

\subsubsection*{$\boldsymbol{\cT^{(2)}_{\varnothing,5}}$}
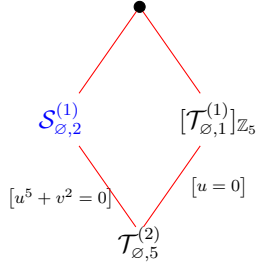
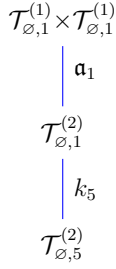
\label{sec:Tvar}

\begin{figure}[h!]
\ffigbox{
\begin{subfloatrow}
\ffigbox[8.5cm][]{
\begin{tikzpicture}[decoration={markings,
mark=at position .5 with {\arrow{>}}}]
\begin{scope}[scale=1.5]
\node[bbc,scale=.5] (p0a) at (0,-1) {};
\node[scale=.5] (p0b) at (0,-3) {};
\node[scale=.8] (t0b) at (0,-3.1) {$\cT^{(2)}_{\varnothing,5}$};
\node[scale=.8] (p1) at (-.7,-2) {$\blue{\cS^{(1)}_{\varnothing,2}}$};
\node[scale=.8] (p2) at (.7,-2) {$[\cT^{(1)}_{\varnothing,1}]_{\Z_5}$};
\node[scale=.8] (t2b) at (-.7,-2.7) {{\scriptsize$\big[u^5+v^2=0\big]$}};
\node[scale=.8] (t3b) at (.7,-2.6) {{\scriptsize$\big[u=0\big]$}};
\draw[red] (p0a) -- (p1);
\draw[red] (p0a) -- (p2);
\draw[red] (p1) -- (p0b);
\draw[red] (p2) -- (p0b);
\end{scope}
\begin{scope}[scale=1.5,xshift=2.5cm]
\node[scale=.8] (p3a) at (0,-1) {$\cT^{(1)}_{\varnothing,1}{\times}\cT^{(1)}_{\varnothing,1}$};
\node[scale=.8] (tp2) at (.2,-1.45) {$\af_1$};
\node[scale=.8] (p2a) at (0,-2) {$\cT^{(2)}_{\varnothing,1}$};
\node[scale=.8] (tp2) at (.2,-2.5) {$k_5$};
\node[scale=.8] (p1a) at (0,-3) {$\cT^{(2)}_{\varnothing,5}$};
\draw[blue] (p1a) -- (p2a);
\draw[blue] (p2a) -- (p3a);
\end{scope}
\end{tikzpicture}}
{\caption{\label{CBhT05}The Coulomb and Higgs stratification of $\cT^{(2)}_{\varnothing,5}$.}}
\end{subfloatrow}\hspace{1cm}%
\begin{subfloatrow}
\capbtabbox[5cm]{%
  \renewcommand{\arraystretch}{1.1}
  \begin{tabular}{|c|c|} 
  \hline
  \multicolumn{2}{|c|}{$\cT^{(2)}_{\varnothing,5}$}\\
  \hline\hline
  $(\D_u,\D_v)$  &\quad $\left(\frac{12}5.6\right)$\quad{} \\
  $24a$ & $\frac{456}5$ \\  
  $12c$ & $\frac{234}5$ \\
$\ff_k$ & $\suf(2)_{14}$ \\ 
$d_{\rm HB}$& 2\\
$h$&2\\
$T({\bf2}\bh)$&1\\
\hline\hline
\end{tabular}
}{%
  \caption{\label{CchT05}Central charges, \gls{CB} parameters and \gls{ECB} dimension.}%
}
\end{subfloatrow}}{\caption{\label{TothT05}Information about $\cT^{(2)}_{\varnothing,5}$.}}
\end{figure}

This theory is curious. It in fact belongs to an infinite series of theories first conjectured in \cite{Ohmori:2018ona} and then shown to be consistent in \cite{Giacomelli:2020gee} which naturally sit in the infinite series of $\cT$ theories \cite{Giacomelli:2020jel}. It is curious since the construction using the twisted compactification of \sd (1,0) theories suggests a brane realization as all other $\cT$ theories. Yet this would imply the presence of a $\Z_5$ S-fold which doesn't seem to be allowed \cite{Aharony:2016kai} for the simple reason that the latter are specified by finite subgroup of $SL(2,\Z)$ and $\Z_5$ simply isn't one. For a more detailed discussion see \cite{Giacomelli:2020gee}. 

The structure of the moduli space of vacua has very much the same features of the remaining $\cT$-theories, which are again discussed in appendix \ref{sec:TandS}. But given the strange nature of the case we will present the CFT data and the depiction of the moduli space stratification separately. All the relevant info are summarized in figure \ref{TothT05}.

\subsubsection*{\boldsymbol{$USp(4)\ \text{w/}\ \frac12\,{\bf 16}$}}\label{sec:Sp416}

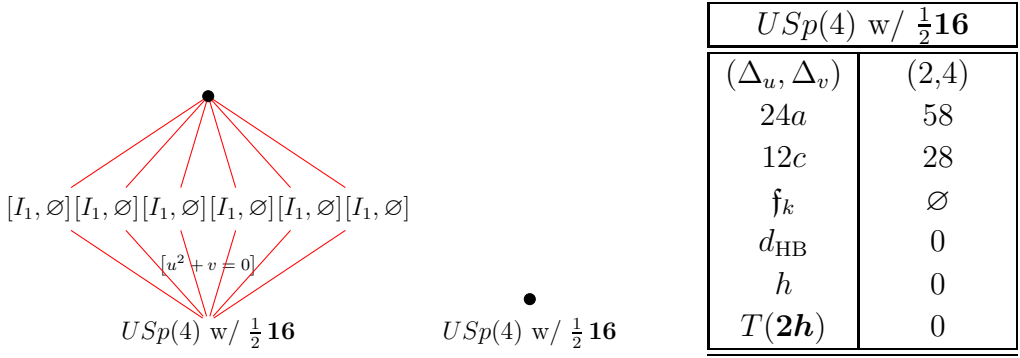
\begin{figure}[h!]
\ffigbox{
\begin{subfloatrow}
\ffigbox[7cm][]{
\begin{tikzpicture}[decoration={markings,
mark=at position .5 with {\arrow{>}}}]
\begin{scope}[scale=1.5]
\node[bbc,scale=.5] (p0a) at (0,0) {};
\node[scale=.5] (p0b) at (0,-2) {};
\node[scale=.8] (t0b) at (0,-2.1) {$USp(4)$ {\rm w/} $\frac12\,{\bf16}$};
\node[scale=.8] (p4) at (-1.5,-1) {$[I_1,\varnothing]$};
\node[scale=.8] (p5) at (1.5,-1) {$[I_1,\varnothing]$};
\node[scale=.8] (p1) at (-.9,-1) {$[I_1,\varnothing]$};
\node[scale=.8] (p2) at (.9,-1) {$[I_1,\varnothing]$};
\node[scale=.8] (p3) at (.3,-1) {$[I_1,\varnothing]$};
\node[scale=.8] (p6) at (-.3,-1) {$[I_1,\varnothing]$};
\node[scale=.8] (t2b) at (0,-1.5) {{\scriptsize$\big[u^2+v=0\big]$}};
\draw[red] (p0a) -- (p1);
\draw[red] (p0a) -- (p6);
\draw[red] (p0a) -- (p2);
\draw[red] (p0a) -- (p3);
\draw[red] (p0a) -- (p4);
\draw[red] (p0a) -- (p5);
\draw[red] (p6) -- (p0b);
\draw[red] (p5) -- (p0b);
\draw[red] (p4) -- (p0b);
\draw[red] (p1) -- (p0b);
\draw[red] (p2) -- (p0b);
\draw[red] (p3) -- (p0b);
\end{scope}
\begin{scope}[scale=1.5,xshift=2.85cm]
\node[bbc,scale=.5] (p0a) at (0,-1.8) {};
\node[scale=.8] (p0b) at (0,-2.1) {$USp(4)$ {\rm w/} $\frac12\,{\bf16}$};
\end{scope}
\end{tikzpicture}}
{\caption{\label{CBSp416}The \gls{Hasse} for the \gls{CB} and the \gls{HB} of the $USp(4)$ gauge theory with a half ${\bf 16}$.}}
\end{subfloatrow}\hspace{1cm}%
\begin{subfloatrow}
\capbtabbox[7cm]{%
  \renewcommand{\arraystretch}{1.1}
  \begin{tabular}{|c|c|} 
  \hline
  \multicolumn{2}{|c|}{$USp(4)$ {\rm w/} $\frac12 {\bf16}$}\\
  \hline\hline
  $(\D_u,\D_v)$  &\quad (2,4)\quad{} \\
  $24a$ &  58\\  
  $12c$ & 28 \\
$\ff_k$ & $\varnothing$ \\ 
$d_{\rm HB}$&0\\
$h$&0\\
$T({\bf2}\bh)$&0\\
\hline\hline
  \end{tabular}
}{%
  \caption{\label{CcSp416}Central charges, \gls{CB} parameters and \gls{ECB} dimension.}%
}
\end{subfloatrow}}{\caption{\label{TotSp416} Information about the $USp(4)$ $\cN=2$ theory with a-half hypermultiplet in the ${\bf 16}$.}}
\end{figure}

The existence of this theory is pointed out in \cite{Bhardwaj:2013qia}. It is important to stress once again that to the author's knowledge, no string theory realization of this theory is known. Let's discuss here how to derive the result depicted in figure \ref{CBSp416}.

The analysis of the \gls{HB} is obvious. Since there is a single half-hypermultiplet no gauge invariant operator can be made from the hypermultiplets and the \gls{HB} is trivial. The analysis of the \gls{CB} is instead more involved.

As mentioned in the discussion of previous cases, in  $USp(4)$ there are two inequivalent $SU(2)$ with rank-1 commutant, one ($a$) with commutant $U(1)$ and one ($b$) with commutant $SU(2)$. By carefully decomposing the ${\bf 16}$ we obtain the following decomposition:
\begin{itemize}

\item[$a$\,-] ${\bf 16}\to ({\bf 3},{\bf 2})\oplus({\bf 2},{\bf 3})\oplus({\bf 2},{\bf 1})\oplus({\bf 1},{\bf 2})$.

\item[$b$\,-] ${\bf 16}\to {\bf4}_{\pm1}\oplus {\bf 2}_{\pm3}\oplus {\bf 2}_{\pm1}$.

\end{itemize}

Since the two $SU(2)$s in $a$ are equivalent, the inequivalent complex co-dimension one directions to ``explore'' correspond to turning on vevs along the Cartan of the $SU(2)$ in $a$, the Cartan of the $SU(2)$ or the $U(1)$ or a combination of both in $b$. For a given vev, only the components of the ${\bf 16}$ which are not charged under the specific direction in $USp(4)$ stay massless. Using the above decomposition, it is therefore straightforward to conclude that the ${\bf 16}$ contributes a single flavor to $a$, and no massless hypers to either single choices in $b$ unless we pick an appropriate linear combination of the Cartan of the $SU(2)$ and the $U(1)$, whose commutant is of course a $U(1)$ gauge theory. This choice lives behind a single massless hyper. 

In summary along $a$ the theory is effectively an $SU(2)$ with a single hypers while along $b$ turning on a vev for the $U(1)$ gives a pure $SU(2)$ $\cN=2$ gauge theory. The linear combination of $U(1)$ plus Cartan of $SU(2)$ gives instead a $U(1)$ gauge theory with a single hypermultiplet. The $SU(2)$ theories are asymptotically free and the flow to strong coupling in the IR causes the \gls{kstr} to spilt. The pure $SU(2)$ theory contributes two $[I_1,\varnothing]$ while the other has an extra $[I_1,\varnothing]$ coming from the massless hyper. The $U(1)$ with a single hyper contributes instead an extra $[I_1,\varnothing]$ for a total of six knotted singularities each supporting an $[I_1,\varnothing]$. To check that this is indeed the correct stratification, we can plug things in the central charge formulae \eqref{actotaint}-\eqref{actotbint} and perfectly reproduce the expected values which are reported in table \ref{CcSp416}.

\subsubsection*{\boldsymbol{\blue{$G_2$}}}\label{sec:N4G2} This is an $\cN=4$ theory. The moduli space of these theories is extremely constrained and it is basically entirely specified by the Weyl group of the gauge algebra which in this case is $D_6$, the dihedral group of order twelve. More details on the moduli space structure of theories with extended supersymmetry can be found in appendix \ref{sec:N34}. The CFT data of this theory, as well as the explicit \gls{Hasse} of both the \gls{CB} and \gls{HB} stratification are depicted in figure \ref{Fig:N4}.

Also, this theory was shown to be realizable as worldvolume theory of two \emph{D3} branes probing a fluxless $\Z_6$ $\cS$-fold \cite{Aharony:2016kai}. Since no exceptional seven brane singularity is compatible with the presence of an $\Z_6$ $\cS$-fold \cite{Apruzzi:2020pmv}, it reasonable that this theory is indeed isolated.

\acknowledgments I would like to thank G. Zafrir for a very enjoyable and fruitful correspondence during which he clarified countless issues for me, many of which made it into this manuscript. I would also like to thank P. Argyres, A. Bourget, J. Distler, J. Grimminger, A. Rocchetto, S. Schafer-Nameki, Y. Tachikawa, M. Weaver and G. Zafrir for comments on the manuscript. Finally I benefited tremendously from many exchanges with P. Argyres, C. Beem, A. Bourget, J. Distler, S. Giacomelli, J. Grimminger, A. Hanany, C. Meneghelli, W. Peelaers, L. Rastelli, S. Schafer-Nameki and Y. Tachikawa. I am extremely grateful for these interactions which illuminated many details of the constructions which made this paper possible. M.M. gratefully acknowledges the Simons Foundation (Simons Collaboration on the Non-perturbative Bootstrap) grants 488647 and 397411, for the support of his work.

\appendix

\section{$\cS$ and $\cT$ theories}\label{sec:TandS}

This set of theories has been introduced recently by generalizing the $\cN=3$ $S$-fold set up \cite{Garcia-Etxebarria:2015wns,Aharony:2016kai} to the $\cN=2$ case \cite{Apruzzi:2020pmv,Giacomelli:2020jel} as well as generalizing the ``classic'' F-theory $\cN=2$ theories \cite{Banks:1996nj,Douglas:1996js,Sen:1996vd,Dasgupta:1996ij}. This construction thus involves considering the \emph{D3} brane worldvolume theory probing an exceptional 7brane in the presence of an $S$-fold. Depending on whether the $S$-fold has fluxes turned on or off, we get $\cS$ or $\cT$ theories respectively \cite{Giacomelli:2020gee}. For a given exceptional 7brane, there is a restricted set of $S$-folds which are allowed. We won't review this discussion here and instead refer the interested reader to the original paper \cite{Apruzzi:2020pmv}. It is important to notice that both the $\cS$ and $\cT$ theories can be also obtained as compactification of $6d$ (1,0) theories with non-commuting holonomies \cite{Ohmori:2018ona,Giacomelli:2020jel}. Since the $(1,0)$ theories are most naturally constructed in $M$-theory, the construction of $\cS$ and $\cT$ theories suggest potentially interesting duality between $F$ and $M$ theory \cite{Giacomelli:2020gee}.

Both the $\cS$ and $\cT$ theories have been studied in depth, particularly recently \cite{Apruzzi:2020pmv,Argyres:2020wmq,Giacomelli:2020jel,Giacomelli:2020gee,Bourget:2020mez,Kimura:2020hgw,Heckman:2020svr}. Rather than literally reproducing here results from other papers, we make the choice of simply refer to the relevant literature. Namely:
\begin{itemize}
\item[1)] For a discussion of the general structure of the moduli space see \cite{Apruzzi:2020pmv,Argyres:2020wmq,Giacomelli:2020jel}. The \gls{HB} \gls{Hasse} has instead been worked out in detail and for general ranks in \cite{Bourget:2020mez}.

\item[2)] For a discussion of the \gls{CB} stratification see \cite{Argyres:2020wmq,Giacomelli:2020gee} (one of the two references also contain a discussion of the mass deformations among these theories).

\item[3)] For a discussion of the generalized free-field VOA construction see \cite{Beem:2019snk,Giacomelli:2020jel}.

\end{itemize}

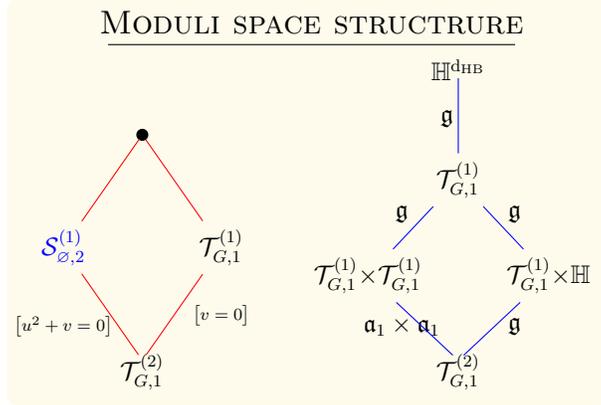
\begin{figure}[h!]
\underline{\Large{\textsc{CFT data of $\cT^{(2)}_{G,1}$ theories\protect\footnotemark}}}\vspace{.5cm}
\ffigbox{
\begin{subfloatrow}
\capbtabbox[5cm]{%
  \renewcommand{\arraystretch}{1.1}
  \begin{tabular}{|c|c|} 
  \hline
  \multicolumn{2}{|c|}{$\cT^{(2)}_{E_8,1}$}\\
  \hline\hline
  $(\D_u,\D_v)$  &\quad $\left(6,12\right)$\quad{} \\
  $24a$ &263\\  
  $12c$ & 161\\
$\ff_k$ & $[\ef_8]_{24}{\times}\suf(2)_{13}$ \\ 
$d_{\rm HB}$& 59\\
$h$&1\\
$T({\bf2}\bh)$&1\\
\hline\hline
\end{tabular}
}{\caption{}\vspace{-.3cm}}
\end{subfloatrow}
\begin{subfloatrow}
\hspace{-.4cm}
\capbtabbox[5cm]{%
  \renewcommand{\arraystretch}{1.1}
  \begin{tabular}{|c|c|} 
  \hline
  \multicolumn{2}{|c|}{$\cT^{(2)}_{E_7,1}$}\\
  \hline\hline
  $(\D_u,\D_v)$  &\quad $\left(4,8\right)$\quad{} \\
  $24a$ &167\\  
  $12c$ & 101\\
$\ff_k$ & $[\ef_7]_{16}{\times}\suf(2)_9$ \\ 
$d_{\rm HB}$& 35\\
$h$&1\\
$T({\bf2}\bh)$&1\\
\hline\hline
\end{tabular}
}{\caption{}\vspace{-.3cm}}
\end{subfloatrow}{}
\begin{subfloatrow}
\hspace{-.4cm}
\capbtabbox[5cm]{%
  \renewcommand{\arraystretch}{1.1}
  \begin{tabular}{|c|c|} 
  \hline
  \multicolumn{2}{|c|}{$\cT^{(2)}_{E_6,1}$}\\
  \hline\hline
  $(\D_u,\D_v)$  &\quad $\left(3,6\right)$\quad{} \\
  $24a$ &119\\  
  $12c$ & 71\\
$\ff_k$ & $[\ef_6]_{12}{\times}\suf(2)_7$ \\ 
$d_{\rm HB}$& 23\\
$h$&1\\
$T({\bf2}\bh)$&1\\
\hline\hline
\end{tabular}
}{\caption{}\vspace{-.3cm}}
\end{subfloatrow}
\begin{subfloatrow}
\capbtabbox[5cm]{%
  \renewcommand{\arraystretch}{1.1}
  \begin{tabular}{|c|c|} 
  \hline
  \multicolumn{2}{|c|}{$\cT^{(2)}_{D_4,1}$/$\spf(4) +4\gls{F}+\gls{V}$}\\
  \hline\hline
  $(\D_u,\D_v)$  &\quad $\left(2,4\right)$\quad{} \\
  $24a$ &71\\  
  $12c$ & 42\\
$\ff_k$ & $\sof(8)_8{\times}\suf(2)_5$ \\ 
$d_{\rm HB}$& 11\\
$h$&1\\
$T({\bf2}\bh)$&1\\
\hline\hline
\end{tabular}
}{\caption{}\vspace{-.3cm}}
\end{subfloatrow}
\begin{subfloatrow}\hspace{-.3cm}
\capbtabbox[5cm]{%
  \renewcommand{\arraystretch}{1.1}
  \begin{tabular}{|c|c|} 
  \hline
  \multicolumn{2}{|c|}{$\cT^{(2)}_{A_2,1}$}\\
  \hline\hline
  $(\D_u,\D_v)$  &\quad $\left(\frac32,3\right)$\quad{} \\
  $24a$ &47\\  
  $12c$ & 26\\
$\ff_k$ & $\suf(3)_6{\times}\suf(2)_4$ \\ 
$d_{\rm HB}$& 5\\
$h$&1\\
$T({\bf2}\bh)$&1\\
\hline\hline
\end{tabular}
}{\caption{}\vspace{-.3cm}}
\end{subfloatrow}
\begin{subfloatrow}
\hspace{-.3cm}
\capbtabbox[5cm]{%
  \renewcommand{\arraystretch}{1.1}
  \begin{tabular}{|c|c|} 
  \hline
  \multicolumn{2}{|c|}{$\cT^{(2)}_{A_1,1}$}\\
  \hline\hline
  $(\D_u,\D_v)$  &\quad $\left(\frac43,\frac83\right)$\quad{} \\
  $24a$ &39\\  
  $12c$ & 21\\
$\ff_k$ &\hspace{-.3cm} ${\suf(2)_{\tiny{\frac{16}3}}{\times}\suf(2)_{\tiny{\frac{11}3}}}$\hspace{-.26cm}  \\ 
$d_{\rm HB}$& 3\\
$h$&1\\
$T({\bf2}\bh)$&1\\
\hline\hline
\end{tabular}
}{\caption{}\vspace{-.3cm}}
\end{subfloatrow}
\begin{subfloatrow}
\capbtabbox[5cm]{%
  \renewcommand{\arraystretch}{1.1}
  \begin{tabular}{|c|c|} 
  \hline
  \multicolumn{2}{|c|}{$\cT^{(2)}_{\varnothing,1}$}\\
  \hline\hline
  $(\D_u,\D_v)$  &\quad $\left(\frac65,\frac{12}5\right)$\quad{} \\
  $24a$ &$\frac{163}5$\\  
  $12c$ & 17\\
$\ff_k$ &\hspace{.55cm} $\suf(2)_{\frac{17}5}$\hspace{.55cm} \\ 
$d_{\rm HB}$& 1\\
$h$&1\\
$T({\bf2}\bh)$&1\\
\hline\hline
\end{tabular}
}{\caption{}\vspace{-.3cm}}
\end{subfloatrow}
\begin{subfloatrow}
\ffigbox[10.2cm][]{
\begin{tikzpicture}[decoration={markings,
mark=at position .5 with {\arrow{>}}}]
\fill[color=Goldenrod!10, rounded corners] (-1.8,1.8) rectangle (6.2,-3.6);
\begin{scope}[scale=1.5]
\node[scale=.5] (p0b) at (1.5,1) {\Huge{\textsc{Moduli space structrure}}};
\node[] (under1) at (-.4,.8) {};
\node[] (under2) at (3.4,.8) {};
\draw[] (under1) -- (under2);
\node[bbc,scale=.5] (p0a) at (0,0) {};
\node[scale=.5] (p0b) at (0,-2) {};
\node[scale=.8] (t0b) at (0,-2.1) {$\cT^{(2)}_{G,1}$};
\node[scale=.8] (p1) at (-.7,-1) {$\blue{\cS^{(1)}_{\varnothing,2}}$};
\node[scale=.8] (p2) at (.7,-1) {$\cT^{(1)}_{G,1}$};
\node[scale=.8] (t2b) at (-.7,-1.7) {{\scriptsize$\big[u^2+v=0\big]$}};
\node[scale=.8] (t3b) at (.7,-1.6) {{\scriptsize$\big[v=0\big]$}};
\draw[red] (p0a) -- (p1);
\draw[red] (p0a) -- (p2);
\draw[red] (p1) -- (p0b);
\draw[red] (p2) -- (p0b);
\end{scope}
\begin{scope}[scale=1.5,xshift=2.8cm]
\node[scale=.5] (p3) at (0,.55) {};
\node[scale=.8] (t0a) at (0,.58) {$\H^{\rm d_{HB}}$};
\node[scale=.8] (tp2) at (-.1,.15) {$\gf$};
\node[scale=.8] (p2) at (0,-.4) {$\cT^{(1)}_{G,1}$};
\node[scale=.8] (tp1) at (-.5,-.7) {$\gf$};
\node[scale=.8] (tp1) at (.5,-.7) {$\gf$};
\node[scale=.8] (p1a) at (-.8,-1.25) {$\cT^{(1)}_{G,1}{\times}\cT^{(1)}_{G,1}$};
\node[scale=.8] (p1b) at (.8,-1.25) {$\cT^{(1)}_{G,1}{\times}\H$};
\node[scale=.8] (tp1) at (-.5,-1.7) {$\af_1\times \af_1$};
\node[scale=.8] (tp1) at (.5,-1.7) {$\gf$};
\node[scale=.5] (p0) at (0,-2) {};
\node[scale=.8] (t0b) at (0,-2.1) {$\cT^{(2)}_{G,1}$};
\draw[blue] (p0) -- (p1a);
\draw[blue] (p0) -- (p1b);
\draw[blue] (p1b) -- (p2);
\draw[blue] (p1a) -- (p2);
\draw[blue] (p2) -- (p3);
\end{scope}
\end{tikzpicture}}
{}
\end{subfloatrow}{}
}{\caption{\label{Fig:T2G1} In this figure we report the relevant CFT data for the $\cT^{(2)}_{G,1}$ theories.}}
\end{figure}

Below we will then simply summarize the results with various tables and the explicit stratification. To better organize the presentation we will collect the relevant CFT data in three figures:
\begin{itemize}
\item[$a)$] In figure \ref{Fig:T2G1} we will collect all the info for the $\cT^{(2)}_{G,1}$. Those correspond to the well-known 4d theories rank-2 theories arising on a worldvolume of two \emph{D3} branes probing an exceptional $G$ \emph{7}brane singularity \cite{Banks:1996nj,Douglas:1996js,Sen:1996vd,Dasgupta:1996ij}. These theories they are all connected to one another by mass deformations.

\item[$b)$] In figure \ref{Fig:T2Gl} we collect the information for the $\cT$ theories which arise when the \emph{D3} probe a $G$ \emph{7}brane singularity plus a \emph{fluxless} $\Z_\ell$ S-fold: $\cT^{(2)}_{G,\ell>1}$. For fixed $\ell$, these theories are connected to one another by mass deformation. They also eventually flow to $\cN=4$ theories with gauge algebra $SU(2){\times}SU(2)$ for $\ell=2$, $SU(3)$ for $\ell=3$ and $USp(4)$ for $\ell=4$ \cite{Giacomelli:2020gee}.

\item[$c)$] Finally figure \ref{Fig:S2Gl} collects the information for the $\cS$ theories which arise instead when two \emph{D3} probe a $G$ \emph{7}brane singularity plus a \emph{fluxfull} $\Z_\ell$ S-fold \cite{Apruzzi:2020pmv}. Again these theories are connected with one another for a given $\ell$ and they flow to $\cN=4$ $USp(4)$ for $\ell=2$ and $\cN=3$ $G(3,1,2)$ and $G(4,1,2)$ theories for $\ell=3$ and $\ell=4$ respectively. 

\end{itemize}

We will separately discuss the case of $\cT^{(2)}_{\varnothing,5}$ in the text because its moduli space structure it does not perfectly fit in the homogenous analysis of the remaining theories. This would also allow us to point out a few interesting features of this case.

\footnotetext{$\cT^{(2)}_{G,1}$ theories correspond to the well-known 4d theories rank-2 theories arising in type \emph{II}B by probing and exceptional $G$ 7brane singularity.}

\begin{figure}[h!]
\underline{\Large{\textsc{CFT data of $\cT^{(2)}_{G,\ell>1}$ theories}}}\vspace{.5cm}
\ffigbox{
\begin{subfloatrow}
\capbtabbox[4.8cm]{%
  \renewcommand{\arraystretch}{1.1}
  \begin{tabular}{|c|c|} 
  \hline
  \multicolumn{2}{|c|}{$\cT^{(2)}_{D_4,3}$}\\
  \hline\hline
  $(\D_u,\D_v)$  &\quad $\left(4,6\right)$\quad{} \\
  $24a$ &120\\  
  $12c$ & 66\\
$\ff_k$ & $[\gf_2]_8{\times}\suf(2)_{14}$ \\ 
$d_{\rm HB}$& 12\\
$h$&2\\
$T({\bf2}\bh)$&1\\
\hline\hline
\end{tabular}}{\caption{}\vspace{-.3cm}}
\end{subfloatrow}
\begin{subfloatrow}
\capbtabbox[4.8cm]{%
  \renewcommand{\arraystretch}{1.1}
  \begin{tabular}{|c|c|} 
  \hline
  \multicolumn{2}{|c|}{$\cT^{(2)}_{A_1,3}$}\\
  \hline\hline
  $(\D_u,\D_v)$  &\quad $\left(\frac83,4\right)$\quad{} \\
  $24a$ &72\\  
  $12c$ & 38\\
$\ff_k$ &\hspace{-.3cm} $\suf(2)_{\frac{16}3}{\times}\suf(2)_{10}$\hspace{-.2cm} \\ 
$d_{\rm HB}$& 4\\
$h$&2\\
$T({\bf2}\bh)$&1\\
\hline\hline
\end{tabular}
}{\caption{}\vspace{-.3cm}}
\end{subfloatrow}{}
\begin{subfloatrow}
\capbtabbox[4.8cm]{%
  \renewcommand{\arraystretch}{1.1}
  \begin{tabular}{|c|c|} 
  \hline
  \multicolumn{2}{|c|}{$\cT^{(2)}_{E_6,2}$}\\
  \hline\hline
  $(\D_u,\D_v)$  &\quad $\left(6,6\right)$\quad{} \\
  $24a$ &156\\  
  $12c$ & 90 \\
$\ff_k$ & $\hspace{-.2cm}[\ff_4]_{12}{\times}\suf(2)_7^2\hspace{-.2cm}$ \\ 
$d_{\rm HB}$& 24\\
$h$&2\\
$T({\bf2}\bh)$&2\\
\hline\hline
\end{tabular}
}{\caption{}\vspace{-.3cm}}
\end{subfloatrow}
\begin{subfloatrow}
\capbtabbox[4.8cm]{%
  \renewcommand{\arraystretch}{1.1}
  \begin{tabular}{|c|c|} 
  \hline
  \multicolumn{2}{|c|}{$\cT^{(2)}_{D_4,2}$}\\
  \hline\hline
  $(\D_u,\D_v)$  &\quad $\left(4,4\right)$\quad{} \\
  $24a$ &96\\  
  $12c$ & 54\\
$\ff_k$ &\hspace{-.2cm} $\sof(7)_8{\times}\suf(2)^2_5$\hspace{-.1cm} \\ 
$d_{\rm HB}$& 12\\
$h$&2\\
$T({\bf2}\bh)$&2\\
\hline\hline
\end{tabular}
}{\caption{}\vspace{-.3cm}}
\end{subfloatrow}
\begin{subfloatrow}
\capbtabbox[4.8cm]{%
  \renewcommand{\arraystretch}{1.1}
  \begin{tabular}{|c|c|} 
  \hline
  \multicolumn{2}{|c|}{$\cT^{(2)}_{A_2,2}$}\\
  \hline\hline
  $(\D_u,\D_v)$  &\quad $\left(3,3\right)$\quad{} \\
  $24a$ &66\\  
  $12c$ & 36\\
$\ff_k$ &\hspace{-.2cm} $\suf(3)_6{\times}\suf(2)_4^2$\hspace{-.1cm} \\ 
$d_{\rm HB}$& 6\\
$h$&2\\
$T({\bf2}\bh)$&2\\
\hline\hline
\end{tabular}
}{\caption{}\vspace{-.3cm}}
\end{subfloatrow}
\begin{subfloatrow}
\hspace{-.3cm}
\capbtabbox[4.8cm]{%
  \renewcommand{\arraystretch}{1.1}
  \begin{tabular}{|c|c|} 
  \hline
  \multicolumn{2}{|c|}{$\cT^{(2)}_{A_2,4}$}\\
  \hline\hline
  $(\D_u,\D_v)$  &\quad $\left(3,6\right)$\quad{} \\
  $24a$ &102\\  
  $12c$ & 54\\
$\ff_k$ & $\hspace{-.2cm}\suf(2)_6{\times}\suf(2)_{14}\hspace{-.2cm}$ \\ 
$d_{\rm HB}$& 6\\
$h$&2\\
$T({\bf2}\bh)$&2\\
\hline\hline
\end{tabular}
}{\caption{}\vspace{-.3cm}}
\end{subfloatrow}
\begin{subfloatrow}
\ffigbox[12cm][]{
\begin{tikzpicture}[decoration={markings,
mark=at position .5 with {\arrow{>}}}]
\fill[color=Goldenrod!10, rounded corners] (-1.8,1) rectangle (9.8,-5);
\node[scale=.5] (p0b) at (4,.8) {\Huge{\textsc{Moduli space structrure}}};
\node[] (under1) at (1.2,.5) {};
\node[] (under2) at (6.8,.5) {};
\draw[] (under1) -- (under2);
\begin{scope}[scale=1.5]
\node[scale=.5] (tit) at (0,-.7) {\LARGE{$\ell=3$}};
\node[bbc,scale=.5] (p0a) at (0,-1) {};
\node[scale=.5] (p0b) at (0,-3) {};
\node[scale=.8] (t0b) at (0,-3.1) {$\cT^{(2)}_{G,{\rm odd}}$};
\node[scale=.8] (p1) at (-.7,-2) {$\blue{\cS^{(1)}_{\varnothing,2}}$};
\node[scale=.8] (p2) at (.7,-2) {$[\cT^{(1)}_{G,\ell}]_{\Z_\ell}$};
\node[scale=.8] (t2b) at (-.7,-2.7) {{\scriptsize$\big[u^2+v=0\big]$}};
\node[scale=.8] (t3b) at (.7,-2.6) {{\scriptsize$\big[u=0\big]$}};
\draw[red] (p0a) -- (p1);
\draw[red] (p0a) -- (p2);
\draw[red] (p1) -- (p0b);
\draw[red] (p2) -- (p0b);
\end{scope}
\begin{scope}[scale=1.5,xshift=2.4cm]
\node[scale=.5] (tit) at (0,-.7) {\LARGE{$\ell=2,4$}};
\node[bbc,scale=.5] (p0a) at (0,-1) {};
\node[scale=.5] (p0b) at (0,-3) {};
\node[scale=.8] (t0b) at (0,-3.1) {$\cT^{(2)}_{G,{\rm even}}$};
\node[scale=.8] (p1) at (-1,-2) {$\blue{\cS^{(1)}_{\varnothing,2}}$};
\node[scale=.8] (p3) at (0,-2) {$\blue{\cS^{(1)}_{\varnothing,2}}$};
\node[scale=.8] (p2) at (1,-2) {$[\cT^{(1)}_{G,\ell}]_{\Z_\ell}$};
\node[scale=.8] (t2b) at (-.7,-2.7) {{\scriptsize$\big[u^2+v=0\big]$}};
\node[scale=.8] (t3b) at (.7,-2.6) {{\scriptsize$\big[v=0\big]$}};
\draw[red] (p0a) -- (p1);
\draw[red] (p0a) -- (p2);
\draw[red] (p0a) -- (p3);
\draw[red] (p3) -- (p0b);
\draw[red] (p1) -- (p0b);
\draw[red] (p2) -- (p0b);
\end{scope}
\begin{scope}[scale=1.5,yshift=-.85cm,xshift=4.55cm]
\node[scale=.5] (p5) at (0.7,.75) {};
\node[scale=.8] (t0a) at (.7,.85) {$\H^{\rm d_{HB}}$};
\node[scale=.8] (tp2) at (.5,.5) {$\gf$};
\node[scale=.8] (p4) at (.7,0) {$\cT^{(1)}_{G,\ell}$};
\node[scale=.8] (tp2) at (1.2,-.4) {$\gf$};
\node[scale=.8] (tp2) at (0.1,-.4) {$h_{m+1,\ell}$};
\node[scale=.8] (p3a) at (1.4,-1.05) {$\cT^{(1)}_{G,\ell}{\times}\cT^{(1)}_{G,\ell}$};
\node[scale=.8] (tp2) at (1.25,-1.75) {$k_\ell$};
\node[scale=.8] (p2b) at (0,-1.05) {$\cS^{(1)}_{G,\ell}$};
\node[scale=.8] (tp2) at (.2,-1.75) {$\gf'$};
\node[scale=.8] (p1a) at (.7,-2.25) {$\cT^{(2)}_{G,\ell}$};
\draw[blue] (p1a) -- (p3a);
\draw[blue] (p1a) -- (p2b);
\draw[blue] (p2b) -- (p4);
\draw[blue] (p3a) -- (p4);
\draw[blue] (p4) -- (p5);
\end{scope}
\end{tikzpicture}}
{}
\end{subfloatrow}{}
}{\caption{\label{Fig:T2Gl} In this figure we report the relevant CFT data for the $\cT^{(2)}_{G,\ell>1}$ theories where the ($m,\gf'$) are equal to ($4,\ff_4$), ($3,\gf_2$), ($2,\sof(7)$), ($2,\suf(2)$), ($1,\suf(3)$) and ($1,\suf(2)$) for $(E_6,2)$, $(D_4,3)$, $(D_4,2)$, $(A_2,4)$, $(A_2,2)$ and $(A_1,3)$ respectively and the $k_\ell$ slice is defined, for example, in \cite[Eq. (C.6)]{Bourget:2020mez}.}}
\end{figure}

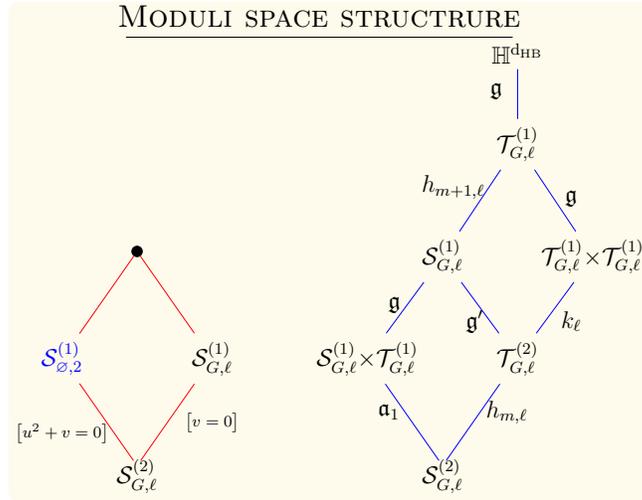
\begin{figure}[h!]
\begin{adjustbox}{center,max width=.95\textwidth}
\ffigbox{
\underline{\Large{\textsc{CFT data of $\cS^{(2)}_{G,\ell}$ theories}}}\\\vspace{.5cm}
\begin{subfloatrow}
\capbtabbox[5.5cm]{%
  \renewcommand{\arraystretch}{1.1}
  \begin{tabular}{|c|c|} 
  \hline
  \multicolumn{2}{|c|}{$\cS^{(2)}_{E_6,2}$}\\
  \hline\hline
  $(\D_u,\D_v)$  &\quad $\left(6,12\right)$\quad{} \\
  $24a$ &130\\  
  $12c$ & 232\\
$\ff_k$ & $\spf(8)_{13}{\times}\suf(2)_{26}$ \\ 
$d_{\rm HB}$& 28\\
$h$&6\\
$T({\bf2}\bh)$&3\\
\hline\hline
\end{tabular}
}{\caption{}\vspace{-.3cm}}
\end{subfloatrow}
\begin{subfloatrow}
\capbtabbox[7cm]{%
  \renewcommand{\arraystretch}{1.1}
  \begin{tabular}{|c|c|} 
  \hline
  \multicolumn{2}{|c|}{$\cS^{(2)}_{D_4,2}$}\\
  \hline\hline
  $(\D_u,\D_v)$  &\quad $\left(4,8\right)$\quad{} \\
  $24a$ &146\\  
  $12c$ & 80\\
$\ff_k$ & $\spf(4)_9{\times}\suf(2)_{16}{\times}\suf(2)_{18}$ \\ 
$d_{\rm HB}$& 14\\
$h$&4\\
$T({\bf2}\bh)$&3\\
\hline\hline
\end{tabular}
}{\caption{}\vspace{-.3cm}}
\end{subfloatrow}{}
\begin{subfloatrow}
\hspace{-.5cm}\capbtabbox[5.8cm]{%
  \renewcommand{\arraystretch}{1.1}
  \begin{tabular}{|c|c|} 
  \hline
  \multicolumn{2}{|c|}{$\cS^{(2)}_{A_2,2}$}\\
  \hline\hline
  $(\D_u,\D_v)$  &\quad $\left(3,6\right)$\quad{} \\
  $24a$ &103\\  
  $12c$ & 55\\
$\ff_k$ & $\hspace{-.1cm}\suf(2)_7{\times}\suf(2)_{14}{\times}\uf(1)\hspace{-.1cm}$ \\ 
$d_{\rm HB}$& 7\\
$h$&3\\
$T({\bf2}\bh)$&3\\
\hline\hline
\end{tabular}
}{\caption{}\vspace{-.3cm}}
\end{subfloatrow}
\begin{subfloatrow}
\hspace{.2cm}\capbtabbox[4.4cm]{%
  \renewcommand{\arraystretch}{1.1}
  \begin{tabular}{|c|c|} 
  \hline
  \multicolumn{2}{|c|}{$\cS^{(2)}_{D_4,3}$}\\
  \hline\hline
  $(\D_u,\D_v)$  &\quad $\left(6,12\right)$\quad{} \\
  $24a$ &219\\  
  $12c$ & 117\\
$\ff_k$ & $\hspace{-.1cm}\suf(3)_{26}{\times}\uf(1)\hspace{-.1cm}$ \\ 
$d_{\rm HB}$& 15\\
$h$&5\\
$T({\bf2}\bh)$&2\\
\hline\hline
\end{tabular}
}{\caption{}\vspace{-.3cm}}
\end{subfloatrow}
\begin{subfloatrow}
\hspace{.2cm}\capbtabbox[4cm]{%
  \renewcommand{\arraystretch}{1.1}
  \begin{tabular}{|c|c|} 
  \hline
  \multicolumn{2}{|c|}{$\cS^{(2)}_{A_1,3}$}\\
  \hline\hline
  $(\D_u,\D_v)$  &\quad $\left(4,8\right)$\quad{} \\
  $24a$ &137\\  
  $12c$ & 71\\
$\ff_k$ & $\hspace{-.1cm}\uf(1)^2\hspace{-.1cm}$ \\ 
$d_{\rm HB}$& 5\\
$h$&3\\
$T({\bf2}\bh)$&-\\
\hline\hline
\end{tabular}
}{\caption{}\vspace{-.3cm}}
\end{subfloatrow}
\begin{subfloatrow}
\vspace{.5cm}
\hspace{-.3cm}
\capbtabbox[5cm]{%
  \renewcommand{\arraystretch}{1.1}
  \begin{tabular}{|c|c|} 
  \hline
  \multicolumn{2}{|c|}{$\cS^{(2)}_{A_2,4}$}\\
  \hline\hline
  $(\D_u,\D_v)$  &\quad $\left(6,12\right)$\quad{} \\
  $24a$ &212\\  
  $12c$ & 110\\
$\ff_k$ & $\suf(2)_{16}{\times}\uf(1)$ \\ 
$d_{\rm HB}$& 8\\
$h$&4\\
$T({\bf2}\bh)$&-\\
\hline\hline
\end{tabular}
}{\caption{}\vspace{-.3cm}}
\end{subfloatrow}
\begin{subfloatrow}
\ffigbox[10.2cm][]{
\begin{tikzpicture}[decoration={markings,
mark=at position .5 with {\arrow{>}}}]
\fill[color=Goldenrod!10, rounded corners] (-1.8,2) rectangle (7.2,-5);
\begin{scope}[scale=1.5]
\node[scale=.5] (p0b) at (1.7,1.2) {\Huge{\textsc{Moduli space structrure}}};
\node[] (under1) at (-.2,1) {};
\node[] (under2) at (3.6,1) {};
\draw[] (under1) -- (under2);
\node[bbc,scale=.5] (p0a) at (0,-1) {};
\node[scale=.5] (p0b) at (0,-3) {};
\node[scale=.8] (t0b) at (0,-3.1) {$\cS^{(2)}_{G,\ell}$};
\node[scale=.8] (p1) at (-.7,-2) {$\blue{\cS^{(1)}_{\varnothing,2}}$};
\node[scale=.8] (p2) at (.7,-2) {$\cS^{(1)}_{G,\ell}$};
\node[scale=.8] (t2b) at (-.7,-2.7) {{\scriptsize$\big[u^2+v=0\big]$}};
\node[scale=.8] (t3b) at (.7,-2.6) {{\scriptsize$\big[v=0\big]$}};
\draw[red] (p0a) -- (p1);
\draw[red] (p0a) -- (p2);
\draw[red] (p1) -- (p0b);
\draw[red] (p2) -- (p0b);
\end{scope}
\begin{scope}[scale=1.5,xshift=2.85cm]
\node[scale=.5] (p5) at (0.7,.75) {};
\node[scale=.8] (t0a) at (.7,.85) {$\H^{\rm d_{HB}}$};
\node[scale=.8] (tp2) at (.5,.5) {$\gf$};
\node[scale=.8] (p4) at (.7,0) {$\cT^{(1)}_{G,\ell}$};
\node[scale=.8] (tp2) at (1.2,-.5) {$\gf$};
\node[scale=.8] (tp2) at (0.1,-.4) {$h_{m+1,\ell}$};
\node[scale=.8] (p3a) at (1.4,-1.05) {$\cT^{(1)}_{G,\ell}{\times}\cT^{(1)}_{G,\ell}$};
\node[scale=.8] (p2b) at (0,-1.05) {$\cS^{(1)}_{G,\ell}$};
\node[scale=.8] (tp2) at (-.45,-1.5) {$\gf$};
\node[scale=.8] (tp2) at (.3,-1.65) {$\gf'$};
\node[scale=.8] (tp2) at (1.2,-1.65) {$k_\ell$};
\node[scale=.8] (p1a) at (.7,-2.0) {$\cT^{(2)}_{G,\ell}$};
\node[scale=.8] (p1b) at (-.7,-2.0) {$\cS^{(1)}_{G,\ell}{\times}\cT^{(1)}_{G,\ell}$};
\node[scale=.8] (tp1) at (-.5,-2.5) {$\af_1$};
\node[scale=.8] (tp2) at (.6,-2.5) {$h_{m,\ell}$};
\node[scale=.5] (p0) at (0,-3) {};
\node[scale=.8] (t0b) at (0,-3.1) {$\cS^{(2)}_{G,\ell}$};
\draw[blue] (p0) -- (p1a);
\draw[blue] (p0) -- (p1b);
\draw[blue] (p1a) -- (p3a);
\draw[blue] (p1a) -- (p2b);
\draw[blue] (p1b) -- (p2b);
\draw[blue] (p2b) -- (p4);
\draw[blue] (p3a) -- (p4);
\draw[blue] (p4) -- (p5);
\end{scope}
\end{tikzpicture}}
{}
\end{subfloatrow}{}
}{\caption{\label{Fig:S2Gl} In this figure we report the relevant CFT data for the $\cS^{(2)}_{G,\ell}$ theories. The $m$, the $k_\ell$ and the $\gf'$ are defined as in figure \ref{Fig:T2Gl}.}}
\end{adjustbox}
\end{figure}

\section{Theories with enhanced supersymmetry}\label{sec:N34}

Theories with enhanced ($\cN\geq3$) supersymmetry have a much tighter moduli space structure. The $\cN=4$ case has been discussed for many decades, \emph{e.g.} \cite{Seiberg:1997ax}, while $\cN=3$ theories have been constructed considerably more recently \cite{Aharony:2015oyb,Garcia-Etxebarria:2015wns,Aharony:2016kai}. The two cases bear many similarities; the metric on the entire moduli space is flat (see \emph{e.g.} \cite{Cordova:2016xhm}) and the $R$-symmetry group enhancement ties in the structure of the \gls{CB} and the \gls{HB} of these theories giving rise to a mathematical structure on the entire moduli space which has been deemed \gls{tsk} \cite{Argyres:2019yyb}. In particular it implies that all the theories which appear on singular starta have to be themselves $\cN\geq3$. We won't delve further into the details of this construction here and only mention that all known $\cN\geq 3$ geometries are orbifold of the type:
\beq\label{orbi}
\cM=\C^{3r}/\Gamma
\eeq
where $r$ is the rank of the theory and $\Gamma$ is a \gls{Ccrg} \cite{Popov:1982,Lehrer:2009} which preserves a principal polarization \cite{Caorsi:2018zsq}. The singular locus of orbifold moduli spaces of vacua \eqref{orbi} can be easily determined by studying the fix locus of the $\G$ action. The case in which $\G$ is a real reflection group gives rise to the $\cN=4$ case with a Lie group $G$ where $\Gamma$ is naturally associated to the Weyl group of $G$. 

It is important to notice that imposing the condition of $\cN\geq3$ supersymmetry leave plenty of space for many new theories, already at rank-2 as:
\begin{itemize}
\item[1)] \gls{tsk} geometry does not imply the orbifold structure in \eqref{orbi}. 

\item[2)] \gls{tsk} geometry does not imply $\Gamma$ being a complex reflection group.

\item[3)] Even within the sub-class of \gls{Ccrg}, there are many $\Gamma$ which give rise to moduli spaces of vacua which do not correspond to any known theories.

\item[4)] As it is well-known for the $\cN=4$ case \cite{Aharony:2013hda}, the structure of the moduli space of vacua alone does not uniquely specify a theory.

\end{itemize} 

Already at rank-2 the situation is rich. A systematic analysis of the allowed orbifold \gls{tsk} geometries at rank-2 was performed not long ago \cite{Argyres:2019ngz} with the result that only a small subset of allowed geometries has been realized as $\cN=3$ geometries. Since the \gls{Ccrg} largely determines the full structure of the moduli space of vacua of $\cN\geq3$ theories, rather than discussing each case individually, we collect all the relevant data in figure \ref{Fig:N4} for $\blue{\cN=4}$ and in figure \ref{Fig:N3} for $\green{\cN=3}$ theories.

\begin{figure}[h!]
\underline{\Large{\textsc{CFT data of \blue{$\cN=4$} theories}}}\vspace{.5cm}
\ffigbox{
\ffigbox[10cm]{
\begin{subfloatrow}

%%beginning \suf(3)
\hspace{-2cm}\ffigbox[7cm][]{
\begin{tikzpicture}[decoration={markings,
mark=at position .5 with {\arrow{>}}}]
\begin{scope}[scale=1.5,xshift=-2.5cm]
\fill[color=Goldenrod!25, rounded corners] (-.8,-1.6) rectangle (.8,-2.4);
\node[scale=.8] (t0b) at (0,-2) {\Large{$\boldsymbol{\G=S_3}$}};
\end{scope}
\begin{scope}[scale=1.5]
\node[bbc,scale=.5] (p0a) at (0,-1) {};
\node[scale=.5] (p0b) at (0,-3) {};
\node[scale=.8] (t0b) at (0,-3.1) {\blue{$\cN=4$\ $SU(3)$}};
\node[scale=.8] (p1) at (0,-2) {\blue{$\blue{\cS^{(1)}_{\varnothing,2}}$}};
\node[scale=.8] (t2b) at (-.5,-2.5) {{\scriptsize$\big[u^2+v^3=0\big]$}};
\draw[red] (p0a) -- (p1);
\draw[red] (p1) -- (p0b);
\end{scope}
\begin{scope}[scale=1.5,xshift=2.5cm]
\node[scale=.8] (p3a) at (0,-1) {$U(1){\times}U(1)$};
\node[scale=.8] (tp2) at (-.2,-1.5) {$\af_1$};
\node[scale=.8] (p2b) at (0,-2) {\blue{$\blue{\cS^{(1)}_{\varnothing,2}}$}};
\node[scale=.8] (p1) at (0,-3) {\blue{$\cN=4$\ $SU(3)$}};
\draw[blue] (p1) -- (p2b);
\draw[blue] (p2b) -- (p3a);
\end{scope}
\end{tikzpicture}}
{}
\hspace{1cm}
\begin{adjustbox}{center,max width=.47\textwidth}
\capbtabbox[3cm]{%
  \renewcommand{\arraystretch}{1.1}
  \begin{tabular}{|c|c|} 
  \hline
  \multicolumn{2}{|c|}{\blue{$\cN=4$ $SU(3)$}}\\
  \hline\hline
  $(\D_u,\D_v)$  &\quad $\left(2,3\right)$\quad{} \\
  $24a$ & 48 \\  
  $12c$ & 24 \\
$\ff_k$ & $\suf(2)_{8}$ \\ 
$d_{\rm HB}$& 2\\
$h$&2\\
$T({\bf2}\bh)$&2\\
\hline\hline
\end{tabular}
}{
}
\end{adjustbox}
\end{subfloatrow}}{}\vspace{-.6cm}

%% end \suf(3)

\begin{center}
\rule[1mm]{2cm}{.4pt}\hspace{1cm}$\circ$\hspace{1cm} \rule[1mm]{2cm}{.4pt}
\end{center}
\vspace{-.3cm}

%%beginning \suf(2)x\suf(2)

\ffigbox[10cm]{
\begin{subfloatrow}
\hspace{-9cm}\begin{adjustbox}{center,max width=.47\textwidth}
\capbtabbox[3cm]{%
  \renewcommand{\arraystretch}{1.1}
  \begin{tabular}{|c|c|} 
  \hline
  \multicolumn{2}{|c|}{\blue{$\cN=4$ $SU(2){\times}SU(2)$}}\\
  \hline\hline
  $(\D_u,\D_v)$  &\quad $\left(2,2\right)$\quad{} \\
  $24a$ & 36 \\  
  $12c$ & 18 \\
$\ff_k$ & $\suf(2)_3\times\suf(2)_3$ \\ 
$d_{\rm HB}$& 2\\
$h$&2\\
$T({\bf2}\bh)$&2\\
\hline\hline
\end{tabular}
}{
}
\end{adjustbox}
\hspace{-1cm}
\ffigbox[4cm][]{
\begin{tikzpicture}[decoration={markings,
mark=at position .5 with {\arrow{>}}}]
\begin{scope}[scale=1.5]
\node[bbc,scale=.5] (p0a) at (0,-1) {};
\node[scale=.5] (p0b) at (0,-3) {};
\node[scale=.8] (t0b) at (0,-3.1) {\blue{$\cN=4$\ $SU(2){\times}SU(2)$}};
\node[scale=.8] (p1) at (-.7,-2) {\blue{$\blue{\cS^{(1)}_{\varnothing,2}}$}};
\node[scale=.8] (p2) at (.7,-2) {\blue{$\blue{\cS^{(1)}_{\varnothing,2}}$}};
\node[scale=.8] (t2b) at (-.7,-2.7) {{\scriptsize$\big[u+v=0\big]$}};
\node[scale=.8] (t3b) at (.7,-2.6) {{\scriptsize$\big[u=0\big]$}};
\draw[red] (p0a) -- (p1);
\draw[red] (p0a) -- (p2);
\draw[red] (p1) -- (p0b);
\draw[red] (p2) -- (p0b);
\end{scope}
\begin{scope}[scale=1.5,xshift=2.5cm]
\node[scale=.8] (p3a) at (0,-1) {$U(1){\times}U(1)$};
\node[scale=.8] (tp2) at (.6,-1.45) {$\af_1$};
\node[scale=.8] (tp2) at (-.6,-1.45) {$\af_1$};
\node[scale=.8] (p2a) at (-.7,-2) {\blue{$\blue{\cS^{(1)}_{\varnothing,2}}$}};
\node[scale=.8] (tp2) at (-.6,-2.5) {$\af_1$};
\node[scale=.8] (p2b) at (.7,-2) {\blue{$\blue{\cS^{(1)}_{\varnothing,2}}$}};
\node[scale=.8] (tp2b) at (.6,-2.5) {$\af_1$};
\node[scale=.8] (p1) at (0,-3) {\blue{$\cN=4$\ $SU(2){\times}SU(2)$}};
\draw[blue] (p1) -- (p2a);
\draw[blue] (p1) -- (p2b);
\draw[blue] (p2a) -- (p3a);
\draw[blue] (p2b) -- (p3a);
\end{scope}
\begin{scope}[scale=1.5,xshift=5cm]
\fill[color=Goldenrod!25, rounded corners] (-1,-1.6) rectangle (1,-2.4);
\node[scale=.8] (t0b) at (0,-2) {\Large{$\boldsymbol{\G=\Z_2{\times}\Z_2}$}};
\end{scope}
\end{tikzpicture}}
{}
\end{subfloatrow}}{}\vspace{-.6cm}

%% end \suf(2)x\suf(2)

\begin{center}
\rule[1mm]{2cm}{.4pt}\hspace{1cm}$\circ$\hspace{1cm} \rule[1mm]{2cm}{.4pt}
\end{center}
\vspace{-.3cm}

%%beginning Sp(4)

\ffigbox[10cm]{
\begin{subfloatrow}
\hspace{-2cm}
\ffigbox[7cm][]{
\begin{tikzpicture}[decoration={markings,
mark=at position .5 with {\arrow{>}}}]
\begin{scope}[scale=1.5,xshift=-2.5cm]
\fill[color=Goldenrod!25, rounded corners] (-.8,-1.6) rectangle (.8,-2.4);
\node[scale=.8] (t0b) at (0,-2) {\Large{$\boldsymbol{\G=D_4}$}};
\end{scope}
\begin{scope}[scale=1.5]
\node[bbc,scale=.5] (p0a) at (0,-1) {};
\node[scale=.5] (p0b) at (0,-3) {};
\node[scale=.8] (t0b) at (0,-3.1) {\blue{$\cN=4$\ $USp(4)$}};
\node[scale=.8] (p1) at (-.7,-2) {\blue{$\blue{\cS^{(1)}_{\varnothing,2}}$}};
\node[scale=.8] (p2) at (.7,-2) {\blue{$\blue{\cS^{(1)}_{\varnothing,2}}$}};
\node[scale=.8] (t2b) at (-.7,-2.7) {{\scriptsize$\big[u^2+v=0\big]$}};
\node[scale=.8] (t3b) at (.7,-2.6) {{\scriptsize$\big[v=0\big]$}};
\draw[red] (p0a) -- (p1);
\draw[red] (p0a) -- (p2);
\draw[red] (p1) -- (p0b);
\draw[red] (p2) -- (p0b);
\end{scope}
\begin{scope}[scale=1.5,xshift=2.5cm]
\node[scale=.8] (p3a) at (0,-1) {$U(1){\times}U(1)$};
\node[scale=.8] (tp2) at (.6,-1.45) {$\af_1$};
\node[scale=.8] (tp2) at (-.6,-1.45) {$\af_1$};
\node[scale=.8] (p2a) at (-.7,-2) {\blue{$\blue{\cS^{(1)}_{\varnothing,2}}$}};
\node[scale=.8] (tp2) at (-.6,-2.5) {$\af_1$};
\node[scale=.8] (p2b) at (.7,-2) {\blue{$\blue{\cS^{(1)}_{\varnothing,2}}$}};
\node[scale=.8] (tp2b) at (.6,-2.5) {$A_3$};
\node[scale=.8] (p1) at (0,-3) {\blue{$\cN=4$\ $USp(4)$}};
\draw[blue] (p1) -- (p2a);
\draw[blue] (p1) -- (p2b);
\draw[blue] (p2a) -- (p3a);
\draw[blue] (p2b) -- (p3a);
\end{scope}
\end{tikzpicture}}
{}
\hspace{1cm}\begin{adjustbox}{center,max width=.47\textwidth}
\capbtabbox[3cm]{%
  \renewcommand{\arraystretch}{1.1}
  \begin{tabular}{|c|c|} 
  \hline
  \multicolumn{2}{|c|}{\blue{$\cN=4$ $USp(4)$}}\\
  \hline\hline
  $(\D_u,\D_v)$  &\quad $\left(2,4\right)$\quad{} \\
  $24a$ & 60 \\  
  $12c$ & 30 \\
$\ff_k$ & $\suf(2)_{10}$ \\ 
$d_{\rm HB}$& 2\\
$h$&2\\
$T({\bf2}\bh)$&2\\
\hline\hline
\end{tabular}
}{
}
\end{adjustbox}
\end{subfloatrow}}{}\vspace{-.6cm}

%% end Sp(4)

\begin{center}
\rule[1mm]{2cm}{.4pt}\hspace{1cm}$\circ$\hspace{1cm} \rule[1mm]{2cm}{.4pt}
\end{center}
\vspace{-.3cm}

%%beginning G2

\ffigbox[10cm]{
\begin{subfloatrow}
\hspace{-9cm}\begin{adjustbox}{center,max width=.47\textwidth}
\capbtabbox[3cm]{%
  \renewcommand{\arraystretch}{1.1}
  \begin{tabular}{|c|c|} 
  \hline
  \multicolumn{2}{|c|}{\blue{$\cN=4$ $G_2$}}\\
  \hline\hline
  $(\D_u,\D_v)$  &\quad $\left(2,6\right)$\quad{} \\
  $24a$ & 84 \\  
  $12c$ & 42 \\
$\ff_k$ & $\suf(2)_{14}$ \\ 
$d_{\rm HB}$& 2\\
$h$&2\\
$T({\bf2}\bh)$&2\\
\hline\hline
\end{tabular}
}{
}
\end{adjustbox}
\hspace{-1cm}
\ffigbox[4cm][]{
\begin{tikzpicture}[decoration={markings,
mark=at position .5 with {\arrow{>}}}]
\begin{scope}[scale=1.5]
\node[bbc,scale=.5] (p0a) at (0,-1) {};
\node[scale=.5] (p0b) at (0,-3) {};
\node[scale=.8] (t0b) at (0,-3.1) {\blue{$\cN=4$\ $G_2$}};
\node[scale=.8] (p1) at (-.7,-2) {\blue{$\blue{\cS^{(1)}_{\varnothing,2}}$}};
\node[scale=.8] (p2) at (.7,-2) {\blue{$\blue{\cS^{(1)}_{\varnothing,2}}$}};
\node[scale=.8] (t2b) at (-.7,-2.7) {{\scriptsize$\big[u^3+v=0\big]$}};
\node[scale=.8] (t3b) at (.7,-2.6) {{\scriptsize$\big[v=0\big]$}};
\draw[red] (p0a) -- (p1);
\draw[red] (p0a) -- (p2);
\draw[red] (p1) -- (p0b);
\draw[red] (p2) -- (p0b);
\end{scope}
\begin{scope}[scale=1.5,xshift=2.5cm]
\node[scale=.8] (p3a) at (0,-1) {$U(1){\times}U(1)$};
\node[scale=.8] (tp2) at (.6,-1.45) {$\af_1$};
\node[scale=.8] (tp2) at (-.6,-1.45) {$\af_1$};
\node[scale=.8] (p2a) at (-.7,-2) {\blue{$\blue{\cS^{(1)}_{\varnothing,2}}$}};
\node[scale=.8] (tp2) at (-.6,-2.5) {$\af_1$};
\node[scale=.8] (p2b) at (.7,-2) {\blue{$\blue{\cS^{(1)}_{\varnothing,2}}$}};
\node[scale=.8] (tp2b) at (.6,-2.5) {$A_5$};
\node[scale=.8] (p1) at (0,-3) {\blue{$\cN=4$\ $G_2$}};
\draw[blue] (p1) -- (p2a);
\draw[blue] (p1) -- (p2b);
\draw[blue] (p2a) -- (p3a);
\draw[blue] (p2b) -- (p3a);
\end{scope}
\begin{scope}[scale=1.5,xshift=5cm]
\fill[color=Goldenrod!25, rounded corners] (-1,-1.6) rectangle (1,-2.4);
\node[scale=.8] (t0b) at (0,-2) {\Large{$\boldsymbol{\G=D_6}$}};
\end{scope}
\end{tikzpicture}}
{}
\end{subfloatrow}}{}

%% end G2

}
{\caption{\label{Fig:N4}CFT data for \blue{$\cN=4$} rank-2 theories.}}
\end{figure}
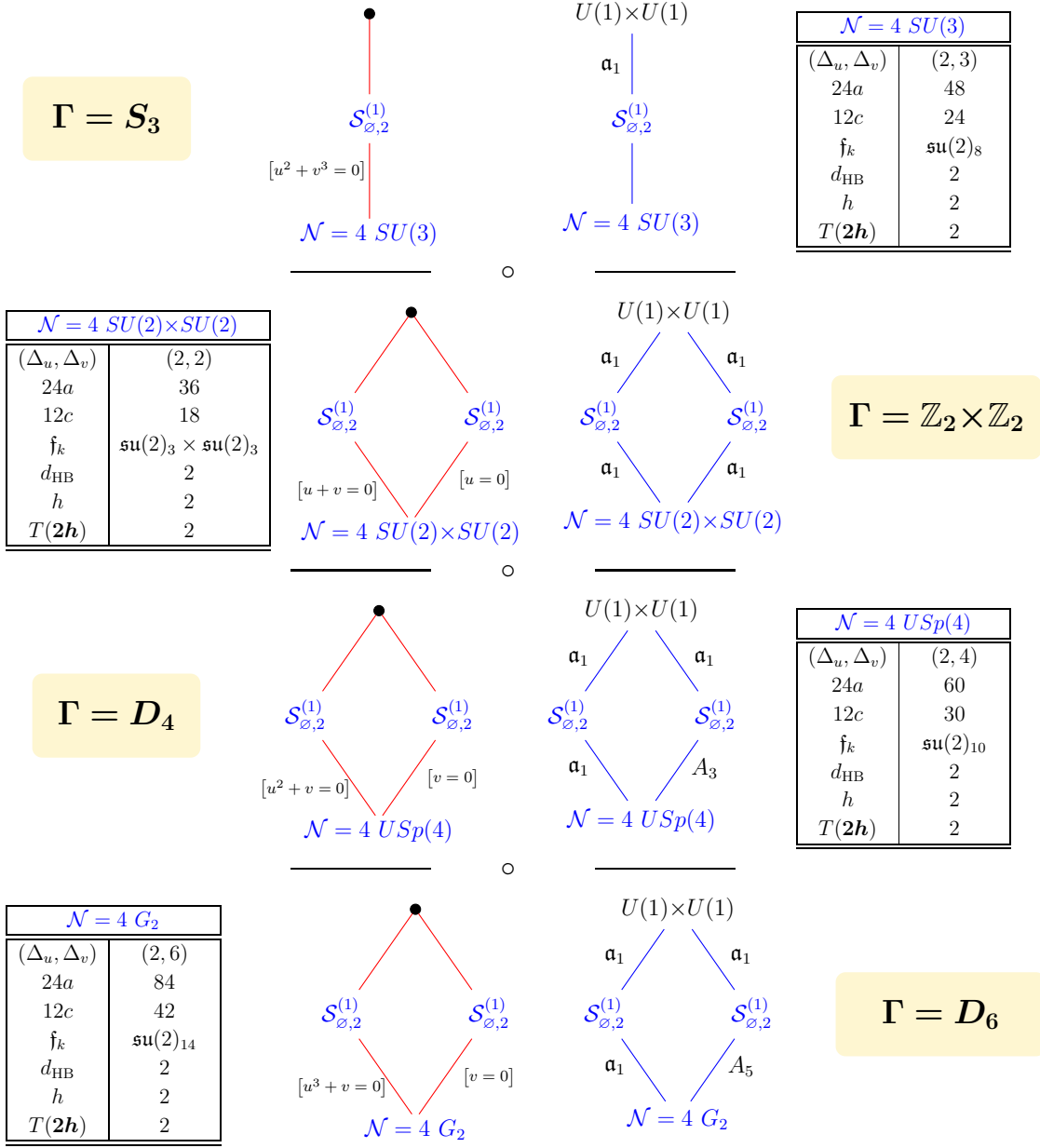

\begin{figure}[h!]
\underline{\Large{\textsc{CFT data of \green{$\cN=3$} theories}}}\vspace{.5cm}
\ffigbox{
\ffigbox[10cm]{
\begin{subfloatrow}

%%beginning G(3,1,2)
\hspace{-2cm}\ffigbox[7cm][]{
\begin{tikzpicture}[decoration={markings,
mark=at position .5 with {\arrow{>}}}]
\begin{scope}[scale=1.5,xshift=-2.5cm]
\fill[color=Goldenrod!25, rounded corners] (-1.2,-1.6) rectangle (1.2,-2.4);
\node[scale=.8] (t0b) at (0,-2) {\Large{$\boldsymbol{\G=G(3,1,2)}$}};
\end{scope}
\begin{scope}[scale=1.5]
\node[bbc,scale=.5] (p0a) at (0,-1) {};
\node[scale=.5] (p0b) at (0,-3) {};
\node[scale=.8] (t0b) at (0,-3.1) {$\green{\cS^{(2)}_{\varnothing,3}}$};
\node[scale=.8] (p1) at (-.7,-2) {$\blue{\cS^{(1)}_{\varnothing,2}}$};
\node[scale=.8] (p2) at (.7,-2) {$\green{\cS^{(1)}_{\varnothing,3}}$};
\node[scale=.8] (t2b) at (-.7,-2.7) {{\scriptsize$\big[u^2+v=0\big]$}};
\node[scale=.8] (t3b) at (.7,-2.6) {{\scriptsize$\big[v=0\big]$}};
\draw[red] (p0a) -- (p1);
\draw[red] (p0a) -- (p2);
\draw[red] (p1) -- (p0b);
\draw[red] (p2) -- (p0b);
\end{scope}
\begin{scope}[scale=1.5,xshift=2.5cm]
\node[scale=.8] (p3a) at (0,-1) {$U(1){\times}U(1)$};
\node[scale=.8] (tp2) at (.6,-1.45) {$A_2$};
\node[scale=.8] (tp2) at (-.6,-1.45) {$\af_1$};
\node[scale=.8] (p2a) at (-.7,-2) {$\blue{\cS^{(1)}_{\varnothing,2}}$};
\node[scale=.8] (tp2) at (-.6,-2.5) {$A_2$};
\node[scale=.8] (p2b) at (.7,-2) {$\green{\cS^{(1)}_{\varnothing,3}}$};
\node[scale=.8] (tp2b) at (.6,-2.5) {$A_2$};
\node[scale=.8] (p1) at (0,-3) {$\green{\cS^{(2)}_{\varnothing,3}}$};
\draw[blue] (p1) -- (p2a);
\draw[blue] (p1) -- (p2b);
\draw[blue] (p2a) -- (p3a);
\draw[blue] (p2b) -- (p3a);
\end{scope}
\end{tikzpicture}}
{}
\hspace{1cm}
\begin{adjustbox}{center,max width=.47\textwidth}
\capbtabbox[3cm]{%
  \renewcommand{\arraystretch}{1.1}
  \begin{tabular}{|c|c|} 
  \hline
  \multicolumn{2}{|c|}{$\green{\cN=3\ \cS^{(2)}_{\varnothing,3}}$}\\
  \hline\hline
  $(\D_u,\D_v)$  &\quad $\left(2,6\right)$\quad{} \\
  $24a$ & 84 \\  
  $12c$ & 42 \\
$\ff_k$ & $\uf(1)$ \\ 
$d_{\rm HB}$& 2\\
$h$&2\\
$T({\bf2}\bh)$&2\\
\hline\hline
\end{tabular}
}{
}
\end{adjustbox}
\end{subfloatrow}}{}\vspace{-.6cm}

%% end G(3,1,2)

\begin{center}
\rule[1mm]{2cm}{.4pt}\hspace{1cm}$\circ$\hspace{1cm} \rule[1mm]{2cm}{.4pt}
\end{center}
\vspace{-.3cm}

%%beginning G(4,1,2)

\ffigbox[10cm]{
\begin{subfloatrow}
\hspace{-9cm}\begin{adjustbox}{center,max width=.47\textwidth}
\capbtabbox[3cm]{%
  \renewcommand{\arraystretch}{1.1}
  \begin{tabular}{|c|c|} 
  \hline
  \multicolumn{2}{|c|}{\green{$\cN=3$ $\cS^{(2)}_{\varnothing,4}$}}\\
  \hline\hline
  $(\D_u,\D_v)$  &\quad $\left(4,8\right)$\quad{} \\
  $24a$ & 132 \\  
  $12c$ & 66 \\
$\ff_k$ & $\uf(1)$ \\ 
$d_{\rm HB}$& 2\\
$h$&2\\
$T({\bf2}\bh)$&2\\
\hline\hline
\end{tabular}
}{
}
\end{adjustbox}
\hspace{-1cm}
\ffigbox[4cm][]{
\begin{tikzpicture}[decoration={markings,
mark=at position .5 with {\arrow{>}}}]
\begin{scope}[scale=1.5]
\node[bbc,scale=.5] (p0a) at (0,-1) {};
\node[scale=.5] (p0b) at (0,-3) {};
\node[scale=.8] (t0b) at (0,-3.1) {$\green{\cS^{(2)}_{\varnothing,4}}$};
\node[scale=.8] (p1) at (-.7,-2) {$\blue{\cS^{(1)}_{\varnothing,2}}$};
\node[scale=.8] (p2) at (.7,-2) {$\green{\cS^{(1)}_{\varnothing,4}}$};
\node[scale=.8] (t2b) at (-.7,-2.7) {{\scriptsize$\big[u^2+v=0\big]$}};
\node[scale=.8] (t3b) at (.7,-2.6) {{\scriptsize$\big[v=0\big]$}};
\draw[red] (p0a) -- (p1);
\draw[red] (p0a) -- (p2);
\draw[red] (p1) -- (p0b);
\draw[red] (p2) -- (p0b);
\end{scope}
\begin{scope}[scale=1.5,xshift=2.5cm]
\node[scale=.8] (p3a) at (0,-1) {$U(1){\times}U(1)$};
\node[scale=.8] (tp2) at (.6,-1.45) {$A_3$};
\node[scale=.8] (tp2) at (-.6,-1.45) {$\af_1$};
\node[scale=.8] (p2a) at (-.7,-2) {\blue{$\blue{\cS^{(1)}_{\varnothing,2}}$}};
\node[scale=.8] (tp2) at (-.6,-2.5) {$A_3$};
\node[scale=.8] (p2b) at (.7,-2) {$\green{\cS^{(1)}_{\varnothing,4}}$};
\node[scale=.8] (tp2b) at (.6,-2.5) {$A_3$};
\node[scale=.8] (p1) at (0,-3) {$\green{\cS^{(2)}_{\varnothing,4}}$};
\draw[blue] (p1) -- (p2a);
\draw[blue] (p1) -- (p2b);
\draw[blue] (p2a) -- (p3a);
\draw[blue] (p2b) -- (p3a);
\end{scope}
\begin{scope}[scale=1.5,xshift=5cm]
\fill[color=Goldenrod!25, rounded corners] (-1.2,-1.6) rectangle (1.2,-2.4);
\node[scale=.8] (t0b) at (0,-2) {\Large{$\boldsymbol{\G=G(4,1,2)}$}};
\end{scope}
\end{tikzpicture}}
{}
\end{subfloatrow}}{}

%% end G(4,1,2)

}
{\caption{\label{Fig:N3}CFT data for \green{$\cN=3$} rank-2 theories.}}
\end{figure}

\section{Flavor structure along the Higgsing and generalized free fields VOA}\label{genfreef}

In this section we quickly sketch how the information in table \ref{Higgs:one}, \ref{Higgs:two} and \ref{Higgs:three} can be leveraged to determine the VOA of the various entries. It is important to stress that we have not performed any in-depth calculation and the presentation in this appendix should be intended as a sketch and not as an actual costruction.

Let's start from the basics. To any four-dimensional $\cN=2$ \gls{SCFT} $\Tf$, one can canonically associate a two-dimensional VOA \cite{Beem:2013sza},
%%%%%%
\begin{equation}
\label{eqn:correspondence}
\chi: \;\; {\rm 4d \;\; \cN=2 \; \; SCFT ~\longrightarrow~ VOA}~.
\end{equation}
%%%%%%
The VOA $\chi[\Tf]$ arises as a cohomological reduction of the full local OPE algebra of a four-dimensional theory $\Tf$ with respect to a certain nilpotent supercharge. We won't review any of the details here but mention that there are numerous indications that $\chi[\Tf]$ is deeply connected with the physics of the \gls{HB} $\bH$. The full extent of the connection remains somewhat elusive but a remarkable fact, observed in many examples and conjectured to be universally true \cite{Beem:2017ooy}, is that $\bH$ can be recovered directly from $\chi[\Tf]$
%%%%%%
\begin{equation}
\label{eqn:M=X}
\bH= X_{\chi [\Tf]}~,
\end{equation}
%%%%%%
where $X_\cV$ denotes a symplectic variety canonically associated to $\cV$ called the \emph{associated variety} of a VOA $\cV$ \cite{Arakawa:2010ni}. It is worth pointing out that using a very nice physical argument \cite[Section 6.2]{Dedushenko:2019mzv}, proving the conjecture in \eqref{eqn:M=X} has been reduced to proving a previous conjecture by Arakawa \cite[Conjecture 1]{Arakawa:2015}.

In \cite{Beem:2019tfp,Beem:2019snk}, striking evidence was provided that data associated with the Higgs branch physics of a theory may be sufficient to determine the full VOA by studying the theory on a higgs stratum $\bbbSf_i$ which supports a non-trivial SCFT $\bTf_i$. The gist of this construction is that the VOA of the initial theory $\Tf$ can be written in terms of free fields which parametrize $\bbbSf_i$ and the VOA generators of $\chi[\bTf_i]$, the VOA of $\bTf_i$. This construction was deemed a generalized free-field construction\footnote{There are simple cases where $\bTf_i$ is itself a theory of free fields, in which case the construction provides an actual free-fields realization.} in \cite{Beem:2019snk,Beem:2019tfp} where the prescription to build $\chi[\Tf]$ from the information about the effective field theory (EFT) $\bTf_i$ was explained. The general picture can be summarized as follows:
%%%%%%
\begin{equation}\label{genconj}
\chi[\Tf]\subset\chi[\bTf_i]\otimes \cV^i_{\rm free}[\bbbSf_i],\qquad \bTf_i=\Tf\wr x,\qquad x\in \bSf_i
\end{equation}
%%%%%%
Here $\cV^i_{\rm free}[\bbbSf_i]$ is a free field VOA for which $X_{\cV^i_{\rm free}[\bbbSf_i]}=\bbbSf_i$. Also we have adopted the notation introduced in \cite{Tachikawa:2017byo}, where $\wr$ signifies ``supported on''.

These generalized free-field realizations are remarkable and handy in many ways, \emph{e.g.}, they realize the \emph{simple quotient} of $\chi[\Tf]$, that is null vectors vanish on the nose when expressed in terms of free fields. They have also been used to characterize many $\cN=2$ \gls{SCFT}s as they give a tool to ``invert the higgsing'' \cite{Giacomelli:2020jel}.  For example the flavor level of a simple factor of the \gls{SCFT} $\Tf$ can be related to properties of the stratum $\bbbSf_\ff$ which arises by spontaneous breaking of the $\ff$, thus the subscript, and the theory supported over it $\bTf_\ff$. The basic formula reads:  
\beq\label{levVOA}
k_{\ff}=\frac{T_2({\bf \bRf})+I_{\ff^\natural\hookrightarrow \ff_{\rm IR}}k_{\ff_{\rm IR}}}{I_{\nff\hookrightarrow \ff}}
\eeq
this of course requires some explanation. $\ff$ is obvious. $I_{\hf_1\hookrightarrow \hf_2}$ indicates the index of embedding of $\hf_1$ into $\hf_2$, $\nff$ is what is left unbroken on the generic point of $\bbbSf_\ff$ of $\ffUV$, the full flavor symmetry of $\Tf$, $\ff_{\rm IR}$ is the subgroup of the $\bTf_{\ff}$ flavor symmetry realized on the component and $T_2({\bf \bRf})$\footnote{We use a somewhat unusual normalization where the ${\bf n}$ of $\suf(n)$ has $T_2({\bf n})=1$.} is Dynkin index of the $\nff$ representation of the goldstone boson associated to the higgsing. 

This data for each one of the theories discussed in this paper, is reported in table  \ref{Higgs:one}, \ref{Higgs:two} and \ref{Higgs:three}. Here we will not provide a careful realization of the VOAs of these theories but only discuss a couple of examples to explain how to check that \eqref{levVOA} applies in all cases using the information in tables. We take this as a suggestion that a generalized free field realization exists for all theories discussed in this paper.

\paragraph{\hyperref[sec:D20E8]{$[\ef_8]_{20}$}} Let's start from a simple case. For this theory there is a single simple factor, $\ef_8$, and the stratum associated with it is its minimal nilpotent orbit, $\bbbSf_{\ef_8}\equiv \ef_8$. From table \ref{NilOrbits} we readily obtain that $\nff=\ef_7$ which will have $k_{\ef_7}=20$ and the goldstone bosons transform in the ${\bf 56}$ of $\ef_7$ with $T_2({\bf 56})=12$, as in fact is reported in the corresponding entry in table \ref{Higgs:one}. The index of embedding of $\ef_7$ inside $\ef_8$ is one as it can be computed by the decomposition of any $\ef_8$ irreducible representations in $\ef_7\oplus\suf(2)$ ones (for example ${\bf 248}\to({\bf 133,1})\oplus ({\bf 1,3})\oplus ({\bf 56,2})$). The theory supported on this stratum is $\bTf_{\ef_8}\equiv\cT^{(1)}_{E_7,1}$ and thus $k_{\ffIR}=8$. This information is enough for the reader to check that \eqref{levVOA} is indeed satisfied.

\paragraph{\hyperref[sec:t2]{$\sof(20)_{16}$}} The matching of the level is not always a simple as the previous example. Consider now another theory of the $\ef_8-\sof(20)$ series, namely theory \hyperref[sec:t2]{$\sof(20)_{16}$}. As discussed in the corresponding section, the \gls{HB} stratum associated to the simple flavor factor is the minimal nilpotent orbit of $\sof(20):\bbbSf_{\sof(20)}\equiv\df_{10}$. From table \ref{NilOrbits} we obtain that $\nff=\sof(16){\times}\suf(2)$, in particular $\nff$ is semi-simple, and that the goldstone bosons transform in the $({\bf2},{\bf16})$. On the other hand the theory supported on this stratum is $\bTf_{\sof(20)}\equiv \cT^{(1)}_{E_8,1}$ which has flavor symmetry $\ef_8$ and we noticed that $\sof(16)$ is a maximal subalgebra of $\ef_8$, thus $\ffIR=\sof(16)$ at level $k_{\sof(16)}=12$. Using the fact that $I_{\sof(16)\hookrightarrow \ef_8}=1$ and that in our normalization $T_2({\bf 16})=2$, we can immediately reproduce the result we are after: $k_{\sof(20)}=16$. But what about the extra $\suf(2)$? It arises from the breaking of the initial $\sof(20)$, so we expect the level of this $\suf(2)$ to be also 16. Since we have ``used up'' all the moment maps of the rank-1 \gls{SCFT} on the stratum, our only hope is that the goldstone bosons alone can make up for that. It is a nice surprise to notice that indeed the goldstone transform as 16 copies of the fundamental of this $\suf(2)$ and we are working in a normalization such that $T_2({\bf 2})=1$.

In a similar manner, the data in tables \ref{Higgs:one}, \ref{Higgs:two} and \ref{Higgs:three} can be leveraged to show the consistency of \eqref{levVOA} of the theories discussed in this paper.

\section{Reading tools}\label{glo}

To ease the reader accessibility to the content of the paper, here we will collect a Glossary and list of Acronyms and Symbols:

\printnoidxglossary[type=acronym]
\printnoidxglossary % default: type=main
\printnoidxglossary[type=symbols]

\bibliographystyle{JHEP}

\end{document}